\documentclass[11pt,a4paper]{article}

\usepackage{setspace}

\usepackage[latin1]{inputenc}
\usepackage{cite}
\usepackage{amsmath,amsfonts,mathtools,mathrsfs}

\usepackage{graphicx}
\usepackage{a4wide}
\usepackage{color}
\usepackage[edge-length=.7cm,root-radius=.1cm]{dynkin-diagrams}
\usepackage{mathdots}

\numberwithin{equation}{section}
\makeatletter 
\@addtoreset{equation}{section} %
\makeatother

\usepackage{titlesec}
\titleformat*{\section}{\large\bfseries}

\usepackage{mdframed}

\usepackage{esint}
\usepackage[hidelinks]{hyperref}

\usepackage{mathrsfs}
\DeclareMathAlphabet{\mathpzc}{OT1}{pzc}{m}{it}
\usepackage[mathscr]{euscript}

\usepackage{tikz}
\usetikzlibrary{trees}
\usetikzlibrary{snakes}
\usetikzlibrary{arrows}

\usepackage{genyoungtabtikz}

\newcount\tableauRow
\newcount\tableauCol

\newenvironment{Tableau}[1]{  \tikzpicture[scale=0.4,draw/.append style={thick,black},
	baseline=(current bounding box.center)]
	\tableauRow=-1.5
	\foreach \Row in {#1} {
		\tableauCol=0.5
		\foreach\k in \Row {
			\draw(\the\tableauCol,\the\tableauRow)rectangle++(1,1);
			\draw(\the\tableauCol,\the\tableauRow)+(0.5,0.5)node{$\k$};
			\global\advance\tableauCol by 1
		}
		\global\advance\tableauRow by -1
	}
}{\endtikzpicture}

\newcommand{\beq}{\begin{equation}}
\newcommand{\eeq}{\end{equation}}
\newcommand{\bea}{\begin{eqnarray}}
\newcommand{\eea}{\end{eqnarray}}

\newcommand{\red}{\color{red}}

\DeclareMathOperator{\Tr}{Tr}

\def\ulmb{{\underline{\smash\lambda}}}
\def\lmb{\lambda}
\def\ue{{\underline{\smash e}}}
\def\umu{{\underline{\smash\mu}}}
\def\unu{{\underline{\smash\nu}}}

\def\udelta{\underline{\delta}}
\def\cO{{\mathcal{O}}}
\def\cN{{\mathcal{N}}}
\def\cC{{\mathcal{C}}}
\def\cD{{\mathbf{D}}}
\def\cS{{\mathcal{S}}}

\def\sdet {\operatorname{sdet}}

\def\JtoB{Jack$\rightarrow$Block\ }
\def\BtoJ{Block$\rightarrow$Jack\ }

\def\BB		 { B}

\def\bslash{\backslash}

\begin{document}

\thispagestyle{empty}

~\\[-1.8cm]

\begin{center}
{ \bf
{\LARGE Superconformal blocks in diverse dimensions \\[.2cm]
		and $BC$ symmetric functions}
	}\\
	\vskip 1truecm
	{\bf Francesco Aprile${}^{1}$ and Paul Heslop${}^{2}$ \\
	}
\end{center}
	
	\vskip 0.4truecm
	
		\centerline{\it ${}^{1}\,$Instituto de Fisica Teorica, UNESP, ICTP South American Institute for Fundamental Research}
		\centerline{\it Rua Dr Bento Teobaldo Ferraz 271, 01140-070, S\~ao Paulo, Brazil} 
		\vskip .4truecm

		\centerline{\it ${}^{2}\,$Mathematics Department, Durham University,} 
		\centerline{\it Science Laboratories, South Rd, Durham DH1 3LE} 
		\vskip .2truecm

\vskip 1.25truecm 
\centerline{\bf Abstract} 
\vskip 0.25truecm

We uncover a precise relation between superblocks for correlators of superconformal field theories (SCFTs) in various  
dimensions and symmetric functions related to the $BC$ root system. The theories we consider are defined by two integers $(m,n)$ together 
with a parameter $\theta$ and they include correlators of all half-BPS correlators in 4d theories with $\cN=2n$ supersymmetry, 
6d theories with $(n,0)$ supersymmetry and 3d theories with $\cN=4n$ supersymmetry, as well as all scalar correlators 
in any non SUSY theory in any dimension,  and conjecturally various 5d, 2d and 1d superconformal theories. The superblocks are eigenfunctions 
of the super Casimir of the superconformal group whose action we find to be precisely that of the $BC_{m|n}$ Calogero-Moser-Sutherland Hamiltonian.  
When $m=0$ the blocks are polynomials, and we show how these relate to $BC_n$ Jacobi polynomials. However, differently from $BC_n$ Jacobi polynomials, 
the $m=0$ blocks possess a crucial stability property that has not been emphasised previously in the literature. 
This property allows for a novel supersymmetric uplift  of the $BC_n$ Jacobi polynomials, which in turn yields 
the $(m,n;\theta)$ superblocks. Superblocks defined in this way are related to Heckman-Opdam hypergeometrics and are non polynomial functions. 
A fruitful interaction between the mathematics of symmetric functions and SCFT follows, and we give a number of new results on both sides. 
One such example is a new Cauchy identity which naturally pairs our superconformal blocks with Sergeev-Veselov  
super Jacobi polynomials and yields the CPW decomposition of any free theory diagram in any dimension.


\medskip

\noindent

\newpage
\setcounter{page}{1}\setcounter{footnote}{0}

\renewcommand{\baselinestretch}{0.75}\normalsize
\tableofcontents
\renewcommand{\baselinestretch}{1.0}\normalsize

\newpage

\section{Introduction}

In this paper we will obtain conformal and superconformal blocks for 
four-point functions of half-BPS scalar operators in diverse dimensions.  
Our construction is based on a  single unified formalism, which uses analytic superspace 
\cite{Galperin:1984av,Howe:1995md,Hartwell:1994rp,hep-th/9408062,hep-th/0008048,Heslop:2001zm,Heslop:2002hp,Heslop:2004du,Doobary:2015gia,Howe:2015bdd} as its starting point, 
but includes various generalisations.  In particular, a parameter $\theta$ which allows us to move 
across dimensions.  By varying the value of $\theta$, we will find superconformal blocks for  
theories  in 1,{2},3,4,6 and conjecturally 5 dimensions.

The superconformal blocks that we present have a beautiful  
interpretation in  the theory of symmetric functions: they are in one-to-one correspondence with
a natural supersymmetric extension  of a class of stable polynomials that we call dual $BC$ Jacobi 
polynomials.  While the appearance of the $BC$ root system in this context is not new,%
\footnote{It was  
pointed out already in the foundational work~\cite{Dolan:2003hv},
and more recently in~\cite{Isachenkov:2016gim,Chen:2016bxc,Isachenkov:2017qgn,Chen:2020ipe}, that bosonic conformal blocks solve 
$BC_2$-type differential equations.} the results here set these observations in a much more general context and apply them directly to the supersymmetric case as well as the bosonic CFT case.

In a quantum field theory, correlation functions can be decomposed by using the operator product expansion (OPE). 
If the theory is a  Conformal Field Theory (CFT), it is well known that the OPE can be organised further
by collecting the contribution of a primary  operator with all its infinite descendants  \cite{Ferrara_1,Ferrara_2,Todorov,Mack_paper}.
The object representing the common OPE between pairs of external operators in a four-point correlator 
is the conformal block. It can be represented 
as the solution of a Casimir operator (for the correlation function under consideration) with 
given boundary conditions~\cite{Dolan:2000ut,Dolan:2001tt,Dolan:2003hv,Nirschl:2004pa,Dolan:2004iy}.  
The conformal block encodes an infinite sum  and is closely related to  hypergeometric functions. 
We will show, more precisely, that the mathematics underlying the physics of superconformal 
blocks is that of the \emph{multivariate} hypergeometric functions developed by Heckman-Opdam  
(applied to the $BC$ root system).

The Heckman-Opdam (HO) hypergeometrics are rigorously defined as  eigenfunctions  
of a system of differential equations \cite{Heckman_1,Heckman_2}. 
In the case of positive weights it was shown  \cite{Opdam}
that the $A$- and $BC$-type solutions reduced to Jack and Jacobi polynomials.
These polynomials, on the other hand, were particularly well known in the literature because 
of their definition as orthogonal polynomials associated to root systems \cite{Macdonald_2,Macdonald_3,Macdonald_book}. 
In fact, the study of orthogonal polynomials had its own independent trajectory, 
perhaps culminating with the introduction of the Koornwinder polynomials  \cite{Koornwinder_BC,Stokman,Rains_1}
and the proof of evaluation symmetry by Okounkov \cite{Okounkov_1}.
In this framework, supersymmetry makes its first appearance in \cite{Veselov_1,Veselov_2,Veselov_Macdonald}, 
where Sergeev and Veselov used the $A_{m-1|n-1}$ root system to construct a supersymmetric version 
of Jack and Macdonald polynomials, i.e.~super Jack and super Macdonald polynomials,
which reduce to the bosonic family for the $A_{n-1}$ root system.

The idea of expanding superconformal blocks in terms of super Jack polynomials  was  developed in \cite{Heslop:2002hp,Dolan:2003hv,Heslop:2004du,Doobary:2015gia}. In particular 
the work of \cite{Doobary:2015gia}
showed that 4d  superconformal blocks on the super Grassmannian  
$Gr(m|n,2m|2n)$ admit a simple representation as a sum over super Schur polynomials \cite{moens}, 
the latter being a particular ($\theta{=}1$) case of super Jack polynomials. 
In this representation each super Schur polynomial contributes with a simple coefficient 
built out of gamma functions. Quite nicely, the whole series was shown to
re-sum into a determinant of a matrix of Gauss hypergeometric functions.

An essential property of the construction in \cite{Doobary:2015gia}  is \emph{stability}, which states 
that the $(m,n)$ superconformal block reduces to the $(m{-}1,n)$ or $(m,n{-}1)$ 
superconformal block when one of its $m$ $x$-variables, or $n$ $y$-variables, is switched off. 
This property is crucial. It implies that
the coefficients of a block expanded in super Schur polynomials are \emph{independent} on $m,n$!
Thus, they can be obtained by examining the $(m,0)$ bosonic conformal blocks, or even simpler,
the $(0,n)$ compact analogues, which are just polynomials. This simple property then leads to  a precise formula of all superblocks (in particular those of short or atypical representations) of half BPS correlators in 4d $\cN=4,2$ supersymmetric theories.

As we will show here, it turns out that the $(0,n)$ polynomial  of \cite{Doobary:2015gia} is actually a rewriting of the $BC_n$ 
Jacobi polynomial $J_{\ulmb}$ with the parameter $\theta=1$. A $BC_n$ Jacobi polynomial is a polynomial in $n$ variables defined by a Young diagram $\ulmb=[\lambda_1,..,\lambda_n]$ with row lengths $\lambda_i\geq \lambda_{i+1} \in \mathbb{Z}^{\geq0}$. We will use the recent definition given by Koornwinder in \cite{Koornwinder}.
Then, the precise relation between the $(0,n)$ block and a Jacobi polynomial is quite interesting as it involves taking a complementary Young diagram, $\beta^n\bslash\ulmb=[\beta{-}\lambda_n,..,\beta{-}\lambda_1]$
and inverting the original variables.  Based on this observation we introduce a new class of polynomials, which we call
the \emph{dual} Jacobi polynomials. These are defined by
\beq\label{formula_Jt_intro}
	\tilde J_{\beta,\ulmb}(y_1,\ldots ,y_n) \equiv   (y_1\ldots y_n)^\beta  J_{\beta^n\bslash\ulmb}(\tfrac{1}{y_1},\ldots ,\tfrac{1}{y_n}) 
\eeq
where the new parameter $\beta$ here is an arbitrary integer, sufficiently large to 
ensure the Jacobi in  inverse variables is again polynomial.  
Remarkably, unlike the Jacobi polynomials themselves, the dual Jacobi 
polynomials are stable! 
This surprisingly simple fact is a key observation,\footnote{The 
importance of stability for the construction of $BC$ symmetric functions from $BC$ polynomials was discussed by E.Rains in \cite{Rains_1}.}  
and opens up the way towards the more general definition of superconformal blocks in diverse dimensions that we
give below. In fact, the above discussion was for $\theta=1$ but can be repeated for any 
value of $\theta$ by going from Schur to Jack polynomials.

We can now define a natural supersymmetric extension of $\tilde J_{\beta,\ulmb}$ using stability in a crucial way:
We simply have to  replace  the expansion in Jack polynomials of $\tilde J_{\beta,\ulmb}$  with super Jack polynomials, 
keeping the same expansion coefficients. The sum over super Jack polynomials now is no longer cut off, 
and  becomes infinite in the direction of the conformal subgroup, since the latter is non compact.  We define in this way the 
$BC_{n|m}$ dual Jacobi functions.

The key claim then is that superconformal blocks $B_{\gamma, \ulmb}({\bf x}|{\bf y})$ in any $(m,n;\theta)$ theory are given by these dual super Jacobi functions 
through a trivial redefinition, normalisation and transposition
\beq
B_{\gamma, \ulmb}({\bf x}|{\bf y})=
\left(\frac{\prod_i x_i^\theta}{\prod_j y_j}\right)^{\!\!\frac{\gamma}{2}} \, 
(-1)^{|\ulmb'|} \Pi_{\ulmb'}(\tfrac{1}{\theta}) \ {
	\tilde J_{\beta,\ulmb'}(  {\bf y}|{\bf x}) }{  }  \ . 
\eeq
The parameter 
$\theta$ will be  related to the spacetime dimension in the four-point correlator, and $\beta$ is given in terms of  the 
parameter $\gamma$, related to the scaling dimension of the operator appearing in the block. Here and throughout the paper we use primed Young diagrams 
to denote transposed Young diagrams, i.e.~the row lengths of $\ulmb'$ are the column lengths of $\ulmb$ and viceversa. 
The factor $ \Pi_{\ulmb'}(\tfrac{1}{\theta})$ is an explicit known function of $\theta$ given later.

The dual Jacobi functions so constructed interpolate between the dual Jacobi polynomials  
for $m=0$, the (infinite series) bosonic conformal blocks of \cite{Dolan:2003hv} for $n=0$ and arbitrary $\theta$,  
and reproduce the $\theta=1$ $(m,n)$ superconformal blocks in 4d. Quite remarkably, another class 
of supersymmetric Jacobi polynomials was constructed by Sergeev and Veselov in~\cite{Veselov_3}, 
however these are polynomial in both directions and are explicitly different from the  super Jacobi functions we defined above.

The correlators in this formalism, which are defined entirely by specifying  $(m,n;\theta)$ and charges for the external operators, include many cases of interest, for example 3d and 6d half BPS superconformal blocks which are notoriously difficult to study. 
Importantly, it will treat long and short representations on the same footing.

The use of symmetric polynomials, in relation with stability, has important payoffs for both the study of CFT and the theory of symmetric functions. 
The first one is that we will  exhibit a formula for the expansion coefficients of our 
superconformal blocks, over the basis of super Jack polynomials, in terms of a (super) 
binomial coefficient \cite{Veselov_3},  after a physically motivated redefinition of the external parameters. 
The second one, which is perhaps the most beautiful, is that we will able to prove, and indeed generalise,  a 
superconformal Cauchy identity, which once properly interpreted  yields the conformal partial 
wave expansion of any free theory propagator structures within the formalism. The objects paired in this 
Cauchy identity are, on the one side the superconformal blocks, on the other side the super Jacobi polynomials 
of Sergeev and Veselov \cite{Veselov_3}, which we mentioned above. 
Another pay off is that we can read off and generalise explicit results derived for Heckman Opdam hypergeometrics with $\theta=1$, allowing us to write down all higher order super Casimir operators for the superblocks.
Finally we will find an interpretation of the coefficients occurring in the decomposition of superblocks into blocks as structure constants for dual Jacobi polynomials.

In the next section we will give a detailed outline of the whole paper before proceeding to the main body.

\section{Overview} \label{overview:sec}

We would like to facilitate the reading of our paper, by presenting an extended overview of our results.

From a physics perspective, we will begin by presenting superconformal blocks  in the formalism of analytic 
superspace as eigenfunctions of the superconformal  Casimir operator which remarkably we will see is equivalent to a $BC_{m|n}$ CMS operator. This will anticipate the more detailed discussion in section~\ref{how_to_read}.
Then, we will outline a practical method to solve the differential equation associated 
to the $BC_{m|n}$ Casimir, by using a recursion relation. This recursion is very important in our story.
Full details will be given in section~\ref{sec_SCBlock}, and an extended discussion about its analytic 
properties will be given in section~\ref{analytic_cont_sec}.

From a more mathematically oriented point of view, we will motivate in section~\ref{dual_Jac_sec} 
our construction of dual BC${}_n$ Jacobi polynomials,  and their supersymmetric extension to dual 
super Jacobi functions. We will show that superblocks are essentially dual 
super Jacobi functions. Properties of these functions, which parallel physical  properties of the 
superconformal blocks, will be reviewed in sections \ref{overv_bin_cauchy} and \ref{overv_conject}, 
and proved later on in the corresponding sections. These have to do with the action of complementation, 
the Cauchy identity, and the structure constants for superconformal blocks.

\subsection{Superconformal  blocks}

We will consider four-point functions of scalar operators living on certain coset spaces  of appropriate superconformal 
groups with coordinates $X$. These operators include a number of cases of interest including half-BPS correlators 
in many superconformal field theories in $1,2,3,4,6$  and conjecturally 5 dimensions,  as well as (non-supersymmetric) 
scalars on Minkowski space of any dimension, as well as scalars in non supersymmetric CFTs and analogous reps 
on purely internal (ie compact) spaces. These cases all belong to the more general family of theories indexed 
by three parameters $(m,n;\theta)$  with $m,n$ non-negative integers. 
The above group theory interpretation exists only for certain values of $(m,n;\theta)$  
and when there is such a group theory interpretation,
the complexified coset space in question will be a maximal (possibly super and/or orthosymplectic) 
Grassmannian. They will all be given by a flag manifold which can be denoted by a (super) Dynkin diagram with a  single marked node (see section~\ref{how_to_read} for more details, in particular section~\ref{summary_list_theory} gives a list of the physical CFTs to which the formalism applies). 
Scalar operators in this space have a weight, $p$, under a 
$\mathbb{C}^*$ subgroup.\footnote{This is the complexification of either a $U(1)$ subgroup of the internal subgroup 
or the group of dilatations when there is no internal subgroup (i.e.~$n=0$) and it corresponds to the single marked Dynkin node.} 
We denote  them by $\mathcal{O}_{p}$, and the four point functions by%
\beq\label{study_4pt}
\langle \mathcal{O}_{p_1}(X_1) \mathcal{O}_{p_2}(X_2) \mathcal{O}_{p_3}(X_3) \mathcal{O}_{p_4}(X_4) \rangle.
\eeq
The operator $\cO_1$ is the basic representation, and corresponds to 
a representation whose highest weight state is a scalar operator 
of scaling dimension $\theta=\frac{d-2}{2}$ and internal charge 1. Then the scaling dimension of  
$\mathcal{O}_p$ is $p\,\theta$ with internal charge $p$.  In the supersymmetric cases $\mathcal{O}_p$ is a half BPS multiplet.
Full details of all theories that fit into this $(m,n,\theta)$ classification can be found in 
section~\ref{how_to_read} and appendix~\ref{app_supecoset}.

In a CFT we can bring two operators close to each other and replace their coincident limit 
with a sum over operators at a single point, as follows,
\beq\label{ope}
\cO_{p_i}(X_1)\cO_{p_j}(X_2)\ =\ \sum_{O^{\gamma,\ulmb}}\ C_{p_i p_j O^{\gamma,\ulmb} } \ 
g_{12}^{\frac{1}{2}(p_i+p_j-\gamma)}\ \mathbb{D}^{\gamma,\ulmb}(X_{12},\partial_2)\ O_{\gamma,\ulmb}(X_2)\ .
\eeq
Formula  \eqref{ope} is known as the Operator Product Expansion or OPE.
As an equality in group theory,  the OPE  corresponds to decomposing the tensor product of two 
(possibly infinite dimensional) representations into its irreducible primary representations, 
denoted above with $O_{\gamma,\ulmb}$, when $X_1\rightarrow X_2$. 
In the $(m,n;\theta)$ theories that we consider here, the coordinates $X_i$ can 
always be written as a square 
(super)-matrix, and the propagator, i.e.~the two point function of two basic 
$\cO_1$ operators, is%
\footnote{\label{f2}The power $\#$ here is defined so that $\cO_1$ has dimension $\theta$ 
and/or internal charge 1. It depends on the theory in question: for 
$\theta=1,2$ we will have $\#=1$ whereas for $\theta=\frac{1}{2}$ we will have 
$\#=\frac{1}{2}$. This precise value won't play any further role in the following. }
\beq\label{2pt}
g_{ij}=\langle \cO_1(X_i)\cO_1(X_j) \rangle
= \sdet(X_i{-}X_j)^{-\#}\ .
\eeq
The operator  $\mathbb{D}$ appearing in \eqref{ope} generates descendants of the exchanged primary operators $O_{\gamma,\ulmb}$,
and $C_{p_i p_j O_{\gamma,\ulmb} }$ is the OPE coefficient of $O_{\gamma,\ulmb}$ w.r.t.~the external fields $\cO_{p_i}(X_1)\cO_{p_j}(X_2)$.

In the OPE of two scalars, as is the case here, the primaries exchanged can always be specified by 
a weight under the afore-mentioned  $\mathbb{C}^*$, which we denote by $\gamma$,  
together with the representation of the remaining isotropy group, $\ulmb$, so $O_{\gamma,\ulmb}$ (and 
 also the external operators $\cO$ can be specified in this way but with $\ulmb$ trivial).  In a free theory,%
 \footnote{We will focus initially on the case where all parameters are integers as for example occurs in a (generalised) free conformal field theory. However the analytic continuation to non integer values (for example to include anomalous dimensions) is considered in detail in section~\ref{analytic_cont_sec}.}
  $\gamma$  varies over the range 
$\max(p_{1}{-}p_2,p_{4}{-}p_3)\leq \gamma\leq \min(p_1+p_2,p_3+p_4)$ in steps $\in 2 \mathbb{Z}$ 
where we will assume $p_1>p_2, p_4>p_3$ without loss of generality.
The representation of the isotropy group is specified via a Young diagram 
$\ulmb$, which will encode the various quantum numbers, for example spin and twist, 
of the operator exchanged. The Young diagram will be consistent with that describing an $SL(m|n)$ representation, meaning it will have at most $m$ rows longer than $n$ and at most $n$ columns taller than $m$. Furthermore the overall height of the Young diagram must be smaller than $\beta\equiv \min\left( \tfrac 1 2 (\gamma {-}  p_{12}), \tfrac 1  2 (\gamma {-}  p_{43}) \right)$. So a valid Young diagram $\ulmb$ must fit inside the red area here
    \beq 
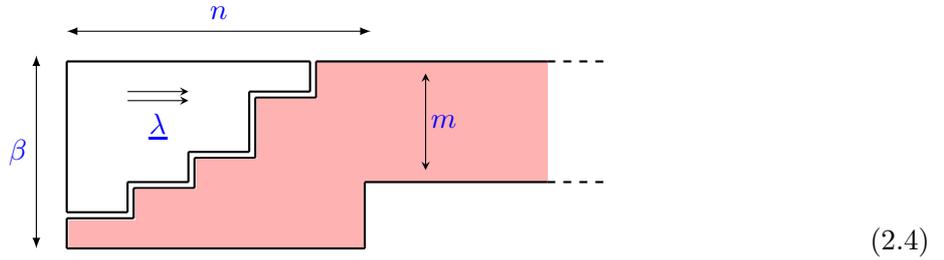
\label{YDpic}
\begin{tikzpicture}[scale=.8]
	
	\def\shift{2.5}
	\def\xuno{2-\shift}
	\def\yuno{2}
	
	\def\ydue{5}
	\def\xtre{6-\shift}
	
	\def\ytre{4.5}
	\def\xquattro{5-\shift}
	
	\def\yquattro{3.5}
	\def\xcinque{4-\shift}
	
	\def\ycinque{3}
	\def\xsei{3-\shift}
	
	\def\step{.5}
	\def\stepp{.1}

	\draw[thick] (\xuno,	\yuno+\step) -- 			(\xuno,	\ydue);
	\draw[thick] (\xuno,	\ydue) -- 				(\xtre,	\ydue);
	\draw[thick] (\xtre,	\ytre) -- 				(\xquattro,	\ytre);
	\draw[thick] (\xquattro,	\yquattro) -- 		(\xcinque,	\yquattro);
	\draw[thick] (\xcinque,	\ycinque) -- 		(\xsei,	\ycinque);
	\draw[thick] (\xsei,	\yuno+\step) -- 			(\xuno, \yuno+\step);
	
	\draw[thick](\xtre,	\ydue)-- 			(\xtre,	\ytre);
	\draw[thick](\xquattro,	\ytre)-- 		(\xquattro,	\yquattro);
	\draw[thick](\xcinque,	\yquattro)-- 	(\xcinque,	\ycinque);
	\draw[thick](\xsei,	\ycinque)-- 		(\xsei,	\yuno+\step);

	\draw[thick] (\xtre+\stepp,	\ytre-\stepp) -- 				(\xquattro+\stepp,	\ytre-\stepp);
	\draw[thick] (\xquattro+\stepp,	\yquattro-\stepp) -- 		(\xcinque+\stepp,	\yquattro-\stepp);
	\draw[thick] (\xcinque+\stepp,	\ycinque-\stepp) -- 		(\xsei+ \stepp,	\ycinque-\stepp);
	\draw[thick] (\xsei+\stepp,	\yuno+\step-\stepp) -- 			(\xuno, \yuno+\step-\stepp);

	\draw[thick](\xtre+\stepp,	\ydue)-- 			(\xtre+\stepp,	\ytre-\stepp);
	\draw[thick](\xquattro+\stepp,	\ytre-\stepp)-- 		(\xquattro+\stepp,	\yquattro-\stepp);
	\draw[thick](\xcinque+\stepp,	\yquattro-\stepp)-- 	(\xcinque+\stepp,	\ycinque-\stepp);
	\draw[thick](\xsei+\stepp,	\ycinque-\stepp)-- 		(\xsei+\stepp,	\yuno+\step-\stepp);

	\filldraw[red!30!white]  (\xtre+\stepp+.025,\ydue) rectangle (\xtre+4-\stepp,\ycinque);
	\filldraw[red!30!white]  (\xquattro+\stepp+.04,	\ytre-\stepp-.025) rectangle(\xtre+1-\stepp,\yuno-.05);
	\filldraw[red!30!white]  (\xcinque+\stepp+.04,	\yquattro-\stepp-.025) rectangle(\xtre+1-\stepp,\yuno-.05);
	\filldraw[red!30!white]  (\xsei+\stepp+.04,	\ycinque-\stepp-.025) rectangle(\xtre+1-\stepp,\yuno-.05);
	\filldraw[red!30!white]  (\xuno+.04,	\yuno+\step-\stepp-.04) rectangle(\xtre+1-\stepp,\yuno-.1);

	\draw[thick]  (\xtre+\stepp,\ydue) -- (\xtre+4-\stepp,\ydue);
	\draw[thick]  (\xuno,\yuno-\stepp) -- (\xuno,\yuno+\step-\stepp);	
	
	\draw[thick]  (\xuno,\yuno-\stepp) -- (\xtre+1-\stepp,\yuno-\stepp);	
	\draw[thick]  (\xtre+1-\stepp,\yuno-\stepp) -- (\xtre+1-\stepp,\ycinque);
	\draw[thick]  (\xtre+1-\stepp,\ycinque) --  (\xtre+4-\stepp,\ycinque);
	
	\draw[thick,dashed] (\xtre+4-\stepp,\ydue)--(\xtre+5-\stepp,\ydue);
	\draw[thick,dashed] (\xtre+4-\stepp,\ycinque)--(\xtre+5-\stepp,\ycinque);
	
	\draw[stealth-stealth]  (\xtre+2-\stepp,	 \ydue-.2)  --     (\xtre+2-\stepp,	 \ycinque+.2) ;
	\draw[]  (\xtre+2.3-\stepp,	 \ycinque+1) 	node [rotate=0]	{ \color{blue} $m$ };

	\draw[latex-latex] (\xuno-.5,\yuno-.1) -- (\xuno-.5,\ydue+.1);
	\draw[latex-latex] (\xuno,\ydue+.5) -- (\xtre+1,\ydue+.5);

	\draw (\xuno-.8,	  3.5) 	node 	{ \color{blue} $\beta$ };
	\draw (4.5-\shift,	  \ydue+.8) 	node 	{ \color{blue} $n$ };
	
	\draw[] (3.5-\shift,	 3.75) 	node [rotate=0]	{ \color{blue} $\ulmb$ };

	\draw[-stealth]  (3-\shift,4.5) -- ( 4-\shift,4.5) ;
	\draw[-stealth]  (3-\shift,4.35) -- ( 4-\shift,4.35);

	\draw[] (8.8,	 2.5) 	node [rotate=0]	{  $\phantom{cazzo}$ };	
	
\end{tikzpicture}
\eeq

Superconformal representations are conventionally specified  via the dilation weight, $\Delta$, various conformal quantum numbers or Dynkin labels (e.g. Lorentz spin), and quantum numbers or Dynkin labels describing the internal representation,  whereas we specify the  the representation  by the Young diagram $\ulmb$ and the parameter $\gamma$. The dilation weight is then 
$\Delta=
\sum_{i=1}^{m}  \max({\lambda_i{-}n},{0}) +m \theta \tfrac{\gamma}{2}$
and the other conformal quantum numbers are specified by differences of the first $m$ row lengths $\lambda_i-\lambda_{i+1}$. The internal group quantum numbers are given as differences of Young diagram column heights $\lambda'_{i}-\lambda'_{i+1}$ as well as the special marked Dynkin label $b=\gamma -2\lambda_1'$.
We refer to section \ref{how_to_read} for a more detailed explanation about how 
to read the Young diagram and relate it to conformal group representations. For now note that whilst all parameters are integers in the free theory, in an interacting quantum theory $\Delta$ alone can become non integer, with all other quantum numbers remaining integer. In particular, this means we can allow for anomalous dimensions by deforming the row lengths $\lambda_i \rightarrow \lambda_i+ \tau$ for  $i=1,..,m$, arbitrary real $\tau$   and/ or $\lambda'_j \rightarrow \lambda'_j+ \tau'$ for $j=1,..,n$, arbitrary real $\tau'$  as long as we also have $\gamma \rightarrow \gamma +2\tau'$ to ensure that $b=\gamma -2\lambda_1'$ remains integer. Furthermore deforming thus,  with $\tau=-\theta \tau'$ leaves the representation unchanged.  We will return to this `shift symmetry'.

The conformal structure of the theory and the use (twice) of the OPE, 
imply that we can decompose a four-point function as follows:
\begin{align}\label{4pt}
&
\!\!\!\langle \mathcal{O}_{p_1}(X_1) \mathcal{O}_{p_2}(X_2) \mathcal{O}_{p_3}(X_3) \mathcal{O}_{p_4}(X_4) \rangle= \notag\\
&
\rule{1cm}{0pt} g_{12}^{ \frac{p_1+p_2}{2} } g_{34}^{ \frac{ p_3+p_4}{2} } \left( \frac{ g_{14}  }{ g_{24} }\right)^{\! \frac{ p_{12} }{2} } \left( \frac{ g_{14} }{ g_{13} } \right)^{\! \frac{ p_{43 } }{2} }  
\sum_{\gamma,\ulmb} \left( \sum_{O} C_{p_1p_2}^{O_{} }  C_{O\, p_2p_3}  \right)
B_{\gamma,\ulmb}^{(m,n)}(X_i;\theta,p_{12},p_{43})
\end{align}
where $B_{\gamma,\ulmb}(X_i)$ is called the (super)conformal  block.

The (super)conformal block $B_{\gamma,\ulmb}$ is the main subject of this paper. It
represents the contribution of an exchanged primary operator  $O_{\gamma, \ulmb}$ with 
its tail of  descendants, which is common  to the OPE of $\mathcal{O}_{p_1}\mathcal{O}_{p_2}$ 
and $\mathcal{O}_{p_3}\mathcal{O}_{p_4}$.%
Thus the block depends on the particular theory of interest  $(m,n;\theta)$, as well as the external operators through the quantities 
\beq
p_{12}\equiv p_1{-}p_2\qquad;\qquad p_{43}\equiv p_4{-}p_3.
\eeq 
Furthermore, because of conformal invariance, $B_{\gamma,\ulmb}(X_i)$ only depends 
on cross ratios built out of the four coordinates $X_{i=1,2,3,4}$. 

The $X_{i=1,2,3,4}$ are square (super) matrices, and the cross 
 ratios are  the $m{+}n$ independent eigenvalues of the (super)-matrix $Z=X_{12}X_{24}^{-1}X_{43}X_{31}^{-1}$
 where $X_{ij}=X_i{-}X_j$. 
These $m{+}n$ independent cross-ratios $z_i$  split into two types, corresponding to whether they arise from 
the non-compact conformal group or the compact internal group, which we denote by $x_i$ and $y_j$ respectively, so 
\begin{align}\label{evalues}
	&B_{\gamma,\ulmb}(X_i)=B_{\gamma,\ulmb}(\bf z)\notag\\
	&{\bf z}=(z_1,\dots,z_m|z_{m+1},\dots ,z_{m+n}), \qquad z_i=x_i, \qquad z_{j+m}=y_j\ .
\end{align}
The details in all cases of physical interest will be given in section \ref{how_to_read}. 
For now note that 
\begin{align}\label{zz}
	\sdet(Z)^\#=\frac{g_{24}g_{13}}{g_{12}g_{34}}=\frac{\prod_{i=1}^m 
	x_i^\theta}{\prod_{j=1}^n y_j}\qquad \qquad 
	\sdet(1-Z)^\#=\frac{g_{14}g_{23}}{g_{12}g_{34}}=\frac{\prod_{i=1}^m 
	(1-x_i)^\theta}{\prod_{j=1}^n (1-y_j)}\ ,
\end{align}
where $\#$ is the power in~\eqref{2pt}.

We can  characterise $B_{\gamma,\ulmb}(\bf z)$ group theoretically, 
as eigenfunctions of the quadratic Casimir  acting at points 1 and 2 on the correlator. Pulling the Casimir 
through the prefactor of~\eqref{4pt} to act on the block itself, it becomes a differential operator in ${\bf z}$. 
This differential operator, denoted hereafter by $\mathbf{C}$, has been computed explicitly in a number of cases\footnote{The 
cases are: $(m,n)=(2,0)$ and $(0,2)$ in~\cite{Dolan:2003hv}.  %
Then, $\theta=1$, $m,n \in \mathbb{Z}^+$,  in~\cite{Doobary:2015gia}, by using a supermatrix formalism,
and partially in the case of $\theta=2$, $m,n \in \mathbb{Z}^+$ in~\cite{Doobary2}  }
and we find, quite remarkably,  that in all these cases it is equivalent to the $BC_{m|n}$ operator of the type of 
Calogero-Moser-Sutherland. We give the details of this identification in appendix 
\ref{app_differential_operators}. We are led to conjecture that such an identification is true for all cases. 
Under this assumption  we  define superconformal blocks group theoretically through the eigenvalue equation: 
 \beq\label{eeq}
\mathbf{C}^{(\theta,-\frac12p_{12},-\frac12 p_{43},0)} B_{\gamma,\ulmb}({\bf 
z})= { E}_{\gamma,\ulmb}^{(m,n;\theta)} B_{\gamma,\ulmb}({\bf z})\ .
\eeq
The Casimir $\mathbf{C}^{(\theta,a,b,c)}$ is a second order differential operator in the variables 
$\bf z$ given explicitly in \eqref{casimir_intro}.  It depends on $ \theta,a,b,c$ and implicitly on $m,n$. 
The eigenvalue is 
\begin{align}\label{ev1}
	{ E}_{\gamma,\ulmb}^{(m,n;\theta)}&={h}_{\ulmb}^{(\theta)}+\theta \gamma |\ulmb|
+\Big[ \gamma\theta|m^n|+ h_{[e^m]}^{(\theta)} - \theta h_{[s^n]}^{(\frac{1}{\theta})}\Big]
\end{align}
where $e=+\frac{\theta\gamma}{2}$, $s=-\frac{\gamma}{2}$, and
\beq\label{intro_eigenv}
{h}_{\ulmb}^{(\theta)}=\sum_{i} \lambda_i (\lambda_i{-}2\theta(i{{-}}1){-}1)\qquad;\qquad |\ulmb|=\sum_i \lambda_i
\eeq
with $\ulmb=[\lambda_1,\ldots ]$ denoting the row lengths of the Young diagram, and 
\beq
 h_{[e^m]}^{(\theta)}=m\theta \tfrac{\gamma}{2}  ( (\tfrac{\gamma}{2}+1-m)\theta-1) \qquad;\qquad \theta h_{[s^n]}^{(\frac{1}{\theta})}= n\tfrac{\gamma}{2}(\theta(\tfrac{\gamma}{2}+1)-1+n)
\eeq
In particular, $[e^{m}]$ and $[s^n]$ are the diagrams 
with $m$ rows of `length' $+\frac{\theta\gamma}{2}$, and $n$ columns of `height' $-\frac{\gamma}{2}$. 

What is important about \eqref{ev1} is that  ${h}_{\ulmb}^{(\theta)}+\theta \gamma |\ulmb|$ does not depend on $m,n$ 
but only depends on the Young diagram $\ulmb$, whereas the other terms clearly have explicit  $m,n$ dependence. 
We will revisit this key point shortly.

\subsection{Solving the Casimir equation}\label{presentation}

From OPE considerations, it has been shown in a number of cases\footnote{For a direct  derivation 
 of this limit from the OPE see the discussion around eq.~(20) in \cite{Doobary:2015gia}  
 for  the $\theta=1$ case and the discussion around (33)-(35) of~\cite{Heslop:2004du}  
 for $\theta=2$. This limit is also implicit in earlier work for $\theta=1$~\cite{Heslop:2001zm} 
 as well as in the bosonic case \cite{Dolan:2003hv}  for any $\theta$.}
 that in the limit ${\bf z}\rightarrow 0$
\beq\label{asym}
B_{\gamma, \ulmb}({\bf z})= \left(\frac{\prod_i x_i^\theta}{\prod_j y_j}\right)^{\!\frac{\gamma}{2}}\!\!\times\big(P_{\ulmb}({\bf z};\theta)\,+\,\ldots\ \big)
\eeq
where $P_{\ulmb}$ is the super Jack polynomial of Young diagram $\ulmb$. (Jack and super 
Jack polynomials are reviewed in appendix \ref{symm_supersym_poly_sec}). The physics 
behind \eqref{asym} is quite simple to understand: When there is a group theory interpretation, 
the prefactor corresponds to the limit $z_i\rightarrow 0$ of a four-point propagator structure 
in which an operator of twist $\theta\gamma$ is the first operator exchanged, and $P_{\ulmb}$ 
accounts for the corresponding superprimary operator in $B_{\gamma, \ulmb}({\bf z})$.

The eigenvalue equation~\eqref{eeq}, together with the asymptotics~\eqref{asym}, 
gives a unique definition of the superblocks $B_{\gamma, \ulmb}({\bf z})$.

For $\theta=1,2$ the whole superblock is known to be representable as a series 
over super Jack 
polynomials~\cite{Heslop:2002hp,Heslop:2004du,Doobary:2015gia}. Thus for arbitrary 
$\theta$ we will seek a representation of $B_{\gamma, \ulmb}({\bf z})$ as such a series, explicitly as, %
\begin{align}\label{Jexp_intro}
B_{\gamma, \ulmb}({\bf z})=
	\left(\frac{\prod_i x_i^\theta}{\prod_j y_j}\right)^{\!\!\frac{\gamma}{2}} \,  F_{\gamma,\ulmb}({\bf z})\qquad;\qquad
	 F_{\gamma,\ulmb}({\bf z};\theta,p_{12},p_{43})=\sum_{\umu \supseteq \ulmb} 
	 { ({ T}_{\gamma;\theta,p_{12},p_{43}})_\ulmb^\umu} \, P_{\umu}({\bf z})\ ,
\end{align}
where the sum is over all Young diagrams $\umu$ which contain $\ulmb$, i.e.~$\ulmb\subseteq \umu$.
The expansion coefficients ${ ({ T}_{\gamma})_\ulmb^\umu}$ will be sometimes referred to as the \JtoB matrix. 
These coefficients depend on $(\theta,p_{12},p_{43})$ 
as well as $\gamma$ and the two Young tableaux $\ulmb,\umu$. In principle they should depend 
also on $(m,n)$, but as we will see quite remarkably  they in fact don't!
The super Jacks themselves vanish if the Young diagram is not of $SL(m|n)$ shape (ie if the $m$th row is bigger than $n$, $\lambda_{m+1}>n$). Furthermore we will see that the coefficients $T_\gamma$ vanish if the height of the Young digram is larger than $\beta$. So the non-vanishing contributions to the sum are only from Young diagrams $\umu$ which fit inside the red area of~\eqref{YDpic}.

From the defining Casimir for the blocks~\eqref{eeq}
we obtain a differential equation for $F_{\gamma,\ulmb}({\bf z})$ by conjugating 
the original Casimir in \eqref{eeq} by the $\gamma$-prefactor in~\eqref{Jexp_intro}. It turns out that 
the result can be written in terms of a shifted version of the same differential operator $\bf C$,   
with a modified eigenvalue:
\beq\label{eeq_F}
\mathbf{C}^{(\theta,\alpha,\beta,\gamma)} F_{\gamma,\ulmb}({\bf z})= (h_{\ulmb}^{(\theta)} +\gamma|\ulmb|) F_{\gamma,\ulmb}({\bf z})\ ,
\eeq
where 
\begin{align}\label{first_def_alpha_beta}
	\alpha &\equiv \max\left( \tfrac 1 2 (\gamma {-}  p_{12}),\tfrac 1  2 (\gamma {-}  p_{43})\right) ,\quad \beta\equiv \min\left( \tfrac 1 2 (\gamma {-}  p_{12}), \tfrac 1  2 (\gamma {-}  p_{43}) \right)\ .
\end{align}

A remarkable outcome of conjugating the Casimir is that the eigenvalue of 
$F_{\gamma,\ulmb}$ 
does not depend explicitly on $(m,n)$, since the $m,n$ dependent term of the 
original eigenvalue~\eqref{ev1},  the one in $[\ldots]$, is now missing.  
In other words, the eigenvalue of $F_{\gamma,\ulmb}$ depends only on $\ulmb$ and $\gamma$.
This fact underlies a property known as `stability' in the maths literature, which means that $F_{\gamma, \ulmb}$ 
depends only on the number of non-vanishing variables of each type. 
So the $F_{\gamma, \ulmb}$ of type $(m+1,n)$ in which one of the $x$'s is set to 
zero reduces to the $F_{\gamma, \ulmb}$ of type $(m,n)$. Similarly if one of the $y$s 
is set to zero. In formulae 
\beq
F_{\gamma,\ulmb}(x_1,\ldots x_{m},0|y_1,\ldots y_{n})=F_{\gamma,\ulmb}(x_1,\ldots x_{m}|y_1,\ldots y_{n},0)=F_{\gamma,\ulmb}(x_1,\ldots x_{m}|y_1,\ldots y_{n})\ .
\eeq
Stability is clear when there is a 
(super) matrix interpretation for the superconformal block (as for $\theta=\frac{1}{2},1,2$), 
but the generalisation to arbitrary $\theta$ is non-trivial.

The next observation is that since super Jack polynomials are also 
stable,\footnote{Notice also that the eigenvalue for 
	$F_{\gamma,\ulmb}$ is the same as for $P_{\ulmb}$, since 
	${\bf H}P_{\ulmb}=h_{\ulmb} P_{\ulmb}$ and $\sum_i z_i\partial_i  P_{\ulmb}=|\ulmb| P_{\ulmb}$. }
 the expansion coefficients   
${ ({ T}_{\gamma})_\ulmb^\umu}$ must be  {\em{independent}} of $(m,n)$!
Indeed this was the main insight of~\cite{Doobary:2015gia}, which reduces the study of 
${ ({ T}_{\gamma})_\ulmb^\umu}$ for $m,n$ superconformal blocks to the study of
the simpler generalised bosonic conformal blocks.

The representation of the super blocks $B_{\gamma, \ulmb}({\bf z})$ as a sum over super 
Jack polynomials, \eqref{Jexp_intro}, is not accidental. In fact, building on the observation that the 
Casimir is a differential operator for the $BC_{m,n}$ root system, as mentioned above, 
we are led to consider writing it in terms of operators ${\bf H}$ and $\sum_i z_i \partial_i$, the $A_{m-1,n-1}$ 
differential operators for which  super Jack polynomials are eigenfunctions.\footnote{More precisely,  ${\bf H}$ is the supersymmetric version of the Calogero-Moser-Sutherland (CMS) 
	Hamiltonian found in \cite{Veselov_1,Veselov_2}. $A$- and $BC$-type differential operators 
	are reviewed in appendix \ref{app_differential_operators}.} The corresponding  decomposition takes the form, 
\begin{align}\label{rewriteCinJack_intro}
	\mathbf{C}^{(\theta,a,b,c)}
	=&\ \mathbf{H}^{(\theta)} +  \theta c\sum_{i=1}^{m+n} \,  z_i \partial_i \notag\\
	&\ -\theta  (a+b) \sum_{i=1}^{m+n}  z_i^2\partial_i -\tfrac{1}{2}\Big[ \mathbf{H}^{(\theta)}, \sum_{i=1}^{m+n} z_i^2 \partial_i \Big]   -  \theta^2 ab \sum_{i=1}^{m+n} z_i (-\theta)^{-\pi_i}\ .
\end{align}
 The operators on the first line thus map Jack polynomials to themselves whereas those on the second line take  a Jack polynomial to another Jack polynomial with an additional box  in its Young diagram and will thus be interpreted as one-box raising operators.
The action of the Casimir on the sum of super Jack polynomials, decomposed as in~\eqref{rewriteCinJack_intro}, thus   turns the 
differential equation~\eqref{eeq_F} into a recursion  relation on the coefficients $(T_{\gamma})^{\umu}_{\ulmb}$. 
A particularly convenient representation of this recursion is
\beq\label{rec}
(T_\gamma)_{\ulmb}^{\umu}= 
\frac{ \sum_{i=1}%
	 \left(\mu_i{-}1{-} \theta(i{-}1{-}\alpha)\right)\left(\mu_i{-}1{-} \theta(i{-}1{-}\beta)\right) \mathbf{f}_{\umu{-}\Box_i}^{(i)} \, (T_\gamma)_{\ulmb}^{\umu{-}\Box_i}  }{
\left( {h}_{\umu}{-}{h}_{\ulmb} {+} \theta \gamma\, (|\umu|{-}|\ulmb| )\right)}.
\eeq
Here $\umu{-}\Box_i$ represents a Young diagram obtained by removing 
the last box at row $i$ from $\umu$, when allowed,  and ${\bf f}$ is a simple function 
given in \eqref{boldf_section}.  Let us emphasise that the recursion~\eqref{rec} is straightforward to 
implement on a computer, and very efficient up to high order.

The recursion \eqref{rec} actually depends only on objects which can be defined combinatorially, 
and admits different equivalent representations, depending on whether we read the Young diagram just
along the rows, or the columns, or we mix rows and columns as it is more appropriate in the supersymmetric theory. 
By the $(m,n)$ independence,
the value of $(T_\gamma)_{\ulmb}^{\umu}$ does not depend on the chosen representation.
This is particularly useful when constructing superconformal blocks for 
short (protected) representations, a case which is notoriously difficult to study with other methods.

As mentioned the above discussion is true for (generalised) free theory blocks where all quantum numbers are integers. In an interacting CFT however, the dilation weight can become non integer. Further it is interesting to consider analytic continuations of other quantum numbers such as spin. 
We then study possible analytic continuations of the recursion in the variables that describe the Young diagrams. 
We do so by promoting the external $\ulmb$ to complex values, and taking $\umu=\ulmb+\vec{n}$ with $\vec{n}\in\mathbb{Z}^+$. 
It turns out that each one of the following representations of the recursion, either row type, column type, or supersymmetric one, 
now gives a distinct analytic continuation (thus analytic continuation breaks the $m,n$ independence), which however coincide with the others when $\ulmb$ reduces to  a valid Young diagram.   

For long (non protected) representations we show in section~\ref{analytic_cont_sec} that a suitable $(m,n)$ 
analytic continuation of  $(T_\gamma)_{\ulmb}^{\umu}$ exists, such that 
the following shift, 
\beq\label{deform_overview_sec}
	\begin{array}{c} \lambda_i \rightarrow \lambda_i-\theta\tau'   \\[.2cm] \mu_i \rightarrow \mu_i-\theta\tau'  \\[.2cm] \!i=1,\ldots m\end{array} \qquad;\qquad 
	\begin{array}{c} \lambda'_j \rightarrow \lambda'_j+\tau' \\[.2cm] \mu'_j \rightarrow \mu'_j+\tau'  \\[.2cm] \!j=1,\ldots n\end{array}  \qquad;\qquad            
	\gamma \rightarrow \gamma+2\tau'\ .
\eeq
is a symmetry of the solution. For integer values this shift symmetry corresponds to the well understood  equivalence  of different Young diagrams describing $SL(m|n)$ reps for $\theta=1$ and similarly for the other group theoretic cases (e.g. when $n=0$ you can see it as the fact that full $m$ columns  correspond to the trivial $SL(m)$ rep and thus can be deleted). It can also be seen directly from the relation between the Young diagram and Dynkin labels (see~\eqref{translation_dynkin_YT}) that there is a redundancy in the Young diagram description of representations which is precisely this shift.

\subsection{Superblocks as dual super Jacobi functions}\label{dual_Jac_sec}

In this section we start from scratch 
and motivate,  from a purely mathematical point of view, the introduction of a certain 
supersymmetric generalisation of $BC$ Jacobi polynomials~\cite{Stokman, 
Koornwinder}. More precisely, 
we will define a family of polynomials that we call {\em dual Jacobi  polynomials}. These are 
closely related to the $BC$ Jacobi polynomials of Koornwinder \cite{Koornwinder}, but
differently from those, they are stable, and thus allow a straightforward supersymmetric generalisation.
The  supersymmetric generalisation of  
a dual Jacobi polynomial is however not  polynomial, in general. We will denote 
them as {\em dual super Jacobi  functions}. 
It will turn out that superblocks are equivalent to these dual super Jacobi  functions.

This section 
can be read largely independently of the previous sections (up to the point where we make the identification with blocks). 

\subsubsection*{Dual Jacobi polynomials}
\label{sec:dualjacobi}

Consider the following operation: take a  $BC_n$ Jacobi polynomial $J_\ulmb()$ in inverse 
variables $y^{-1}_i$, and   multiply by a sufficiently high power of $(y_1...y_n)^\beta$ for 
$\beta\in \mathbb{N}$ in order to ensure the result is polynomial again in the $y_i$. We will call this a 
{\em dual Jacobi polynomial}. 

Note that when the above operation is performed on a {\em Jack} polynomial the 
result is again a Jack polynomial of the {\em complementary} Young diagram 
$\beta^n \backslash \ulmb$:
\begin{align}\label{complement_on_jack}
	(y_1 \ldots y_n)^\beta P_\ulmb(\tfrac{1}{y_1},\ldots ,\tfrac{1}{y_n})= P_{\beta^n\bslash\,\ulmb}(y_1,\ldots,y_n)\ ,
\end{align} 
where the  {\it complement} of $\ulmb$ in $\beta^n$,  i.e.~$\beta^n\bslash\,\ulmb$, is 
defined as the Young diagram with row lengths 
\beq\label{lambdap_compl}
\begin{tikzpicture}%
	
	\def\shift{2.5}
	\def\xuno{2-\shift}
	\def\yuno{2}
	
	\def\ydue{5}
	\def\xtre{6-\shift}
	
	\def\ytre{4.5}
	\def\xquattro{5-\shift}
	
	\def\yquattro{3.5}
	\def\xcinque{4-\shift}
	
	\def\ycinque{3}
	\def\xsei{3-\shift}
	
	\def\step{.5}
	\def\stepp{.1}

	\draw[thick] (\xuno,	\yuno+\step) -- 			(\xuno,	\ydue);
	\draw[thick] (\xuno,	\ydue) -- 				(\xtre,	\ydue);
	\draw[thick] (\xtre,	\ytre) -- 				(\xquattro,	\ytre);
	\draw[thick] (\xquattro,	\yquattro) -- 		(\xcinque,	\yquattro);
	\draw[thick] (\xcinque,	\ycinque) -- 		(\xsei,	\ycinque);
	\draw[thick] (\xsei,	\yuno+\step) -- 			(\xuno, \yuno+\step);
	
	\draw[thick](\xtre,	\ydue)-- 			(\xtre,	\ytre);
	\draw[thick](\xquattro,	\ytre)-- 		(\xquattro,	\yquattro);
	\draw[thick](\xcinque,	\yquattro)-- 	(\xcinque,	\ycinque);
	\draw[thick](\xsei,	\ycinque)-- 		(\xsei,	\yuno+\step);

	\draw[thick] (\xtre+\stepp,	\ytre-\stepp) -- 				(\xquattro+\stepp,	\ytre-\stepp);
	\draw[thick] (\xquattro+\stepp,	\yquattro-\stepp) -- 		(\xcinque+\stepp,	\yquattro-\stepp);
	\draw[thick] (\xcinque+\stepp,	\ycinque-\stepp) -- 		(\xsei+ \stepp,	\ycinque-\stepp);
	\draw[thick] (\xsei+\stepp,	\yuno+\step-\stepp) -- 			(\xuno, \yuno+\step-\stepp);

	\draw[thick](\xtre+\stepp,	\ydue)-- 			(\xtre+\stepp,	\ytre-\stepp);
	\draw[thick](\xquattro+\stepp,	\ytre-\stepp)-- 		(\xquattro+\stepp,	\yquattro-\stepp);
	\draw[thick](\xcinque+\stepp,	\yquattro-\stepp)-- 	(\xcinque+\stepp,	\ycinque-\stepp);
	\draw[thick](\xsei+\stepp,	\ycinque-\stepp)-- 		(\xsei+\stepp,	\yuno+\step-\stepp);

	\draw[thick]  (\xuno,\yuno-\stepp) -- (\xtre+1-\stepp,\yuno-\stepp);	
	\draw[thick]  (\xtre+1-\stepp,\yuno-\stepp) -- (\xtre+1-\stepp,\ydue);
	\draw[thick]  (\xtre+\stepp,\ydue) -- (\xtre+1-\stepp,\ydue);
	\draw[thick]  (\xuno,\yuno-\stepp) -- (\xuno,\yuno+\step-\stepp);

	\draw[very thick, red]  (\xuno-.1,\ydue+.1) rectangle (\xtre+1,\ycinque-2*\step-.2);
	
	\draw[latex-latex] (\xuno-.5,\yuno-.1) -- (\xuno-.5,\ydue+.1);
	\draw[latex-latex] (\xuno,\ydue+.5) -- (\xtre+1,\ydue+.5);

	\draw (\xuno-.8,	  3.5) 	node 	{ \color{blue} $n$ };
	\draw (4.5-\shift,	  \ydue+.8) 	node 	{ \color{blue} $\beta$ };
	
	\draw[] (3.5-\shift,	 4.) 	node [rotate=0]	{ \color{blue} $\ulmb$ };
	\draw[] (5.9-\shift,	 2.9) 	node [rotate=0]	{ \color{blue} $\beta^n\bslash\,\ulmb$ };

	\draw[-stealth]  (6.5-\shift,	 2.2)  --     (5-\shift,	 2.2) ;
	\draw[-stealth]  (6.5-\shift,	 2.35)  --     (5-\shift,	 2.35) ;
	
	\draw[-stealth]  (3-\shift,4.5) -- ( 4-\shift,4.5) ;
	\draw[-stealth]  (3-\shift,4.65) -- ( 4-\shift,4.65) ;

	\draw[] (-5,	 3.5) 	node [rotate=0]	{  $\displaystyle (\beta^n\bslash\,\ulmb)_j= \beta- (\ulmb)_{n+1-j}\qquad;\qquad$ };	
	\draw[] (+4,	 3.5) 	node [rotate=0]	{  $\phantom{cc}$ };	
	
\end{tikzpicture}
\eeq
The integer  $\beta$ should be large enough to take the complement, 
i.e.~$\beta\ge \lambda_1$.

The aforementioned operation maps a BC${}_n$ Jacobi polynomial, 
$J({\bf y};\theta,p_-,p_+)$ to a different polynomial, which we thus
define as the {\em  dual Jacobi polynomial}, 
\beq\label{dual_jacobi}
\tilde J_{\beta,\ulmb}(y_1,\ldots ,y_n) \equiv   (y_1\ldots y_n)^\beta  
J_{\beta^n\bslash\,\ulmb}(\tfrac{1}{y_1},\ldots ,\tfrac{1}{y_n}) \ .
\eeq
Crucially, dual Jacobi polynomials turn out to be stable (meaning that 
switching off one of the variables reduces the polynomial to the polynomial 
with one variable fewer)
\begin{align}\label{dual_Jacobi_stable}
	\tilde J_{\ulmb}(y_1,..,y_{n-1},0)= \tilde J_{\ulmb}(y_1,..,y_{n-1})\ .
\end{align}
This is the same stability property possessed by  Jack polynomials, $
	P_{\ulmb}({\bf y},0)=P_{\ulmb}({\bf y})$, but  which is absent for the original  
 Jacobi polynomials themselves: 
$	J_{\ulmb}({\bf y},0)\neq J_{\ulmb}({\bf y})$.

This stability property  is key to  a direct supersymmetric uplift of
the dual Jacobi polynomials to $BC_{n|m}$ functions. As we will see, this uplift lands 
precisely on our superconformal blocks!

The  $BC_n$ Jacobi polynomial has as an explicit expansion 
in Jack polynomials (just like the blocks)\footnote{We have momentarily suppressed the parameters 
	in $J_{\ulmb}(;\theta,p^-,p^+)$, since they do not play an immediate role.
	Note that $P_{\umu}(y_1,\ldots,y_n)$ here is $P^{(n,0)}_{\umu}(y_1,\ldots,y_n|;\theta)$ 
	in supersymmetric notation, not to be confused with the $P^{(0,n)}_{\umu}(|y_1,\ldots,y_n;\theta)$ 
	polynomials. See appendix \ref{symm_supersym_poly_sec} and \ref{intro_sjack_sec} for further details.}  
\begin{align}\label{Paul_Jacobi_1}
J_{\ulmb}({\bf y};\theta,p^-,p^+) = \sum_{\umu \subseteq \ulmb}  
(S^{(n)}_{\theta,p^-,p^+})_{\ulmb}^{\umu} 
\,P_{\umu}({\bf y};\theta)\ .
\end{align}
The coefficients, $(S^{(n)})_{\ulmb}^{\umu}$ are not stable, meaning they have 
explicit $n$ dependence and don't just depend on the Young diagrams $\ulmb,\umu$. 
Indeed, this is what  prevents stability of the Jacobi polynomial $J$. 
 Let us note that the $(S^{(n)})_{\ulmb}^{\umu}$  can be computed quite explicitly, either through 
a recursion, investigated by Macdonald \cite{Macdonald_3},  or independently by using a 
binomial formula due to Okounkov  \cite{Okounkov_1,Okounkov_FactoShur,Koornwinder}. 
In the latter case, $S^{(n)}$,  is  written in terms of $BC_n$ interpolation polynomials (IPs). 
We will discuss  this in much more detail in section~\ref{sec_guideline}.

We can understand the stability of the dual Jacobi polynomials by considering their defining 
differential equation. The original eigenvalue equation for 
the Jacobi polynomials, translated  to the dual Jacobi polynomials,  can be written in terms of the Casimir~\eqref{rewriteCinJack_intro} as
\begin{gather}\label{diff_def_dualJ}
\ \ \ \ \mathbf{C}_{}^{(\frac{1}{\theta},a,b,c)}(|{\bf y})\ \tilde{ J}_{\beta,\ulmb}(y_1,\ldots y_n; \theta,p^-,p^+)= 
e_{\beta,\ulmb}^{(\theta)}\    \tilde{ J}_{\beta,\ulmb}(y_1,\ldots y_n; \theta,p^-,p^+) \\[.2cm]
a=\beta\qquad;\qquad b=\beta+p^-\qquad;\qquad c=2\beta+p^-+p^+\ . \notag
\end{gather}
where the eigenvalue is
\beq
e_{\beta,\ulmb}^{(\theta)}=-\tfrac{1}{\theta} h_{\ulmb}^{(\theta)}+\tfrac{1}{\theta} (2\beta+p^-+p^+) |\ulmb|
\eeq
Both  $h_{\ulmb}^{(\theta)}$ and $|\ulmb|$, given already in \eqref{intro_eigenv},
 are functions of the Young diagram $\ulmb$ only, (unlike the corresponding 
 eigenvalue for the Jacobi polynomial which has explicit $n$ dependence) and 
 stability~\eqref{dual_Jacobi_stable} follows.
In our conventions, the normalisation of $\tilde J$ will be chosen so that the leading
 term of $\tilde J_{\beta,\ulmb}$ is $P_{\ulmb}$.

Dual Jacobi polynomials  depend on $\theta,p^{\pm}$, as the Jacobi polynomials do, 
and in addition depend on $\beta$, which sets the boundary for the 
Young 
diagram in ~\eqref{lambdap_compl}.
From the definition of $\tilde J_{\beta,\ulmb}$, in terms of the Jacobi 
polynomials~\eqref{dual_jacobi}, 
the expansion of Jacobi polynomials~\eqref{Paul_Jacobi_1}, and the relation~\eqref{complement_on_jack},  
we  obtain the following expansion of the dual Jacobi polynomials in Jack polynomials with 
coefficients given by binomial coefficients, $S$, in complemented Young diagrams
\begin{align}\label{Paul_Jacobi_22}
	\tilde J_{\beta,\ulmb}( y_1,\ldots,y_n) = 
	\sum_{\umu:\, \ulmb \subseteq \umu}  \big(\widetilde 
	S_{\beta}\big)_{\ulmb}^{\umu}\,P_\umu(y_1,\ldots,y_n) \qquad ; \qquad 
	\big(\widetilde S_{\beta}\big)_{\ulmb}^{\umu} = 
	\big(S^{(n)}\big)_{\beta^n\bslash\,\ulmb}^{\beta^n\bslash\,\umu} \ .
\end{align}
Note that if the Young diagram  $\umu$ is wider than $\beta$ (ie $\mu_1 > \beta$) the 
coefficient $\tilde S$ will vanish and so the sum is finite, $\umu\subseteq 
\beta^n $ giving a polynomial.

Stability of dual Jacobi polynomials implies that the coefficients $(\widetilde S_{\beta})_{\ulmb}^{\umu}$  
are independent of $n$. Looking at the way this is related to $S^{(n)}$~\eqref{Paul_Jacobi_22}, we see that even though 
$S^{(n)}$ is $n$ dependent, remarkably it becomes independent of $n$ when specified by complemented  Young diagrams as is the case for $(\widetilde S_{\beta})_{\ulmb}^{\umu}$.

\subsubsection*{Dual super Jacobi functions}

The supersymmetric extension of the dual Jacobi polynomial which we call {\em dual super Jacobi functions}
is now immediate from stability:   replace the expansion over Jack polynomials with super Jack polynomials, 
keeping the same expansion coefficients $\widetilde{S}$.  This operation defines the $n|m$ dual super Jacobi function
\begin{align}\label{sJ}
	\tilde J_{\beta,\ulmb}( {\bf y}|{\bf x}) = \sum_{ \umu \supseteq \ulmb } 
	(\widetilde S_{\beta})_{\ulmb}^{\umu}\,P_\umu( {\bf y}|{\bf x}) \ .
\end{align}
When $m>0$, the sum is over all Young diagrams $\umu$ such that $\ulmb\subseteq 
\umu$. As before,  if $\umu$ is wider than $\beta^n$, the coefficient $\tilde 
S$ will vanish and so  $\mu_1\leq \beta$. But now $\umu$ can have arbitrary height, unlike 
in the non  supersymmetric case~\eqref{Paul_Jacobi_22} where it was naturally cut-off by $n$ 
since the Jack polynomials would vanish for $\mu'_1>n$. This means that the sum is \emph{infinite} in the vertical Young diagram direction,
and this is why we call $\tilde J_{\beta,\ulmb}$ a  function rather than a polynomial. 
As a result, the dual Jacobi functions $\tilde J_{\beta,\ulmb}$ are explicitly different from the super Jacobi polynomials introduced 
in~\cite{Veselov_3}. The latter also provide a supersymmetric generalisations of Jacobi 
polynomials but the generalisation is polynomial in both $x$ and $y$ variables and not stable. 
We will have something to say about  these other supersymmetric Jacobi polynomials in section~\ref{sec_guideline}.

\subsubsection*{Blocks as dual super Jacobi functions}

The claim is then that the dual super  Jacobi 
functions are precisely the superconformal blocks up to changes of conventions. 
Explicitly the relation is
\beq\label{Block_dualJacobi}
	F_{\gamma,\ulmb}({\bf x}|{\bf y};\theta,p_{12},p_{43})=
	(-1)^{|\ulmb'|} \Pi_{\ulmb'}(\tfrac{1}{\theta}) \ {
	\tilde J_{\beta,\ulmb'}(  {\bf y}|{\bf x};\tfrac{1}{\theta},p^{-},p^+) }{  }   
	\eeq
where\footnote{In other words, $\beta = \min\left( \tfrac 1 2 (\gamma {-}  p_{12}), \tfrac 1  2 (\gamma {-}  p_{43}) \right)$.}
\beq\label{bppm}
\gamma=2\beta+p^++p^-
\qquad;\qquad
p^{\pm}=\tfrac{1}{2}|p_{12}\pm p_{43}|\ 
\eeq
then $\Pi(\theta)$ is a numerical  factor, given explicitly later on in 
\eqref{def_Pi_new}, and $p_{43}$ and $p_{12}$ can be solved as linear combinations of $p^+$ and $p^-$. 
Specialising to the internal case $(0,n)$ we have a direct relation between internal blocks and Jacobi polynomials
\beq\label{BCJacobi_internalBlock}
\prod_{i=1}^n 
{y_i^{\frac{1}{2}(p^++p^-) } }
\!\times\!{\BB}_{\gamma=2\beta+p^++p^-,\,\ulmb}^{}({\bf
 y};\theta,p_{12},p_{43})\,=\,
(-)^{|\ulmb|} \Pi_{\ulmb'}(\tfrac{1}{\theta})\times 
{J}_{\beta^n-\ulmb'}^{}\,\left(\tfrac{1}{{\bf 
y}};\tfrac{1}{\theta},p^-,p^+\right)  
\eeq
where the original Jacobi polynomial $J$ appears on the RHS~with complemented, transposed Young diagram.

Equation \eqref{BCJacobi_internalBlock}, is an equation purely between symmetric polynomials, 
and provides  the cleanest way to prove the more general supersymmetric 
relation~\eqref{Block_dualJacobi}. Indeed it was the discovery of this relation which lead us to dual Jacobi polynomials and their supersymmetric generalisation. This polynomial  identification~\eqref{BCJacobi_internalBlock} 
can be proved by simply showing they obey the same Casimir equation. Then the general 
supersymmetric relation~\eqref{Block_dualJacobi} follows directly because both sides can be  
uplifted supersymmetrically in the same way, i.e.~by expanding in Jack polynomials, and uplifting 
the bosonic polynomials to super Jack polynomials thanks to stability. 

The relation between dual super Jacobi functions and superconformal blocks gives a way of seeing why  there must be an infinite expansion in the ${\bf x}$ variable, at least in cases 
where there is a  supergroup interpretation. This infinite sum is a consequence of the non 
compactness  of the conformal subgroup.

To summarise this subsection then,  we have  a link between internal blocks and Jacobi polynomials, 
both of which, through stability, naturally uplift to superblocks and dual 
super Jacobi functions respectively:
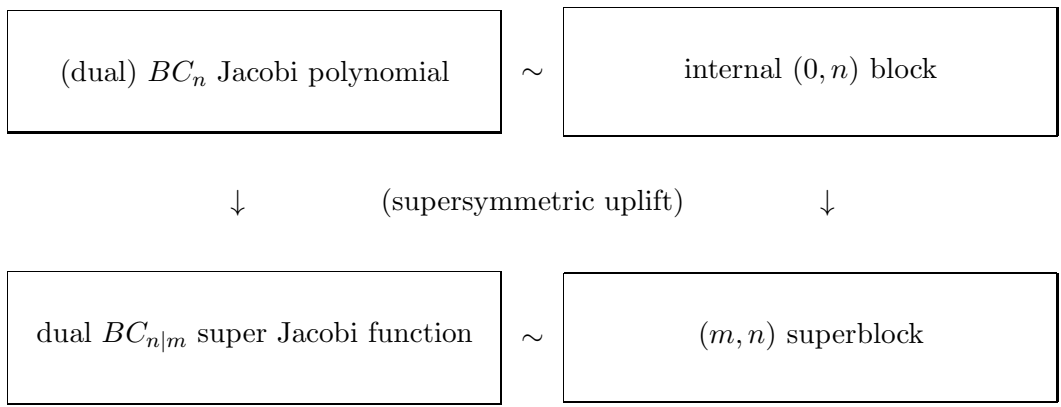
\begin{center}
	\begin{minipage}{6.5cm}
		\begin{mdframed}
			\begin{center}
				\rule{0pt}{.25cm}\\
			(dual) $BC_n$ Jacobi polynomial \\
				\rule{0pt}{.045cm}
			\end{center}
		\end{mdframed}
	\end{minipage}
\ $\sim$ \ 
	\begin{minipage}{6.5cm}
		\begin{mdframed}
			\begin{center}
				\rule{0pt}{.25cm}\\
				internal $(0,n)$ block\\
				\rule{0pt}{.45cm}
			\end{center}
		\end{mdframed}
	\end{minipage}

	\begin{minipage}{2.5cm}
		\begin{center}
			\rule{0pt}{.25cm}\\
		$\downarrow$ \\
			\rule{0pt}{.45cm}
		\end{center}
\end{minipage}
	\begin{minipage}{5cm}
	\begin{center}
		\rule{0pt}{.25cm}\\
		(supersymmetric uplift) \\
		\rule{0pt}{.45cm}
	\end{center}
\end{minipage}
	\begin{minipage}{2.5cm}
	\begin{center}
		\rule{0pt}{.25cm}\\
		$\downarrow$ \\
		\rule{0pt}{.45cm}
	\end{center}
\end{minipage}

	\begin{minipage}{6.5cm}
	\begin{mdframed}
		\begin{center}
			\rule{0pt}{.25cm}\\
			dual $BC_{n|m}$ super Jacobi function \\
			\rule{0pt}{.45cm}
		\end{center}
	\end{mdframed}
\end{minipage}
\ $\sim$ \ 
\begin{minipage}{6.5cm}
	\begin{mdframed}
		\begin{center}
			\rule{0pt}{.25cm}\\
			$(m,n)$ superblock\\
			\rule{0pt}{.45cm}
		\end{center}
	\end{mdframed}
\end{minipage}
\end{center}

\subsection{Binomial coefficient and Cauchy identities}\label{overv_bin_cauchy}

The previous subsection outlined the relation between Jacobi polynomials and blocks. In this section  we consider some consequences of this relation, summarising the results of section~\ref{sec_guideline}.

After computing the \JtoB matrix  $(T_\gamma)_{\ulmb}^{\umu}$ in many specific cases by solving the  recursion~\eqref{rec} we find 
that,
  for generic $\theta$, it is a rational functional with an increasingly 
complicated  numerator as a function of $\gamma$. To understand this non-trial dependence on  
$\gamma$ we  use  the relation to Jacobi polynomials of the $m=0$ theory described in the previous section.

The relation between blocks and dual Jacobis~\eqref{Block_dualJacobi} leads to a relation between the coefficients, $\tilde S$ and  $T_\gamma$  in their respective Jack expansions~\eqref{Jexp_intro},\eqref{Paul_Jacobi_22}, explicitly:%
\footnote{To derive this relation one also needs to use the fact that super Jack polynomials are well behaved under transposition
\begin{align}\label{sJduality1}
	P_{\ulmb'}({\bf y}|{\bf x};\tfrac{1}{\theta}) = (-1)^{|\ulmb|} \Pi_{\ulmb}(\theta) P_{\ulmb}({\bf x}|{\bf y};\theta)\ 
\end{align}  
For more details see the discussion around~\eqref{sJduality}.}
\begin{align}\label{TtoTtilde}
	(T_{\gamma;\theta,p_{12},p_{43}})_{\ulmb}^\umu = 	
	(\widetilde S_{\beta;\frac{1}{\theta},p^-,p^+})_{\ulmb'}^{\umu'} \times 
	\frac{(-1)^{|\umu|} \Pi_{\umu}(\theta)}{(-1)^{|\ulmb|} 
	\Pi_{\ulmb}(\theta)}\ .
\end{align}
Hence from~\eqref{Paul_Jacobi_22} the  coefficients   $T_\gamma$ are related to  the binomial coefficients $S$~\eqref{Paul_Jacobi_1}
\begin{align}
 (T_{\gamma;\theta,p_{12},p_{43}})_{\ulmb}^{\umu} \ &=\ 
  \frac{(-)^{|\umu|} \Pi_{\umu}(\theta) }{(-)^{|\ulmb|} \Pi_{\ulmb}(\theta)}\times
(S^{(n)}_{\frac{1}{\theta};\,p^{-},p^+})_{\beta^n\bslash\,\ulmb'}^{ 
\beta^n\bslash\,\umu'}
  \ ,
 \label{binomial_f_intro}
\end{align}
where $\beta, p^\pm$ are read off  $\gamma,p_{12},p_{43}$ through~\eqref{bppm}.

Now the $BC_n$ Jacobi polynomials are very well studied objects. In particular, 
the coefficients in the expansion over Jack polynomials can be computed by Okounkov \emph{binomial formula}.
Okounkov's binomial formula for $S^{(n)}$ (see \cite{Okounkov_1,Okounkov_FactoShur,Rains_1,Koornwinder}) is 
computed via objects called $BC$ interpolation  polynomials (IPs) 
evaluated on  partitions. 
The combinatorics  
is thus completely different from the combinatorics induced by the recursion and so the above rewriting of $T_{\gamma}$~\eqref{binomial_f_intro} is quite non trivial.

In fact, a very non trivial feature of this relation~\eqref{binomial_f_intro}  
is the way the RHS~depends on $\gamma$. In the binomial coefficient we have  $\gamma=2\beta+p^++p^-$
where $\beta$ is an integer specifying the complemented Young diagram. 
On the other hand  in the recursion, $\gamma$ is a free parameter, and the 
solution of the recursion is a rational function of $\gamma$ and thus straightforwardly 
analytically continued. This should also be the case for the binomial coefficient then. 
In the process of understanding how this works we find that the complicated $\gamma$ dependence of 
$(T_{\gamma})_{\ulmb}^{\umu}$ is precisely captured by the  interpolation polynomial.
We also find that the inverse of the \JtoB matrix (i.e.~the \BtoJ matrix) $T^{-1}_{\gamma}$, has an even more direct 
characterisation in terms of the binomial coefficient.

Note that  since 
$(T_\gamma)_{\ulmb}^{\umu}$ does not depend on $(m,n)$, for integer quantum numbers, the above characterisation via the binomial coefficient, through the $(0,n)$ case,  
can be used to compute any superconformal block for arbitrary $(m,n)$. 
However, from this insight we understand that the binomial coefficient can also be  upgraded to a super binomial coefficient which uses the super interpolation polynomials introduced by 
Sergeev and Veselov~\cite{Veselov_3}.
We will investigate this whole story in detail in  section~\ref{sec_guideline}, and then in appendix \ref{know_solu}  we collect a number of explicit  solutions for $T_\gamma$ found by solving the recursion,  and  we show explicitly 
	how the two sides of \eqref{binomial_f_intro} match.

We will conclude the main body of the paper by giving two nice applications of the relation between blocks and Jacobi 
polynomials, together with stability. Firstly, in section~\ref{sc_Cauchy} we see how upon re-interpreting a Cauchy identity for Jacobi  polynomials, 
given by Mimachi in~\cite{Mimachi}, we can obtain (essentially with no effort) a formula for 
the conformal partial wave expansion of any generalised free theory within the class analysed 
here, i.e.~whose supergroup description fits $(m,n,\theta)$ for specific values. In the process
of investigating this  we discover a new non-trivial double  uplift of Mimachi's 
Cauchy identity to a doubly supersymmetric Cauchy identity involving both dual super Jacobi functions and Sergeev and Veselov's super Jacobi polynomials.
Secondly in section~\ref{sec_hoc} we use results from the study of $BC$ hypergeometric functions in the special $\theta=1$ case of Shimeno~\cite{Shimeno} to give simple explicit formulae for all higher order super Casimirs.

\subsection{Superblock to block decomposition: a conjecture}\label{overv_conject}

Our formalism for constructing superconformal blocks might be classified as \emph{top-down}, 
because we start from a supergroup perspective in which superconformal symmetry is built-in, 
and we derive its consequences for the superconformal blocks. Instead, a \emph{bottom-up} approach 
constructs superconformal blocks starting from an ansatz made of a
sum of products of conformal and internal blocks, and afterwards imposes 
the constraints of superconformal symmetry, i.e.~the superconformal Ward identity 
and the Casimir equation \cite{Dolan:2002zh,Dolan:2004mu,Beem:2014zpa,Chester:2014fya,Beem:2016wfs,Lemos:2016xke,Liendo:2018ukf,Alday:2020tgi}.

The top-down and the bottom-up approaches should obviously give the same final result. However, 
the two paths are quite different, and   the crucial point is that decomposing a superconformal blocks, 
say for example on a basis of conformal and internal blocks, is quite hard. In fact, we will now show that
 all the nice properties about stability and the \JtoB matrix, explained in the previous sections, become hidden in the details.

From the top-down approach we are able to provide an implicit formula for decomposing a $(m,n)$ superconformal 
in subgroups. Mathematically, this formula is the equivalent of decomposing a super Jack polynomial $(m+m',n+n')$ 
into sum of products of two super Jack polynomials for $(m,n)$ and $(m',n')$.  This decomposition is achieved by using
the structure constants for super Jack polynomials $({\cS}_{})_\ulmb^{\umu\unu}$, 
\beq\label{above_LR}
J^{(m+m',n+n')}_{\ulmb}=\sum_{\umu,\unu}({\cS}_{})_\ulmb^{\umu\unu}J_{\umu}^{(m,n)} J_{\unu}^{(m',n')}
\eeq
which by stability are the same as those for bosonic Jack polynomials.   For $\theta=1$ these are the Littlewood-Richardson 
coefficients. Combining \eqref{above_LR} with the expansion of superconformal blocks in super Jack polynomials we  arrive at  
\begin{align}\label{superblockdecomp_intro}
		F^{(m+m',n+n')}_{\gamma,\ulmb}= \sum_{\umu,\unu} F_{\gamma,\umu}^{(m,n)}\, ({\cS}_{\gamma})_\ulmb^{\umu\unu}\, F_{\gamma,\unu}^{(m',n')}\ ,
\end{align}
where the block structure constants ${\cS}_{\gamma}$ depend on $\theta,p_{12},p_{43}$ and are related 
to the Jack structure constants via the matrices $T_\gamma$ and $T^{-1}_{\gamma}$,
	\begin{align}\label{blockstructure_intro}
		 ({\cS}_{\gamma})_{\ulmb}^{\umu\unu} = \sum_{\substack{\tilde{\ulmb}\supseteq{\ulmb}\\\tilde{\umu}\subseteq {\umu}\\\tilde{\unu}\subseteq{\unu}}}{\cS}_{\tilde{\ulmb}}^{\tilde{\umu}\tilde{\unu}}\, (T_{\gamma})_{{\ulmb}}^{\tilde{\ulmb}}   (T^{-1}_{\gamma})_{\tilde{\umu}}^{{\umu}}  (T^{-1}_{\gamma})_{\tilde{\unu}}^{{\unu}}\  . 
	\end{align}

Let us point out that \cite{Lassalle_2,Lassalle_3,Schlosser} obtained a recursive formula for 
the Jack structure constants  ${\cal S}$,   but we are not aware of a more explicit formula, and 
to the best of our knowledge there is no closed formula expression for it,  and therefore for ${\cal S}_{\gamma}$. 
The bottom-up construction of superblocks in terms of conformal and internal blocks  mentioned above is a special case of the more general decomposition \eqref{superblockdecomp_intro} when $m'=n=0$, 
and the ignorance about ${\cal S}$ is what makes it complicated.

Not only is  ${\cal S}$ not known explicitly, but a very non-obvious fact about  \eqref{blockstructure_intro} 
is the following: the \JtoB matrix  and its inverse are infinite size triangular matrices, however
$({\cS}_{\gamma})_{\ulmb}^{\umu\unu}$ should truncate, because,  for example, superconformal group 
theory for specific values of $\theta=\frac{1}{2},1,2$ tells us that \eqref{superblockdecomp_intro} is a finite sum. 
Consider again the case of conformal block $\times$ internal block decomposition, $m'=n=0$. Whilst a truncation in the number of rows of $\unu$ is expected, since 
$F_{\gamma,\ulmb}$ is polynomial in the $\bf{y}$, the truncation over finitely many conformal block 
$F_{\gamma,\umu}$ is not at all manifest.

We conjecture that the subgroup decomposition in \eqref{superblockdecomp_intro} is a finite sum. 
In section \ref{section_mn_deco} we discuss in more detail various aspects of this conjecture, relating it to a more specific conjecture for a general decomposition 
and providing a number of examples.  We have checked the conjecture with computer algebra, in many cases, 
and find that it holds for generic $\theta$, even beyond the cases $\theta=\frac{1}{2},1,2$ related to superconformal groups.  
Perhaps the connection with the binomial coefficient will provide a simple mechanism to prove it in the future.

That concludes our extended outline of  the paper. 
The methods employed throughout the paper strongly rely on the use of Young diagrams and
symmetric polynomials. To help the reader familiarise themself with 
the relevant mathematics, we have summarised in appendix \ref{symm_supersym_poly_sec}  
a number of results on the theory of symmetric polynomials.

\section{Supergroups, physical theories, and beyond}\label{how_to_read}

In this section we detail the theories and the external operators for which we are computing 
superconformal blocks, as a function of $(m,n;\theta)$. We also explain the field theory 
interpretation of the labels assigned to the superconformal  block $B_{\gamma,\ulmb}$,  
namely $\gamma$ and $\ulmb$.

A unified description of all relevant theories can be achieved by using the formalism based on harmonic/analytic superspace, 
and building on previous literature. In particular, 4d superconformal theories with $\theta=1$ discussed 
in~\cite{Galperin:1984av,Hartwell:1994rp,Howe:1995md,hep-th/0105218,Heslop:2001zm,Heslop:2002hp,hep-th/0307210,Heslop:2004du,Doobary:2015gia,Howe:2015bdd}, 
6d superconformal field theories with $\theta=2$ discussed in~\cite{hep-th/9812133,hep-th/0008048,Heslop:2004du}, and
3d superconformal field theories with $\theta=\frac{1}{2}$ discussed in some detail in~\cite{hep-th/9408062,hep-th/9812133}.
This is explained in section \ref{supergroup_sec}, with supplementary material given in appendix \ref{app_supecoset}.

Let us emphasise that we will be able to construct $B_{\gamma,\ulmb}$ for any $(m,n;\theta)$ (for
$m,n$ positive integer), but only {\em some} values of $(m,n;\theta)$ appear to have a group theoretic 
meaning (and even fewer will correspond to a physical CFT). The spacetime dimension $d$ and the parameter $\theta$ are identified as $\theta=\frac{d-2}{2}$,
 apart for special cases, that we will explicitly mention. A useful summary of the different cases is given here below.

\subsection{List of theories and their superconformal blocks}\label{summary_list_theory}

The main examples to have in mind are three complete families of supergroups with  
$\theta=1,2,\frac{1}{2}$ and arbitrary $m,n$. These include the cases of four-point functions of half-BPS 
operators in  4d,6d,3d superconformal field theories respectively:\\[.2cm]

\begin{minipage}{15cm}

$\bullet$ $SU(m,m|2n)$: $(m,n)$ arbitrary and  $\theta=1$. 
For $(m,n)=(2,1)$ and $(2,2)$ we compute $\cN= 2,4$ superblocks, respectively, in four-dimensions. 
For $(m,n)=(1,1)$  we compute $\cN=4$ superblocks in one-dimension.\\ 

$\bullet$ $OSp(4m^*|2n)$: $(m,n)$ arbitrary and $\theta=2$. For $m=2$ we compute ${\cal N}=(n,0)$ superblocks in six-dimensions.\\

$\bullet$ $OSp(4n|2m)$: $(m,n)$ arbitrary and $\theta=\frac{1}{2}$. For $m=2$ we compute $\cN= 4n$ superblocks in three-dimensions.\\
\end{minipage}\\[.2cm]
Alternatively, fixing $(m,n)=(2,0)$ or $(0,2)$ with arbitrary (half-integer) $\theta$ gives blocks in purely bosonic theories\\

\begin{minipage}{15cm}

	$\bullet$  $SO(2,2\theta+2)$: $(m,n)=(2,0)$ with arbitrary $\theta$: 
	We compute standard conformal blocks for scalar correlators in dimensions $d=2\theta+2$. \\

	$\bullet$ $SO(2/\theta+4)$: $(m,n)=(0,2)$  with arbitrary $\theta$: 
	We compute ``compact blocks'', i.e. blocks where the external operators are finite dimensional representations. These are the
	dual Jacobi polynomials.\\  
\end{minipage}\\[.2cm]
Note the apparent duality 
\begin{align}
	m \leftrightarrow n \qquad ;\qquad \theta\leftrightarrow \frac{1}{\theta}.
\end{align}
It means that whenever there is a group theory interpretation, there should be a corresponding interpretation for the dual case, 
although this exchanges non-compact groups with compact groups.

There are other theories, in addition to the ones listed above, which we would like to point out. 
We have not spelled out the details and therefore the discussion here will be in a sense speculative. 
Nevertheless we will provide evidence that these other cases appear as special cases in our formalism:\footnote{We 
thank Tadashi Okazaki for pointing out $D(2,1;\alpha)$ in the list.}\\

\begin{itemize}
\item  {$m=1,n=1$, and arbitrary $\theta$.}
There is a one-parameter family of one-dimensional $\cN=4$ superconformal groups called $D(2,1;\alpha)$~\cite{1503.03906}.
For special values of $\alpha$, $D(2,1;\alpha)$ is isomorphic (at the level of the algebra at least) to the following   
	\begin{itemize}
		\item 	$OSp(4^*|2)$, for  $\alpha=-2,1$  
		\item $SU(1,1|2)\times SU(2)$, for $\alpha=-1,0$, 
		\item $OSp(4|2)$ for $\alpha=-\frac{1}{2}$  
	\end{itemize}
Noting the relation with the supergroups discussed above for general $m,n$, we relate 
 $\theta=-\alpha$ (or  $\theta = \alpha+1$)  and speculate that for $m=1,n=1$, arbitrary 
 $\theta$ we obtain superblocks for $D(2,1;\alpha)$ superconformal theories.\\

\item   {$m=1,n=2$, for various $\theta$.} There are four different 
possibilities for a one-dimensional $\cN=8$ CFT~\cite{1503.03906}: 
\beq\label{F4primo}
OSp(8|2),\ SU(1,1|4),\ Osp(4^*|4)\qquad {\rm and}\qquad F(4)\,.
\eeq 
The first three correspond to supergroup theories discussed above for $\theta=\frac{1}{2},1,2$ and $m=1,n=2$. 
To understand $F(4)$ note first that $F(4)$ has R-symmetry $SO(7)$.  The blocks 
for this group belong to the family of $SO(2/\theta+4)$ bosonic blocks mentioned above, 
and would correspond to $m=0,n=2,\theta=\frac{2}{3}$. Thus we conjecture 
that the $F(4)$ supergroup corresponds to $m=1,n=2,\theta=\frac{2}{3}$.\\
	 
\item {Five-dimensional blocks.} 
	There is also an $F(4)$ superconformal group in five dimensions, and 
	since the 5d bosonic blocks correspond to $m=2,n=0$ and $\theta=\frac{3}{2}$, 
	we conjecture that the 5d superblocks correspond to $m=2,n=1$ and $\theta=\frac{3}{2}$.
	Note that, nicely,  upon considering the duality $m \leftrightarrow n$ and 
	$\theta\leftrightarrow \frac{1}{\theta}$, we obtain the other $F(4)$ group in 
	\eqref{F4primo}.\\

\item  {Two-dimensional blocks.} In this case holomorphic/anti-holomorphic 
	factorisation suggests  that superconformal blocks are obtained from a basis constructed 
	by first taking products of one-dimensional blocks, i.e.~the ones that appeared at the previous points, 
 	and then by rearranging the product basis according to the relevant representations. 
	An example of this has been worked out in \cite{Aprile:2021mvq}. 
\end{itemize}

\subsection{Coset space formalism}\label{supergroup_sec}

In the previous section we introduced the groups corresponding to superconformal blocks with various  
values of $(m,n;\theta)$.  In order to make our discussion self-contained we will now specify the space on 
which the superconformal reps lie, both the external and the exchanged states in the four-point function.

The external states we consider are scalar (super)fields (line bundles) on a certain coset space of the group.
For example, it is well known that complexified Minkowski space in $d=2\theta+2$ dimensions, $M_d$, can 
be viewed as a coset of the complexified conformal group $SO(d+2;\mathbb{C})$ divided by the subgroup 
consisting of Lorentz transformations, dilatations and special conformal transformations. This corresponds to 
the case $(m,n)=(2,0)$ with arbitrary half integer $\theta$. More generally, the coset space for arbitrary $(m,n;\theta)$ 
will always be a special type of (super) flag manifold, which can be specified by a (super) Dynkin diagram 
with a single marked node, also known as generalised (super) Grassmannian spaces.

A beautiful classification and description of flag manifolds, together with representations on them, 
can be found in~\cite{bastoneastwood}.  A flag manifold and its field content, i.e.~the irreducible representations of the group, 
can be read off from a marked Dynkin diagram and its associated  Dynkin labels~\cite{bastoneastwood}.
From a root system and a Dynkin diagram one can construct the corresponding  Lie algebra. 
Similarly, from a root system and a {\em marked} Dynkin diagram one constructs a parabolic subalgebra 
by simply omitting the positive  simple roots corresponding to the marked nodes and retaining all the 
negative roots. The  coset space is then the group $G$ divided by this parabolic subgroup $H$. 
Irreducible representations are defined by Dynkin labels giving the transformation properties of the 
highest weight state. 

 The generalisation of Dynkin diagrams to the supersymmetric case is well known (we recommend~\cite{Cornwell} for a nice introduction) and 
the generalisation of the flag manifold techniques of~\cite{bastoneastwood}  
to supergroups proceeds fairly straightforwardly and has been considered in the  $\theta =1,2$ cases explicitly in~\cite{Heslop:2001zm,Heslop:2002hp,hep-th/0307210,Heslop:2004du}.
We should point out that, compared to the bosonic cases,
when dealing with supergroups there is no longer a unique (super) Dynkin diagram.\footnote{An easy way 
		to understand the existence of inequivalent Dynkin diagrams  for the case of $SL(M|N)$ 
		is to notice that a change of basis (in particular swapping even and odd basis elements) 
		can not always be achieved via an $SL(M|N)$ transformation  (unlike in the bosonic case). 
		For each inequivalent change of basis there is a different Dynkin diagram. 	
		In the case of interest here, for $\theta=1$ the standard basis for the complexified superconformal group 
		$SL(2m|2n)$ would give a Dynkin diagram with  $2m-1$ even nodes then one odd node then $2n-1$ even nodes. 
		If instead we choose the basis $(m|2n|m)$ we end up with the diagram~(\ref{dynkin}a) involving two odd nodes.}
A super Dynkin diagram will have even (black) and odd (white) nodes, and then the coset space of this superspace will be indicated by marked nodes (represented by crosses).   Remarkably all unitary superconformal representations 
of the superconformal group are obtained as {\em unconstrained} analytic superfields on the the coset superspaces 
we use here. This is proven for 4d superconformal  theories $\theta=1,m=2$ in~\cite{Heslop:2002hp} 
and is conjectured in other supersymmetric cases.\footnote{These results suggest the existence of a supersymmetric generalisation 
of the Bott-Borel-Weil theorem, which would be interesting to explore further.}

Various details about the coset construction for the specific cases listed in section~\ref{summary_list_theory} 
are given in appendix \ref{app_supecoset}. We repeat the main points here.

In the following will consider the {\em complexified}, e.g. $SU(m,m|2n)\ \xrightarrow{ \mathbb{C}} SL(2m|2n;\mathbb{C})$, of the superconformal groups. One  can return to the real case by choosing the appropriate real coordinates at the end.
The marked super Dynkin diagrams specifying the relevant coset space 
for the three families $\theta=1,2,\frac{1}{2}$ are\footnote{Other cases  will be discussed in appendix \ref{app_supecoset}.}
\beq
\begin{array}{c}\label{dynkin}
\begin{tikzpicture}
\draw (0,0)       node 		         {$\begin{dynkinDiagram}{A}{*.*o*.*x*.*o*.*} \dynkinBrace[m-1]{1}{2}\dynkinBrace[n-1]{4}{5}\dynkinBrace[n-1]{7}{8}\dynkinBrace[m-1]{10}{11}  \end{dynkinDiagram}  $}; 
\draw (7,0.2)  node		           {$SL(2m|2n)$};

\draw (-5,-1) -- (9,-1) ;

\draw (0,-2)       node 		 {$\begin{dynkinDiagram}{B}{*.*o*.*x} \dynkinBrace[2m-1]{1}{2} \dynkinBrace[n-1]{4}{5}   \end{dynkinDiagram} $};
							
\draw (7,-2+0.2)  node		 {$OSp(4m|2n)$};	

\draw (-5,-2.75) -- (9,-2.75) ;				

\draw (-.15,-4)  node                    {$\,\begin{dynkinDiagram}{D}{*.*o*.**x}   \dynkinBrace[m-1]{1}{2} \dynkinBrace[2n-2]{4}{5}	 \end{dynkinDiagram}$};
\draw (7,-4.2+0.2)  node		 {$OSp(4n|2m)$};								

\draw  (5.2,1) --( 5.2,-5);
\end{tikzpicture}
\end{array}
\eeq
We consider a matrix representation for the (complexified) (super)group such that all positive roots correspond to upper triangular matrices.
Crucially, in all cases the basis can be chosen such that (super)coset space $H  \backslash G$ has the following block $2\times 2$ structure 
\beq \label{hg}
\ \ G\ \, =\  \, \left\{\left(\begin{array}{cc}\hat a^A_{\ B}&\,  \hat b^{AB'}\\\hat c_{A'B}&\ \hat d_{A'}^{\ B'}
\end{array}\right)\right\}
\qquad \qquad
\ \ H\ \, =\  \, \left\{\left(\begin{array}{cc}a^A_{\ B}&\, 0\\c_{A'B}&\ d_{A'}^{\ B'}
\end{array}\right)\right\}\ .
\eeq
where $\hat a,\hat b,\hat c,\hat d,a,c,d$ are square (super-)matrices of equal dimensions (in general they will have some constraints).
For $\theta=1$ they are $(m|n)\times (m|n)$ matrices; 
for $\theta=2$ they are $(2m|n)\times (2m|n)$ matrices; 
for $\theta=\frac{1}{2}$ they are $(m|2n)\times (m|2n)$ matrices.\footnote{More 
		precisely,  the second half of the basis is reversed compared to the first half. 
		Taking $\theta=1$ as the illustrative example, then $a$ is $(m|n)\times (m|n)$, 
		$X$ is $(m|n)\times (n|m)$, $c$ is $(n|m)\times (m|n)$ and $d$ is $(n|m)\times (n|m)$. However when referring 
		to $a,c,d,X$ we will consider them with the bases rearranged, so they all have the same dimensions $(m|n)\times (m|n)$. \label{fbasis}}
For example,  the special  case $(m,n;\theta)=(2,0;1)$ corresponds to 4d Minkowski space viewed as the coset space 
of the complexified conformal group $SL(4)$  modded out by dilatations, Lorentz and special conformal transformations. 
Here the $2\times2$ matrices $a,b$ give Lorentz and dilatations whereas $c$ represents special conformal transformations.

Fields living on the coset space transform non-trivially under the block diagonal part of the parabolic subgroup 
$H$ (known as the  Levi subgroup, $L$). This subgroup consists of  the matrices $a,d$ (e.g. in the Minkowski space  example
these correspond to Lorentz and dilatations under which conformal operators transform). The Levi subgroup is read off from the same Dynkin diagrams \eqref{dynkin} 
upon deleting the  marked node (and replacing it with $\mathbb{C}^*$). In the above cases they are as follows,
\beq\label{levi}
\begin{array}{|c|c|c|c|}
	\hline & & &\\
	\theta&1&2&\frac{1}{2}\\ & & &\\
	\hline & & & \\
G&SL(2m|2n) 						      &  Osp(4m|2n) 						&  Osp(4n|2m) \\ 
& & &\\
\hline & & & \\
L& SL(m|n)\otimes SL(m|n) \otimes \mathbb{C}^* & \quad SL(2m|n)\otimes  \mathbb{C}^*       \quad          & \qquad SL(m|2n)\otimes  \mathbb{C}^* \quad\\
 & & &\\
 \hline
\end{array}
\eeq
There is an overall constraint to take into account: for $h\in H$, $\sdet(h)=\sdet(a)\sdet(d)=1$. The $\mathbb{C}^*$ 
subgroup of the parabolic group is then identified  with $\sdet(a) = 1/(\sdet(d))$. 

The coset space $H \backslash G$ is finally the collection of all the orbits under the 
equivalence $g \sim hg$ for $g\in G$ and $h\in H$. A representative for each orbit is\footnote{A 
		toy model for the coset construction, corresponding to $(m,n;\theta)=(0,1;1)$ is just the Riemann sphere, realised by taking 
		$g=\big(\,^{\hat{a}\ \hat{b}\,}_{\hat{c}\ \hat{d}\,}\big)$ in $SL(2,\mathbb{C})$ and $h=\big(\,^{a\,\, 0\,}_{c\ d\,}\big)$. 
		Then $\big(\,^{1\,\, x\,}_{0\ 1\,}\big)$ is the coset representative and 
		$\big(\,^{1\,\, x\,}_{0\ 1\,}\big) g=h \big(\,^{1\, f(x)}_{0\ 1\,}\big) \sim \big(\,^{1\, f(x)}_{0\ 1\,}\big)$ 
		where $f(x)=({ \hat{d}x + \hat{b}})/({\hat{c}x + \hat{a}})$ and $a=\hat{a}+x \hat{c}, c=\hat{c},d=\hat d-\hat{c} f(x)$.
		From $f$ we recognise M\"obius transformations acting on the Riemann sphere 
		as the coset  $ H \backslash SL(2,\mathbb{C})$. By taking just the top row of the coset representative, 
		$(1,x)$, we recognise this construction to be completely equivalent to the projective space $P^1$.  
		Similarly,  the general $(m,n;\theta)$ construction is equivalent to a Grassmannian space.}
\begin{align}\label{cosetgen}
g \sim s(X)\ =\ \left(\begin{array}{cc}\, 1&X^{AA'}\\ \, 0&\ 1\ \ \end{array}\right).
\end{align}
This supermatrix $X$ are then  coordinates for the coset space, which we write as
\beq
\ X=
\left( \begin{array}{c|c} x & \rho  \\\hline   \overline{\rho} & y   \end{array}\right)\,.
\eeq
For  $\theta=1$, $x$ is $m\times m$, $y$ is $n\times n$ and both are bosonic, whereas $\rho$ is $m\times n$, 
$\bar \rho$ is $n\times m$ and they are fermionic. The other cases are similar except $m\rightarrow 2m$ when 
$\theta=2$ and $n\rightarrow 2n$ when $\theta=\frac{1}{2}$ (and the supermatrix is generalised (anti-)symmetric 
in its indices, see appendix~\ref{app_supecoset} for more details). For $\theta=1$ the cosets are equivalent to 
super Grassmannians which can be seen by considering the upper half of the group matrix
$(1,X)\sim a(1,X)$, which  is indeed $Gr(m|n,2m|2n)$, the space of $m|n$ planes in $2m|2n$ dimensions. 
Then the $\theta=2,\frac{1}{2}$ cases are generalised Grassmannians. For $\theta=1,2,\frac{1}{2}$ note that 
$x^{\alpha\dot{\alpha}}$ represents coordinates for Minkowski space in dimensions $M_{d=4,6,3}$ respectively 
written in a (Weyl) spinor notation.

Now we can connect with the superblocks. The  four-point superconformal invariant combinations of four coordinates $X_1,X_2,X_3,X_2$ can be viewed as  the independent eigenvalues of the cross-ratio matrix:
\begin{align}\label{Zmat}
	Z=X_{12}X_{23}^{-1}X_{34}X_{41}^{-1}\qquad ;\qquad X_{ij}=X_i-X_j\ .
\end{align}
These independent eigenvalues then correspond to the arguments of the blocks, $\bf z$~\eqref{evalues}.
Further the building block two-point function $g_{ij}$ in~\eqref{2pt} is 
\begin{align}
	g_{ij} = \sdet^{-\#}(X_i{-}X_j).
\end{align}
where $sdet$ is the superdeterminant (Berezinian) and the exponent $\#$ is discussed in footnote~\ref{f2}.

The external representations appearing in the four-point function~\eqref{4pt} are scalars, $\cO_p(X)$, on this coset space, 
thus they transform non-trivially, with a certain weight $p$ under  $\mathbb{C}^*$ and are invariant under the 
rest of the parabolic group~\eqref{levi}. In terms of the Dynkin labels, the label above a marked node gives 
the weight under the $\mathbb{C}^*$ subgroup of the parabolic subgroup  associated with that node. 
The Dynkin labels next to the unmarked nodes give the representation under non-trivial subgroups of the parabolic 
subgroup.\footnote{Although we will be dealing with infinite dimensional representations, the representation of the 
			      parabolic subgroup itself will always be finite dimensional.}  
The external operators ${\cal O}_p$ are therefore written as, 
\beq\label{dynkin_summaryhbps}
\begin{array}{c}
\begin{tikzpicture}
\draw (0,0)       node 			{$\begin{dynkinDiagram}[labels={,,,,,p,,,,,}]{A}{*.*o*.*x*.*o*.*}	\end{dynkinDiagram}$}; 
\draw (7,0.2)  node		 	{$SL(2m|2n)$};

\draw (-5,-1) -- (9,-1) ;

\draw (0,-2)       node 		{$ \begin{dynkinDiagram}[labels={,,,,,,,,,,p}]{B}{*****.**o*.*x} \end{dynkinDiagram} $};
							
\draw (7,-2+0.2)  node		 {$OSp(4m|2n)$};	

\draw (-5,-2.75) -- (9,-2.75) ;				

\draw (-.15,-4)  node 			{$\begin{dynkinDiagram}[labels={,,,,,,,,,,p}]{D}{*.*o**.*****x} \end{dynkinDiagram}$};
\draw (7,-4.2+0.2)  node		 {$OSp(4n|2m)$};								

\draw  (5.2,1) --( 5.2,-5);

\end{tikzpicture}
\end{array}
\eeq
where all unlabelled nodes are understood to have the label $0$.
These fields transform under a general superconformal transformation $g$ as follows:
\begin{align}
	s(X)\,g= h\, s(X') \qquad \qquad \cO_p(X)\rightarrow \cO'_p(X')=\sdet(a)^p \cO_p(X)\ .
\end{align}

The superblock gives the contribution of an exchanged  representation  appearing 
in the common OPE (or tensor product)  between the first two, $\cO_1\cO_2$, and last two, $\cO_3\cO_4$, 
external operators (see~\eqref{ope}). The representations which can appear in the OPE are summarised 
by the labelled Dynkin diagrams given below, for $n>0$,
\beq\label{dynkin_summary}
\begin{array}{c}
	\begin{tikzpicture}
		\draw (0,0)       node 			{$\begin{dynkinDiagram}[labels={l_1,l_{m-1},\delta,a_1,a_{n-1}\ ,b,\ a_{n-1},a_1,\delta,\ l_{m-1},l_1}]{A}{*.*o*.*x*.*o*.*}	\end{dynkinDiagram}$}; 
		\draw (7,0.2)  node		 	{$SL(2m|2n)$};
		
		\draw (-6.5,-1) -- (8.5,-1) ;
		
		\draw (0,-2)       node 		{$ \begin{dynkinDiagram}[labels={,l_1,,l_2,,l_{m-1},,\delta,2a_1,2a_{n-1}\ ,b}]{C}{*****.**o*.*x} \end{dynkinDiagram} $};
		
		\draw (7,-2+0.2)  node		 {$OSp(4m|2n)$};	
		
		\draw (-6.5,-2.75) -- (8.5,-2.75) ;				
		
		\draw (-.15,-4)  node 			{$\begin{dynkinDiagram}[labels={2l_1,2l_{m-1},2\delta,,a_1,,a_{n-2},,a_{n-1},,b}]{D}{*.*o**.*****x} \end{dynkinDiagram}$}; %
		\draw (7,-4.2+0.2)  node		 {$OSp(4n|2m)$};								
		
		\draw  (5.2,1) --( 5.2,-5);
		\draw  (-5,1) --( -5,-5);
			\draw (-6,0.2)  node		 	{$\theta=1$};
			\draw (-6,-2+0.2)  node		 {$\theta=2$};
			\draw (-6,-4.2+0.2)  node		 {$\theta=\frac{1}{2}$};
	\end{tikzpicture}
\end{array}
\eeq
The Dynkin labels $l_{i=1,\ldots, m-1}$ and $a_{j=1,\ldots n-1}$, are all positive integers, and  denote 
generalised spins of the conformal and compact subgroups respectively, of the superconformal groups. 
Note that for $OSp(4m|2n)$, the Dynkin labels for the $SO(2m)$ conformal subgroup are alternately vanishing. 
Similarly, for $OSp(4n|2m)$ the internal spins are alternately vanishing.  All of the above representations are specified 
by the same set of labels $l_i,\delta,a_j,b$ even though they are describing representations of different groups. 

The special cases of $m=1,0$ and $n=1,0$ require a small discussion.
The Dynkin diagrams above indeed include the special cases:  $m=1$, which corresponds to having external 
nodes with $l_i$ absent;  $m=0$, which corresponds to having external and odd nodes absent; $n=1$, which 
corresponds to having internal uncrossed nodes labelled by $a_j$ absent. However, the case $n=0$, which 
corresponds to a bosonic scalar conformal field theory, can not be read off directly from the above diagrams, 
even though it follows directly from the same matrix construction. The relevant Dynkin diagrams for $n=0$ is instead
\beq\label{dynkin_summaryn0}
\begin{array}{c}
	\begin{tikzpicture}
		\draw (0,0)       node 			{$\begin{dynkinDiagram}[labels={l_1,l_{m-1}\ ,-b',\ l_{m-1},l_1}]{A}{*.*x*.*}	\end{dynkinDiagram}$}; 
		\draw (7,0.2)  node		 	{$SL(2m)$};
		
		\draw (-6.5,-1) -- (8.5,-1) ;
		
		\draw (0,-2)       node 		{$ \begin{dynkinDiagram}[labels={,l_1,,l_2,l_{m-1}\!\!\!\!\!,,-b'}]{D}{****.**x} \end{dynkinDiagram} $};
		
		\draw (7,-2+0.2)  node		 {$OSp(4m)$};	
		
		\draw (-6.5,-2.75) -- (8.5,-2.75) ;				
		
		\draw (-.15,-4)  node 			{$\begin{dynkinDiagram}[labels={2l_1,2l_{m-1},-b'}]{C}{*.*x}  \end{dynkinDiagram}$};
		\draw (7,-4.2+0.2)  node		 {$Sp(2m)$};								
		
		\draw  (5.2,1) --( 5.2,-5);
		\draw  (-5,1) --( -5,-5);
		\draw (-6,0.2)  node		 	{$\theta=1$};
		\draw (-6,-2+0.2)  node		 {$\theta=2$};
		\draw (-6,-4.2+0.2)  node		 {$\theta=\frac{1}{2}$};
	\end{tikzpicture}
\end{array}
\eeq

From all of the above Dynkin digrams we can immediately read off the corresponding field in the appropriate  coset space.
The weight under the $\mathbb{C}^*$ is the label on the marked node, and the non trivial transformation under the Levi subgroup, i.e.~under the  supermatrices $a,d$ in~\eqref{hg} is that dictated by the rest of the Dynkin diagram.
All of these reps can equivalently be given by a parameter $\gamma$ and a Young diagram $\ulmb$ consistent with an $SL(m|n)$ rep. 
We will give the precise equations relating $\gamma,\ulmb$ to the above  Dynkin labels shortly, but first we explain how the Young diagram arises as it slightly different  in the three cases $\theta=1,2,\frac{1}{2}$. 

First note that the representations appearing in the OPE of two scalars, $\cO_{p_1} \cO_{p_2}$, 
are not the most general possible (indeed these  would have arbitrary Dynkin labels above every node in the diagrams above). Rather the representation of $a$ must be identical to  that of $d$. 
This statement can be understood 
most directly in the free theory, where the operators appearing in the OPE  have the form 
\beq\label{free_theory_repOPE}
\cO_{\underline A \underline A'} = \cO_{p_1-w} (\partial^k)_{\underline A \underline A'} \cO_{p_2-w}+..\qquad;\qquad (\partial^k)_{\underline A \underline A'}:=(\delta_R)_{\underline A}^{A_1..A_k}(\delta_R)_{\underline A'}^{A'_1..A'_k} \prod_{i=1}^k \frac{\partial}{\partial X^{A_iA'_i}}
\eeq 
 with $\underline A$ and $\underline A'$  multi-indices and $A_i,A'_i$ (anti-)symmetrised into irreps $R$ via the invariant tensors $\delta_R$. 
 Since  partial derivatives $\partial_{A A'}$ commute with each other, the representation of $\underline A$ 
 and $\underline A'$ must  be the same as each other for this to be non-vanishing. 
 Here the $A$ indices carry the rep of $a$ and $A'$ indices carry the rep of $d$ so we see these must be the same.
Now for $\theta=1$,  $a$ and $d$ are independent and form the two $SL(m|n)$ subgroups of $L$ (
see~\eqref{dynkin_summary}). Associating this common representation with its Young diagram, we define $\ulmb$.
For $\theta=2,\frac{1}{2}$, the matrices  $a$ and $d$ are essentially equal to each other (see appendix~\ref{appth2})
 and this means that the representation of $SL(2m|n)$ and $SL(m|2n)$ has a particular special form.
For $\theta=2$, the Young diagram of the  $SL(2m|n)$ rep, $[r_1,\ldots$], will  have 
duplicated rows so that $r_{2i-1}= r_{2i}$.  Thus, the only relevant information is given in a Young diagram $\ulmb$ 
defined by $\lambda_i= r_{2i}$, i.e. half the height, and this $\ulmb$ will label the blocks. Furthermore since the Young diagram $[r_1,\ldots$] is consistent with $SL(2m|n)$ the resulting Young diagram $\ulmb$ will be consistent with $SL(m|n)$. 
For $\theta=\frac{1}{2}$ on the other hand,
the resulting $SL(m|2n)$ Young diagram  will always have duplicated columns so $r'_{2j-1}= r'_{2j}$. 
Thus,we will label the blocks with the Young diagram $\ulmb$ defined to have column lengths $\lambda'_j= r'_{2j}$.

Summarising: In all cases with a group theory interpretation we have specified a representation exchanged in the 
OPE of two scalars, through a Young diagram, $\ulmb$. Thus  $\ulmb$, 
together with a parameter $\gamma$, is the data we will use to specify the superconformal blocks $B_{\gamma,\ulmb}$. 
It will turn out that this same data is very natural from the point of view of $BC_{m|n}$ functions!

A closing remark regarding the non-supersymmetric degeneration  for either $n=0$ or $m=0$, which we discuss 
in appendix \ref{app_supecoset}. This has a group theory interpretation for any half integer (or inverse of a half integer) $\theta$, 
and it is interesting to mention that the group theoretic interpretation we give here meshes closely with a previously 
understood interpretation of the $BC_n$  Heckman Opdam hypergeometric functions  as spherical functions 
on certain coset spaces~\cite{0907.2447}. The latter are Grassmannians in $SU(p,q), SO(p,q)$ and $Sp(p,q)$, 
for $\theta=1,2,\frac{1}{2}$ respectively. Upon setting $p=q=m$ (or $n$) they look very reminiscent of the ones that 
we are using and it would be interesting to pursue this connection further.

\subsubsection*{Young diagrams vs representation labels}

A Young diagram $\ulmb$ with a $GL(m|n)$ structure has a hook shape in which 
the row lengths $\lambda_{i=1,2,\ldots} $ are such that $\lambda_{m+1}\leq n$.  Equivalently, the 
column heights $\lambda'_{j=1,2,\ldots}$ are such  that $\lambda'_{n+1}\leq m$. Graphically, the diagram 
fits inside the red dashed lines:\footnote{In this specific drawing $m=4$, $n=7$, then $\lambda_1=20$ 
and $\lambda_9=1$, while $\lambda'_1=9$ and $\lambda'_{20}=1$. }
\beq\label{yd}
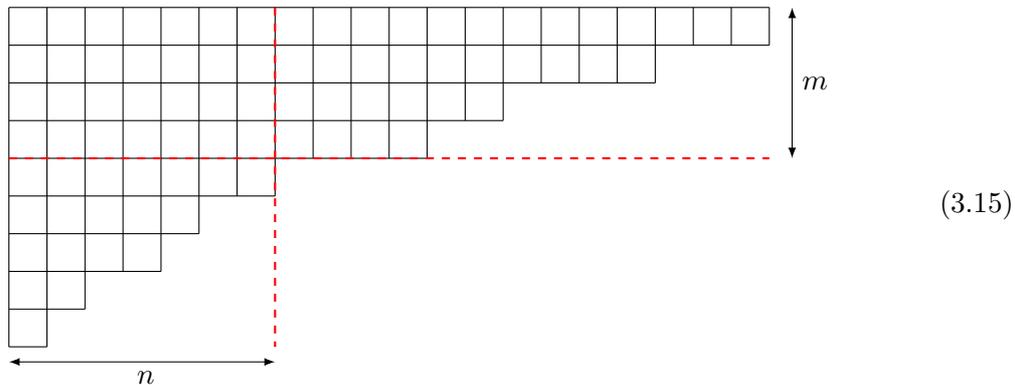
\begin{array}{c}
\begin{tikzpicture}

\foreach \y / \x  in  { 0 / 10,    0.5/ 10,    1 / 8.5,     1.5 / 6.5 ,   2 /5.5 ,    2.5 / 3.5,    3 /2.5 ,    3.5 /2 ,    4 /1,     4.5 /.5}
\draw (0,-\y) -- (\x,-\y);

\foreach \x / \y  in  { 0 /4.5, 0 .5 /4.5,   1 /4 , 1.5 /3.5,       2 /3.5,      2.5 /3,      3 / 2.5 ,  3.5 / 2.5,  4 /2 , 4.5 /2, 5 /2, 5.5 /2  , 6 / 1.5 , 6.5 /1.5, 7 /1, 7.5 /1, 8 /1, 8.5 /1 , 9 /.5 , 9.5 / .5 ,10 /.5 }
\draw (\x,0) -- (\x,-\y);

\draw [red,thick,dashed] (3.5,0 ) -- (3.5, -4.5);
\draw [red,thick, dashed] (0,-2 ) -- (10, -2);

\draw [latex-latex]   (0,-4.7)--(3.5, -4.7);
\draw (1.8,-4.9)	node 	{$n$};

\draw [latex-latex]   (10.3,0 ) -- (10.3, -2);
\draw (10.6, -1)	node 	{$m$};

\end{tikzpicture}
\end{array}
\eeq
We shall say that a box $\Box\in\ulmb$ has integer coordinates $(i,j)$, where $i$ corresponds 
to the row index and $j$ to the column index. For example, the rightmost boxes of $\ulmb$ from 
top to bottom have coordinates $(i,\lambda_i)$ for $i=1,2,\ldots$. Equivalently, from bottom 
to top they have coordinates $(\lambda'_j,j)$ for $j=1,2,\ldots$. 

Young diagrams $\ulmb$ with a hook shape might or might not contain the rectangle $n^m$.  
In the supersymmetric case, a Young diagram $\ulmb$ which contains the box with 
coordinates $(m,n)$, and therefore contains the full rectangle $n^m$, will correspond to 
a typical or long representation.  Otherwise the representation is atypical. We will discuss concrete 
examples related to physical theories later on (see discussion around ~\eqref{typical} and \eqref{atypical} for examples and a fuller description).

The Dynkin labels given in \eqref{dynkin_summary} (and \eqref{dynkin_summaryn0} for the case $n=0$) translate to the data specifying 
a Young diagram $\ulmb$, and a parameter $\gamma$, for $\theta=1,2,\frac{1}{2}$ as follows
\beq\label{translation_dynkin_YT}
\begin{array}{clccl}
	l_{m-i}&=(\lambda_i{-}n)^+-(\lambda_{i+1}{-}n)^+  & i=1,\ldots, m-1   &\  ;\quad &		{\delta}=(\lambda_1{-}n)^+ +\theta\lambda'_n \\[.2cm]
	a_{n-i}&=(\lambda'_i{-}\lambda'_{i+1}) & i=1,\ldots,n	-1		   &\  ;\quad &               b=\gamma-2\lambda_1' \\[.2cm]
	b'&=\gamma+\frac2\theta \lambda_1
\end{array} 
\eeq
where $(x)^+=\max(x,0)$.  
In physical applications it is useful to read off the dilation weight, especially as this is the only quantum number which can become non integer in an interacting CFT. The following equality gives the dilation weight in terms of the Dynkin labels, and then $\gamma,\ulmb$
 \beq
  \Delta=-\sum_{i=1}^{m-1}il_i+m\left( {\delta}+ \theta \sum_{j=1}^{n-1}a_j +  \theta \tfrac{b}{2} \right)=
 \sum_{i=1}^{m}  (\lambda_i{-}n)^+ +m \theta \tfrac{\gamma}{2}\ .
 \eeq
Note now that  for long representations (those for which the box with coordinates  $(m,n)$ lies in the Young diagram $\ulmb$) the dictionary 
\eqref{translation_dynkin_YT} is invariant under
\begin{align}\label{shift}
	\begin{array}{c} \lambda_i \rightarrow \lambda_i-\theta\tau'   \\[.2cm]  \!i=1,\ldots,m\end{array} \qquad;\qquad 
	\begin{array}{c} \lambda'_j \rightarrow \lambda'_j+\tau' \\[.2cm] \!j=1,\ldots,n\end{array}  \qquad;\qquad            
	\gamma \rightarrow \gamma+2\tau'\ .
\end{align}
We will refer to this redundancy as the \emph{shift invariance}. 

The shift invariance \eqref{shift} arises from the fact that $SL(m|n)$ reps are not uniquely specified by a 
Young diagram. This is very familiar in the bosonic case $n=0$.  In fact, for  $SL(m)$ reps one can add 
arbitrarily many height $m$ columns to a Young diagram without changing the rep. This leads directly to~\eqref{shift} for $n=0$. 
In the same way, $SL(n)$ reps have a shift invariance, and therefore when $m=0$ one can add arbitrarily many length $n$ rows 
(since the Young diagram in $SL(0|n)$ is transposed w.r.t.~$SL(n)$).  For the supergroup $SL(m|n)$ 
what happens is that if there is a full width $n$ row below the $n^m$ rectangle in the Young diagram, this can be 
removed and replaced  by a full height $m$ column to the right of the rectangle without changing the representation. 
The presence of $\theta$ in the shift~\eqref{shift} reflects the fact that this shift applies to the reps of $SL(2m|n)$ 
or $SL(m|2n)$ with double numbers of rows or columns around \eqref{free_theory_repOPE}.
The implications of this shift for $\ulmb$ then give the $\theta$ dependence in \eqref{shift}.

Let us exemplify the relevant Young diagrams in some cases of physical interest.  
Blocks for maximally supersymmetric theories 
\beq
\left.\begin{array}{c}
\cN=4, 4d \ {\rm for}\ \theta=1\\[.15cm]
\cN=(2,0), 6d\ {\rm for}\ \theta=2\\[.15cm]
\cN=8, 3d\ {\rm for}\ \theta=\frac{1}{2}
\end{array}\right\}\quad\ {\rm all\ correspond\ to}\quad m=2,n=2
\eeq
These blocks are labelled by $\gamma$ and a Young diagram with at most two rows of length $\lambda_1,\lambda_2$ 
and two columns of length $\lambda_1',\lambda_2'$.  There are two  quantum numbers for the conformal group, 
dilation weight $\Delta=(\lambda_1{-}2)^++(\lambda_2{-}2)^+ +\theta \gamma$  and spin $l_1=(\lambda_1{-}2)^+-(\lambda_2{-}2)^+$ (only symmetrised Lorentz indices appear in these four-point functions) and two analogous  quantum 
numbers for the internal subgroup $a_1,b$.\footnote{It might be instructive to compare with~\cite{Dolan:2004mu}, 
which also treats all maximally supersymmetric  cases together. They use internal labels $(a_{\tt there},b_{\tt there})$ 
where $a_{there}=\tfrac12 \gamma{-}\lambda'_2, \quad  b_{there}=\tfrac12 \gamma{-}\lambda'_1$.} 
When $\lambda_2\geq 2$ the representation exchanged in the OPE (assuming integer dilation weight) is long or typical . 
If $\lambda_2=1,0$ the diagram is a thin hook, with at most a single row and a single column, 
and the representation is short.  When the Young diagram is empty it is half BPS. We emphasise 
that in our formalism there is no need to distinguish the different shortening conditions!

Another case of physical interest is given by blocks with $m=2,n=1$, which includes
$\mathcal{N}=2$ superconformal theories in 4d, the $\mathcal{N}=(0,1)$ superconformal theories 
in 6d, and also $\cN=4$ in 3d. In this case the Young diagram has at most a single long column $\lambda_1'$ 
and therefore a single integer classifies the internal representations that can appear in the OPE. 
The reps are often defined by $\Delta, l$ and $b$. This time the atypical diagrams can only be the single 
row diagrams, since these do not contain the rectangle $[1^2]$. The rest of the allowed diagrams, 
$[\lambda_1,\lambda_2,1^{\lambda'_1-2}]$, are necessarily long/typical.

\section{Rank-one invitation}%

Before presenting the most general $(m,n)$ superconformal block, we describe in some detail the
bosonic $(1,0)$ and $(0,1)$ blocks, since these are simple enough to familiarise with the formalism 
based on super Jack polynomials.

\subsection{Rank-one bosonic blocks}\label{sec_rank1}

\subsubsection*{$(1,0)$ blocks}

Consider first the $(1,0)$ blocks, i.e.~a single $x$ variable and no $y$ variables. The only allowed 
Young diagrams have a single row shape $\ulmb=[\lambda]$. The Casimir~\eqref{casimir_intro} reduces 
to a one-variable Gauss hypergeometric equation, from which follows the solution (normalised according to ~\eqref{asym}),
\begin{equation}
	\label{10block}
	\begin{split}
		\mathbf{C}^{} B^{}_{\gamma,[\lambda]}(x|)&=(\lambda{+}\tfrac{\theta\gamma}2)(\lambda{+}\tfrac{\theta\gamma}2{-}1) B_{\gamma,\ulmb}(x|) \\[.2cm]%
		{\BB}_{\gamma,[\lambda]}(x|)&= x^{\frac{\theta \gamma}2+\lambda}~{}_2F_1\big(\lambda{+}{\theta \alpha},\lambda{+}{\theta \beta};2\lambda+{\theta \gamma};x\big) %
	\end{split}
\end{equation}
where recall $
	\alpha=\max\left( \tfrac 1 2 (\gamma {-}  p_{12}), \tfrac 1  2 (\gamma {-}  p_{43}) \right)$, $ \beta=\min\left( \tfrac 1 2 (\gamma {-}  p_{12}), \tfrac 1  2 (\gamma {-}  p_{43}) \right)$~\eqref{first_def_alpha_beta}.
The Casimir and the solution is symmetric in $\alpha$ and $\beta$. 
Furthermore, let us
point out that out of the five parameters 
$\lambda,\gamma,\theta,p_{12},p_{43}$ only the three quantities $\lambda+\frac{\theta\gamma}{2}$, 
$p_{12}\theta$ and $p_{43}\theta$ appear.\footnote{The combination $\lambda+\frac{\theta\gamma}{2}$  appears
in the Dynkin diagram \eqref{dilbos} for the $(1,0)$ bosonic theory.} 

Writing the $(1,0)$ block as a series over superJack polynomials is  very 
simple since the $(1,0)$ Jack is simply $P^{}_{[\lambda]}(x|)=x^\lambda$, therefore the \JtoB matrix 
is just the coefficient in the  Taylor expansion of the Gauss hypergeometric:
\bea\label{10blockexp}
	{\BB}_{\gamma,[\lambda]}(x|)=\,  
	x^{ \theta\frac{\gamma}{2}} \sum_{\mu=\lambda}^\infty  
	(T_{\gamma})^{[\mu]}_{[\lambda]}  P_{[\mu]}(x|) \ ,\qquad
	(T_{\gamma})^{[\mu]}_{[\lambda]}=
	\frac{(\lambda+{\theta \alpha})_{\mu-\lambda}(\lambda+{\theta \beta})_{\mu-\lambda}}{(\mu{-}\lambda)!(2\lambda+{\theta \gamma})_{\mu-\lambda}}\ .
\eea

\subsubsection*{$(0,1)$ blocks}

Consider now the $(0,1)$ blocks, i.e.~no $x$ variables and a single $y$ variable. 
Here the only allowed Young diagrams have a single column shape $\ulmb=[1^{\lambda'}]$. 
The Casimir reduces again to a hypergeometric equation, and is solved by:
\beq
\label{internal}
\begin{split}
	\!\!\!\!\mathbf{C}^{} B_{\gamma,[1^{\lambda'}]}(|y)&= -\theta (\lambda'{-}\tfrac\gamma2)(\lambda'{-}\tfrac\gamma2{-}1) B_{\gamma,[1^{\lambda'}]}(|y) %
	\\[.2cm]
	{\BB}_{\gamma,[1^{\lambda'}]}(|y)&= \tfrac{(-1)^{\lambda'}(\frac{1}{\theta})_{\lambda'}}{\lambda'!}\,  y^{-\frac\gamma2 + \lambda'} {}_2F_1\big(\lambda'{-}\alpha,\lambda'{-}\beta;2\lambda'{-}\gamma;y\big)\,.
\end{split}
\eeq
Notice that the entries of the$~_2F_1$ this time are independent of $\theta $. {Then,  
$\lambda'$ and $\gamma$ only enter through the combination  $\frac{\gamma}{2}- \lambda'$.\footnote{This combination 
is the value appearing in \eqref{p_dynkin_label} sitting on crossed through Dynkin node. %
}

The normalisation of ${\BB}_{\gamma,[1^{\lambda'}]}(|y)$ has been chosen as for the 
the $(0,1)$ superJack polynomials (in agreement with~\eqref{asym}). In fact, the series expansion in superJack polynomials 
this time reads
\begin{align}\label{01blockexp}
	{\BB}^{}_{\gamma,[1^{\lambda'}]}(|y)=\,  y^{-\frac{\gamma}{2} }  
	\sum_{\mu'=\lambda'}^{\infty}  (T_{\gamma})_{[1^{\lambda'}]}^{[1^{\mu'}]}
	P_{[1^{\mu'}]}^{}(|y)\ ,\qquad
	P^{}_{[1^{\mu'}] }(|y)=\frac{ (\frac{1}{\theta})_{\mu'} } {\mu'!} (-y)^{\mu'} %
\end{align}
where 
\begin{align}
(T_{\gamma})_{[1^{\lambda'}]}^{[1^{\mu'}]}= 
\frac{(\lambda'{-}\alpha)_{\mu'-\lambda'} 
		(\lambda'{-}\beta)_{\mu'-\lambda'}} 
		{(\mu'-\lambda')!(2\lambda'{-}\gamma)_{\mu'-\lambda'}}\ 
	\frac{ \mu'! (\frac{1}{\theta})_{\lambda'}}{\lambda'! (\frac{1}{\theta})_{\mu'} }\, (-1)^{ \mu'-\lambda'}, %
\end{align}
An important observation to make at this point is that this series truncates when the arguments 
of the Pochhammers are negative. Thus  it truncates for values of $\mu'$ greater than  
$ \beta$.

\subsubsection*{Properties of $T_{\gamma}$}

We can already with this information perform do a very rudimentary check  that the coefficients $T_{\gamma}$ are $(m,n)$ independent
i.e.~they only depend on the Young diagrams, $\theta$, and $p_{12},p_{43}$. %
This implies that  when the Young diagrams coincide, evaluating \eqref{10blockexp} and \eqref{01blockexp} 
should give the same result.  In the $(1,0)$ case the Young diagram has a single row, and 
in $(0,1)$ case a single column, thus they can only be compared when the Young diagram  is empty 
or consists of a single box i.e.~the cases 
$(\ulmb,\umu)=([0],[0]),([1],[1]),([0],[1])$. We indeed see that for both $(1,0)$ and $(0,1)$ we obtain:
\beq
\begin{array}{ccl}
([0],[0]) & ;  & (T_{\gamma})_{[0]}^{[0]}=1\\[.2cm]
([1],[1]) & ; &(T_{\gamma})_{[1]}^{[1]}=1\\[.2cm]
([0],[1]) &; &(T_{\gamma})_{[0]}^{[1]}=\theta\frac{\alpha \beta}{\gamma}\ .
\end{array}
\eeq
The above computation is limited, but nevertheless it is nice to see that~\eqref{10blockexp} and~\eqref{01blockexp} do indeed give the same results where they overlap.  This illustrates the more general $(m,n)$ independence of $T_\gamma$  which we will consider further in  section \ref{sec_SCBlock}.

\subsection{Relation with the Heckman-Opdam hypergeometrics}
\label{sec:HO}

Despite the fact that both $(1,0)$ and $(0,1)$ blocks are given by the same building block, 
a Gauss hypergeometric, there is an important difference, which we emphasised already: 
the $(0,1)$ block is a polynomial, while the $(1,0)$ block is an infinite series. 
However, since both external and internal blocks are based on the same$~_2F_1$, comparing the two expressions~\eqref{10block} and~\eqref{internal},
we read off the following formal relation between them:
\beq\label{ancont}
	{\BB}^{(1,0)}_{\gamma,\, [\lambda] }(y|;\theta ,p_{12},p_{43})
	=
	\frac{\lambda!}{(-1)^{\lambda}({\theta})_{\lambda} }	{\BB}^{(0,1)}_{-\theta\gamma, \,[1^\lambda]}(|y;\tfrac1\theta,-\theta p_{12},-\theta p_{43})\,.
\eeq
So the $(1,0)$ and $(0,1)$ blocks can be viewed as analytic continuations of each other. In particular, 
if $\alpha,\beta,\gamma$  are assumed to be positive on the $(1,0)$ side, they become 
negative on the $(0,1)$ side,    specifically, $(\alpha,\beta,\gamma)\rightarrow -\theta (\alpha,\beta,\gamma)$.  
This can be also understood from the general relation satisfied by the defining Casimir 
\beq
{\bf C}^{(\theta,a,b,c)}({\bf x}|{\bf y})=-\theta\, {\bf C}^{(\frac{1}{\theta}, -a \theta,-b \theta,-c \theta)}({\bf y}|{\bf x})\ .
\eeq
Let us then try and understand this analytic continuation in the parameters and relate it to the Heckman Opdam hypergeometric function.

In section \ref{dual_Jac_sec} we identified the internal $(0,n)$ block with 
the dual ${BC}_n$ Jacobi polynomial, up to a normalisation, and we showed how the latter 
is obtained from the ${BC}_n$ Jacobi polynomial (in particular see \eqref{BCJacobi_internalBlock}). Now the ${BC}_n$ Jacobi polynomials  
are also Heckman Opdam (HO) hypergeometric functions for positive weights \cite{Opdam}, but the ${BC}_n$ HO 
hypergeometric are defined more generally for arbitrary parameters.  It is therefore interesting to compare 
these functions with our $(1,0)$ and $(0,1)$ blocks, and discuss in this context the analytic continuation 
w.r.t.~the external parameters.

Both the $BC_1$ Jacobi polynomial and the  dual Jacobi polynomial 
can be given explicitly in terms of ${}_2F_1$ Hypergeometrics as follows,\footnote{These can be obtained by solving their defining differential equations. In the dual 
case this is \eqref{diff_def_dualJ}. For the Jacobi polynomial we refer to \cite{Macdonald_3}. 
Alternatively, the 1d combinatorial formula in~\cite{Koornwinder} straightforwardly gives the result. 
The normalisation of the Jacobi is taken such that $J_{[\lambda]}=y^\lambda+\ldots$, and this gives the prefactor w.r.t. to the$~_2F_1$ series. }
\begin{align}\label{dJ2}
	J_{[\lambda]}(y;\theta,p^-,p^+)&=%
	\frac{(-)^\lambda (1+p_-)_{\lambda} }{(1+\lambda{+}p^+{+}p^-)_{\lambda}}
	\  {}_2F_1(1+\lambda{+}p^+{+}p^-,-\lambda;1{+}p^-;y)\qquad \lambda \in \mathbb{Z}^{\geq 0} \\[.2cm]
	\tilde J_{\beta,[\lambda]}(y;\theta,p^-,p^+)&=y^\lambda\, {}_2F_1(\lambda{-}\beta{-}p^-,-(\beta-\lambda);2(\lambda{-}\beta){-}p^+{-}p^-;y)	\qquad \beta{-}\lambda \in \mathbb{Z}^{\geq 0}\notag\\
	&= y^\lambda {}_2F_1(\lambda{-}\alpha,\lambda{-}\beta;2\lambda{-}\gamma;y) 
\end{align}
where $\alpha=\beta+p^-$ and $\gamma=2\beta+p^++p^-$. 
The use of the Gauss hypergeometric to write the polynomials has the bonus that 
they give a natural  analytic  continuation to arbitrary values of the parameters, i.e. away from the polynomial restriction  $\lambda\in \mathbb{Z}^+$ for $J$ and $\beta-\lambda\in \mathbb{Z}^+$ for $\tilde J$. 
The key point here is that even though the Jacobi polynomial and the dual Jacobi polynomial are directly related to each other, the above expressions in terms of Gauss hypergeometrics  yield two {\em inequivalent} analytic continuations of the parameters. The first is the HO hypergeometric, the second is the one relevant for blocks. In both cases the analytic continuation gives a function that is regular at $y=0$ with a branch cut between $y=1$ and $y=\infty$.

Let us explore how this works in more detail. The point is that 
when $\beta-\lambda \in {\mathbb{Z}}^+$, the hypergeometrics are polynomial and there is a hypergeometric identity~\cite{wolfram1}
\beq\label{id1}
	\frac{(-1)^{\lambda-\beta}}{
		c(\alpha,\beta,\gamma,\lambda)} \ {}_2F_1(1{-}\lambda{+}\gamma{-}\beta,-(\beta-\lambda);1{-}\beta{+}\alpha;\tfrac1y)
	= 
	y^{\lambda-\beta} 
	{}_2F_1(\lambda{-}\alpha,-(\beta-\lambda);2\lambda{-}\gamma;y)\ ,
\eeq	
where
\beq
	c(\alpha,\beta,\gamma,\lambda)= \frac{\Gamma[1 {{+}}\alpha {-}\beta] \Gamma\left[1+\gamma {-}2
		\lambda \right]  \, }{\Gamma\left[1+\alpha {-}\lambda \right] \Gamma\left[1{-}\beta {+}\gamma {-}\lambda \right]}\,. \ \ 
\eeq
This is precisely the identity between the Jacobi and dual Jacobi polynomials $J_{[\beta-\lambda]}(1/y)=y^{-\beta}\tilde J_{\beta,[\lambda]}(y)$.

But as soon as we continue away   from $\beta-\lambda \in {\mathbb{Z}}^+$, the above identity 
relaxes to  a three term identity~\cite{wolfram2}: 
\begin{align}\label{paul_formula_analitic_c}
	\!\!\!\!	\,    {}_2F_1(1{-}\lambda{+}\gamma{-}\beta,\lambda{-}\beta;1{-}\beta{+}\alpha;\tfrac1y)=&
		\phantom{+}\ (-1)^{\lambda {-}\beta}c(\alpha,\beta,\gamma,\lambda)
		\times  y^{\lambda {-}\beta}{}_2F_1\left(\lambda {-}\alpha ,\lambda {-}\beta ;2 \lambda {-}\gamma ;{y}\right)\notag
		\\
		&
		+(-1)^{\tilde \lambda {-}\tilde\beta}c(\tilde\alpha,\tilde\beta,\tilde\gamma,\tilde\lambda)
		\times  y^{\tilde\lambda {-}\tilde\beta}{}_2F_1(\tilde\lambda {-}\tilde\alpha ,\tilde\lambda {-}\tilde\beta ;2 \tilde\lambda {-}\tilde\gamma ;{y})
\end{align}
where $\tilde \alpha=\alpha-\gamma, \tilde \beta=\beta-\gamma,\tilde\gamma=-\gamma, \tilde \lambda =1-\lambda$. 
This is a well know connection formula for the $\,_2F_1$.
Notice in particular that $c(\tilde\alpha,\tilde\beta,\tilde\gamma,\tilde\lambda)\rightarrow 0$, as  $\beta-\lambda \rightarrow n\in {\mathbb{Z}}^+$, 
therefore we recover the previous identity~\eqref{id1}.

Now the combination on the  RHS~of \eqref{paul_formula_analitic_c} is precisely the 
expression for the $BC_1$ HO hypergeometric (in $1/y$) given as a specific sum of independent solutions of its defining equation (which is equivalent to the Casimir equation). 
In general the $BC_n$  HO hypergeometric is defined as a sum over $W= S_n\ltimes (\mathbb{Z}_2)^n$ of building block functions called Harish Chandra functions~\cite{Heckman_1,Heckman_2} each of which are independent solutions of the same defining equation. For $BC_1$ this sum is just a sum over $\mathbb{Z}_2$ and the two functions on the RHS of~\eqref{paul_formula_analitic_c} are precisely these building block   Harish Chandra functions (the normalisations are known as the Harish-Chandra $c$ functions). More details for this $BC_1$ case can be found for example in \cite{Shimeno}.

In our context however on the RHS~of \eqref{paul_formula_analitic_c},  the first contribution is (up to a normalisation)  
the $(1,0)$ block in \eqref{10block} with parameters ${B}^{}_{-\theta\gamma, \,[\lambda]}(y|;\tfrac1\theta,-\theta p_{12},-\theta p_{43})$. 
The second contribution is 
$B_{-\theta\gamma;[1-\lambda]}$, and corresponds to 
an independent solution of the same Casimir equation~\eqref{10block}, with different properties in the 
small $y$ expansion. In the CFT context this is also an important object known as the shadow block.

So we see therefore that  the block and the shadow block are both Harish-Chandra functions\footnote{In fact twisted Harish 
Chandra functions since we exclude the $(-1)^{\lambda-\beta}$ in the definition of the block.}. 
The blocks have different analyticity properties to the Heckman Opdam hypergeometrics, explaining why they are not the same  even though they are trivially related for certain integer values of their parameters.
In particular, in the variables we are using the blocks are regular at $y\rightarrow 0$, whereas instead the Heckman Opdam hypergeometrics are regular at $1/y \rightarrow 0$. For a more detailed analysis of the analytic structure of the standard bosonic conformal $m=2,n=0$ blocks and the relation to Harish Chandra functions see~~\cite{Isachenkov:2016gim,Isachenkov:2017qgn}.

Summarising. The  $BC_1$ HO hypergeometric is the analytic continuation of the $BC_1$ Jacobi 
polynomial when $\beta-\lambda\in\mathbb{C}$ which is regular when its argument approaches zero. 
The dual Jacobi polynomials however have a different natural analytic continuation which is regular when its argument (which is the inverse of that of that of the Jacobi polynomial) approaches zero.
This latter analytic continuation gives the Harish Chandra functions, 
the building blocks in the sum over $W={\pm1}$ giving the  HO hypergeometric. Furthermore this latter analytic continuation interpolates between 
the $(1,0)$ or $(0,1)$ blocks as we vary the  parameters. 

This story would appear to have a natural generalisation to higher $n$. The $BC_n$ Jacobi  polynomials have a natural analytic continuation in their parameters as $BC_n$ HO Hypergeometrics.  The dual Jacobi polynomials (internal $(0,n)$ blocks) however  will have a natural analytic continuation as $S_n \subset W= S_n\times (\mathbb{Z}_2)^n $ invariant combinations of Harish Chandra functions. This combination  will then coincide with external $(n,0)$ blocks for certain values of their parameters. Indeed this latter story can be seen explicitly in the $\theta=1$ case from the results of~\cite{Doobary:2015gia} where both external and internal blocks are given in terms of a determinant of Gauss Hypergeometric functions which indeed coincide on mapping the parameters appropriately.  It is then interesting to consider the generalisation to the supersymmetric $(m,n)$ case and if there is a relation between the dual super Jacobi functions and super Jacobi polynomials via analytic continuation. 
We consider $(m,n)$  analytic continuations further in section~\ref{analytic_cont_sec}

\section{Superconformal blocks (I): recursion}\label{sec_SCBlock}

Having considered in detail the $(1,0)$ and $(0,1)$ blocks in the previous section we return to the general case.
In this section we explain how to solve the Casimir equation~\eqref{eeq} for the  superconformal blocks,  
defined via the expansion
\beq\label{another_superb_def}
B_{\gamma,\ulmb}=\left(\frac{\prod_i x_i^\theta}{\prod_j y_j}\right)^{\!\!\frac{\gamma}{2}} \, F_{\gamma,\ulmb}\qquad;
\qquad F_{\gamma,\ulmb}=\sum_{\umu \supseteq \ulmb} { ({ T}_{\gamma})_\ulmb^\umu} \, P_{\umu}({\bf z})\ \quad;
\eeq
for any  values of the $m|n$ variables ${\bf z}$ as well as any $\theta,p_{12},p_{43}$ and 
$\gamma$. We will use a recursion, as anticipated in section \ref{overview:sec}, and to find it we will first derive the representation of the Casimir~\eqref{rewriteCinJack_intro} suited for acting on  the  basis of super Jack polynomials,  $P_\umu$ and then use this to turn the differential equation into a recurrence relation for the coefficients  
$(T_{\gamma})_{\ulmb}^{\umu}$.  

We will write the recursion in a way which 
is manifestly independent of the dimension of the blocks, and the $m,n$ labels for the super 
Jack polynomials, purely in terms of the Young diagram~\eqref{rec2}. Later this is made very 
explicit in a manner prepared for different possible analytic continuations, see for example \eqref{rec_section_scblocks}, \eqref{south_recursion}. 
Appendix \ref{symm_supersym_poly_sec} provides a concise summary of 
properties and definitions  of Jack and super Jack polynomials which  might serve as a helpful 
guide for the reader.

Various properties of the recursion, and therefore of its solution, the \JtoB matrix $T_\gamma$, will then be discussed. 
First we will exemplify the special case of the half-BPS block $\ulmb=[\varnothing]$, which 
admits a simple hypergeometric solution. This is instructive because the details are still quite 
non trivial, despite the final simplicity of the solution. 
Then, we will consider  the {most} general case. We do not obtain an explicit solution for the general case,  however, it is very 
efficient to implement the recursion on a computer, and we use this to explore properties of  $(T_{\gamma})_{\ulmb}^{\umu}$, 
for example its dependence on $\theta$, $\gamma$, $p_{12}$ and $p_{43}$.
The most intriguing is the dependence on $\gamma$,  which becomes highly non-trivial and takes the form of a more and more complicated mostly non factorisable polynomial.  In  section \ref{sec_guideline} we will find a surprising  interpretation  for the $\gamma$ dependence, 
by elaborating on the  connection between superconformal blocks and Jacobi polynomials.

\subsection{Derivation of the recursion and higher order Casimirs}\label{casimir_deco_sec}

Recall briefly our construction. Superconformal blocks as eigenfunctions of the Casimir 
operator introduced in~\eqref{eeq}-\eqref{ev1}. On $F_{\gamma,\ulmb}$, the eigenvalue problem reduces to 
\beq\label{eeqF2}
\mathbf{C}_{}^{(\theta,\alpha,\beta,\gamma)}   F_{\gamma,\ulmb} = ({h}_{\ulmb}^{(\theta)}+\theta \gamma |\ulmb|) F_{\gamma,\ulmb}\ ,
\eeq
where
\begin{equation}\label{hevalue}
	{h}_{\ulmb}^{(\theta)}=\sum_{i} \lambda_i (\lambda_i{-}2\theta(i{{-}}1){-}1)
\end{equation}
 and $|\ulmb|=\sum_i \lambda_i$, with $\ulmb=[\lambda_1,\ldots ]$.  As highlighted already, 
 the eigenvalue for $F_{\gamma,\ulmb}$ does not  depend explicitly on $(m,n)$ unlike the original eigenvalue 
 for $B_{\gamma,\ulmb}$ in~\eqref{ev1}. Recall also 
 the definitions of  $\alpha$ and $\beta$ in terms of $p_{12},p_{43},\gamma$ given in \eqref{first_def_alpha_beta}
\begin{align}\label{second_def_alpha_beta}
	\alpha &\equiv \max\left( \tfrac 1 2 (\gamma {-}  p_{12}),\tfrac 1  2 (\gamma {-}  p_{43})\right) ,
			\quad \beta\equiv \min\left( \tfrac 1 2 (\gamma {-}  p_{12}), \tfrac 1  2 (\gamma {-}  p_{43}) \right)
\end{align}

 As discussed above~\eqref{eeq}, the Casimir  can be nicely related to the $BC_{m|n}$ CMS operator, 
 and we do so in appendix~\ref{app_differential_operators}. The explicit result in our notation takes the form,
\begin{align}\label{casimir_intro}
 \mathbf{C}_{}^{(\theta,a,b,c)}= 
 \sum_{I=1}^{m+n}  (-\theta)^{\pi_I} \mathbf{D}_I^{(\theta,a,b,c)}  + 2\theta \sum_{I=1}^{m+n}  \sum_{J\neq I} \frac{ z_I z_J}{z_I-z_J}\, (-\theta)^{-\pi_J}\mathbf{d}_I
 \end{align}
where the two operators ${\bf d}_I$ and $\mathbf{D}_{I}$ are defined as
 \begin{align}
 \mathbf{d}_I&=(1-z_I) \partial_I\\[.2cm]
 \mathbf{D}^{(\theta,a,b,c)}_I &=  z^2_I \partial_{I}(1-z_I)\partial_I -  (-\theta)^{1-\pi_I}(c-(a +b )z_I)  z_I\partial_I -(-\theta)^{2-2\pi_I}  a b  z_I\ .
 \label{operator_D}
 \end{align}
We use the vector $\mathbf{z}=(z_1,\ldots,z_m|z_{m+1},\ldots ,z_{m+n})$ to label the
variables $z_i=x_{i=1,\ldots m}$ and $z_{m+j}=y_{j=1,\ldots n}$. The parity $\pi_{i}=0$ 
is assigned to the non-compact direction $i=1,..,m$,  otherwise $\pi_{j}=1$ is assigned to 
the compact direction $j=m+1,..,m+n$.

The identification of the Casimir with a $BC_{m|n}$  differential operator helps deducing  
the decomposition~\eqref{rewriteCinJack_intro} of the Casimir  into operators with well defined actions on $P_{\umu}$:
\begin{align}\label{rewriteCinJack}
	\mathbf{C}^{(\theta,a,b,c)}
	=& \mathbf{H}^{(\theta)} +  \theta c\sum_{i=1}^{m+n} \,  z_I \partial_I \notag\\
	& -\theta  (a+b) \sum_{I=1}^{m+n}  z_I^2\partial_I -\tfrac{1}{2}\Big[ \mathbf{H}^{(\theta)}, \sum_{I=1}^{m+n} z_I^2 \partial_I \Big]   -  \theta^2 ab \sum_{I=1}^{m+n} z_I (-\theta)^{-\pi_I}
\end{align}
where $\mathbf{H}^{(\theta)}$ is the $A_{m|n}$ CMS Hamiltonian 
(see appendix~\ref{app_differential_operators})\footnote{Notice that 
	$\mathbf{H}_{}$ shows up directly in \eqref{rewriteCinJack}, since it is found by replacing 
	$\mathbf{D}_I \rightarrow z_I^2 \partial_I^2 $ and $\mathbf{d}_I\rightarrow \partial_I$. Terms of the Casimir in 
	which  $\mathbf{D}_I \rightarrow -z_I^3 \partial_I^2$ and $\mathbf{d}_I\rightarrow -z_I\partial_I$, 
	are generated by the commutator. }
\begin{align}\label{hh}
	\mathbf{H}^{(\theta)}=&
	\sum_{I=1}^{m+n} (-\theta)^{\pi_I}z_I^2\partial_I^2\ +\ 2\theta \sum_{I\neq J}  \frac{  z_I z_J }{z_I-z_J}(-\theta)^{-\pi_J} \partial_I\ .
\end{align}

The operator decomposition in \eqref{rewriteCinJack} is perfectly suited to acting on the expansion 
$F_{\gamma,\ulmb}=\sum (T_{\gamma})_{\ulmb}^{\umu} P_{\umu}$ because
super Jack polynomials are eigenfunctions of both $\sum_{I} z_I \partial_I$ 
(reflecting the fact that they are homogeneous polynomials)  and $\mathbf{H}^{(\theta)}$,\footnote{Indeed 
these are the first two Hamiltonians  of a tower found in \cite{Veselov_1}, which establishes 
the classical  integrability of the system.} and moreover, the other two operators  
are themselves related to the simplest one-box  super Jack polynomial, $P_\Box$
\begin{align}\label{offdiag}
	\sum_{I} z_I (-\theta)^{-\pi_I} = P_\Box({\bf z}), \qquad  \text{and} \qquad \sum_{I} z_I^2 \partial_I=\tfrac12 [{\bf H}^{(\theta)},P_{\Box}(\bf z)]\ .
\end{align}

The action of all the operators $z_I \partial_I, \ {\bf H},\  P_\Box$ on super Jack polynomials yields 
a linear combination of Jack polynomials whose coefficients are independent of  $m,n$. 
This will imply that the recursion for $T_{\gamma}$, which we derive from the Casimir, 
will be independent of $m,n$ at every stage.

 Let us begin with the simplest 
operator in~\eqref{rewriteCinJack}, i.e.~the momentum operator,
\bea
\sum_{I=1}^{m+n} z_I \partial_I\, {P}_{\umu }{({\bf z})}= |\umu| {P}_{\umu }{({\bf z})}\ .
\label{superMeigen}
\eea
The eigenvalue here simply counts the number of boxes of the Young diagram $\umu$ 
(which obviously does not depend on $(m,n)$).  This operator just reflects the fact that super Jack polynomials 
are homogeneous functions of the $z_I$ of degree $|\mu|$.

For the Hamiltonian,  we have
\bea\label{ham}
\mathbf{H}^{(\theta)}\, {P}_{\umu }{({\bf z})}   &=&   {h}^{(\theta)}_{\umu} {P}_{\umu }{({\bf z})}\ .
\eea
This equation is part of the standard definition of the super Jack polynomials 
and the eigenvalue  ${h}^{(\theta)}_{\umu}$ in~\eqref{hevalue}  has  no explicit 
dependence  on  $(m,n)$.

Next, the two `off-diagonal' operators in~\eqref{offdiag}. The first of the two is simply the 
multiplicative operator acting with $P_\Box$. In general the product $P_{\umu_1}P_{\umu_2}$ 
gives a linear combination of super Jacks whose coefficients, ${\cal C}_{\umu_1 \umu_2}^{\umu_3}(\theta)$, known 
as structure constants, are independent of $m,n$. %
 When $\theta=1$ Jack polynomials reduce to Schur polynomials and the structure constants 
become precisely the standard Littlewood-Richardson coefficients. Finding an explicit formula for the 
structure constants for general $\theta$ however is a difficult problem/task in modern mathematics \cite{Schlosser}. 
Here however we only need the case in which one of the Young diagrams is a single box. This case 
reduces to the simplest case of the Pieri rule for Jack and super Jack polynomials,
and ${\cal C}_{ \Box\, \umu}^\unu$ {\em is} known combinatorially~\cite{Stanley,Lassalle_1}. It is perhaps less known 
that this combinatorial formula can be made fully explicit, 
as we will show shortly.

So the action of the first off-diagonal operator in \eqref{offdiag} coincides with 
the product  $P_{\Box}P_{\umu}$ and returns
 \beq\label{superPieri1}
P_\Box({\bf z})\,  {P}_{\umu }{({\bf z})} =  
\ \sum_{\unu \in \{\umu+\Box\}} {\cal C}_{ \Box\, \umu}^\unu  P_{  \unu }{({\bf z})}\ . 
\eeq
where the RHS~spans all polynomials labelled by the Young diagram $\unu$ 
obtained by adding an extra box to the Young tableau $\umu$, in all possible ways.

The action of the second off-diagonal operator in \eqref{offdiag}, $\sum_i z_i^2 \partial_i$, 
can then be obtained straightforwardly from its representation as a commutator together 
with~\eqref{ham},\eqref{superPieri1}. We have
\begin{align}
	\sum_{I} z_I^2 \partial_I P_{\umu}({\bf z})&=\tfrac12 [{\bf H}^{(\theta)},P_{\Box}({\bf z})]P_{\umu}({\bf z}) = 
	\sum_{\unu \in \{\umu+\Box\}} \tfrac12(h_{\unu}-h_{\umu}){\cal C}_{ \Box\, \umu}^\unu P_{  \umu+\Box_i }{({\bf z})}\ . 
\end{align}

Having understood the action of all the operators appearing in the operator decomposition 
of the Casimir, and  having observed that all of them decompose into Jack polynomials 
with coefficients which only depend on the Young diagram, we can now write the representation 
of the Casimir itself in the basis of superJack polynomials.  The result is quite neat and nicely 
generalises the structure noticed for $\theta=1$ in \cite{Doobary:2015gia},
\begin{align}\label{ConJack}
	\rule{0pt}{.8cm}
	\mathbf{C}_{}^{(\theta,a,b,c)}  P_{  \umu }&=\left( {h}_{\umu } + \theta c|\umu|\right) P_{  \umu }
	\ -\ \sum_{\unu \in \{\umu+\Box\}} \left(\tfrac12(h_{\unu}{-}h_{\umu})+\theta a \right)\left(\tfrac12(h_{\unu}{-}h_{\umu})+\theta b \right) {\cal C}_{ \Box\, \umu}^\unu\  P_{  \unu }\ .
\end{align}
From this and the definition $F_{\gamma,\ulmb}=\sum_{\umu} (T_{\gamma})_{\ulmb}^{\umu} P_{\umu}$, 
then the Casimir equation~\eqref{eeqF2} gives  the following recursion equation for the coefficients $T_\gamma$
\begin{mdframed}
\begin{gather}\label{rec2}
(T_\gamma)_{\ulmb}^{\umu}=  \frac{ \sum_{\unu \in \{\umu-\Box\}} 
	\left(\tfrac12(h_{\umu}{-}h_{\unu})+\theta \alpha \right)\left(\tfrac12(h_{\umu}{-}h_{\unu})+\theta \beta \right) {\cal C}_{ \Box\, \unu}^\umu \, (T_\gamma)_{\ulmb}^{\unu}  }{
	\left( {h}_{\umu}{-}{h}_{\ulmb} {+} \theta \gamma\, (|\umu|{-}|\ulmb| )\right)} \\ \notag
\end{gather}
\end{mdframed}
with $\alpha,\beta$ as defined earlier in terms of $\gamma,p_{12},p_{43}$~\eqref{second_def_alpha_beta}.
Here we see that the recursion initiates from the Young diagram $\umu$ (which contains $\ulmb$) 
and goes back to  $\ulmb$ (for which we know $(T_\gamma)_\ulmb^\ulmb=1$) recursively by subtracting boxes. Thus if $\umu$ is  compatible  
with the $(m,n)$ structure (i.e. $\mu_{m+1}\leq n$) then $\unu\in\{\umu-\Box\}$ is also compatible with it and the recursion always remains inside the $(m,n)$ Young diagram. 
This means the recursion can be solved 
quite efficiently on a computer.

It is also interesting to consider the inverse of $T_\gamma$, i.e.~the \BtoJ matrix, $T_\gamma^{-1}$. 
This gives Jack polynomials as a sum of blocks, via the inverse of~\eqref{another_superb_def},  
\beq
P_{\ulmb}=\sum_{\umu} 
(T_{\gamma}^{-1})_{\ulmb}^{\umu}\, F_{\gamma,\umu}. 
\eeq
Note that $F_{\gamma,\umu}$, when $m>0$ is an infinite series, and the sum, an infinite sum. Many non trivial cancellations take place in order to recover a polynomial. 
From~\eqref{ConJack} and~\eqref{eeqF2} 
we obtain a recursion for $T_\gamma^{-1}$ which is very similar looking to that of $T_\gamma$
\begin{gather}\label{rec2inv}
	(T_\gamma^{-1})_{\ulmb}^{\umu}=  \frac{ \sum_{\unu \in \{\ulmb+\Box\}} 
		\left(\tfrac12(h_{\unu}{-}h_{\ulmb})+\theta \alpha \right)\left(\tfrac12(h_{\unu}{-}h_{\ulmb}+\theta \beta \right) {\cal C}_{ \Box\, \ulmb}^\unu\, (T_\gamma^{-1})_{\unu}^{\umu}  }{
		\left( {h}_{\ulmb}{-}{h}_{\umu} {+} \theta \gamma\, (|\ulmb|{-}|\umu| )\right)} \\ \notag
\end{gather}
The main difference is that here the recursion gives $(T_\gamma^{-1})^\umu_\ulmb$  in terms 
of $(T_\gamma^{-1})^\umu_{\ulmb+\Box}$, obtained by {\em adding} boxes to $\ulmb$, whereas   
the recursion for $(T_\gamma)^\umu_\ulmb$ is in terms of $(T_\gamma)^{\umu-\Box}_\ulmb$, 
is obtained by {\em subtracting} boxes from $\umu$.

We will return to the relation between $T$ and $T^{-1}$ later on in section~\ref{Jack_to_BlockMatrix}.

\subsubsection*{Higher order Casimirs for $\theta=1$}
\label{sec_hoc}

Before considering the recursion in more detail we would like to discuss higher order Casimirs in the CFT. 
These arise from the relation between the Casimir and CMS system with $BC$ root system. In fact, higher order Casimirs correspond 
to the defining higher order differential operators of the Heckman Opdam $BC$ hypergeometric functions. This is a general statement, but here
we will specialise to $\theta=1$, where formulae are both explicit and simple. 
Generalising the work of~\cite{Shimeno} (who focussed on the purely bosonic case $n=0$) we now show that higher order Casimirs in the supersymmetric case are given by simply 
replacing the variables $z_I$ in a super Schur with the corresponding second order operator $\cD_I$ of~\eqref{operator_D} and conjugating with the Vandermonde (super-) determinant 
\begin{align}\label{hocs2}
	\mathbf{C}_{\umu}^{(\theta=1,a,b,c)}= V^{-1}(z_I)\,P_\umu\Big(\cD_{I}^{(\theta=1;a,b,c)};\theta=1\Big)\,V(z_I)
\end{align}
where the super van der Monde determinant is
\begin{align}
	V(z_I) &= \frac{\prod_{1\leq i<j \leq m}(x_i^{-1}-x_j^{-1})\prod_{1\leq i<j \leq n}(y_i^{-1}-y_j^{-1})}{ \prod_{\substack{1\leq i\leq m\\ 1 \leq j\leq n}}(x_i^{-1}-y_j^{-1})}\ ,
\end{align}
and where $\mathbf{D}_I$ is given in~\eqref{operator_D}.
The quadratic Casimir of~\eqref{casimir_intro} correspond to the above for the one-box super Schur, i.e.~$\umu=\Box$. The latter is simply the sum $\sum_{i} x_i -\sum_j y_j$.

The superblocks $B_{\gamma,\ulmb}$ are eigenvalues of $\mathbf{C}_\ulmb^{(\theta,-\frac12p_{12},-\frac12 p_{43},0)}$ for {\em all} Young diagrams $\umu$ 
\beq\label{eeqh}
\mathbf{C_{\umu}}^{(\theta,-\frac12p_{12},-\frac12 p_{43},0)} B_{\gamma,\ulmb}({\bf 
	z})= { (E_\umu)}_{\gamma,\ulmb}^{(m,n;\theta=1)} B_{\gamma,\ulmb}({\bf z})\ ,
\eeq
The corresponding eigenvalue is
\begin{align}
	{(E_\umu)}_{\gamma,\ulmb}^{(m,n;\theta=1)} = 	P_\umu( 	(E_I)_{\gamma,\ulmb}  ) -\tfrac13(m-n-1)_3
\end{align}
obtained by replacing $z_I$ by $E_I$ in the corresponding Schur polynomial $P_\umu$ where
\begin{align}
	(E_I)_{\gamma,\ulmb}^{(m,n;\theta=1)} = \left\{  \begin{array}{ll}
		(\lambda_i-i+\tfrac12 \gamma +1)(\lambda_i-i+\tfrac12 \gamma)\qquad & I=i=1,..,m\\
		(\lambda'_j-j-\tfrac12 \gamma +1)(\lambda'_j+j-\tfrac12 \gamma)\qquad & I-m=j=1,..,n
	\end{array}       \right. \ .
\end{align}
The case $\umu=\Box$ is the one we discussed in our overview:
one can check that putting the single box Young diagram, this agrees with~\eqref{ev1} for $\theta=1$.  

Note that an eighth order super Casimir in $\cN=4$ SYM considered in~\cite{Caron-Huot:2017vep} corresponds to~\eqref{hocs2}
with $m{=}n{=}2$ and $\mu{=}[2,2]$ for which $P_\umu(z_I){=}(x_1{-}y_1)(x_1{-}y_2)(x_2{-}y_1)(x_2{-}y_2)$.
Higher order Casimirs both in 1d CFTs ($m{=}n{=}\theta{=}1$) and in $\cN{=}4$ SYM ($m{=}n{=}2, \theta{=}1$) are considered in~\cite{arthur}.

\subsection{Explicit formulae for the recursion} \label{explicit_form_sec}

So far the recursion~\eqref{rec2} is entirely combinatoric, written in terms of Young 
diagrams, and our main concern has been to 
emphasize this property, that it depends only on objects that are $(m,n)$ independent.  We will now make the discussion more concrete by considering explicit formulae. 
Remarkably, there are different explicit forms that the recursion can take.
The difference is simply a preference for rows over columns when reading the Young 
diagrams, or vice-versa, or even mixing rows and columns as is more natural in the supersymmetric $(m,n)$ case.

\subsubsection*{Eigenvalue}

Let us begin from $h_{\ulmb}^{(\theta)}$ which in~\eqref{hevalue} we wrote as
\beq\label{eveast}
h_{\ulmb}^{(\theta)}=\sum_{i=1,2,..} \lambda_i (\lambda_i{-}2\theta(i{{-}}1){-}1)\ .
\eeq
It is simple to see that it admits an equivalent representation in which instead we take the sum to run over columns,
\beq
{h}_{\ulmb}^{(\theta)}=-\theta\sum_{j=1,2,..} \lambda'_j ( \lambda'_j-1-\tfrac{2}{\theta} (j-1) )=-\theta h_{\ulmb'}^{{(\frac{1}{\theta})}}\ ,
\label{superHeigen_22}
\eeq
or if $\ulmb$ is a typical $(m,n)$ Young diagram (ie it has $\lambda_m\geq n, \lambda_{n+1}\leq n$ see~\eqref{typical}) 
then it can be rewritten as a combination of these row and column formulae:
\beq\label{evboth}
h_{\ulmb}^{(\theta)}=\sum_{i=1}^m \lambda_i (\lambda_i{-}2\theta(i{{-}}1){-}1)  -\theta\sum_{j=1}^n \lambda'_j ( \lambda'_j-1-\tfrac{2}{\theta} (j-1) ) -h_{n^m}^{({\theta})}\ .
\eeq
where  $h_{n^m}^{(\theta)}= nm((n-1)-\theta(m-1))$. 
All these forms for the eigenvalue can be easily derived from the following combinatoric formula 
\begin{align}\label{hbox}
	{h}^{(\theta)}_{\ulmb} = 2\sum_{(i,j) \in \ulmb}\Big((j{-}1)-\theta(i{-}1)\Big)\ .
\end{align}
where the eigenvalue is given as a sum over boxes in the Young tableau where each box 
is understood to have row $i$ column $j$. 
Then one obtains the equivalent formulae~\eqref{hevalue} or~\eqref{superHeigen_22} by simply 
performing the first sum  in each of the following ways of writing the sum over Young diagram boxes explicitly
\begin{align}\label{resum}
	\sum_{(i,j) \in \ulmb} = \sum_{i=1,2,..} \sum_{j=1}^{\lambda_i} = \sum_{j=1,2,..} \sum_{i=1}^{\lambda'_i} = \sum_{i=1}^m \sum_{j=1}^{\lambda_i} + \sum_{j=1}^n \sum_{i=1}^{\lambda'_i}-   \sum_{i=1}^m \sum_{j=1}^{n}\ .
\end{align}
with the last formula for typical (long) $(m,n)$ Young diagrams only.

In the numerator of the recursion~\eqref{rec2} we have the expression 
\beq
\left(\tfrac12(h_{\umu}{-}h_{\unu})+\theta \alpha \right)\left(\tfrac12(h_{\umu}{-}h_{\unu})+\theta \beta \right) \qquad \unu \in \{\ulmb-\Box\}\,.
\eeq
Since  $\unu$ is related to $\umu$ by subtracting a box in position $(i,j)$, 
 clearly from~\eqref{hbox} we have $\frac{1}{2}(h_{\umu} - h_{\unu})=
(j{-}1)-\theta(i{-}1)$.

\subsubsection*{Structure constants}

Let us come to ${\cal C}_{ \Box\, \umu}^{\umu+\Box}$ in~\eqref{rec2}. The 
corresponding combinatorial formula is well known and can be found in \cite{Stanley}. 
It can be rearranged quite explicitly 
and  we find\footnote{See also \cite{Lassalle_1}. We actually first derived this 
formula using computed algebra, by seeking a generalisation of the formula for two-row polynomials, 
which can be derived with pencil and paper. Then, we confirmed its expression by rewriting the combinatorial 
formula coming from the Pieri rule \cite{Macdonald_book}.}  
\begin{align}\label{boldf_section}
	{\cal C}_{ \Box\, \umu}^{\umu+\Box_{ij}} = \mathbf{f}_{\umu}^{(i)}(\theta)=
	& \
	\prod_{k=1}^{i-1} \frac{
		\left( \theta{(i{-}k{+}1)}{-}1{+}\mu_{k}{-}\mu_i\right) \left( \theta{(i{-}k{-}1)} {+}\mu_{k}{-}\mu_i \right)}{ 
		\left( \theta{(i{-}k)}{-}1{+}\mu_{k}{-}\mu_i \right)\left( \theta {(i{-}k)}{+}\mu_{k}{-}\mu_i \right) }\ .
\end{align}%
where we are using the notation $\umu+\Box_{ij}$ to mean that we add a box to row $i$ and column 
$j$ of the Young tableau so that we have $j=\mu_i+1$ and $i=\mu'_j+1$.
  The above formula is written in terms of Young diagram row lengths $\mu_i$. Alternatively, we might prefer a rewriting of ${\cal C}_{ \Box\, \umu}^{\umu+\Box_{ij}}$ over columns. 
 
 To do so we define first, $\mathbf{g}^{(i)}_{\umu}$,  obtained by multiplying  $\mathbf{f}^{(i)}_{\umu}$ 
 by the following normalisation in terms of $C$ symbols (defined more generally later in~\eqref{recap_Csymbol}) 
\begin{align}\label{boldg_section}
	\mathbf{g}^{(i)}_{\umu}(\theta) &\equiv 	\	\frac{ \Pi_{\umu }(\theta) }{ \Pi_{\umu+\Box_{ij}}(\theta) } \ \mathbf{f}^{(i)}_{\umu}(\theta)\\
	& \qquad \Pi_{\underline{\kappa} }(\theta)=
		\frac{C_{\underline{\kappa}}^-(\theta;\theta)}{C_{\underline{\kappa}}^-(1;\theta)} =
		 \frac{ \prod_{(ij)\in\underline{\kappa} } \left( \kappa_i{-}j+\theta(\kappa'_j{-}i)+\theta \right) }{ \prod_{(ij)\in\ulmb } \left( \kappa_i{-}j+\theta(\kappa'_j{-}i)+1 \right)}\ .
\end{align}
Note that this can be rewritten as an explicit function of the row lengths $\mu_i$ ($i=1,..,n$) only%
\footnote{\label{fn27}To do this we
	 use  explicit formulae for $C^-_{\underline{\kappa}}(w;\theta)$%
from~\cite{Opdam},
	\begin{align}	
	C^{-}_{\umu}(\theta;\theta)&= \frac{ \prod_{k=1}^{\mu'_1} \frac{ (\mu_k+\theta(\mu'_1-k)+\theta-1)!  }{ (\theta-1)!} }{ \prod_{1\leq k_1<k_2\leq \mu'_1} (\mu_{k_1}{-}\mu_{k_2}{+}\theta(k_2{-}k_1) )_\theta }   \qquad 
	C^{-}_{\umu}(1;\theta)=\frac{ \prod_{k=1}^{\mu'_1} (\mu_k+\theta(\mu'_1-k))!  }{ \prod_{1\leq k_1<k_2\leq \mu'_1} (\mu_{k_1}{-}\mu_{k_2}{+}\theta(k_2{-}k_1)-(\theta{-}1) )_\theta }\, . \notag
	\end{align}
after which  the majority of contributions to ${ \Pi_{\umu }(\theta) }\big/{ \Pi_{\umu+\Box_{ij}}(\theta) }$ simplify in the ratio.}
\begin{align}\label{explicit_boldg_section}
{\bf g}^{(i)}_{\umu}(\theta)&
=\frac{\mu_i-\mu_{i+1}+1}{\mu_i-\mu_{i+1}+\theta}\prod_{k=i+1}^n  \frac{ 
	(\theta(k-i+1)+\mu_i-\mu_k) ( \theta(k-i)+\mu_i-\mu_{k+1}+1 ) }{  (\theta(k-i+1)+\mu_i-\mu_{k+1}) ( \theta(k-i)+\mu_i-\mu_k+1) }\notag\\
\end{align}
(with $\mu_{n+1}$ defined to vanish).

The structure constant~\eqref{boldf_section} then has an alternative writing in terms of column  lengths using  $\mathbf{g}^{(j)}_{\umu'}$
as follows%
\footnote{A posteriori, 
	${\bf f}$ can also be written in terms of the
	$A$-type binomial coefficient, built out of the $A$-type interpolation polynomials of Okounkov \cite{Okounkov_Olshanski}, 
	which can be seen as the limit $u=\infty$ in \eqref{BCpsi}. The duality ${\bf f}\leftrightarrow {\bf g}$ can be proven using the identification with the $A$-type binomial coefficient.} 
\begin{align}\label{bosoniz_fandg}
{\cal C}_{ \Box\, \umu}^{\umu+\Box_{ij}}=	\mathbf{f}^{(i)}_{\umu}(\theta)=\tfrac{1}{\theta}\, \mathbf{g}^{(j)}_{\umu'}(\tfrac{1}{\theta})\,,\qquad \mu'_j= i-1\,,\quad \mu_i=j-1\ .
\end{align}
where recall that $(i,j)$ are the coordinates of $\Box_{ij}$.

\subsubsection*{Equivalent forms of the recursion} \label{equiv}

Choosing all row-type formulae is  the most natural from a bosonic CFT point 
of view, since in a physical theory with conformal symmetry, we would be summing over $P_{\umu}$  
with a finite  number of rows but an arbitrary number of columns such that $\ulmb\subseteq\umu$, 
in order to construct $B_{\gamma,\ulmb}$. With this choice then
we use~\eqref{eveast} and~\eqref{boldf_section} to  write the recursion~\eqref{rec2} 
in terms of row lengths explicitly as 
\bea\label{rec_section_scblocks}
\rule{0pt}{1cm}
(T_\gamma)_{\ulmb}^{\umu}= 
\frac{ \sum_{i}^{} \left(\mu_i{-}1{-} \theta(i{-}1{-}\alpha)\right)\left(\mu_i{-}1{-} \theta(i{-}1{-}\beta)\right) \mathbf{f}^{(i)}_{\umu{-}\Box_i} (\theta) \, (T_\gamma)_{\ulmb}^{\umu{-}\Box_i}  }{
\sum_i (\mu_i- \lambda_i)(\lambda_i+\mu_i+\theta \gamma -2\theta(i-1)-1)}  \\ \notag
\eea
where $\umu{-}\Box_i:=[\mu_1,..,\mu_{i-1},\mu_i-1,\mu_{i+1},...]$.
Note that ${\bf f}$ imposes that $\mu_i>\mu_{i+1}$ automatically. 
In fact if it was the case that $\mu_{i}=\mu_{i-1}+1$ for some $i$,
then second term in the denominator of $\mathbf{f}^{(i)}_{\umu{-}\Box_i}$ 
in \eqref{boldf_section} would vanish for $k=i-1$.

If instead we choose all column-type explicit formulae
then using~\eqref{superHeigen_22} and~\eqref{boldg_section} we write the  recursion~\eqref{rec2}  in terms of column  lengths explicitly as
\bea			\label{south_recursion} 
	\  \  (T_\gamma)_{\ulmb}^{\umu}
	= 
	\frac{\sum_{j}\left(\mu'_j {-}1{-}\alpha{-} \frac{j{-}1}{\theta} \right)\left(\mu'_j {-}1{-}\beta{-}\frac{j{-}1}{\theta}\right) 
	\mathbf{g}^{(j)}_{\umu'{-}\Box_j}(\tfrac1\theta)  (T_\gamma)_{\ulmb}^{(\umu'{-}\Box_j)'}}
{
	\sum_j (\lambda'_j-\mu'_j)(\lambda'_j+\mu'_j- \gamma -\frac{2}{\theta}(j-1)-1)}  \rule{.5cm}{0pt} \\ \notag
\eea
where $\umu'{-}\Box_j:=[\mu'_1,..,\mu'_{j-1},\mu'_j-1,\mu'_{j+1},...]$.

Finally we could also split the sum $\sum_{\unu\in\{\umu-\Box\}}$ in \eqref{rec2} 
into east and south part arbitrarily.  The outcome for $(T_{\gamma})_{\ulmb}^{\umu}$ will remain unchanged. 
This is most natural for an $(m,n)$ supersymmetric theory and indeed is crucial if we wish to analytically continue consistently with the $(m,n)$ structure. We will discuss this supersymmetric splitting in the section \ref{analytic_cont_sec}.

\subsection{Special solutions: the half-BPS superconformal block}

There is a special case in which we can solve the recursion for all coefficients explicitly with a particularly simple solution.
It is the case corresponding to an empty diagram $\ulmb=[\varnothing]$, 
and therefore corresponding   to a half-BPS superconformal block exchanged 
in a propagator structure with $\gamma$ propagators going from ${\cal O}_{p_1}{\cal O}_{p_2}$ 
to ${\cal O}_{p_3}{\cal O}_{p_4}$. The solution is
\beq\label{solu_halfBPS}
(T_{\gamma})^{\umu}_{[\varnothing]}= \frac{ C^0_{\umu}(\theta\alpha;\theta)C^0_{\umu}(\theta\beta;\theta) }{ C^0_{\umu}(\theta \gamma;\theta) }\frac{1}{C^-_{\umu}(1;\theta)}
\eeq
 where the combinatorial symbols are
\bea
C_\ulmb^0(w;\theta)&=&\ \,\prod_{(ij)\in\ulmb } \left(j{-}1-\theta(i{-}1) +w\right)\notag\\
C_\ulmb^-(w;\theta)&=&\ \,\prod_{(ij)\in\ulmb } \left( \lambda_i{-}j+\theta(\lambda'_j{-}i)+w \right)\ .
\label{C0symbol_sec5}
\eea

Let us  verify that this solution indeed satisfies the recursion using the row-type recursion~\eqref{rec_section_scblocks}. We will see that  despite the final simplicity of the result, the recursion is solved in a very non trivial way. 
Afterwards we will present a more direct argument based on the fact that the Casimir annihilates the 
half-BPS superconformal block, therefore it can be split into simpler independent hypergeometric equations
studied by Yan in \cite{Yan}.

Consider  the recursion~\eqref{rec_section_scblocks} with $\ulmb=[\varnothing]$
\beq\label{half_bps_rec}
(T_\gamma)_{[\varnothing]}^{\umu}= 
\frac{ \sum_{i}^{} \left(\mu_i{-}1{-} \theta(i{-}1{-}\alpha)\right)\left(\mu_i{-}1{-} \theta(i{-}1{-}\beta)\right) \mathbf{f}^{(i)}_{\umu{-}\Box_i} (\theta) \, (T_\gamma)_{[\varnothing]}^{\umu{-}\Box_i}  }{
\left( {h}_{\umu}+ \theta \gamma\, |\umu| \right)}\ .
\eeq
Inserting the solution~\eqref{solu_halfBPS} the terms   $\left(\mu_i{-}1{-} \theta(i{-}1{-}\alpha)\right)$ and $C^0_{\umu{-}\Box_i}(\theta\alpha;\theta)$ combine and 
simplify with the LHS~$C^0_{\umu}(\theta\alpha;\theta)$. Similarly for the terms with $\beta$.  We are left with needing to prove the non trivial statement
\beq
\ 1=\frac{ 1}{ \left( {h}_{\umu}+ \theta \gamma\, |\umu| \right)}\sum_i \left( {\bf f}^{(i)}_{\umu{-}\Box_i} \frac{ C^-_{\umu}(1;\theta) }{ C^-_{\umu-\Box_i}(1;\theta) } \right)
( (\mu_i-1)-\theta(i-1)+\theta\gamma)\ .
\eeq
We can deal with $C^-_{\umu}$ by using the explicit  formulae of footnote~\ref{fn27}
and find that 
\beq
\ 
{\bf f}^{(i)}_{\umu-\Box_i} \frac{ C^-_{\umu}(1;\theta) }{ C^-_{\umu-\Box_i}(1;\theta) } = (\mu_i+\theta( \mu'_1-i) )\prod_{k\neq i} \frac{ \mu_i-\mu_k+\theta(k-i-1) }{ \mu_i-\mu_k+\theta(k-i) }\ .
\eeq
Proving the recursion is now equivalent to proving that
\begin{align}
&\sum_i (\mu_i+\theta( \mu'_1-i) )\prod_{k\neq i} \frac{ \mu_i-\mu_k+\theta(k-i-1) }{ \mu_i-\mu_k+\theta(k-i) } = |\umu| \\
&\sum_i (\mu_i+\theta( \mu'_1-i) )\prod_{k\neq i} \frac{ \mu_i-\mu_k+\theta(k-i-1) }{ \mu_i-\mu_k+\theta(k-i) } ( \mu_i-\theta i)= 2\sum_{(ij)\in \umu }(j-\tfrac{1}{2})-\theta(i-\tfrac{1}{2}) \notag
\end{align}
where the RHS~corresponds to a decomposition of ${h}_{\umu}+ \theta \gamma$. 
Note that even the first one of these relations is quite non trivial, since the common 
denominator in the LHS~has to cancel out in the end, and this only happens after taking the sum. 
These two relations are proven in a beautiful way in \cite{Lassalle_nice} by using the Lagrange Lemma for alphabets.

The superconformal block $B_{\gamma,[\varnothing]}$ itself is now obtained by summing up 
$(T_{\gamma})^{\umu}_{[\varnothing]}$ with super Jack polynomials.  It takes the form of a multivariate$~_2F_1$ series, 
\beq
B_{\gamma,[\varnothing]}=\left(\frac{\prod_i x_i^\theta}{\prod_j y_j}\right)^{\!\!\frac{\gamma}{2}}\sum_{\umu}
 \frac{ C^0_{\umu}(\theta\alpha;\theta)C^0_{\umu}(\theta\beta;\theta) }{ C^0_{\umu}(\theta \gamma;\theta) } \frac{P_{\umu}({\bf z};\theta)}{C^-_{\umu}(1;\theta)}. 
\eeq
In the bosonic $(m,0)$ theory this  is indeed equal to a well known multivariate hypergeometric function  \cite{Yan}.%
\footnote{As discussed at the end of section~\ref{sec:HO} we believe this is also a certain $S_n$ combination  of Harish Chandra contributions to the corresponding  Heckman-Opdam hypergeometric.}
In particular, it is known to satisfy a set of $m$ hypergeometric equations, which in our notation take the form
\beq
\!\!\!\!\!\!
\left[ z_I^{{-}1} {\bf D}_I^{(\theta,\alpha,\beta,\gamma)}{-}\theta(m{-}1)(1{-}z_I) {+} \theta \sum_{J\neq I} \frac{ z_I (1{-}z_I)  \partial_I {-}z_J(1{-}z_J) \partial_J}{z_I{-}z_J}\right]B_{\gamma,[\varnothing]}=0 \quad;\quad I=1,\ldots m
\label{yan_eq}
\eeq
where ${\bf D}_I$ is the same operator introduced in \eqref{operator_D}. The trick to recover our 
Casimir ${\bf C}$ is to sum $\sum_I z_I\times \eqref{yan_eq}$  and put together
$
{-}(m-1){+}\sum_{J\neq I}\frac{z_I}{z_I-z_J} = +\sum\frac{ z_J}{z_I-z_J}
$.

Since $h_{\varnothing}+\gamma|\varnothing|=0$, the half-BPS block is annihilated by the Casimir, 
and therefore the system of equations \eqref{yan_eq} is equivalent to the original Casimir eigenvalue problem.

\subsection{General features and non-trivial $\gamma$-dependence} \label{example_Acoeff}

The solution of the recursion for $T_\gamma$ reveals many non trivial features. 
In order to systematise the discussion, let us first show that the $\alpha$ and $\beta$ dependence of 
$(T_{\gamma})_{\ulmb}^{\umu}$ always takes the following form
\beq\label{attempt_gen_sol}
(T_{\gamma})_{\ulmb}^{\umu} = {C^0_{\umu/\ulmb}(\theta\alpha;\theta) C^0_{\umu/\ulmb}(\theta\beta;\theta) }\times (T^{\tt\, rescaled}_\gamma)_{\ulmb}^{\umu} 
\eeq
with $(T^{\tt\, rescaled}_\gamma)_{\ulmb}^{\umu}$ then independent of $\alpha,\beta$. 
Here by definition $C^0_{\umu/\ulmb}\equiv C^0_{\umu}/C^0_{\ulmb}$, thus\footnote{The $C_\umu^0(w;\theta)$ was defined 
in \eqref{C0symbol_sec5}. Here we are giving the final result for ${C^0_{\umu/\ulmb}(\theta\alpha;\theta) C^0_{\umu/\ulmb}(\theta\beta;\theta) }$.} 
\beq\label{C0ontheeast}
{C^0_{\umu/\ulmb}(\theta\alpha;\theta) C^0_{\umu/\ulmb}(\theta\beta;\theta) }
=\prod_{(ij)\in\, \umu/\ulmb} (j{-}1{-}\theta(i{-}1{-}\alpha))(j{-}1{-}\theta(i{-}1{-}\beta)).
\eeq
The proof of \eqref{attempt_gen_sol} is immediate. In fact, the ratio of $C^0$ satisfies the relation
\beq
C^0_{\umu/\ulmb}(\theta w;\theta)=\left(\mu_i-1- \theta(i-1-w)\right)C^0_{\umu-\Box_i/\ulmb}(\theta w;\theta)
\eeq
and therefore in the row-type formulae \eqref{rec_section_scblocks} for the recursion, 
the $\alpha$ and $\beta$ dependence drops.  This is indeed the same argument we saw at work 
in the half-BPS solution, around \eqref{half_bps_rec}, with $\ulmb=[\varnothing]$.
Then, it follows from \eqref{attempt_gen_sol} that $(T^{\tt\, rescaled}_\gamma)_{\ulmb}^{\umu}$ satisfies a simpler recursion, 
\beq
(T^{\tt\, rescaled}_\gamma)_{\ulmb}^{\umu}= 
\frac{ \sum_{i}^{} \mathbf{f}^{(i)}_{\umu-\Box_i} (\theta) \, (T^{\tt \, rescaled}_{\ulmb})_{\ulmb}^{\umu-\Box_i}  }{\sum_i (\mu_i- \lambda_i)(\lambda_i+\mu_i+\theta \gamma -2\theta(i-1)-1)},
\eeq
and $(T^{\tt\, rescaled}_\gamma)_{\ulmb}^{\umu}$ is  \emph{only} a function of $\gamma,\theta$ and the Young diagrams $\umu$ and $\ulmb$.

The recursion for $(T^{\tt\, rescaled}_\gamma)_{\ulmb}^{\umu}$ can be turned into column or mixed 
type formulae, exactly as for $(T_\gamma)_{\ulmb}^{\umu}$ itself, as we did in section~\ref{equiv}.
The $C^0$ rescaling will change accordingly, see in particular section \ref{alphabetarevisited}.

Given the various explicit formulae provided in the previous section \ref{explicit_form_sec}, 
it is not difficult to implement the recursion on a computer
and solve it in many cases. Let us look at a couple of examples to illustrate the general features:
\beq\label{example_gamma}
\!\!\!\!
(T^{\tt\,rescaled}_{\gamma})_{[3,1]}^{[5,3]}\ =
\frac{ 162 +222\theta+67 \gamma\theta-168 \theta^2 +109 \gamma\theta^2 +7\gamma^2\theta^2 +24 \theta^3-36 \gamma\theta^3 +13\gamma^2\theta^3 }{ 
2(1+\theta)(3+\theta)_2 (6+\gamma\theta)_2 (2-2\theta+\gamma\theta)_2(4-\theta+\gamma\theta)_2 }
\eeq
The dependence on $\gamma$ in the denominator as a product of linear factors in $\gamma$ is simple to understand: 
it can only come from putting together all denominators encountered upon solving the recursion, 
and these are all of the form $(\theta\gamma-\theta \mathbb{N}+ \mathbb{N})$ 
with integers coming from Young diagrams.  
The numerator on the other hand is fairly complicated, degree two in $\gamma$ and degree three in $\theta$.
This degree depends on the Young diagrams.
Another example shows this more clearly,  
\begin{align}
\!\!\!\! %
(T_{\gamma}^{\tt rescaled})_{[3,2,1]}^{[6,5,3]}\ \sim\ & +21208 +6 (1861 \gamma +7012) \theta +\left(1943 \gamma ^2+26524 \gamma -56312\right) \theta ^2\notag\\
&
+\left(111 \gamma ^3+5312 \gamma ^2-17918 \gamma +11112\right) \theta ^3\notag\\
&
+\left(339 \gamma ^3-965 \gamma ^2-796 \gamma +2560\right) \theta ^4 \notag\\
&
+\left(66 \gamma ^3-410 \gamma ^2+800 \gamma -480\right) \theta ^5\ .
\end{align}
(we only display the complicated polynomial factor, and we omitted the various linear factors  of the form $(\theta\gamma-\theta 
\mathbb{N}+ \mathbb{N})$.)

The point is that guessing the $\gamma$ dependence of $(T_{\gamma})_{\ulmb}^{\umu}$, for arbitrary $\ulmb$, $\umu$, and $\theta$, 
would be quite challenging. However, as anticipated in the introduction, 
due to the relation with Jacobi polynomials we will find in section \ref{SCBinomialCoeff_sec} an alternative 
description for this complicated factor  as 
an interpolation polynomial.

\section{Superconformal blocks (II): analytic continuation}\label{analytic_cont_sec}

We have understood in the previous section how to construct a superconformal block $B_{\gamma,\ulmb}$
given a Young diagram $\ulmb$ for which e.g. its row lengths   $\lambda_i \in \mathbb{N}$ 
with $\lambda_i\geq \lambda_{i+1}$.

For physical applications in an interacting CFT however, non-protected operators exchanged in the OPE acquire an anomalous dimension and thus it is 
crucial to be able to understand non-integer quantum numbers, at least for the dilation weight. Even though 
the other quantum numbers remain integer in a unitary CFT,  it is nevertheless also very 
useful to consider  analytically continuing these too (for example the spin, see~\cite{Caron-Huot:2017vep} for example).

For general $\theta$, we do not have explicit solutions for either $B_{\gamma,\ulmb}$ or $(T_{\gamma})_{\ulmb}^{\umu}$ 
to look at, in order to discuss analytic continuation beyond Young diagrams.
In fact, only the case $\theta=1$ has been solved for both $B_{\gamma,\ulmb}$ and $(T_{\gamma})_{\ulmb}^{\umu}$ 
in \cite{Doobary:2015gia}. We will then have to look at the recursion. 
The combinatorial form given in~\eqref{rec2} 
is not immediately useful, since does not lend itself to analytic continuation. However, from the explicit 
forms of the recursion~\eqref{rec_section_scblocks},\eqref{south_recursion} we can obtain analytic results  for the coefficients $T_\gamma$ as rational functions of the row lengths $\lambda_i$ or column lengths $\lambda'_i$ respectively.

The logic will be quite simple. For the row type recursion~\eqref{rec_section_scblocks} we will think of $(T_{\gamma})_{\ulmb}^{\umu}$ as a function of 
$\ulmb\in \mathbb{C}$, with
$\umu$ given by $\umu=\ulmb+\vec{n}$ with $n_i\in \mathbb{Z}^+$. The recursion 
takes $\umu$ and goes back to $\ulmb$ by negative integer shifts, regardless of whether the $\lambda_i$ 
are integer or not. In this framework, the various representations of the recursion, 
i.e.~whether row-type or vertical-type or a mixed one, lead to different ways of analytically continuing 
the $(T_{\gamma})_{\ulmb}^{\umu}$ in the variables describing the Young diagrams.
Crucially we know that all these analytic continuations of $T_\gamma$ must 
coincide in the cases when $\ulmb$ and $\umu$ return to values such that they represent a valid Young diagrams.

\label{sec_various_analcont}

\subsection{Row-type representation on the east}

The row-type representation of the recursion on the east, which we derived in the previous section using Young diagrams technology, 
is given in \eqref{rec_section_scblocks}.  This has the following $(m,0)$ analytic continuation in the row lengths $\lambda_i$, 
\begin{gather}\label{rec_section_scblocks_analyt}
\!\!\!\!\!\!(T_\gamma)_{\ulmb}^{\umu}= 
\frac{ \sum_{i=1}^{m} \left(\mu_i{-}1{-} \theta(i{-}1{-}\alpha)\right)\left(\mu_i{-}1{-} \theta(i{-}1{-}\beta)\right) \mathbf{f}^{(i)}_{\umu{-}\ue_i} (\theta) \, (T_\gamma)_{\ulmb}^{\umu{-}\ue_i}  }{
\sum_{i=1}^m (\mu_i- \lambda_i)(\lambda_i+\mu_i+\theta \gamma -2\theta(i-1)-1)} \\[.2cm]
\lambda_i\in\mathbb{C},\qquad\mu_i=\lambda_i +n_i, \qquad n_i \in \mathbb{Z}^+,\qquad (T_\gamma)_{\ulmb}^{\ulmb} =1,\qquad (T_\gamma)_{\ulmb}^{\umu}=0 \text{ if } n_i<0 \notag
\end{gather}
Here  $\umu$ and $\ulmb$ are parameters such that  $\mu_i-\lambda_i=n_i$ is a positive integer, 
but $\lambda_i\in \mathbb{C}$ and $\ue_i$ is the usual basis vector with a 1 in position $i$ and zeroes elsewhere.  
As we discussed below \eqref{rec_section_scblocks} this recursion {\em automatically} gives 
$(T_\gamma)_{\ulmb}^{\umu}=0$ if $\mu_{i}=\mu_{i-1}+1$.  Thus if $\ulmb$ corresponds to a Young diagram, 
meaning that $\lambda_i-\lambda_{i-1} \in \mathbb{Z}^+$, then  $(T_\gamma)_{\ulmb}^{\umu}=0$ unless  $\umu$ 
also corresponds to the row lengths of a Young diagram.  This is a key point which allows the analytic continuation 
away from the Young diagram: if this was not the case, then the solution to the above recursion would not recover the correct solution when $\ulmb$ becomes a Young diagram.
The expression for $\mathbf{f}_{\umu}^{(i)}(\theta)$ is given in~\eqref{boldf_section} and is rational in $\lambda_i$ and $\theta$.
For fixed, integer $\mu_i-\lambda_i$,  the recursion~\eqref{rec_section_scblocks_analyt} has a solution which is manifestly 
rational in $\lambda_i,\theta,\alpha,\beta,\gamma$. We can thus view the solution of \eqref{rec_section_scblocks_analyt} 
as an analytic continuation of the solution of~\eqref{rec_section_scblocks} which is valid only when $\lambda_i$ are the 
row lengths of Young diagrams.

An important bonus of the solution of \eqref{rec_section_scblocks_analyt} is the manifest shift symmetry~\eqref{deform_overview_sec} (for any value of $\theta$)
\beq\label{shift_east_pre}
\lambda_i \rightarrow \lambda_i{-}\theta\tau' \,, \qquad \mu_i \rightarrow \mu_i-\theta\tau'  \,, 
	\qquad \gamma \rightarrow \gamma+2\tau'\,, \qquad 
 \!i=1,\ldots m\ .
\ 
\eeq
We can see this invariance  immediately. First,  the combinations $(\mu_i+\theta\alpha)$, $(\mu_i+\theta\beta)$, 
are manifestly invariant, since both  $\alpha$ and $\beta$ have the form $\frac{\gamma-p_{ij}}{2}$. 
Then, the combination of eigenvalues appearing in the denominator, i.e.~$(\mu_i- \lambda_i)(\lambda_i+
\mu_i+\theta \gamma -2\theta(i-1)-1)$,  is also manifestly invariant. 
Finally, $\mathbf{f}_{\umu-\ue_i}$ only depends on differences $\mu_i-\mu_j$ and so  is invariant.

The simplest analytic continuation we can think of
is the $(1,0)$ solution given in section \ref{sec_rank1},   which we repeat here for convenience,
\beq
(T_{\gamma})^{[\mu]}_{[\lambda]}=
	\frac{(\lambda+{\theta \alpha})_{\mu-\lambda}(\lambda+{\theta \beta})_{\mu-\lambda}}{(\mu{-}\lambda)!(2\lambda+{\theta \gamma})_{\mu-\lambda}}\ .
\eeq
We see that indeed, for fixed integer $\mu-\lambda$ this is a rational function of $\lambda,\theta,\alpha,\beta,\gamma$.
(Of course, having an explicit solution we can now also see how to analytically continue in $\umu$ if we desired, just by changing the Pochhammers to Gamma functions.)

Summarising, we have shown that defining  $(T_{\gamma})_{\ulmb}^{\umu}$ via \eqref{rec_section_scblocks_analyt} 
produces a rational (and hence) analytic function of $\ulmb,\theta,\alpha,\beta,\gamma$, which is also invariant under the shift symmetry~\eqref{shift_east_pre}.

\subsection{Column-type representation on the south}

An alternative analytic continuation of $(T_\gamma)_{\ulmb}^{\umu}$ can be obtained by viewing 
the Young diagram as function of the column lengths $\lambda'_j$, rather than the row lengths $\lambda_i$.
The column-type representation of the recursion on the south was derived in \eqref{south_recursion}. 
It has the following  $(0,n)$ analytic continuation,
\begin{gather}		\label{south_recursion_nonshift} 
	\!\!\!\!\!\!\!\!(T_\gamma)_{\ulmb}^{\umu}
	= 
	\frac{\sum_{j=1}^n\left(\mu'_j {-}1{-}\alpha{-} \frac{j{-}1}{\theta} \right)\left(\mu'_j {-}1{-}\beta{-}\frac{j{-}1}{\theta}\right) 
	\mathbf{g}^{(j)}_{\umu'{-}\ue_j}(\tfrac1\theta)  (T_\gamma)_{\ulmb}^{(\umu'{-}\ue_j)'}}
{
	\sum_{j=1}^n (\lambda'_j-\mu'_j)(\lambda'_j+\mu'_j- \gamma -\frac{2}{\theta}(j-1)-1)} \\[.2cm]
\lambda'_j\in\mathbb{C},\qquad\mu'_j=\lambda'_j +n'_j, \qquad n'_j \in \mathbb{Z}^+,\qquad (T_\gamma)_{\ulmb}^{\ulmb} =1,\qquad (T_\gamma)_{\ulmb}^{\umu}=0 \text{ if } n'_j<0 \notag
\end{gather}
where the explicit formula for ${\bf g}$ was given already in \eqref{explicit_boldg_section}
Similarly to the recursion on the east, for fixed integer $\mu'_j-\lambda'_j$, the solution is clearly a rational function of $\lambda'_j,\theta,\alpha,\beta,\gamma$.

The simplest solution of \eqref{south_recursion_nonshift} we can look at 
is  the $(0,1)$ solution studied in section \ref{sec_rank1}. %
Namely
\beq\label{example_01_analcont}
	(T_{\gamma})^{[1^{\mu'}]}_{[1^{\lambda'}]}=(-1)^{ \mu'-\lambda'}
	\frac{(\lambda'{-}\alpha)_{\mu'-\lambda'}  
		(\lambda'{-}\beta)_{\mu'-\lambda'}}{(\mu'-\lambda')! 
		(2\lambda'{-}\gamma)_{\mu'-\lambda'}}\ 
	\frac{(\lambda'+1)_{\mu'-\lambda'}}{(\lambda'+\tfrac1\theta)_{\mu'-\lambda'}}\
	 .
\eeq

Comparing with the row-type solution from \eqref{rec_section_scblocks_analyt} there is one small difference, 
which can appreciated also in \eqref{example_01_analcont}. This $(T_{\gamma})^{\umu}_{\ulmb}$ 
is not quite invariant under the $n=1$ case of the shift symmetry~\eqref{shift_east_pre}
\beq\label{shift_south_pre}
 \lambda'_j \rightarrow \lambda'_j{+}\tau' \,, \qquad \mu'_j \rightarrow \mu'_j{+}\tau' \,, \qquad 	\gamma \rightarrow \gamma+2\tau'\,, \qquad \!j=1,\ldots n\ .
\eeq
The problem is a simple normalisation issue, due to the fact that we are expanding in 
$BC_{0,n}$ super Jack polynomials rather than directly in $BC_n$ Jack polynomials.  These two are essentially equal but have  a  different 
normalisation, (e.g see~\eqref{sJduality} with $m=0$)
\beq
\label{sJduality2}
P^{(n)}_{\ulmb'}({\bf y}; \tfrac{1}{\theta}) = (-1)^{|\ulmb|} \Pi_{\ulmb}(\theta) P^{(0,n)}_{\lambda}(|{\bf y};\theta)\ .
\eeq
Thus the combination of superJack $\times$ coefficient $T_\gamma$ does satisfy shift invariance.

\subsection{Supersymmetric representation}

We now come to a form of the recursion which gives analytic (indeed rational) results in the 
variables $\lambda_i,\lambda'_j$,  the first $m$ row lengths and first $n$ column lengths, 
for long (or typical,  the see discussion around~\eqref{typical} for these meanings) $(m,n)$ Young diagrams $\ulmb$. 
This is  suited for supersymmetric theory described 
by a $(m,n)$ theory. It  includes the previous cases as special cases by taking $m=0$ or $n=0$.

We first give a version of the recursion~\eqref{rec2} for {\em long} reps $\ulmb$ in the $(m,n)$ theory, which gives 
rational results in $\lambda_i,\lambda'_j$.  To do this we split the RHS~of the recursion into south 
and east components and define it as follows, 
\begin{align}
	\label{super_recursion2} 
&
\!\!\!\Big({h}_{\umu}{-}{h}_{\ulmb} {+} \theta \gamma\, (|\umu|{-}|\ulmb| )\Big)  (T_\gamma)_{\ulmb}^{\umu}
						=  \notag\\[.2cm]
&
\rule{1cm}{0pt}
\sum_{i=1}^{m } \left(\mu_i{-}1{-} \theta(i{-}1{-}\alpha ) \right)\left(\mu_i{-}1{-} \theta(i{-}1{-}\beta) \right) 
					\mathbf{f}^{(i)}_{\umu{-}\ue_i}(\theta) (T_\gamma)_{\ulmb}^{\umu{-}\ue_i} {+}%
\notag
\\
& 
\rule{.5cm}{0pt}
{+}\theta\sum_{j=1}^{n }\left(\mu'_j {-}1{-}\alpha{-} \tfrac{j{-}1}{\theta} \right)\left(\mu'_j {-}1{-}\beta{-}\tfrac{j{-}1}{\theta}\right) 
					\mathbf{g}^{(j)}_{\umu'{-}\ue_j}(\tfrac1\theta)  (T_\gamma)_{\ulmb}^{(\umu'{-}\ue_j)'} %
					\\[.2cm]
					&\lambda_i,\lambda'_j\in\mathbb{C},\quad\mu'_j=\lambda'_j +n'_j,\mu_i=\lambda_i +n_i \quad n_i,n'_j \in \mathbb{Z}^+,\quad (T_\gamma)_{\ulmb}^{\ulmb} =1,\quad (T_\gamma)_{\ulmb}^{\umu}=0 \text{ if } n_i,n'_j<0 \notag
\end{align}
where~\eqref{evboth} is used for ${h}_{\umu},{h}_{\ulmb}$ in the above.
As in previous cases, for given integers $n_i,n'_j$ this recursion yields  a solution which is  
manifestly rational in $\lambda_i, \lambda'_j, \theta,\alpha,\beta,\gamma$  and therefore gives  
a supersymmetric-type analytic continuation in the variables $\lambda_i,\lambda'_j$ of $T_\gamma$.

For example the explicit solution for the $(1,1)$ theory (relevant for 1d and 2d supersymmetric theories)
is
\begin{align}
	&
	(T^{}_\gamma)^{\umu}_{\ulmb}= %
	\frac{  (\mu'_1-1)! (\frac{1}{\theta})_{\lambda'_1-1} }{ (\lambda'_1-1)!  (\frac{1}{\theta})_{\mu'_1-1}  } 
	\frac{
		\theta^{2\lambda'_1-2\mu'_1}\, {C^0_{\umu/\ulmb}(\theta\alpha;\theta) C^0_{\umu/\ulmb}(\theta\beta;\theta) } 
	}{  (\mu'_1-\lambda'_1)! (\mu_1-\lambda_1)!  }\ \times \notag
	\\[.2cm]
	&\rule{1.5cm}{0pt}  \frac{
		(\lambda_1+\theta\lambda'_1-\theta)( \mu_1+\theta\mu'_1-1) +(1-\theta)(\lambda_1-\mu_1) \frac{ (\theta\gamma+\lambda_1-\theta\mu'_1)}{ (\theta\gamma+\lambda_1-\theta\lambda'_1) }  }{
		(\lambda_1+\theta\lambda'_1-1)(\mu_1+\theta \mu'_1-\theta) (2\lambda_1+\theta\gamma)_{\mu_1-\lambda_1}  (2\lambda'_1-\gamma)_{\mu'_1-\lambda'_1} (-1)^{\mu'_1-\lambda'_1}  } %
	\label{nonoffendingline11main}
\end{align}
where $\umu=[\mu_1,1^{\mu'_1-1}]$ and $\ulmb=[\lambda_1,1^{\lambda'_1-1}]$. 
Note that this $(1,1)$ coefficient reduces to the $(1,0)$ case for $\mu'_1=\lambda'_1=1$ and to the $(0,1)$ case for $\mu_1=\lambda_1=1$.

In appendix~\ref{sec_shift}  we discuss in detail the shift symmetry of this form of the recursion relation and various other features.

We conclude with some comments.

Note that the recursion has been explicitly solved analytically for arbitrary $\lambda_i$ and/or $\lambda'_j$ in only  a few cases: using both row and column variables, the 
determinantal formula of \cite{Doobary:2015gia} for $\theta=1$, which we 
repeat in appendix \ref{D1508}, and  the $(1,1)$ solution presented above, with arbitrary $\theta$. In this latter case the solution is precisely a
formula depending on one row length and one column height.

In the  next section \ref{sec_guideline} we will give a completely different description of the coefficients 
$T_{\gamma}$ and $T_{\gamma}^{-1}$, closely related to  
the $BC$ binomial formula of Okounkov \cite{Okounkov_1} involving  evaluating $BC$ interpolation polynomials at partitions arising via the identification between $(0,n)$ blocks and $BC_n$ Jacobi polynomials. 
It is tempting to speculate that   this form should also 
have a supersymmetric representation, more suited for a $(m,n)$ theory and  indeed we will find that the super interpolation 
polynomials of Sergeev 
and Veselov \cite{Veselov_3} do the job, but only for $T_{\gamma}^{-1}$ not $T_\gamma$ itself.

\section{SCFT perspective on the binomial coefficient}\label{sec_guideline}

In previous sections we defined superconformal blocks as an expansion in super Jack polynomials and 
solved for the coefficients of this expansion, $T_{\gamma}$, by setting up a recursion 
relation. As we pointed out in the overview, see section~\ref{dual_Jac_sec}, the  
blocks can be viewed as supersymmetric generalisations of the $BC$ Jacobi 
polynomials defined in~\cite{Stokman,Koornwinder}, which we called dual super 
Jacobi functions.  Then, the claim about stability of superconformal blocks implies that the coefficients $T_\gamma$ are 
directly related to the coefficients in the Jack expansion of $BC_N$ 
Jacobi polynomials, $(S^{(N)}$. %
In our notation the explicit relation reads
\begin{align}
	(T_{\gamma})_{\ulmb}^{\umu} \ &=\ 
	\frac{(-)^{|\umu|} \Pi_{\umu}(\theta) }{(-)^{|\ulmb|} 
	\Pi_{\ulmb}(\theta)}\times
	(S^{(N)}_{\frac{1}{\theta};\,p^{\pm}})^{\beta^N\bslash\,\ulmb'}_{ 
		\beta^N\bslash\,\umu'}
	\ .
	\label{startingBC}
\end{align}
Specifically, the starting relation is the one between $(0,N)$ blocks and $BC_N$ Jacobi 
polynomials (see~\eqref{BCJacobi_internalBlock}), and therefore \eqref{startingBC} is valid for {\em any}  
integer $N\geq \mu_1$.  Here we will use $N$ rather than $n$ to avoid confusion with the $(m,n)$ labels of the
superblock. Indeed, as emphasised
in previous sections, we know that the $(T_{\gamma})_{\ulmb}^{\umu}$ on the LHS~of \eqref{startingBC} do not depend on the particular $(m,n)$ theory (when $\ulmb$ is a strict Young diagram), 
and therefore do not depend on the number of variables of the superblock. %
The crucial point about \eqref{startingBC} is that the coefficients $S$ can be given explicitly in terms of the $BC_N$ interpolation 
polynomials $P^*$ of Okounkov 
\cite{Okounkov_1,Okounkov_FactoShur}.\footnote{In these refs, the more general 
Koornwinder polynomials are discussed, but the coefficients we are interested in are obtained 
by taking a limit on the Koornwinder polynomials as explained in \cite{Stokman,
Koornwinder}} evaluated at partitions. These polynomials 
are an important class of symmetric polynomials in $N$ variables, whose definitions and properties are given explicitly below and in the appendices.

To begin appreciating the subtle aspects of \eqref{startingBC} it is useful to picture how the Young diagrams involved change from LHS~to RHS. 
On the LHS~we have $\lambda,\mu$ and on the RHS~we 
have their transposed complements in $\beta^N$
\beq
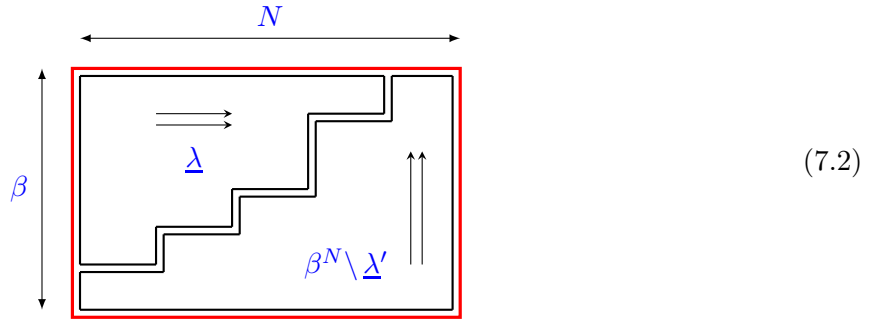
\begin{array}{c}
	\begin{tikzpicture}
		
		\def\shift{2.5}
		\def\xuno{2-\shift}
		\def\yuno{2}
		
		\def\ydue{5}
		\def\xtre{6-\shift}
		
		\def\ytre{4.5}
		\def\xquattro{5-\shift}
		
		\def\yquattro{3.5}
		\def\xcinque{4-\shift}
		
		\def\ycinque{3}
		\def\xsei{3-\shift}
		
		\def\step{.5}
		\def\stepp{.1}

		\draw[thick] (\xuno,	\yuno+\step) -- 			(\xuno,	\ydue);
		\draw[thick] (\xuno,	\ydue) -- 				(\xtre,	\ydue);
		\draw[thick] (\xtre,	\ytre) -- 				(\xquattro,	\ytre);
		\draw[thick] (\xquattro,	\yquattro) -- 		(\xcinque,	\yquattro);
		\draw[thick] (\xcinque,	\ycinque) -- 		(\xsei,	\ycinque);
		\draw[thick] (\xsei,	\yuno+\step) -- 			(\xuno, \yuno+\step);
		
		\draw[thick](\xtre,	\ydue)-- 			(\xtre,	\ytre);
		\draw[thick](\xquattro,	\ytre)-- 		(\xquattro,	\yquattro);
		\draw[thick](\xcinque,	\yquattro)-- 	(\xcinque,	\ycinque);
		\draw[thick](\xsei,	\ycinque)-- 		(\xsei,	\yuno+\step);

		\draw[thick] (\xtre+\stepp,	\ytre-\stepp) -- 				(\xquattro+\stepp,	\ytre-\stepp);
		\draw[thick] (\xquattro+\stepp,	\yquattro-\stepp) -- 		(\xcinque+\stepp,	\yquattro-\stepp);
		\draw[thick] (\xcinque+\stepp,	\ycinque-\stepp) -- 		(\xsei+ \stepp,	\ycinque-\stepp);
		\draw[thick] (\xsei+\stepp,	\yuno+\step-\stepp) -- 			(\xuno, \yuno+\step-\stepp);

		\draw[thick](\xtre+\stepp,	\ydue)-- 			(\xtre+\stepp,	\ytre-\stepp);
		\draw[thick](\xquattro+\stepp,	\ytre-\stepp)-- 		(\xquattro+\stepp,	\yquattro-\stepp);
		\draw[thick](\xcinque+\stepp,	\yquattro-\stepp)-- 	(\xcinque+\stepp,	\ycinque-\stepp);
		\draw[thick](\xsei+\stepp,	\ycinque-\stepp)-- 		(\xsei+\stepp,	\yuno+\step-\stepp);

		\draw[thick]  (\xuno,\yuno-\stepp) -- (\xtre+1-\stepp,\yuno-\stepp);	
		\draw[thick]  (\xtre+1-\stepp,\yuno-\stepp) -- (\xtre+1-\stepp,\ydue);
		\draw[thick]  (\xtre+\stepp,\ydue) -- (\xtre+1-\stepp,\ydue);
		\draw[thick]  (\xuno,\yuno-\stepp) -- (\xuno,\yuno+\step-\stepp);

		\draw[very thick, red]  (\xuno-.1,\ydue+.1) rectangle (\xtre+1,\ycinque-2*\step-.2);
		
		\draw[latex-latex] (\xuno-.5,\yuno-.1) -- (\xuno-.5,\ydue+.1);
		\draw[latex-latex] (\xuno,\ydue+.5) -- (\xtre+1,\ydue+.5);

		\draw (\xuno-.8,	  3.5) 	node 	{ \color{blue} $\beta$ };
		\draw (4.5-\shift,	  \ydue+.8) 	node 	{ \color{blue} $N$ };
		
		\draw[] (3.5-\shift,	 3.75) 	node [rotate=0]	{ \color{blue} $\ulmb$ };
		\draw[] (5.5-\shift,	 2.5) 	node [rotate=0]	{ \color{blue} 
		$\beta^N\bslash\,\ulmb'$ };

		\draw[-stealth]  (6.5-\shift,	 2.5)  --     (6.5-\shift,	 4) ;
		\draw[-stealth]  (6.35-\shift,	 2.5)  --     (6.35-\shift,	 4) ;
		
		\draw[-stealth]  (3-\shift,4.5) -- ( 4-\shift,4.5) ;
		\draw[-stealth]  (3-\shift,4.35) -- ( 4-\shift,4.35);

		\draw[] (5,	 2.5) 	node [rotate=0]	{  $\phantom{space}$ };	
		
	\end{tikzpicture}
\end{array}
\eeq
where $\beta^N\bslash\,\ulmb'$ is to be read along the columns, and bottom-up 
w.r.t.~$\ulmb$.

At this point the relation~\eqref{startingBC} is quite intriguing because superficially the LHS~appears to have many properties that are not manifest for the binomial 
coefficient on the RHS~:
\label{points}
\begin{enumerate}
\item We know from the recursion that we can obtain the LHS~as an explicit 
rational function of $\gamma$, whereas the RHS~depends on $\gamma$ only through 
the integer $\beta\equiv \min\left( \tfrac 1 2 (\gamma {-}  p_{12}), \tfrac 1  
2 (\gamma {-}  p_{43}) \right)$.
\item To obtain superblocks we sum over Young diagrams with arbitrarily large 
width $\mu_1$. On the RHS~this translates to arbitrarily large $N$ 
and so we need interpolating polynomials with arbitrarily large number of 
variables which is very inefficient.
\item Moreover, the LHS~depends mostly on the skew diagram 
$\umu/\ulmb$ (indeed the recursion of the previous sections is solved along 
this skew diagram),
the RHS~instead depends on the increasing large (as $N$ gets large, and also when $\beta$ 
is large) Young diagrams $\beta^N \backslash \ulmb'$ and $\beta^N \backslash 
\umu'$. Again this looks very inefficient. 

\end{enumerate}
 From the above considerations we expect the 
binomial coefficient, $S^{(N)}$,  to have various non-trivial hidden properties, in 
order to recover those of $(T_{\gamma})_{\ulmb}^{\umu}$. 

Our plan is thus the following: After giving a detailed definition of the binomial coefficient in 
terms of interpolating polynomials we will  examine the above properties from 
this perspective. In the process we will obtain a rewriting of  the binomial 
coefficient which deals with the first two of these points.
 Furthermore we will
find that the inverse coefficient $(T_{\gamma}^{-1})_{\ulmb}^{\umu}$ i.e.~the 
coefficient in the expansion of a super Jack polynomial in terms of blocks, has a form which 
deals with all three points.

\subsection{The binomial coefficient}

Recall~\eqref{Paul_Jacobi_1}, we wrote the coefficients of a $BC_N$ Jacobi polynomial, 
expanded in terms of Jack polynomials, %
\begin{align}\label{Paul_Jacobi_2}
	J_{\ulmb}({y}_1,\ldots,y_N;\theta,p^-,p^+) = \sum_{\umu \subseteq \ulmb}  
	(S^{(N)}_{\theta,p^-,p^+})_{\ulmb}^{\umu} 
	\,P_{\umu}({\bf y};\theta)\ .%
\end{align}
Now these binomial coefficients $S^{(N)}$ have been well studied and are known to be given, up to normalisations,  by $BC_N$ interpolation 
polynomials $P^*_\umu$ evaluated on $\ulmb$. The interpolation 
polynomials themselves are symmetric polynomials,  
depending on a single parameter $u$, and are uniquely 
defined\cite{Okounkov_1,Rains_1} by the property that they vanish when 
evaluated on a partition ${\ulmb}$ 
which is strictly contained in $\umu$, i.e. 
\beq
P^*_{\umu}(\ulmb;\theta,u)=0 \qquad \text{if} \quad \ulmb \subset \umu\ .
\eeq
Quite remarkably, the interpolation polynomials admit a combinatorial definition which generalises the one of Jack polynomials 
in a simple way. 
See appendix~\ref{sec_interpolation}. 

Now for some normalisation of the Jacobi polynomial, say $\hat J_{\ulmb}({\bf 
y})=J_\ulmb({\bf y})/N_\ulmb$, and some normalisation of the Jack polynomials, say
$\hat P_{\umu}({\bf y})=P_{\umu}({\bf y})/M_\umu$, the corresponding binomial 
coefficients in the expansion of $\hat J_\ulmb$ in the $\hat P_\umu$ really are simply 
interpolating polynomials with the variables taking the values of the Young 
diagram $\ulmb$
\begin{align}
	{\hat J_{\ulmb}(y_1,\ldots,y_N;\theta,A_1,A_2)} = \sum_{\umu \subseteq 
	\ulmb}  P^{*(N)}_\umu(\ulmb;\tfrac{A_1+A_2+1}2) %
	\,{\hat P_{\umu}({\bf y};\theta)}\ .
\end{align}
Returning to  the standard choice of normalisations (such that the 
coefficient of $P_{\ulmb}$ in the expansion~\eqref{Paul_Jacobi_1} is 1 and that 
$P_\umu(\underline 0)=\delta_{\umu,\varnothing}$) %
we can see that the normalisation of $S$ will  have the more complicated looking form:
\bea\label{FABC_binomial_def}
\qquad\qquad
(S^{(N)}_{\theta,\vec A})_{\ulmb}^{\umu} 
= 
\frac{ J_{\ulmb}(\underline{0};\theta,A_1,A_2) 
}{J_{\umu}(\underline{0};\theta,A_1,A_2)}  
\frac{ P_{\umu}^{*(N)}( \ulmb; \theta,u ) }{  P_{\umu}^{*(N)}( \umu;\theta,u )  
}\Bigg|_{u=\frac{A_1+A_2+1}{2} }\ .
\eea
However, the only non trivial part of \eqref{FABC_binomial_def} is still $P_{\umu}^{*(N)}( \ulmb; \theta,u )$!
The remaining objects entering this formula 
are explicitly known formulae which are all products 
of linear factors in the parameters
\begin{align}
		P_{\umu}^{*(N)}( \umu;\theta,u)&={C}_{\umu}^+(2u{-}1{+}2\theta N ;\theta)  
		{C}^-_{\umu}(1;\theta) \label{evaluationBCinterp}\\
		J^{(N)}_{\ulmb}( \underline{0};\theta,A_1,A_2)&=
		\frac{(-)^{|\ulmb|} \
			{C}^0_{\ulmb }(A_1{+}1{+}\theta (N-1); \theta) {C}^0_{\ulmb}(\theta 
			N ;\theta) }{ 
			P_{\ulmb }^*( \ulmb;\theta,u ) \Pi_{\ulmb}(\theta) 
			}\Bigg|_{u=\frac{A_1+A_2+1}{2} }\ ,
		\label{Jacobi_evaluation_in0}
\end{align}
where the combinatorial symbols $C^{0,\pm}$ are
\bea
C_\ulmb^0(w;\theta)&=&\ \,\prod_{(ij)\in\ulmb } \left(j{-}1-\theta(i{-}1) 
+w\right)\notag\\
C_\ulmb^-(w;\theta)&=&\ \,\prod_{(ij)\in\ulmb } \left( 
\lambda_i{-}j+\theta(\lambda'_j{-}i)+w \right)\notag\\
C_\ulmb^+(w;\theta)&=&\ \, 
\prod_{(i,j)\in\ulmb}(\lambda_i+j-\theta(\lambda'_j+i)+w), \label{recap_Csymbol}
\eea
and %
\beq\label{def_Pi_new}
\Pi_{\ulmb }(\theta)\equiv { C^{-}_{\ulmb }(\theta;\theta) }\big/{ C^{-}_{\ulmb 
}(1;\theta) }\qquad;\qquad
\Pi_{\ulmb}(\tfrac{1}{\theta} )= 1\big/\Pi_{\ulmb' }(\theta).
\eeq
Although these evaluation formulae may look 
complicated at first sight, they  consist simply  of products of linear factors.
For later use we will  sometimes use the notation 
$C^0_{\ulmb}(A,B;\theta)=C^0_{\ulmb}(A;\theta)C^0_{\ulmb}(B;\theta)$ and and 
$C^0_{\umu/\ulmb}\equiv C^0_{\umu}/C^0_{\ulmb}$ whenever $\ulmb\subseteq\umu$.

Putting the above definitions in we arrive at the expression for the binomial coefficient
\beq\label{FABC_binomial_def2}
\qquad\qquad
\big(S^{(N)}_{\theta,\vec A}\big)_{\ulmb}^{\umu} 
= 
{(-)^{|\ulmb|-|\umu|} \frac{\Pi_\umu(\theta)}{\Pi_\ulmb(\theta)} 
C^0_{\ulmb / \umu}(A_1{+}1{+}\theta(N{-}1),\theta N;\theta) 
}{}\times   
\frac{ P_{\umu}^{*(N)}( \ulmb; \theta,u ) }{  P_{\ulmb}^{*(N)}( \ulmb;\theta,u 
)  
}\Bigg|_{u=\frac{A_1+A_2+1}{2} }\ .
\eeq
Note that the $P^*$ in the denominator now involves $P^*_{\ulmb}$.

\subsection{Binomial representation of the \JtoB matrix}%
\label{SCBinomialCoeff_sec}

From this expression~\eqref{FABC_binomial_def2} we can obtain an expression for $T_\gamma$ in terms of interpolation polynomials via~\eqref{startingBC}.
Manipulating this with the help of some identities for interpolation polynomials of Rains \cite{Rains_1} (the details will be given shortly), 
we obtain:
\beq
(T_{\gamma})_{\ulmb}^{\umu}= 
(\mathcal{N}_{}\,)^{\umu}_{\ulmb}\times
\frac{ P^{*(M)}_{N^M\,\bslash\umu}(  N^M\,\bslash\ulmb ;\theta,u ) }{  P^{*(M)}_{ 
N^M\,\bslash\ulmb}( N^M\,\bslash\ulmb ;\theta,u) 
}\Bigg|_{u=\tfrac{1}{2}-\theta\tfrac{\gamma}{2} -N } \ .
 \label{SCbinomial2} 
\eeq
The normalisation is
\begin{gather}\label{nn}
(\mathcal{N}_{}\,)^{\umu}_{\ulmb}=
C^0_{\umu/\ulmb}\left(\theta \alpha,\theta \beta;\theta\right)\times 
\frac{(-)^{|\umu|} \Pi_{\umu}(\theta) }{(-)^{|\ulmb|} \Pi_{\ulmb}(\theta)}\ 
\frac{C^0_{\umu/\ulmb}(1{+}\theta(M-1);\theta)}{C^0_{\umu/\ulmb}(M\theta;\theta)
 } 
\end{gather}
where $\alpha,\beta$ are as defined in~\eqref{alphabetarevisited}, and the symbols $\Pi$ 
and $C^0$ are given in \eqref{recap_Csymbol}. %

Crucially
the whole expression is independent of $M,N$ as long as the rectangle $N^M$ 
contains the Young diagram $\umu$ (which in turn contains $\ulmb$). For 
practical computations we would always choose the 
minimal case $M = \mu'_1$ 
and $N= \mu_1$.

Note that the manipulations we performed   have produced a formula which solves the first two problems listed on page~\pageref{points}.
Firstly, the non factorisable $\gamma$ 
dependence of $T_{\gamma}$ now appears entirely in the parameter of $P^*_{N^M\,\bslash\umu}(  N^M\,\bslash\ulmb ;\theta,u )$, and can thus
be continued to arbitrary non-integer values. This polynomial is precisely
the complicated polynomial in $\gamma$ in 
$(T_{\gamma})_{\ulmb}^{\umu}$ discussed in section~\ref{example_Acoeff}.
Secondly, %
when the Young diagrams $\ulmb$ and $\umu$ are arbitrarily wide with limited height, i.e.~less than or equal to $\beta$ for superblocks, the dimension of the interpolating 
polynomials $M$ is limited (since we can choose $M=\beta$ for example) and does not 
grow indefinitely as for the original formula.

Finally,  note that the bosonic $(m,0)$ case of the shift 
symmetry~\eqref{deform_preview} is 
manifest in this formula: choosing $N=\mu_1$, we see that  $\gamma$ only appears 
through the combination $\mu_1+\theta\frac{\gamma}{2}$ and the row lengths 
$\lambda_i, \mu_i$  only appear as differences with $\mu_1$ via 
$\mu_1^M\,\bslash\ulmb$ and 
 $\mu_1^M\,\bslash\umu$ all of which  leave the shift invariant.

\subsubsection*{Proof of~\eqref{SCbinomial2}}

To show that~\eqref{startingBC} is equivalent to~\eqref{SCbinomial2} we 
will need two formulae which can both be found in~\cite{Rains_1}, by taking the appropriate classical limits. 
The first relates $BC_N$ interpolation polynomials with a Young 
diagram $\underline{\kappa}$ to one with a Young diagram with additional $\nu$  
full (ie height $N$) columns 
\begin{align}
	P^{*(N)}_{\underline{\kappa}+\nu^N}(w_1\ldots w_N;\theta,s)&=
	\prod_{i=1}^{N }\prod_{j=1}^{\nu} (w_i^2-(s+j-1)^2)\ 
	P^{*(N)}_{\underline{\kappa}}(w_1\ldots w_N;\theta,s+\nu)\ .
	\label{rectangle_relationPstar}
\end{align}
The second relation is an identity between a $BC_N$ interpolation polynomial with Young 
diagram $\umu$ evaluated on a Young diagram $ \underline{\kappa}$ and a $BC_M$ 
interpolation polynomial with transposed Young diagram $\umu'$ evaluated on the 
transposed  Young diagram $ \underline{\kappa}'$. Here $M,N$ can be arbitrary 
(this freedom to choose $M,N$ in fact arises from the previous relation) as 
long  as they are big enough to contain the relevant Young diagrams, so $M\geq 
\mu'_1$ and $N\geq \mu_1$. It reads
\bea
\label{rains_duality_step1}
\frac{ P^{*(N)}_{\umu}( \underline{\kappa};\theta^{},u) }{ P^{*(N)}_{\umu}( 
\umu;\theta^{},u) }&=&
\frac{ P^{*(M)}_{\umu'}(  \underline{\kappa}';\frac{1}{\theta},u') }{ P^{*(M)}_{\umu'}( 
\umu';\frac{1}{\theta},u') }\\
\text{where} \qquad \theta u'+   (M-\tfrac{1}{2} ) &=&- u 
-\theta(N-\tfrac{1}{2} )\ . \notag
\eea
We can now use~\eqref{rectangle_relationPstar} to remove 
the explicit $\beta$ dependence in the Young diagram of the non-trivial $P^*$ 
term in~\eqref{startingBC} with~\eqref{FABC_binomial_def2}. Taking  $\nu= M-\beta$ positive  we find
\begin{align}
	{ P^{*(N)}_{\beta^N\bslash\umu'}( 
	\beta^N\bslash\ulmb';\theta^{-1},\tfrac12(p_+{+}p_-{+}1)) }= { 
	P^{*(N)}_{M^N\bslash\umu'}(M^N\bslash\ulmb';\theta^{-1},\tfrac12(\gamma+1)-M) } 
	\times \cN_1  
\end{align}
where we recalled the relation between $\beta, 
\gamma,p_{\pm}$ in~\eqref{bppm} to get this. Note however that this equation is true 
for both $\nu$ positive or negative and indeed is 
valid for any $M\geq \mu_1$. The normalisation $\cN_1$ is 
given in terms of the known evaluation 
formula~\eqref{evaluationBCinterp} 
for $P^*_\mu(\mu)$ as
\begin{align}\label{n1}
	\cN_1&=\frac{  P^{*(N)}_{\beta^N\bslash\ulmb'}( 
		\beta^N\bslash\ulmb';\frac{1}{\theta},\tfrac12(p_+{+}p_-{+}1))  } 
	{  P^{*(N)}_{M^N\bslash\ulmb'}( 
		M^N\bslash\ulmb';\frac{1}{\theta},\tfrac12(\gamma+1)-M)  }
	\end{align}

Next we use the relation~\eqref{rains_duality_step1} to transpose the Young 
diagrams. Thus obtaining
\begin{align}
{ 
P^{*(N)}_{M^N\bslash\umu'}(M^N\bslash\ulmb';\,\tfrac{1}{\theta},\tfrac12(\gamma{+}1){-}M) 
}
 =  { 
 	P^{*(M)}_{N^M\bslash\umu}(N^M\bslash\ulmb;\,\theta,\tfrac12(1{-}\gamma\theta){-}N)}
 	 \times \cN_2  \ ,
\end{align}
with the normalisation $\cN_2$ also given in terms of explicitly known formulae
as
\begin{align}
\cN_2 &=\frac{  P^{*(N)}_{M^N\bslash\umu'}( 
	M^N\bslash\umu';\,\frac{1}{\theta},\tfrac12(\gamma{+}1){-}M)  }{  
	P^{*(M)}_{N^M\bslash\umu}( 
	N^M\bslash\umu;\,\theta,\tfrac12(1{-}\gamma\theta){-}N)  }\ .
\end{align}

So inserting these into the  formula  for $T_\gamma$~\eqref{startingBC} 
using  the definition of the binomial 
coefficient~\eqref{FABC_binomial_def} we obtain~\eqref{SCbinomial2} with the 
normalisation:
 \begin{align}	
 	(\mathcal{N}_{}\,)^{\umu}_{\ulmb}= 
 	\frac{(-)^{|\umu|} \Pi_{\umu}(\theta) }{(-)^{|\ulmb|} 
 		\Pi_{\ulmb}(\theta)} \times \frac{J_{\beta^N\backslash 
 		\ulmb'}(\underline 0;\frac{1}{\theta},p^-,p^+)}{J_{\beta^N\backslash 
 		\umu'}(\underline 0;\frac{1}{\theta},p^-,p^+)} \times \cN_1 \cN_2 \times 
 		\frac{  P^{*(M)}_{ 
 			N^M\,\bslash\ulmb}( N^M\,\bslash\ulmb 
 			;\theta,\tfrac{1}{2}(1-\gamma\theta)-N) 
 	}{  P^{*(N)}_{\beta^N\bslash\umu'}( 
 	\beta^N\bslash\umu';\frac{1}{\theta},\tfrac12(p_+{+}p_-{+}1)))  } \ .
 \end{align}
This  normalisation, despite its complicated appearance, is just made of known linear factors. 
Moreover, it dramatically simplifies!!  We know this in advance because it is clear from the recursion for $T_{\gamma}$ 
that only the $\gamma$ dependence is non factorisable and complicated.
In particular all factors of $C^+$ occurring in the evaluation 
formulae for \eqref{evaluationBCinterp} appear in ratios with transposed Young 
diagrams and appropriate parameters such that they cancel using the relations
\bea
{C}^0_{\umu}(x;\theta)&=&(-\theta)^{|\umu|}\ 
{C}^0_{\umu'}(-\tfrac{x}{\theta};\tfrac{1}{\theta} )\label{C0transposo}\\
{C}^-_{\umu}(x;\theta)&=&(+\theta)^{|\umu|}\ 
{C}^-_{\umu'}(+\tfrac{x}{\theta};\tfrac{1}{\theta} )\notag\\
{C}^+_{\umu}(x;\theta)&=&(-\theta)^{|\mu|}\ 
{C}^+_{\umu'}(-\tfrac{x}{\theta};\tfrac{1}{\theta} )\notag\ ,
\eea
together with the identities
\begin{align}\label{complement_identity_C01}
	\frac{ {C}^0_{N^M\backslash\umu}(x;\theta)}{ {C}^0_{N^M}(x;\theta) }&=\frac{(-)^{|\umu|} }{{C}^0_{\umu} 
	(1-N+\theta(M{-}1)-x;\theta)}\ ,
\end{align}
and
\begin{gather}\label{complement_identity_C02}
  \frac{\Pi_{N^M \backslash \umu}(\theta) 
		\Pi_{N^\beta \backslash \ulmb}(\theta) }{\Pi_{N^M \backslash 
			\ulmb}(\theta) \Pi_{N^\beta \backslash \umu}(\theta)} 
=
	\frac{C^0_{\umu/\ulmb}\left(\beta\theta 
	;\theta\right)}{C^0_{\umu/\ulmb}(M\theta;\theta)}\times
	 {}\ 
	\frac{C^0_{\umu/\ulmb}(1{+}\theta(M-1);\theta)}{C^0_{\umu/\ulmb} 
	(1{+}\theta(\beta-1);\theta)}
\end{gather}
The remaining products of linear factors give the much simpler expression quoted in~\eqref{nn}.

\subsubsection*{Comparing the recursion formula with the binomial coefficient}

The binomial formula for $T_\gamma$ in \eqref{SCbinomial2} can be tested, on a computer, against 
the recursion of section~\ref{sec_SCBlock}, very much exhaustively. 
It can also be tested analytically in all known cases in which the recursion has been solved explicitly.
These are: rank-one $(1,0)$ and $(0,1)$ cases for any $\theta$; rank-two $(2,0)$ for any $\theta$ 
and finally $(m,n)$ for $\theta=1$. We explicitly check all of them in appendix \ref{know_solu}, 
where we refer the reader for a detailed discussion.  The computation is instructive since, 
as we pointed out in various occasions, the interpolation polynomials and the 
recursion are based on two very different combinatorics, 
therefore there will be many non trivial operations involved in order for the two to agree.

\subsection{Inverse \JtoB matrix and complementation}\label{Jack_to_BlockMatrix}

We now have two completely different methods for computing the \JtoB matrix 
$T_\gamma$, via the binomial formula~\eqref{SCbinomial2} in terms of interpolation polynomials or via 
the recursion of 
section~\ref{sec_SCBlock}. 
The recursion %
has the advantage that in order to compute $(T_{\gamma})_{\ulmb}^{\umu}$, 
we only run over Young diagrams in between $\umu/\ulmb$, and therefore we use 
the minimum 
number of steps. Moreover, in the super case the steps  respect the $(m,n)$ 
hook structure of the Young diagrams. This structure is broken in the binomial 
formula~\eqref{SCbinomial2} %
because we take the complement Young diagrams $N^M\,\bslash\ulmb$. 
For physical applications, where $m,n$ 
are small, this is not efficient because the diagrams generically will have a slim hook structure, 
but in order to compute the $P^*$ we would need to consider the red diagram in the 
following picture, 
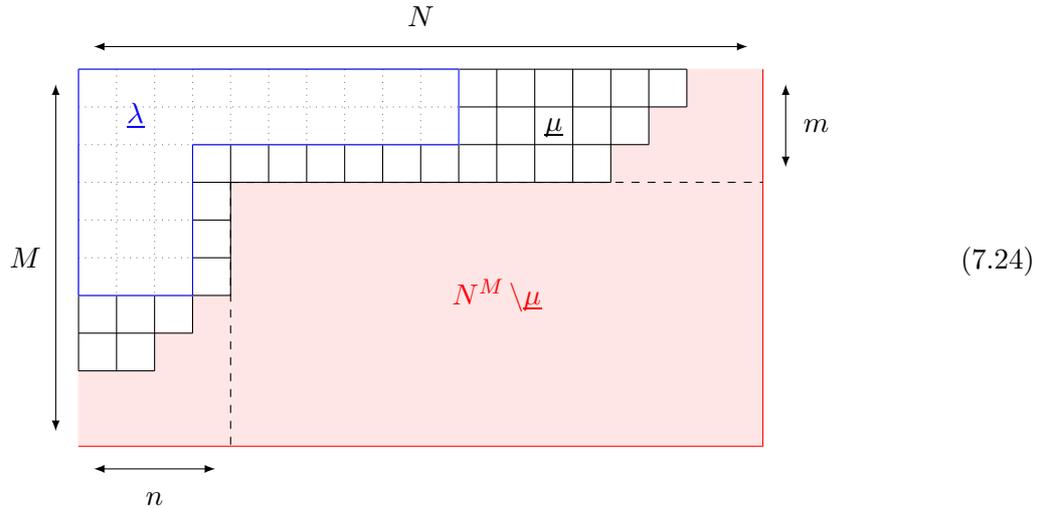
\beq\label{disegno_complemetation}
\!\!\!\!\!\!\!\!\!\!\begin{array}{c}
	\begin{tikzpicture}

		\filldraw[red!10] (2,-1.5) rectangle (8,-4) ;
		\filldraw[red!10] (8,-4) rectangle (7.5, -0.5);
		\filldraw[red!10] (8,-4) rectangle (7, -1);
		
		\filldraw[red!10] (8,-4) rectangle (1,-3.5);
		\filldraw[red!10] (8,-4) rectangle (1.5,-3);
		\filldraw[red!10] (9,-5) rectangle (0,-4);
		\filldraw[red!10] (9,-5) rectangle (8,0);
		
		\draw[red] (9,-5) -- (9,0);
		\draw[red] (9,-5) -- (0,-5);

			\draw[dashed] (2,-1.5) -- (2,-5);
		\draw[dashed] (2,-1.5) -- (9,-1.5);

		\foreach \y / \x  in  { 0 / 8,    0.5/ 8,  1 / 7.5,     1.5 / 7 ,   2 
			/2 ,    2.5 / 2,    3 /2 ,    3.5 /1.5 ,    4 /1}
		\draw (0,-\y) -- (\x,-\y);

		\foreach \x / \y  in  { 0 / 4,  0.5 / 4 , 1 / 4 , 1.5 / 3.5 ,  2 / 3 ,  
			2.5 / 1.5 ,  3  / 1.5,   3.5 / 1.5 ,  4  / 1.5 ,   4.5  / 1.5,   5  
			/ 
			1.5 ,  5.5  / 1.5,  6  / 1.5 ,  6.5  / 1.5 , 7 /1.5,  7.5 /1 , 8 / 
			0.5}
		\draw (\x,0) -- (\x,-\y);	
		
		\filldraw [white!25]  (0,0) rectangle (5,-1);	
		\filldraw [white!25]  (0,0) rectangle (1.5,-3);
		
		\foreach \y / \x  in  {   0.5/ 5,  1 / 1.5, 1.5 /1.5  ,  2 / 1.5  , 2.5 
			/1.5 }
		\draw[dotted,gray] (0,-\y) -- (\x,-\y);
		
		\foreach \x / \y  in  { 0.5 / 3 , 1/ 3, 1.5/ 1, 2/1 , 2.5 /1, 3/1 , 3.5 
			/1, 4/1, 4.5/1 }
		\draw[dotted,gray] (\x,0) -- (\x,-\y);

		\draw[blue] (0,0) -- (5,0);
		\draw[blue] (5,0) -- (5,-1);
		\draw[blue] (5,-1) -- (1.5,-1);
		\draw[blue] (1.5,-1) -- (1.5,-3);
		\draw[blue] (1.5,-3)-- (0,-3);
		\draw[blue] (0,-3) -- (0,0);

		\draw[thin, latex-latex]   (0+.2,	0+.3) -- 			(9-.2,	 +.3);
		\draw[thin, latex-latex]   (0-.3,	0-.2) -- 			(0-.3,	 -5+.2);
		\draw[thin, latex-latex]   (0+.2,	-5-.3) -- 			(2-.2,	 -5-.3);
		\draw[thin, latex-latex]   (9+.3,	0-.2) -- 			(9+.3,-1.5+.2);
		
		\draw (.75,-.75) node {\color{blue} $ \ulmb$};
			\draw (6.25,-.85) node { $ \umu$};
		\draw (-.7,	 -2.5) 	node 	{ $M$};
		\draw (4.5,	 +.7) 	node 	{ $N$};
			\draw (9+.7,	 -.75) 	node 	{ $m$};
		\draw (1,	-5-.7) 	node 	{ $n$};
		\draw (5.5,	 -3) 	node 	{ $ \red N^M\,\bslash \umu$};

	\end{tikzpicture}
\end{array}	   
\eeq
which can become quite `fat', compared to the skew shape $\umu/\ulmb$.

In a number of applications, $(T_{\gamma}^{-1})$, the inverse 
of the \JtoB matrix,  shows up and plays an important role.
This matrix is such that
\begin{align}\label{formula_jack_block}
	P_{\ulmb}({\bf z})= 
	\sum_{\umu: \ulmb\subseteq\umu} (T_{\gamma}^{-1})_{\ulmb}^{\umu}\  
	{F}_{\gamma,\umu }({\bf z})%
\end{align}
where as usual we understand the dependence on $\theta,p_{12},p_{43}$ in 
$F_{\gamma,\umu}$. Note that when $m>0$ the Jack polynomials is represented as an infinite sum. 
We will now see that the inverse 
of the \JtoB matrix, the \BtoJ matrix, has a very nice interpretation in terms of interpolation polynomials, and avoids the above mentioned problems with complementation.
  
An illuminating way of obtaining the \BtoJ matrix is to 
apply the Casimir~\eqref{ConJack} and obtain a  recursion very similar to the recursions for $T_\gamma$, see~\eqref{rec2inv}. 
By carefully examining and comparing the two recursions~\eqref{rec2} and~\eqref{rec2inv}, in particular the denominator term in each,   
one can deduce  that there is a direct relation between $T_\gamma^{-1}$ and $T_{\tilde \gamma}$ with complemented 
Young diagrams. This complementation then accounts for the difference that in~\eqref{rec2} we sum over 
Young diagrams with one subtracted box, whereas in~\eqref{rec2inv} we sum over Young diagrams with one added box. The precise relation is then
\begin{align}
	\!\! \left( {T}_{\gamma}^{-1} \right)_{\ulmb}^{\umu}(\theta,p_{12},p_{43} ) =\   
	\frac{ C^0_{(N^M\bslash\ulmb)/(N^M\bslash\umu)}(M\theta;{\theta} ) }{   C^0_{(N^M\bslash\ulmb)/(N^M\bslash\umu)}(1+\theta(M-1);{\theta} )  } \times \frac{ \Pi_{N^M\bslash \umu }(\theta) }{ \Pi_{N^M\bslash \ulmb }(\theta) }
	\left({T}_{\tilde{\gamma} }\right)_{N^M\bslash\ulmb}^{N^M\bslash\umu}( {\theta}, -p_{12} ,-p_{43} )
	\label{complement_inverse_prima}
\end{align}
where 
\beq 
\tilde \gamma =  -\gamma+2(M-1)-\tfrac{2}{\theta} (N-1)\ .
\eeq 
This identity is valid for any $\gamma$, $p_{12},p_{43}$ and $N\ge \mu_1$, $M\ge \mu'_1$. 
The intuition behind \eqref{complement_inverse_prima} is that $\umu/\ulmb$ 
and $(N^M\backslash\ulmb)/(N^M\backslash\umu)$ represent the same skew diagram, seen from different orientations. Then, the map $\gamma \rightarrow \tilde \gamma$
is the only real non trivial transformation in going from left to right in \eqref{complement_inverse_prima}, since the $\gamma$ dependence is the non factorisable one.
Analyticity in $\gamma$ of $T_{\gamma}$ is also crucial, since $\gamma$ might be an integer but $\tilde\gamma$ is certainly not. 
The factors of $C^0$ can be eventually rewritten/simplified by using \eqref{complement_identity_C01}-\eqref{complement_identity_C02}.\footnote{For example, the RHS~can be transposed by using
	$$
	(T_{-\theta\gamma})^{\umu'}_{\ulmb'}(\tfrac{1}{\theta},-\theta p_{12},-\theta p_{43} )=\frac{(-)^{|\ulmb|} \Pi_{\ulmb}(\theta) }{(-)^{|\umu| }\Pi_{\umu}(\theta) }(T_{\gamma})^{\umu}_{\ulmb}(\theta,p_{12},p_{43})
	$$
	which follows from~${\bf C}^{(\theta,a,b,c)}({\bf x}|{\bf y})=-\theta\, {\bf C}^{(\frac{1}{\theta}, -a \theta,-b \theta,-c \theta)}({\bf y}|{\bf x})$, and properties of the (super) Jack polynomials under transposition.}
The form given above is useful for the  computation that follows.

Now notice that  the binomial formula for the \JtoB matrix, $T_\gamma$~\eqref{SCbinomial2} involves complemented 
Young diagrams, as does the RHS formula~\eqref{complement_inverse_prima} for  $T_\gamma^{-1}$ 
in terms of $T_\gamma$. We thus conclude that   the  {\em inverse} matrix $T_{\gamma}^{-1}$   is 
very naturally computed by a 
binomial coefficient,  given by interpolation polynomials directly in $\umu$ and 
$\ulmb$! In fact, by using \eqref{SCbinomial2} on the RHS of \eqref{complement_inverse_prima}, a number of  simplifications take place,\footnote{
Upon writing $\left({T}_{\tilde{\gamma} }\right)_{N^M\bslash\ulmb}^{N^M\bslash\umu}( {\theta}, -p_{12} ,-p_{43} )$ note that 
 $C^0_{\umu/\ulmb}\left( \theta \alpha, \theta \beta;\theta\right)$ is recovered by using \eqref{complement_identity_C01}.}  
and the final formula is very clean and reads: 
\begin{align}\label{inverse_T_formula}
	(T_{\gamma}^{-1})_{ \ulmb}^{\umu }&= \frac{(-)^{|\umu| } }{(-)^{|\ulmb|} } 	
C^0_{\umu/\ulmb}\left( \theta \alpha, 
\theta \beta;\theta\right)
	\left.\frac{ P^{*(M)}_{\ulmb}( \umu ; \theta,v) }{  P^{*(M)}_{\umu}( \umu; 
	\theta,v) 
	}\right|_{	v=-\tfrac{1}{2}+\theta\tfrac{\gamma}{2}-\theta( M-1) }
\end{align}
where $P^*_{\ulmb}$ are $BC_M$ interpolation polynomials with any $M \geq \mu'_1$ 
and $\alpha,\beta$, as always, given by~\eqref{alphabetarevisited}. Note, the formula above is 
independent of $M$ with the minimal choice $M=\mu'_1$.
One can also deduce this relation from formulae in Rains~\cite{Rains_1}, together with the identification \eqref{SCbinomial2},  
where it is seen that both the binomial coefficient and its inverse are given by interpolation polynomials 
with complemented Young diagrams (see~\cite{Rains_1} (4.1),(4.2),(4.22),(4.23)).
Observe that this expression for $T_\gamma^{-1}$ is very similar  
to the expression 
for the binomial coefficients $S$ written in~\eqref{FABC_binomial_def2}. We will make use 
of this observation when examining the Cauchy identity in the next section.

The nice feature of \eqref{inverse_T_formula} is that the Young diagrams 
labelling the interpolation 
polynomials are now $\ulmb$ and $\umu$ rather than their complements. 
As a result, computing $T^{-1}_{\gamma}$ by using the binomial coefficient is 
also quite efficient.
Finally, the evaluation of the interpolation polynomial in the numerator,  
$P^*_{\ulmb}( \umu ; \theta,v)$,  is 
again the only term not involving products of linear factors, and depends non 
trivially on $\gamma$.

Even though \eqref{inverse_T_formula}  has nice properties, it still 
has the inconvenience that the number of variables in the interpolation polynomials 
grows with $M$.
Therefore, if we are considering blocks with large $\beta$, since $\mu_1\leq \beta$ and $M\ge\mu'_1$, with $M=\mu'_1$ being the minimal choice,
we will will have to consider interpolation polynomials with fixed but large number of variables.
In fact, \eqref{inverse_T_formula}  is still a bosonic formula, which reads the Young diagram only by rows, 
whereas in the super block the typical Young diagram is a $(m,n)$ hook. We can then ask if there is an interpolation polynomial type formula which respects the $(m,n)$ structure. 
Nicely enough, this problem has a simple solution 
which is suggested by the independence of
the 
dimension of the interpolation polynomials, $M$, of the expression 
$T^{-1}$~\eqref{inverse_T_formula}. This stability property 
immediately 
 suggests that one might be able to replace the bosonic interpolation polynomials with %
{\em super}-interpolation polynomials, where the latter have a natural hook-type parametrisation of their variables.

Sergeev 
and Veselov  introduced  the $BC_{M|N}$ super interpolation 
polynomials in~\cite{Veselov_3} that we need. They first modified the 
free parameter on which the interpolation polynomials depend, to $h=v+\theta M$ and 
defined the modified interpolation polynomial $\tilde P^{*(M)}_{\ulmb}$ as
\begin{align}\label{tildepstar}
	\tilde P^{*(M)}_{\ulmb}( {\bf x} ; \theta,h):= 	 P^{*(M)}_{\ulmb}( 
	{\bf x} ; \theta,h {-} \theta M)\ . 
\end{align}
They then showed that these modified IPs  have  a natural 
supersymmetric 
generalisation, denoted by $\tilde P^{*(M|N)}_{\ulmb}( {\bf x} | {\bf y}; \theta,h)$, where taking $N=0$ returns the bosonic interpolation polynomial 
\begin{align}\label{our_veselov}
	\tilde P^{*(M)}_{\ulmb}(  {\bf x}; \theta,h)=\tilde P^{*(M|0)}_{\ulmb}( {\bf x}|\ ; \theta,h)\ .
\end{align}
We spell out the conventions~\cite{Veselov_3} in appendix \ref{rewriting_Veselov}.

In~\cite{Veselov_3} these super polynomials were used to obtain  super binomial coefficients, 
i.e. the coefficients of the expansion of super Jacobi polynomials in terms of super 
Jacks. We will return to this point in the next section.
Here we notice that stability of $T_\gamma^{-1}$ 
in~\eqref{inverse_T_formula} gives further insight into these objects. 
Indeed, we understand from stability that the ratio of interpolation polynomials in~\eqref{inverse_T_formula} is 
independent of $M$,
 \begin{align}\label{Ptilde_indepM}
	\frac{ \tilde P^{*(M)}_{\ulmb}( \umu ; \theta,h) }{  \tilde 
		P^{*(M)}_{\umu}( \umu; 
		\theta,h) 
	}=\frac{ \tilde P^{*(\mathsf{M})}_{\ulmb}( \umu ; \theta,h) }{  \tilde 
	P^{*(\mathsf{M})}_{\umu}( \umu; 
	\theta,h) 
}\ 
\end{align}
for any $\mathsf{M},M\ge \mu'_1$.
This in turn suggests that one can further replace this ratio with the 
appropriate ratio of {\em super} 
interpolation polynomials evaluated on Young diagrams. Namely, 
 \begin{align}\label{ptildeinv}
 	\frac{ \tilde P^{*(\mathsf{M})}_{\ulmb}( \umu ; \theta,h) }{  \tilde 
 	P^{*(\mathsf{M})}_{\umu}( \umu; 
 		\theta,h) 
 	}=\frac{ \tilde P^{*({M}|{N})}_{\ulmb}( \umu_e|\umu_s ; \theta,h) }{  \tilde 
 	P^{*({M}|{N})}_{\umu}( \umu_e|\umu_s; \theta,h) }\ .
 \end{align}
where $\umu_s=[\mu'_1,.., \mu'_{{N}}]$ and $\umu_e\equiv\umu/\umu'_s=[(\mu_1 {-} {N})_+,..,(\mu_{{M}} {-} {N})_+]$ with $(x)_+\equiv\max(x,0)$. 

An example illustrating the decomposition of $\umu$ into $\umu_s$ and $\umu_e$ relevant here is the following,
    \beq
    \begin{array}{c}
	\begin{tikzpicture}
		\def\xuno{2}
		\def\yuno{2}
		\def\ydue{5}
		\def\xtre{7}
		\def\ytre{4.5}
		\def\xquattro{5}
		\def\yquattro{3.5}
		\def\xcinque{4}
		\def\ycinque{3}
		\def\xsei{3}
		\def\step{.5}
		\draw[thick,black!20] (\xuno,	\yuno) -- 			(\xuno,	\ydue);
		\draw[thick,black!20] (\xuno,	\ydue) -- 			(\xtre,	\ydue);
		\foreach \x in {1, ..., 10}
		\draw[thick,black!20] (\xuno+\x*.5,	\ydue) -- 			(\xuno+\x*.5,	
		\ydue-\step);
		\draw[thick,black!20] (\xuno,	\ydue-\step) -- 			(\xtre,	
		\ydue-\step);
		\draw[thick,black!20] (\xquattro,	\ydue-\step) -- 			(\xtre,	
		\ydue-\step);
		\foreach \x in {1, ..., 3}
		\draw[thick,black!20] (\xuno+\x*.5,	\ytre) -- 			(\xuno+\x*.5,	
		\ytre-2*\step);
		\foreach \x in {4, ..., 7}
		\draw[thick,black!20] (\xuno+\x*.5,	\ytre) -- 			(\xuno+\x*.5,	
		\ytre-1*\step);
		\draw[thick,black!20] (\xuno,	\ytre-\step) -- 			
		(\xquattro+0.5,		\ytre-\step);
		\foreach \x in {1, ..., 3}
		\draw[thick,black!20] (\xuno+\x*.5,	\yquattro) -- 			
		(\xuno+\x*.5,	
		\yquattro-\step);
		\draw[thick,black!20] (\xuno,	\yquattro-\step) -- 			
		(\xcinque-.5,	\yquattro-\step);
		\draw[thick,black!20] (\xuno,	\ycinque) -- 			(\xsei,	
		\ycinque);
		\foreach \x in {1, ..., 2}
		\draw[thick,black!20] (\xuno+\x*.5,	\ycinque) -- 			
		(\xuno+\x*.5,	
		\ycinque-2*\step);
		\draw[thick,black!20] (\xuno,	\ycinque-\step) -- 			(\xsei,	
		\ycinque-\step);
		\draw[thick,black!20] (\xuno,	\ycinque-2*\step) -- 			(\xsei,	
		\ycinque-2*\step);
		\draw[thick,black!20] (\xuno,	\ycinque+.5) -- 			(\xsei+.5,	
		\ycinque+.5);
		\draw (\xtre+4,	\yquattro+.5) 	node 	{ 
			$\umu_{\color{red}}=[10,7,3,3,2,2]$};
		\draw (\xtre+4,	\yquattro) 	node 	{ 
			$\umu_{s}=[6,6,4,2,2]$};
		\draw (\xtre+4,	\yquattro-.5) 	node 	{ 
			$\umu_{e}=[5,2,0]$};
		\draw (\xtre-4.2,	\yquattro+.25) 	node 	{ 
			$\umu'_{s}$};
		\draw (\xtre-2,	\yquattro+.9) 	node 	{ 
			$\umu_{e}$};
		\draw[very thick, black]  (\xuno,\ydue) -- 
		(\xcinque+.5-.05,\ydue) -- (\xcinque+.5-.05,\yquattro+.5)-- 
		(\xcinque-.5,\yquattro+.5)-- (\xcinque-.5,\yquattro-.5)-- 
		(\xcinque-1,\yquattro-.5)-- 
		(\xcinque-1,\yquattro-1.5)-- 
		(\xcinque-2,\yquattro-1.5)-- (\xuno,\ydue);
		\draw[very thick, black]	(\xcinque+.5+.05,\ydue) -- 
		(\xcinque+.5+.05,\yquattro+.5)-- (\xcinque+1.5,\yquattro+.5)-- 
		(\xcinque+1.5,\yquattro+1)-- (\xcinque+3,\yquattro+1)-- 
		(\xcinque+3,\ydue)--	(\xcinque+.5+.05,\ydue)
		;

\draw[-latex] (\xuno,	\ydue+.25) -- (\xcinque+.5,	\ydue+.25);
\draw (\xuno+1.2,\ydue+.5) node[scale=.8] {$N=5$};

\draw[-latex] (\xuno-.25,	\ydue) -- (\xuno-.25,	\ydue-1.5);
\draw (\xuno-1,	\ydue-.5) node[scale=.8] {$M=3$};
	\end{tikzpicture}
	 \end{array}
\eeq
We have checked extensively that \eqref{ptildeinv} equality is true.

Putting the above reasoning together, we conclude that the inverse block coefficients can be written in terms of super-interpolation polynomials as
\begin{align}\label{inverse_T_formula_susy}
	(T_{\gamma}^{-1})_{ \ulmb}^{\umu }&= \frac{(-)^{|\umu| } }{(-)^{|\ulmb|} } 	
	C^0_{\umu/\ulmb}\left( \theta \alpha, 
	\theta \beta;\theta\right)
	\left.\frac{ \tilde P^{*(M|N)}_{\ulmb}( \umu_e|\umu_s ; \theta,v) }{  
	\tilde 
	P^{*(M|N)}_{\umu}( \umu_e|\umu_s; 
		\theta,v) 
	}\right|_{	v=-\tfrac{1}{2}+\theta\tfrac{\gamma}{2}+\theta}\ ,
\end{align}
and this expression is independent of $M,N$.

Thus in the expansion of an $(m,n)$ superblock we are free to take $M=m$ and $N=n$, 
and then we will only ever need to evaluate 
super interpolation polynomials in a fixed number ($m+n$) of  variables.

An  interesting application of \eqref{inverse_T_formula_susy} for the inverse matrix
will involve a transformation of the dual Cauchy identity. This is explained in the next section, and will lead to 
a formula 
for decomposing \emph{any} free theory diagram into superconformal blocks for 
any $\theta$, therefore any dimension.

\section{Generalised free theory %
from a Cauchy identity 
}\label{sc_Cauchy}

The most straightforward application of the identification of superblocks with dual super Jacobi functions
consists in borrowing from the maths literature a Cauchy identity for Jacobi polynomials, found by 
Mimachi~\cite{Mimachi}, and re-interpreting it to yield a 
formula for {\em every}  coefficient in the superblock expansion of {\em any} 
generalised 
free theory in {\em any} theory of the type described by this formalism 
(summarised 
in section~\ref{summary_list_theory}).
Inspired by this  we also conjecture the uplift of this Cauchy identity to a new dually supersymmetric 
Cauchy identity.

In the following we will have in mind the following picture, 
\beq\label{disegno_complemetation_Jacobi}
\!\!\!\!\!\!\!\!\!\!\begin{array}{c}
	\begin{tikzpicture}

		\filldraw[red!10] (2,-1.5) rectangle (8,-4) ;
		\filldraw[red!10] (8,-4) rectangle (7.5, -0.5);
		\filldraw[red!10] (8,-4) rectangle (7, -1);
		
		\filldraw[red!10] (8,-4) rectangle (1,-3.5);
		\filldraw[red!10] (8,-4) rectangle (1.5,-3);
		\filldraw[red!10] (9,-5) rectangle (0,-4);
		\filldraw[red!10] (9,-5) rectangle (8,0);
		
		\draw[red] (9,-5) -- (9,0);
		\draw[red] (9,-5) -- (0,-5);

		\foreach \y / \x  in  { 0 / 8,    0.5/ 8,  1 / 7.5,     1.5 / 7 ,   2 
			/2 ,    2.5 / 2,    3 /2 ,    3.5 /1.5 ,    4 /1}
		\draw (0,-\y) -- (\x,-\y);

		\foreach \x / \y  in  { 0 / 4,  0.5 / 4 , 1 / 4 , 1.5 / 3.5 ,  2 / 3 ,  
			2.5 / 1.5 ,  3  / 1.5,   3.5 / 1.5 ,  4  / 1.5 ,   4.5  / 1.5,   5  
			/ 
			1.5 ,  5.5  / 1.5,  6  / 1.5 ,  6.5  / 1.5 , 7 /1.5,  7.5 /1 , 8 / 
			0.5}
		\draw (\x,0) -- (\x,-\y);	
		
		\filldraw [white!25]  (0,0) rectangle (5,-1);	
		\filldraw [white!25]  (0,0) rectangle (1.5,-3);
		
		\foreach \y / \x  in  {   0.5/ 5,  1 / 1.5, 1.5 /1.5  ,  2 / 1.5  , 2.5 
			/1.5 }
		\draw[dotted,gray] (0,-\y) -- (\x,-\y);
		
		\foreach \x / \y  in  { 0.5 / 3 , 1/ 3, 1.5/ 1, 2/1 , 2.5 /1, 3/1 , 3.5 
			/1, 4/1, 4.5/1 }
		\draw[dotted,gray] (\x,0) -- (\x,-\y);

		\draw[blue] (0,0) -- (5,0);
		\draw[blue] (5,0) -- (5,-1);
		\draw[blue] (5,-1) -- (1.5,-1);
		\draw[blue] (1.5,-1) -- (1.5,-3);
		\draw[blue] (1.5,-3)-- (0,-3);
		\draw[blue] (0,-3) -- (0,0);

		\draw[thin, latex-latex]   (0+.2,	0+.3) -- 			(9-.2,	 +.3);
		\draw[thin, latex-latex]   (0-.3,	0-.2) -- 			(0-.3,	 -5+.2);
		
		\draw (.75,-.75) node {\color{blue} $ \ulmb$};
			\draw (6.25,-.85) node { $ \umu$};
		\draw (-.9,	 -2.5) 	node 	{ $M,m$};
		\draw (4.5,	 +.7) 	node 	{ $N,n$};
		\draw (5.5,	 -3) 	node 	{ $ \red N^M\,\bslash \umu$};

	\end{tikzpicture}
\end{array}	   
\eeq
as in the previous section.

\subsection{Super Cauchy identity}

Let us consider eq.~(4.2) of Mimachi \cite{Mimachi}, which  gives a Cauchy identity involving  BC${}_n$ 
and  BC${}_M	$  Jacobi polynomials. Looking at that identity, we immediately see that out of the two Jacobi polynomials entering the formula, one depends on 
a complemented  Young diagram $M^n\backslash \ulmb'$ and is more 
naturally written  in terms of  our {\em dual} Jacobi 
polynomial~\eqref{dual_jacobi}. Mimachi's Cauchy formula then reads
\begin{align}\label{cauchyjacobi}
	\prod_{i}^M\prod_{j}^n (1-y_iy'_j) &= \sum_{\ulmb} 
	(-1)^{|\ulmb|} { J}^{(M)}_\ulmb({\bf y};\theta,\tilde p^-,\tilde p^+) {
		\tilde J}^{(n)}_{M,\ulmb'}({\bf y}';\tfrac{1}{\theta},p^-,p^+) \\
		\tilde p^-&=  \theta p^-+\theta- {1}{} \qquad;\qquad  \tilde p^+=  \theta 
	p^++\theta-{1}{} \ .\label{ptilde}
\end{align}
Note that the sum on the RHS~is automatically cut off, so that $\ulmb  
\subseteq 
n^M$. Indeed, the $BC_M$ Jacobi polynomial, ${ J}_\ulmb$ vanishes if $\ulmb$ has 
more than $M$ rows
and the $BC_n$ dual Jacobi ${
	\tilde J}_{M,\ulmb'} $ vanishes if $\ulmb'$ has more than $n$ rows.
Also note that the LHS~is manifestly stable in both $M$ and $n$: setting 
$y_M=0$ or $y_n=0$ yields the same  formula  with $M \rightarrow M{-}1$ or $n 
\rightarrow n{-}1$. The RHS~is also manifestly stable in $n$ since, as we know, the 
dual Jacobi's $\tilde J$ are stable. However the RHS~is not manifestly stable 
in $M$. It depends on $M$ both in the Jacobi $J^{(M)}$, which is not stable, and in ${\tilde J}_{M,\ulmb'} $. 
Remarkably these two $M$ dependencies must precisely cancel each other out! 
To understand how this is possible, in the next subsection we will re-derive \eqref{cauchyjacobi}
 starting from the simpler and  
well known Cauchy identity involving Jack polynomials. The derivation will then
explain the stability property in $N$ and also 
explain  the  appearance of the modified parameters 
$\tilde p^{\pm}$.

But first let us motivate various supersymmetric   generalisations of this Jacobi Cauchy identity which we will explicitly prove in the next section.
Through stability, %
we know that dual Jacobi polynomials have a 
natural uplift to dual super Jacobi functions. %
Thus we can uplift ${\tilde J}^{(n)}_{M,\ulmb'}({\bf y'})$ on the RHS~of the previous formula \eqref{cauchyjacobi} to $(n,m)$
variables. %
Regarding the LHS of  \eqref{cauchyjacobi},~the eigenvalue interpretation of the variables %
in the super matrix formalism for $\theta=1,2,\frac{1}{2}$ (see appendix~\ref{app_supecoset})  
provides intuition for the corresponding uplift. Namely, we interpret the LHS~as the product of 
determinants $\prod_i\det(1-y_i Z)$ where $Z$ is the diagonal matrix with 
eigenvalues $y'_j$, and therefore we simply lift the matrix $Z$ to the supermatrix 
(see~\eqref{th1ev},\eqref{th2ev},\eqref{ev2}).

This then  leads to propose the following supersymmetric 
Jacobi Cauchy  identity 
\begin{align}\label{superCauchy1}
	\prod_{i}^M\frac{\prod_{j}^n (1-y_iy'_j)}{\prod_{k}^m 
	(1-y_ix'_k)^\theta} &= 
	\sum_{\ulmb } 
	(-1)^{|\ulmb|} { J}^{(M)}_\ulmb({\bf y};\theta,\tilde p^-,\tilde p^+) {
		\tilde J}^{(n|m)}_{M,\ulmb'}({\bf y}'|{\bf x}';\tfrac{1}{\theta},p^-,p^+)\ ,
\end{align}
involving $BC_N$ Jacobi polynomials and $BC_{n|m}$ super dual 
Jacobi functions. 

Now, the dual super Jacobi functions on the RHS~of \eqref{superCauchy1} are 
simply superblocks \eqref{Block_dualJacobi}.
Thus, redefining variables this becomes
\beq\label{scid_intro}
\prod_{r=1}^\beta \frac{ \prod_{j=1}^n (1-s_r y_j) }{ \, \prod_{i=1}^m (1-s_r 
	x_i)^{\theta} }= 
\sum_{\ulmb} {J}^{(\beta)}_{\ulmb}({\bf s};\theta,  \tilde p^- , \tilde p^+ )\ 
{\Pi_{\ulmb}(\theta) }\, F_{\gamma,\ulmb}({\bf z};\theta,p_{12},p_{43})\ ,
\eeq
where $M=\beta$ and therefore $\gamma=2\beta+p^++p^-$, as usual. 
Upon setting $s_{r=1,\ldots, \beta}$ to the values $0$ or $1$, the LHS~of this formula can  be 
interpreted as a diagram contributing to a  generalised free theory correlator in a $(m,n,\theta)$ SCFT as we show in section~\ref{freetheory}. The RHS then gives the expansion in superconformal blocks of this diagram and we see that the coefficients in a superblock decomposition of any generalised free theory diagram are simply bosonic Jacobi polynomials 
${J}^{(\beta)}_{\ulmb}({\bf s};\theta,  \tilde p^- , \tilde p^+ )$ evaluated on 0s and 1s,  (up to multiplication by an explicitly known factor ${\Pi_{\ulmb}(\theta) }$).

Indeed, if we now specialise back to $m=0$,  the coefficients of this expansion don't change and  we get
\beq\label{scid_intro_bos}
\prod_{r=1}^\beta { \prod_{j=1}^n (1-s_r y_j) }= 
\sum_{\ulmb} {J}^{(\beta)}_{\ulmb}({\bf s};\theta, \tilde p^- , 
\tilde p^+  
)\ {\Pi_{\ulmb}(\theta) }
\, F_{\gamma,\ulmb}(|{\bf y};\theta,p_{12},p_{43}) \ .
\eeq
which is just Mimachi's formula~\eqref{cauchyjacobi}. The above is a formula involving only polynomials! Nevertheless
it gives the same information, the superblock coefficients i.e.~$ {J}^{(\beta)}_{\ulmb}({\bf s};\theta, \tilde p^- , 
\tilde p^+)\Pi_\ulmb$, as \eqref{scid_intro}, which instead contains the full (infinite series) superconformal blocks.

But before discussing this generalised free CFT interpretation of the Cauchy formula let us try to first generalise the formula further.   Having lifted the dual Jacobi polynomials to supersymmetric functions, 
it is natural to also lift the Jacobi polynomials! We will do so in two steps. By looking at \eqref{superCauchy1}, we first consider that $BC_M$ Jacobi polynomials have an 
uplift to $BC_{M|N}$ super Jacobi polynomials, ${ J}^{(M|N)}_\ulmb({\bf y}|{\bf 
x})$ discussed below~\eqref{tildepstar}. %
Once we disentangle the various pieces, we arrive at the doubly supersymmetric Cauchy 
identity:
\begin{align}\label{superCauchy2}
&
\frac{\prod_{i,j}^{M,n} (1-y_iy'_j).\prod_{l,k}^{N,m} 
		(1-x_lx'_k)}{\prod_{i,k}^{M,m} 
		(1-y_ix'_k)^\theta . \prod_{l,j}^{N,n} 
		(1-x_ly'_j)^{\frac1\theta}} =  \\
&\rule{2cm}{0pt}	
	\sum_{\ulmb } 
	(-1)^{|\ulmb|} J^{(M|N)}_\ulmb({\bf y}|{\bf x};\theta,\tilde 
	p^-,\tilde p^+) {
		\tilde J^{(n|m)}}_{M-\frac{1}{\theta}N,\,\ulmb'}({\bf y}'|{\bf 
		x}';\tfrac{1}{\theta},p^-,p^+)\notag
\end{align}
which is therefore the master formula for all Cauchy identities previously discussed.%

In the next section we will prove the supersymmetric Cauchy 
identities~\eqref{superCauchy1} and~\eqref{superCauchy2} and in the process, 
illuminate the appearance of the parameters $\tilde p^\pm$ as well as 
explaining how stability manifests in all parameters. %

\subsection{Proof of the super Cauchy identity}

The idea of the proof is to start from analogous super 
Cauchy identities involving superJack polynomials~\cite{matsumoto} and by using properties of the \JtoB matrix and its inverse, 
derive from them the super Jacobi Cauchy identities~\eqref{superCauchy1} 
and~\eqref{superCauchy2}. We will first use formulae involving interpolation polynomials, in order to have a direct connection with mathematics. 
Then we will give a perspective based on our \JtoB matrix $T$ and its inverse $T^{-1}$.

It is useful to begin with the internal case, namely derive~\eqref{cauchyjacobi}, the Jacobi  Cauchy identity in Mimachi~\cite{Mimachi}, from 
the Cauchy identity for Jack polynomials, 
\beq\label{cauchyjack}
\prod_{i=1}^M \prod_{j=1}^n (1-y_iy'_j) = \sum_{\ulmb} 
(-)^{|\ulmb|}  P^{(M)}_{\ulmb}({\bf y};\theta) P^{(n)}_{\ulmb'}({\bf 
	y}';\tfrac{1}{\theta})\ . 
\eeq
The rough idea is  to change basis from Jacks to Jacobis and dual Jacobis, noting that the change of basis matrices are inverses of each other.
  
So to convert this into~\eqref{cauchyjacobi} recall first the 
relevant definitions for $\tilde S,S$~\eqref{Paul_Jacobi_22},
\begin{align}%
	J^{(M)}_{\umu} &= \sum_{\ulmb \subseteq \umu}  
	(S^{(M)}_{\theta,p^-,p^+})^{\ulmb}_{\umu} 
	\,P_{\ulmb}(y_1,\ldots, y_M;\theta)\\ %
	\tilde J^{(n)}_{\beta,\ulmb} &= 
	\sum_{ \ulmb \subseteq \umu}  \big(\widetilde 
	S_{\beta;\theta,p^-,p^+}\big)_{\ulmb}^{\umu}\,P_\umu(y_1,\ldots,y_n;\theta) \qquad ; \qquad 
	\big(\widetilde S_{\beta}\big)_{\ulmb}^{\umu} = 
	\big(S^{(n)}\big)_{\beta^n\bslash\,\ulmb}^{\beta^n\bslash\,\umu} \ .
\end{align}	
We will proceed by using 
the observation that %
the inverse  of the dual  Jacobi coefficient $\tilde S_\beta^{-1}$, once written via interpolation polynomials %
is very similar to the binomial coefficient $S$ for Jacobi polynomials. Given the precise relation $S \sim {\tilde S}^{-1}$, we will use it in  \eqref{cauchyjack} to obtain the desired 
identity~\eqref{cauchyjacobi}.

Let us go into details. Continuing from \eqref{inverse_T_formula}, 
with the $\tilde P$ given in \eqref{tildepstar}-\eqref{Ptilde_indepM}, and using \eqref{TtoTtilde}, we find %
\begin{align}\label{stildeinverse}
(\tilde S^{-1}_{\beta;\frac{1}{\theta},p^-,p^+})^{\umu'}_{\ulmb'} 
=  \frac{ \Pi_{\ulmb}(\theta) }{
		\Pi_{\umu}(\theta)}  	
	C^0_{\umu/\ulmb}\left( \theta \alpha, 
	\theta \beta;\theta\right)
	\left.\frac{ \tilde P^{*(\mathsf{M})}_{\ulmb}( \umu ; \theta,v) }{  \tilde 
	P^{*(\mathsf{M})}_{\umu}( \umu; 
		\theta,v) 
	}\right|_{	v=-\frac{1}{2}+\theta(\frac{p^++p^-}{2}+1+\beta)}
\end{align}
where on the RHS~we are
assuming the usual relations 
\begin{align}
	\alpha&=\max\left( \tfrac 1 2 (\gamma {-}  p_{12}), \tfrac 
1  2 (\gamma {-}  p_{43}) \right),\quad \beta=\min\left( \tfrac 1 2 (\gamma 
{-}  p_{12}), \tfrac 1  2 (\gamma {-}  p_{43}) \right), \quad 
p^{\pm}=\tfrac{1}{2}|p_{12}\pm p_{43}|,\notag%
\end{align}
and therefore $\gamma=2\beta+p^++p^-$ and $\alpha = \beta{+}p^-$. 

Recall that $\mathsf{M}$ is arbitrary in \eqref{stildeinverse} as long as $\mathsf{M} \geq \mu'_1$. %
Now compare this with the binomial coefficient for the Jacobi 
polynomial~\eqref{FABC_binomial_def2}
\bea\label{FABC_binomial_def2_repeat}
\big(S^{(M)}_{\theta,\vec A}\big)_{\umu}^{\ulmb} 
= 
{(-)^{|\umu|-|\ulmb|} \frac{\Pi_\ulmb(\theta)}{\Pi_\umu(\theta)} 
	C^0_{\umu / \ulmb}(A_1{+}1{+}\theta(M{-}1),\theta M;\theta) 
}{}\times   
\frac{ \tilde P_{\ulmb}^{*(\mathsf{M})}( \umu; \theta,u ) }{  \tilde P_{\umu}^{*(\mathsf{M})}( 
\umu;\theta,u 
	)  
}\Bigg|_{u=\frac{A_1+A_2+1}{2}+\theta M }\ .
\eea
where the $\mathsf{M}$-dimensional $\tilde{P}$ are introduced to display 
the same $\mathsf{M}$ independence as~\eqref{stildeinverse}. 
Comparing~\eqref{FABC_binomial_def2_repeat} with~\eqref{stildeinverse} we see 
that if we match the parameters as 
\beq\label{identify_labels}
\beta\leftrightarrow M\qquad;\qquad A_1\leftrightarrow \tilde p^-\qquad;\qquad A_2 \leftrightarrow \tilde p^+
\eeq
as defined in $\eqref{ptilde}$, i.e.~$\tilde p^{\pm}=  \theta p^{\pm}+\theta- {1}{}$, there is a precise matching (up to a sign) 
between $S^{(M)}$ and $\tilde 
S^{-1}_M$, namely   
\begin{align}\label{stostildeinv}
	\big(S^{(M)}_{\theta,\tilde p^-, \tilde p^+}\big)_{\umu}^{\ulmb} 
	=(-)^{|\umu|-|\ulmb|} \big(\tilde 
	S^{-1}_{M;\frac{1}{\theta},p^-,p^+}\big)^{\umu'}_{\ulmb'} \ .
\end{align}

If we now expand  the Jack polynomial $P^{(n)}_{\ulmb'}({\bf 
	y},\tfrac{1}{\theta})$ on the RHS~of the Cauchy 
	identity~\eqref{cauchyjack} in terms of dual Jacobi polynomials, %
	and then use the   
	relation~\eqref{stostildeinv}, we will recognize a 
	Jacobi 
	polynomial~\eqref{Paul_Jacobi_2}, for specific values of the parameters. 
	Let us explain the reasoning step by step: First
\begin{align}
\prod_{i=1}^M \prod_{n=1}^n (1-y_iy'_j) &= \sum_{\ulmb} 
(-)^{|\ulmb|}  P^{(M)}_{\ulmb}({\bf y};\theta) P^{(n)}_{\ulmb'}({\bf 
	y}',\tfrac{1}{\theta})\notag \\ 
	&=
\sum_{\ulmb} 
(-)^{|\ulmb|}   P^{(M)}_{\ulmb}({\bf y};\theta) \sum_{\umu} (\tilde 
		S^{-1}_{\beta;\frac{1}{\theta},p^-,p^+})^{\umu'}_{\ulmb'} \,  {
			\tilde J}^{(n)}_{\beta,\umu'}({\bf y}';\tfrac{1}{\theta},p^-,p^+) \label{initial_step_proofC} 
\end{align}
where $\beta$ is still arbitrary, since it was not entering the Cauchy identity for Jack polynomials we started with. Then, 
we fine tune $\beta=M$ and $A_{1,2}$ as in \eqref{identify_labels}, so to use \eqref{stostildeinv}
\begin{align}
\prod_{i=1}^M \prod_{n=1}^n (1-y_iy'_j) %
		&=
		 \sum_{\umu}\sum_{\ulmb} 
		(-)^{|\umu|} \big(S^{(M)}_{\theta,\tilde p^-, \tilde 
		p^+}\big)_{\umu}^{\ulmb}   P^{(M)}_{\ulmb}({\bf y};\theta) 
		 \,  {
		 	\tilde J}^{(n)}_{M,\umu'}({\bf y}';\theta^{-1},p^-,p^+) \notag \\
		&=
		\sum_{\umu}
		(-)^{|\umu|}  { J}^{(M)}_\umu({\bf y};\theta,\tilde p^-,\tilde p^+)
		\,  {
			\tilde J}^{(n)}_{M,\umu'}({\bf y}';\tfrac{1}{\theta},p^-,p^+)\ .
		\label{cauchyjacktojacobi}
\end{align}
To obtain the second line we recognised that the sum over $\ulmb$ gives the Jacobi polynomial.%
\footnote{The sum over $\ulmb$  in \eqref{initial_step_proofC} we kept unbounded (since it automatically truncates to $M^n$)  will now also automatically truncate since  $(S^{(M)})_{\umu}^{\ulmb}=0$  when $\umu\subset\ulmb$. The result is therefore a polynomial, i.e.~the Jacobi polynomial. }

Our proof of~\eqref{cauchyjacobi} is thus concluded.
Our derivation here explains both the appearance of the parameter combinations 
$\tilde p^\pm$ in the Cauchy identity, needed to obtain the 
relation~\eqref{stostildeinv}, as well as its stability in $M$. The latter follows because the  
Cauchy identity for Jack polynomials we started with is manifestly stable in $M$.

A bonus of our discussion is that following the logic that leads to \eqref{cauchyjacktojacobi} we can prove other supersymmetric 
generalisations of the Cauchy identity for Jacobi and dual Jacobi polynomials, starting from known 
Cauchy identity involving supersymmetric Jack polynomials~\cite{matsumoto}. 

There is an immediate supersymmetrisation of $P^{(n)}({\bf y}')\rightarrow P^{(n|m)}_{}({\bf 
		y}'|{\bf x}')$ that we can perform and will lead to ~\eqref{superCauchy1}. 
The known Cauchy identity for this case is~\cite{matsumoto} %
\begin{align}
	\prod_{i}^M\frac{\prod_{j}^n (1-y_iy'_j)}{\prod_{k}^m 
		(1-y_ix'_k)^\theta} &= 
	\sum_{\ulmb } 
	(-1)^{|\ulmb|}  P^{(M)}_{\ulmb}({\bf y};\theta)
	P^{(n|m)}_{\ulmb'}({\bf 
		y}'|{\bf x}';\tfrac{1}{\theta}) \ ,
\end{align}
The key point is that the super expansion of $P^{(n|m)}_{\ulmb'}$ in terms of ${\tilde J}^{(n|m)}_{M,\ulmb'}$ depends exactly on the same coefficients $(\tilde 
S^{-1}_{M})^{\umu'}_{\ulmb'}$ as in the bosonic case, because $\tilde J$ is stable. Thus, 
by the very same manipulations as in~\eqref{cauchyjacktojacobi}, 
we arrive at~\eqref{superCauchy1}, 
\begin{align}%
	\prod_{i}^M\frac{\prod_{j}^n (1-y_iy'_j)}{\prod_{k}^m 
	(1-y_ix'_k)^\theta} &= 
	\sum_{\umu } 
	(-1)^{|\umu|} { J}^{(M)}_\umu({\bf y};\theta,\tilde p^-,\tilde p^+) {
		\tilde J}^{(n|m)}_{M,\umu'}({\bf y}'|{\bf x}';\tfrac{1}{\theta},p^-,p^+)\ .
\end{align}
For completeness we point out that the sum over $\umu$ is now unbounded on the east, 
since it only truncates to $M$ rows due to $J^{(M)}$. The $J^{(n|m)}$ is instead an infinite series in small ${\bf x}'$, as it is the LHS.

Next, we want to supersymmetrise $ P^{(M)}_{}({\bf y})\rightarrow P^{(M|N)}_{}({\bf y}|{\bf x})$ and obtain the doubly supersymmetric Cauchy identity~\eqref{superCauchy2}.
As before, the starting point is the doubly supersymmetric Cauchy identity for super Jack polynomials~\cite{matsumoto}
\begin{align}\label{superCauchyjack2}
	\frac{\prod_{i,j}^{M,n} (1-y_iy'_j).\prod_{l,k}^{N,m} 
		(1-x_lx'_k)}{\prod_{i,k}^{M,m} 
		(1-y_ix'_k)^\theta . \prod_{l,j}^{N,n} 
		(1-x_ly'_j)^{\frac1\theta}} &= 
	\sum_{\ulmb } 
	(-1)^{|\ulmb|} P^{(M|N)}_\ulmb({\bf y}|{\bf x};\theta) 
		P^{(n|m)}_{\ulmb'}({\bf y}'|{\bf 
			x}';\tfrac{1}{\theta})\ .
\end{align} 
and the initial step is the same as~\eqref{initial_step_proofC}, for arbitrary $\beta$. But this time we will need to carefully understand how to fine tune $\beta$ with both $M$ and $N$.

In~\cite{Veselov_3} super Jacobi polynomials were defined and shown to have 
an expansion in terms of super Jacks with coefficients given in terms of super 
interpolation polynomials:  
\begin{align}\label{Paul_super_Jacobi}
	J^{(M|N)}_{\umu}({\bf y}|{\bf x};\theta,p^-,p^+) = \sum_{\ulmb \subseteq 
	\umu}  
	(S^{(M|N)}_{\theta,p^-,p^+})_{\umu}^{\ulmb} 
	\,P^{(M|N)}_{\ulmb}({\bf y}|{\bf x};\theta)\ ,%
\end{align}
where
\begin{align}
\label{FABC_binomial_super}
\big(S^{(M|N)}_{\theta,A_1,A_2}\big)_{\umu}^{\ulmb} 
&= 
{(-)^{|\umu|{-}|\ulmb|} \frac{\Pi_\ulmb(\theta)}{\Pi_\umu(\theta)} 
	C^0_{\umu / \ulmb}(a,b;\theta) 
}{}\times   
\frac{ \tilde P_{\ulmb}^{*(M|N)}( \umu_e|\umu_s; \theta,u ) }{  \tilde 
P_{\umu}^{*(M|N)}( \umu_e|\umu_s;\theta,u 
	)  
}\notag\\
a&= A_1{+}1{+}\theta(M{-}1){-}N,\quad b=\theta M{-}N,\quad u=\tfrac{A_1{+}A_2{+}1}{2}{+}\theta M{-}N
\ ,
\end{align}
with $\umu_s,\umu_e$ defined below~\eqref{ptildeinv}.

In the above formula the dimensions of the interpolation polynomials $(M|N)$ is tied directly to the 
dimension of the super Jacobi. However in~\eqref{ptildeinv} we saw that the dimensions of the interpolation polynomials can be arbitrary 
(as long as they are big enough to contain the corresponding Young diagram).
So we can write the more general expression:
\bea\label{FABC_binomial_super_prime}
\big(S^{(M|N)}_{\theta,A_1,A_2}\big)_{\umu}^{\ulmb} 
= 
{(-)^{|\umu|{-}|\ulmb|} \frac{\Pi_\ulmb(\theta)}{\Pi_\umu(\theta)} 
	C^0_{\umu / \ulmb}(a,b;\theta) 
}{}\times   
\left.\frac{ \tilde P_{\ulmb}^{*(\mathsf{M}|\mathsf{N})}( \umu_e|\umu_s; \theta,u ) }{  \tilde 
	P_{\umu}^{*(\mathsf{M}|\mathsf{N})}( \umu_e|\umu_s ;\theta,u 
	)}\right|_{u=\tfrac{A_1{+}A_2{+}1}{2}{+}\theta M{-}N}  
\ 
\eea
with $a,b$ as above, any $\mathsf{M}|\mathsf{N}$. (In particular, we could choose $\mathsf{N}=0$ and obtain the super Jacobi 
coefficients in terms of non supersymmetric interpolation polynomials).

Compare \eqref{FABC_binomial_super_prime} with 
$\tilde S_\beta^{-1}$, in which we use the analogue of the supersymmetric formula~\eqref{inverse_T_formula_susy}  
for  $T_\gamma^{-1}$, namely, 
\begin{align}%
(\tilde S^{-1}_{\beta;\frac{1}{\theta},p^-,p^+})^{\umu'}_{\ulmb'} 
=  \frac{ \Pi_{\ulmb}(\theta) }{
		\Pi_{\umu}(\theta)}  	
	C^0_{\umu/\ulmb}\left( \theta \alpha, 
	\theta \beta;\theta\right)
\left.\frac{ \tilde P^{*(\mathsf{M}|\mathsf{N})}_{\ulmb}( \umu_s|\umu_e ; \theta,v) }{  
	\tilde 
	P^{*(\mathsf{M}|\mathsf{N})}_{\umu}( \umu_s|\umu_e; 
		\theta,v) 
	}\right|_{	v=-\frac{1}{2}+\theta(\frac{p^++p^-}{2}+1+\beta)}
\end{align}
we see that
\begin{align}\label{stostildeinv_plus}
	\big(S^{(M|N)}_{\theta,\tilde p^-, \tilde p^+}\big)_{\umu}^{\ulmb} 
	=(-)^{|\umu|-|\ulmb|} \big(\tilde 
	S^{-1}_{M-\frac{1}{\theta} N;\frac{1}{\theta},p^-,p^+}\big)^{\umu'}_{\ulmb'} \ .
\end{align}
We can now apply manipulations similar to the ones performed 
in~\eqref{cauchyjacktojacobi} to derive the doubly supersymmetric Jacobi 
Cauchy~\eqref{superCauchy2} from the doubly supersymmetric Jack 
Cauchy~\eqref{superCauchyjack2}.

\subsection{Free theory block coefficients}
\label{freetheory}

We now turn to our main interest in the Cauchy formulae, namely, we will compute conformal partial wave (CPW) coefficients, in  
generalised free theories, with no effort, just using Cauchy identities and nothing else. 

Four-point functions of generalised free theories are given simply as sums of 
products of propagators with the correct weights: 
\begin{align}\label{intro_4pt0}
	&
	\langle \mathcal{O}_{p_1} \mathcal{O}_{p_2} \mathcal{O}_{p_3} 
	\mathcal{O}_{p_4} 
	\rangle_{\text{free}}= \sum_{\{b_{ij}\}}
	a_{\{b_{ij}\}} \prod_{1\leq i<j \leq 4} g_{ij}^{b_{ij}}, \qquad  \qquad 
	\sum_{i} 
	b_{ij}=p_j\ .
\end{align}
Diagrammatically each term in the sum looks as
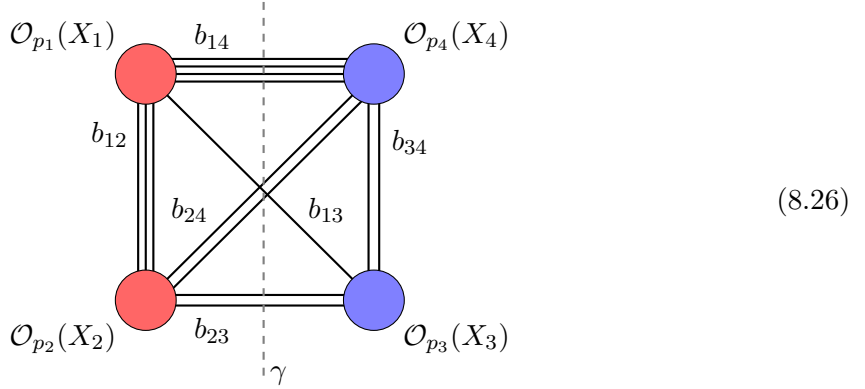
\beq\label{diagramma_figure}
\begin{array}{c}
	\begin{tikzpicture}
		\def\lato 	{3}
		
		\def\xuno	{2}
		\def\yuno	{2}
		
		\def\xdue {2}
		\def\ydue {2-\lato}
		
		\def\xtre {2+\lato}
		\def\ytre {2-\lato}
		
		\def\xquat {2+\lato}
		\def\yquat {2}
		
		\foreach \x in {-1,0,1,2}
		\draw[thick]    (\xuno,\yuno+\x*.1 ) --  (\xquat,\yquat+\x*.1 );
		
		\foreach \x in {-1,0,1}
		\draw[thick]    (\xuno+\x*.1,\yuno) --  (\xdue+\x*.1,\ydue 
		);				
		
		\foreach \x in {-1,1}
		\draw[thick]    (\xdue,\ydue+\x*.07) --  (\xtre,\ytre+\x*.07 );	
		
		\foreach \x in {-1,1}
		\draw[thick]    (\xtre+\x*.07,\ytre) --  (\xquat+\x*.07,\yquat );	
		
		\foreach \x in {-1,0}
		\draw[thick]    (\xdue-\x*.1,\ydue+\x*.1) --  (\xquat-\x*.1,\yquat 
		+\x*.1);		
		
		\foreach \x in {0}
		\draw[thick]    (\xtre-\x*.1,\ytre+\x*.1) --  (\xuno-\x*.1,\yuno 
		+\x*.1);

		\draw[gray, thick, dashed] 		(\xtre*.5+.05, \ydue-1 ) -- 	
		(\xtre*.5+.05, \yuno+1);
		
		\draw  (\xtre*.5, \ydue-1 )			node[right] 	{$\gamma$};
		
		\draw (\xuno-.25,\yuno+.5) 			 node[left] 
		{$\mathcal{O}_{p_1}(X_1)$};
		\draw (\xdue-.25,\ydue-.5)  			node[left] 
		{$\mathcal{O}_{p_2}(X_2)$};
		\draw (\xtre+.25,\ytre-.5)  			node[right] 
		{$\mathcal{O}_{p_3}(X_3)$};
		\draw (\xquat+.25,\yquat+.5)  		node[right] 
		{$\mathcal{O}_{p_4}(X_4)$};

		\draw (\xuno-.1,\yquat-.8) 				node[left] 		{$b_{12}$};
		\draw (\xtre+.1,\yuno-.9) 				node[right] 	{$b_{34}$};
		\draw (\xuno+.5,\yquat+.5) 			node[right] 	{$b_{14}$};
		\draw (\xuno+.5,\ytre-.4) 				node[right] 	{$b_{23}$};	
		\draw (\xuno+.2,\ytre+1.2) 			node[right] 	{$b_{24}$};		
		\draw (\xuno+2,\ytre+1.2) 			node[right] 	{$b_{13}$};

		\draw[fill=red!60,draw=black] (\xuno,\yuno) circle (.4cm);
		\draw[fill=red!60,draw=black] (\xdue,\ydue) circle (.4cm);
		\draw[fill=blue!50,draw=black] (\xtre,\ytre) circle (.4cm);
		\draw[fill=blue!50,draw=black] (\xquat,\yquat) circle (.4cm);

	\end{tikzpicture}
\end{array}
\eeq
Here there are six $b_{ij}$ connecting 
operators between insertion points, 
and four constraints  $\sum_{i} b_{ij}=p_j$ leaving two degrees of freedom. 
We parametrise these with $\gamma:=p_1+p_2-2b_{12}$ and $k:= b_{23}$.%
\footnote{This then gives
	$$
	\rule{.8cm}{0pt}
	\begin{array}{lll}
		b_{12}=\tfrac12(p_1+p_2-\gamma)\quad&;\quad 
		b_{34}=\tfrac12(p_3+p_4-\gamma) \quad&;\quad   
		b_{13}=\tfrac12(p_3-p_4+\gamma)-k \\[.2cm]
		b_{24}=\tfrac12(p_2-p_1+\gamma)-k \quad&;\quad b_{23}=k \quad&;\quad 
		b_{14}=\tfrac12(p_1-p_2-p_3+p_4)+k\ .
	\end{array}
	$$
	An equivalent parametrisations, $\gamma_{ij}=p_i-p_j-2b_{ij}$,
	counts the number of bridges 
	going from $\cO_{p_i}\cO_{p_j}$ to the opposite pair. 
	$\gamma_{12}+\gamma_{13}+\gamma_{23}=\sum_i p_i$, 
	is a Mandelstam-type constraint for free theory four-point correlators. }
The  free four-point 
correlator 
can then be re-written as 
\begin{align}\label{intro_4pt}
	&
	\langle \mathcal{O}_{p_1} \mathcal{O}_{p_2} \mathcal{O}_{p_3} 
	\mathcal{O}_{p_4} 
	\rangle_{\text{free}}=
	g_{12}^{ \frac{p_1+p_2}{2} } g_{34}^{\frac{ p_3+p_4}{2} } \left( \frac{ 
	g_{24}  }{ g_{14} }\right)^{\!\!\frac{ p_{21} }{2} } \left( \frac{ g_{14} 
	}{ g_{13} } \right)^{\!\!\frac{ p_{43 } }{2} }  %
	\!\!\sum_{\gamma,k }  a_{\gamma,k}
	\left(  \frac{ g_{13} g_{24} }{ g_{12} g_{34} 
	}\right)^{\frac\gamma2 }
	\left(\frac{ g_{14}  g_{23} }{ g_{13} g_{24} }\right)^{\!\! 
	k}\quad
\end{align}
where $ \gamma \in\{ \gamma_m,\gamma_m+2,\dots,\gamma_M \}$ with 
$\gamma_m=\max(|p_{12}|,|p_{43}|)$, $\gamma_M=\min( 
p_1+p_2, 
p_3+p_4 )$ and $k=0,1,..,\tfrac12(\gamma-\gamma_m)$.
So the propagator structures  are classified 
firstly by the total 
number of bridges $\gamma$ going from the pair of operators 
$\mathcal{O}_{p_1}\mathcal{O}_{p_2}$ 
to the pair $\mathcal{O}_{p_3}\mathcal{O}_{p_4}$,  and secondly by the the 
number of 
bridges $k$ connecting 
$\mathcal{O}_{p_2}$ with $\mathcal{O}_{p_3}$. 
The parameters $a_{\gamma,k}$ are arbitrary coefficients, determined by the 
particular microscopic theory in question.

The form of~\eqref{intro_4pt}, has exactly the same prefactor as 
the superconformal blocks~\eqref{4pt}. Moreover,  %
changing from $B_{\gamma,\ulmb}$ to $F_{\gamma,\ulmb}$, as in~\eqref{Jexp_intro},
we can drop the first $\gamma$-dependent prefactor.
Thus the superconformal block decomposition  of the above  propagator 
structure~\eqref{diagramma_figure} reduces to 
 \begin{align}\label{def_calA_k}
 	\left(\frac{ g_{14}  g_{23} }{ g_{13} g_{24} 
   }\right)^{\!\! k} =
 \sum_{ \ulmb}
 { A }_{\gamma,k, \ulmb}
 F_{\gamma,\ulmb}
\end{align}
where the expansion coefficients ${ A }_{\gamma,k, \ulmb}$ also depend on $\theta;p_{12},p_{43}$, as $B_{\gamma,\ulmb}$ does.

 The LHS~of \eqref{def_calA_k} is simple to understand, using from~\eqref{zz} we have that:
 \begin{align}
	\left(\frac{ g_{14}  g_{23} }{ g_{13} g_{24} }\right)^{\!\! k} =
 \sdet (1{-}Z)^{-k\#}=
 	 \prod_{r=1}^\beta \frac{ \prod_{j=1}^n (1-s_r y_j) }{ \, \prod_{i=1}^m 
 	 (1-s_r x_i)^{\theta} } \qquad;\qquad
 	\begin{array}{l} s_{i=1,\ldots ,k}=1\\[.2cm] s_{i=k+1,\ldots, 
 	\beta}=0\end{array}\
 	\label{from_prop_to_Cauchy}
 \end{align}
where $\beta=\frac{1}{2}\min(\gamma-p_{43},\gamma-p_{12})$ as usual. 
We recognise here the LHS~of the superconformal 
Cauchy identity \eqref{scid_intro}. Thus comparing~\eqref{def_calA_k} with this 
Cauchy identity~\eqref{scid_intro} we can immediately identify the CPW 
coefficients
${ A }_{\gamma,k,\ulmb}$ as an evaluation formula for a 
Jacobi polynomial:
\beq
A_{\gamma,k,\ulmb;\theta,p_{12},p_{43}}= \ {\Pi_{\ulmb}(\theta) }\, 
{J}^{(\beta)}_{\ulmb}({\bf s};\theta,  \tilde p^- , \tilde p^+ )|_{{\bf 
s}=(1^k,0^{\beta-k})}
\eeq
with $ \tilde p^\pm$ defined in the same way, i.e.~$\tilde p^\pm=  \theta 
p^\pm+\theta- {1}{}$ and $p^{\pm}=\tfrac{1}{2}|p_{12}\pm p_{43}|$.

Finally then, through Okounkov's binomial formula \eqref{Paul_Jacobi_2}, the free theory 
CPW coefficients, $A_{\gamma,k,\ulmb}$, can be written as a sum over 
interpolation polynomials evaluated at partitions 
multiplied by Jack polynomials evaluated at 1 and 0, 
$P_{\umu}(1^k,0^{\beta-k};\theta)$.
Note that such a Jack polynomial, $P_{\umu}(1^k,0^{\beta-k};\theta)$,  truncates to a Young 
diagram with just $k$ rows out of $\beta$, 
$P_{[\mu_1,\ldots,\mu_k]}(1^k;\theta)$, from stability.
The latter is a known explicit evaluation formula due to Stanley \cite{Stanley}.
Putting this together then we arrive at an explicit formula for the CPW coefficients as a sum of binomial coefficients,
\begin{align}\label{acoeff}
{ A }_{\gamma,k, \ulmb;\theta,p_{12},p_{43}}=  {\Pi_{\ulmb}(\theta) }\ 
\sum_{\umu\subseteq \ulmb} 
	(S^{(\beta)}_{\theta,\tilde p^-,\tilde p^+})_{\ulmb}^{\umu}  \ \times
\prod_{1\leq i<j\leq k} \frac{ (\theta (j{-}i{+}1))_{\mu_i-\mu_j} }{ (\theta 
(j{-}i))_{\mu_i-\mu_j} }\ .
\end{align}
This formula provides the decomposition of any free theory diagram in superconformal blocks. 
It depends on $\gamma,k$, then $\theta$ and the external charges $p_{12}$ and 
$p_{43}$. It does not of course depend on $(m,n)$ by stability of  the  super 
Cauchy identity. Indeed, because of stability, the derivation of~\eqref{acoeff} 
doesn't rely on any supersymmetry at all, and can be derived 
directly from the original bosonic Cauchy identity of 
Mimachi~\eqref{cauchyjacobi}.

We will investigate this formula for free theory CPW coefficients further in a 
separate publication~\cite{freeblockcoeff} giving even simpler and completely explicit 
formulae for the $\theta=1$ case.

\section{Decomposing superblocks} \label{section_mn_deco}

The  superconformal block (or equivalently dual super Jacobi functions) $B_{\gamma,\ulmb}$ has been so far defined as the multivariate series 
\beq\label{an_an_def_B}
B_{\gamma,\ulmb}=\left(\frac{\prod_i x_i^\theta}{\prod_j y_j}\right)^{\!\!\frac{\gamma}{2}} \, F_{\gamma,\ulmb}\ ,
\qquad F_{\gamma,\ulmb}=\sum_{\umu \supseteq \ulmb} { ({ T}_{\gamma})_\ulmb^\umu} \, P_{\umu}({\bf z};\theta)\ 
\eeq
with $({T}_{\gamma})_\ulmb^\umu$ computed either from the recursion of section \ref{sec_SCBlock} or by the binomial coefficient of section~\ref{sec_guideline}. 

In this section we  want to relate this to a more common approach where instead $B_{\gamma,\ulmb}$ is decomposed into its constituent bosonic blocks. 
It is useful to  view this in a more general context of
decomposing superblocks into smaller {\em super}blocks, 
\beq\label{an_ecomposi}
(m+m'|n+n') \rightarrow (m|n) \otimes (m'|n')\ .
\eeq  
Then we will specialise to the case of decomposing into bosonic blocks an internal blocks by considering $m'=n=0$.

\subsection{General decomposition of blocks}

The analogous decomposition \eqref{an_ecomposi} for superJack polynomials is given in~\eqref{superdecomp}, and it involves the structure constants ${\cal S}$. Note that these 
structure constants are independent of the dimensions $m,n,m',n'$, because of stability!
Combining this with the expansion of Blocks into Jacks~\eqref{an_an_def_B} we arrive at the   decomposition of superblocks as 
\begin{align}\label{superblockdecomp}
	F^{(m+m',n+n')}_{\gamma,\ulmb}= \sum_{\umu,\unu} F_{\gamma,\umu}^{(m,n)}\, ({\cS}_{\gamma})_\ulmb^{\umu\unu}\, F_{\gamma,\unu}^{(m',n')}\ ,
\end{align}
where the block structure constants ${\cS}_{\gamma}$ depend on $\theta,p_{12},p_{43}$ and are related to the Jack structure constants via the (inverse) block to Jack matrices $T_\gamma$ as
\begin{align}\label{blockstructure}
	({\cS}_{\gamma})_{\ulmb}^{\umu\unu} = \sum_{\substack{\tilde{\ulmb}\supseteq{\ulmb}\\\tilde{\umu}\subseteq {\umu}\\\tilde{\unu}\subseteq{\unu}}}{\cS}_{\tilde{\ulmb}}^{\tilde{\umu}\tilde{\unu}}\, (T_{\gamma})_{{\ulmb}}^{\tilde{\ulmb}}   (T^{-1}_{\gamma})_{\tilde{\umu}}^{{\umu}}  (T^{-1}_{\gamma})_{\tilde{\unu}}^{{\unu}}\  . 
\end{align}
Note we distinguish the Jack structure constants $\cS$ from the block structure constants $\cS_\gamma$ purely by the presence of the  subscript for the latte.r 
Since we know $T_{\gamma}$ infinite triangular matrices, the issue in this formula is to understand when and how the $\cS_\gamma$ are non vanishing.  

For the structure constants of  {\em Jack} polynomials, since these are homogeneous, we clearly require $|\tilde \ulmb|=|\tilde \umu|+|\tilde \unu|$ for ${\cS}_{\tilde{\ulmb}}^{\tilde{\umu}\tilde{\unu}} \neq 0$. A stronger statement conjectured by Stanley (conjecture 8.4 in~\cite{Stanley}) is that the Jack structure constants are non vanishing if and only if the corresponding Schur structure constants (given by the Littlewood-Richardson rule) are non vanishing
\begin{align}\label{conjecturestanley}
	{\cS}_{\tilde{\ulmb}}^{\tilde{\umu}\tilde{\unu}}(\theta) \neq 0 \qquad \Leftrightarrow \qquad 	{\cS}_{\tilde{\ulmb}}^{\tilde{\umu}\tilde{\unu}}(1)\neq 0\ .
\end{align} 
In any case, clearly for a given $\tilde \ulmb$ there are only a finite number of non vanishing Jack structure constants ${\cS}_{\tilde{\ulmb}}^{\tilde{\umu}\tilde{\unu}}$ and thus a finite number of terms in the decomposition of $P_{\tilde \ulmb}$ into smaller Jacks.

What about the block structure constants? Since the \JtoB matrix $T_\gamma$ is triangular we see that~\eqref{blockstructure} gives the  $({\cS}_{\gamma})_{\ulmb}^{\umu\unu}$ in terms of	${\cS}_{\tilde{\ulmb}}^{\tilde{\umu}\tilde{\unu}}$ with $\tilde \ulmb \supseteq \ulmb$. Thus  $\tilde{\ulmb}$ can become arbitrarily large and the decomposition of a superblock into smaller superblocks might  involve an infinite number of terms. However let us consider more carefully the structure of the Young diagrams which can survive.

First note that the decomposition~\eqref{superblockdecomp} has a vertical cut-off: the height of the Young diagram $\tilde \lambda$ can be no larger than $\beta=\min\left( \tfrac 1 2 (\gamma {-}  p_{12}), \tfrac 1  2 (\gamma {-}  p_{43}) \right)$, since $T_\gamma$ has such a vertical cut-off (see discussion above~\eqref{example_picture_block}). 

Now we claim that if $\umu$ is much wider than $\ulmb$ (i.e.~$\mu_1>>\lambda_1$ ) then $\unu$ will also have to be very wide in order for $(\cS_\gamma)_\ulmb^{\umu\unu} \neq 0$.
More precisely we make the following conjecture:
\begin{align}\label{conjecturestructure}
	(\cS_\gamma)_\ulmb^{\umu\unu} \neq 0 \qquad \Rightarrow \qquad \exists\,  \tilde \umu,\tilde \unu,\underline{\kappa}, \text{ such that } \cS_\ulmb^{\tilde \umu \tilde \unu}\neq 0, \ \cS_\umu^{\tilde \umu \underline{\kappa}}\neq 0,\  \cS_\unu^{\tilde \unu \underline{\kappa}}\neq 0\ .
\end{align}
To get a feel for  this claim, consider the case $\ulmb=\varnothing $ (corresponding to half BPS operators in a superconformal theory). Then the above conditions will require $\tilde \umu=\tilde \unu=\varnothing$ and so $\umu=\unu(=\underline{\kappa})$. Therefore the conjecture implies the following in this case
\begin{align}
	(\cS_\gamma)_{\varnothing}^{\umu\unu} \neq 0 \qquad \Rightarrow \qquad \umu=\unu\ ,
\end{align} 
so the decomposition of a superblock with a trivial Young diagram is diagonal. This is not obvious from~\eqref{blockstructure} and requires non trivial relations between the Jack structure constants $\cS$ and the block to Jack matrix $T_\gamma$. 

More generally, the conjecture~\eqref{conjecturestructure} can be proven in the case $\theta=1$ where there is a group theory interpretation. 
In the $\theta=1$ case the decomposition of the superblock $F_{\gamma,\ulmb}^{(m+m'|n+n')}$ in~\eqref{superblockdecomp} has the group theoretic interpretation of decomposing reps of $U(2m{+}2m'|2n{+}2n')\rightarrow U(2m|2n)\otimes U(2m'|2n')$ and taking only the states which are diagonal under the decomposition $U(2m{+}2m'|2n{+}2n')\rightarrow U(m+m'|n+n')\otimes U(m+m'|n+n')$ ( equivalently taking only those reps whose Dynkin labels are symmetric as only such states contribute to the block).%
\footnote{Note the difference  with the corresponding group theory interpretation of the Jack decomposition~\eqref{superdecomp} which is instead equivalent to decomposing  reps of $U(m+m'|n+n')\rightarrow U(m|n)\otimes U(m'|n')$ in the $\theta=1$ case.}
Let us perform this decomposition $U(2m+2m'|2n+2n')\rightarrow U(2m|2n)\otimes U(2m'|2n')$ for the induced $U(2m+2m'|2n+2n')$ representations $\cO_{\gamma,\ulmb}(X^{A A'})$ (described in section~\ref{supergroup_sec}. We must first decompose the $U(m{+}m'|n{+}n')$ rep $\ulmb$ into $U(m|n)\otimes U(m'|n')$ reps. In fact we do this decomposition twice but then take the same reps for both copies of  $U(m{+}m'|n{+}n')$.
Under the decomposition let the $U(m{+}m'|n{+}n')$ index $A$ split into the $U(m|n)$ indices $(\alpha,a)$ and similarly $A' \rightarrow (\dot \alpha,a')$.
So $$\cO_{\gamma,\ulmb(A)\ulmb(A')}(X^{AA'})\rightarrow \bigcup_{\tilde \umu \tilde \unu} \cO_{\gamma,\tilde \umu(\alpha)\tilde \unu(a)\tilde \umu(\dot \alpha)\tilde \unu(a')}(X^{AA'})$$
(this decomposition is dictated by $\cS_\ulmb^{\tilde \umu \tilde \unu}$ and explains the appearance of this  in~\eqref{conjecturestructure}). 
We must now also consider the derivatives $\partial_{A A'}$ acting on these. In particular the off-diagonal derivatives  $\partial_{\alpha a'}$ and $\partial_{a \dot \alpha}$ will yield new induced representations of the subgroup $U(2m|2n)\otimes U(2m'|2n')$. Since the derivatives commute (and we again project onto the diagonal rep) we arrive at
$$\cO_{\gamma,\tilde \umu(\alpha)\tilde \umu(\dot \alpha)\unu(a)\unu(a')}(X^{AA'}) \rightarrow 
\bigcup_{\underline 
	\kappa} \partial^{|\underline{\kappa}|}_{\underline{\kappa}(\alpha)\underline{\kappa}(\dot \alpha)\underline{\kappa}(a)\underline{\kappa}(a')}\cO_{\gamma,\tilde \umu(\alpha)\tilde \umu(\dot \alpha)\unu(a)\unu(a')}(X^{\alpha \dot \alpha},X^{aa'})\ ,$$ 
and arrive at the tensor products $\underline{\kappa}\otimes \tilde \umu$ and  $\underline{\kappa}\otimes \tilde\unu$ which explains the structure constants $\cS_\umu^{\tilde \umu \underline{\kappa}}, \cS_\unu^{\tilde \unu \underline{\kappa}}$ appearing in~\eqref{conjecturestructure}.

We believe a similar argument for~\eqref{conjecturestructure} can be made for the other group theoretic cases $\theta=2,\frac{1}{2}$ although we haven't gone carefully  through the details.

\subsection{Superblock to block decomposition}

We now focus on the main interest,  the case most relevant for physics application, namely that of decomposing a superblock into its corresponding bosonic blocks. This corresponds to taking   $m'=n=0$ in the general case described in the previous section. 
Specialising to this case the decomposition~\eqref{superblockdecomp} becomes
\begin{align}
	F^{(m,n)}_{\gamma,\ulmb}({\bf x}|{\bf y})= \sum_{\umu,\unu} F_{\gamma,\umu}^{(m,0)}({\bf x})\, ({\cS}_{\gamma})_{\ulmb}^{\umu\unu}\, F_{\gamma,\unu}^{(0,n)}({\bf y})\ .
\end{align}
Now note that the conjecture~\eqref{conjecturestructure} together with Stanley's conjecture~\eqref{conjecturestanley} implies that the sum on the rhs is finite in this case. 
First, as already noted, the Young diagrams $\umu,\unu$ can not extend vertically below row $\beta$.  Since $\unu$ is a $(0,n)$ rep it can not be wider than $n$, $\nu_1\leq n$ (and cannot have no more than $n$ columns, since this is a $U(n)$ rep transposed). Thus $\unu$ is restricted to sit inside the rectangle $[n^\beta]$.
But what about $\umu$? If this was much wider than $\ulmb$ it would be impossible to satisfy the conjecture~\eqref{conjecturestructure}. 
To see this, first note that the Littlewood-Richardson rule (and Stanley's conjecture~\eqref{conjecturestanley}) means that $\cS_\ulmb^{\tilde \umu \tilde \unu}\neq 0$ implies $\tilde \mu_1,\tilde\nu_1 \leq \lambda_1\leq \tilde\mu_1+\tilde\nu_1$.
Then this together with conjecture~\eqref{conjecturestructure} means that
$(\cS_\gamma)_\ulmb^{\umu\unu} \neq 0$ only if  $\mu_1-\lambda_1 \leq \mu_1-\tilde \mu_1 \leq \kappa_1 \leq \nu_1\leq n$. So in particular the width $\mu_1$ of $\umu$ can not be bigger than $\lambda_1+n$, i.e.~the width of $\ulmb + n$. There are  tighter bounds but this is already enough to show that there are a finite number of terms.
Note that specialising to $m=n=2, \theta=1$ the decomposition should reproduce  the diagonal component reps in the decomposition of ${\cal N}=4$ superconformal multiplets which were written out in great detail in appendix B of \cite{Dolan:2002zh}.

\subsection{Examples of the superblock to block decomposition}
\label{sec_examples}
To ground the discussion about the conformal and internal block decomposition of a superconformal block, we give some explicit  examples here.

Let us fix the external data to be $\frac{\gamma}{2}=\alpha=\beta=5$, $\theta=\frac{1}{2}$, and $\ulmb=[3,1]$,  
and consider different values of $(m,n)$. This also illustrates the $m,n$ independence.

In the  $(1,1)$ theory, the decomposition of $F_{\gamma=10,[3,1]}(x|y;\frac{1}{2})$ gives
\beq
(\cS_{\gamma=10})_{[3,1]}^{\unu\umu}: \qquad  \begin{array}{c}
	\begin{tikzpicture}[scale=.8]
		
		\foreach \x / \y  in  { -1/$[1]$,  0/ $[2]$ , 1/ $[3]$,  2/$[4]$  } %
	\draw (\x,0) node {\y};

	\foreach \y / \x  in  { -1/$[0]$,  -2/ $[1]$ , -3/ $[2]$,  -4/$[3]$ ,  -5/{$[4]$}  ,-6/{$[5]$}  }
	\draw (-2,\y) node {\x};	
	
	\draw (-2,-.5) -- (4.5,-.5);
	\draw (-1.9, 0) node {$\umu$};
	
	\draw (-1.5,.5) -- (-1.5,-6.5);
	\draw (-2.5,-.3) node[scale=.9] {$\unu'$};
	
	\foreach \x / \y  in  {  -1/0,  0/ 0 , 1/ 0,  2/0 }%
\draw (\x,-1) node {$\y$};

\foreach \x / \y  in  {  -1/0,  0/ 0 , 1/ {\color{red} 1},  2/0 }%
\draw (\x,-2) node {$\y$};

\foreach \x / \y  in  {  -1/0,  0/ {\color{red} \frac{7}{9}} , 1/ 0,  2/{\color{red} \frac{1573}{40320}} }%
\draw (\x,-3) node {$\y$};

\foreach \x / \y  in  {  -1/0,  0/0 , 1/ {\color{red} \frac{39}{1120}}, 2/0 },  %
\draw (\x,-4) node {$\y$};

\foreach \x / \y  in  {  -1/0,  0/0 , 1/ {0},  2/{0} }%
\draw (\x,-5) node {$\y$};

\foreach \x / \y  in  {  -1/0,  0/0 , 1/ {0},  2/{0} }%
\draw (\x,-6) node {$\y$};

\end{tikzpicture}
\end{array}
\eeq
Now let us look at the same external data but in the $(m,n)=(2,1)$ theory, thus considering the decomposition of $F_{\gamma=10,[3,1]}(x_1,x_2|y;\frac{1}{2})$. We find
\beq \label{11ex}
\!\!\!\!\!\!\!\!\!
(\cS_{\gamma=10})_{[3,1]}^{\umu \unu}: \ \ \ 
\begin{array}{c}
\!\!\!\!\!\!\begin{tikzpicture}[scale=.9]

\foreach \x / \y  in  { -1/$[1]$,  0/ $[2]$ , 1/ $[3]$,  2/$[4]$ ,  3/{$[1,1]$},  4/ {$[2,1]$} , 5/{$[3,1]$},   6/{$[4,1]$} ,  7/{$[2,2]$},   8/{$[3,2]$},    9/{$[4,2]$} ,   10/{$[3,3]$},    11/{$[4,3]$},   12/{$[4,4]$}   }
\draw (\x,0) node {\y};

\foreach \y / \x  in  { -1/$[0]$,  -2/ $[1]$ , -3/ $[2]$,  -4/$[3]$ ,  -5/{$[4]$}  ,-6/{$[5]$}  }
\draw (-2,\y) node {\x};	

\draw (-2,-.5) -- (12.5,-.5);
\draw (-1.9, 0) node {$\umu$};

\draw (-1.5,.5) -- (-1.5,-6.5);
\draw (-2.5,-.3) node[scale=.9] {$\unu'$};

\foreach \x / \y  in  {  -1/0,  0/ 0 , 1/ 0,  2/0 ,  3/0,  4/ 0 , 5/{\color{blue} 1},   6/0 ,  7/0,   8/0,    9/0 ,   10/0,    11/0,   12/0   }
\draw (\x,-1) node {$\y$};

\foreach \x / \y  in  {  -1/0,  0/ 0 , 1/ {\color{red} 1},  2/0 ,  3/0,  4/ {\color{blue} \frac{28}{45}} , 5/0,   6/{ \color{blue} \frac{1573}{53550}} ,  7/0,   8/{\color{blue} \frac{4096}{133875}},    9/0 ,   10/0,    11/0,   12/0   }
\draw (\x,-2) node {$\y$};

\foreach \x / \y  in  {  -1/0,  0/ {\color{red} \frac{7}{9}} , 1/ 0,  2/{\color{red} \frac{1573}{40320}} ,  3/0,  4/ 0 , 5/{ \color{blue} \frac{93428}{1461915}},   6/{0} ,  7/{\color{blue} \frac{64}{2025}},    8/0 ,   
9/{\color{blue} \frac{37752}{33632375}},    10/0,   11/0, 12/0   }
\draw (\x,-3) node {$\y$};

\foreach \x / \y  in  {  -1/0,  0/0 , 1/ {\color{red} \frac{39}{1120}},  2/{0} ,  3/0,  4/ { \color{blue} \frac{8}{225}} , 5/{0},   6/{ \color{blue} \frac{1573}{937125}} ,  7/{0},    8/{\color{blue} \frac{1664}{1561875}} ,   
9/{0},    10/0,   11/0, 12/0   }
\draw (\x,-4) node {$\y$};

\foreach \x / \y  in  {  -1/0,  0/0 , 1/ {0},  2/{0} ,  3/0,  4/ {0} , 5/{ \color{blue} \frac{13}{7875}},   6/{0} ,  7/{0},    8/{0} ,   9/{0},    10/0,   11/0, 12/0   }
\draw (\x,-5) node {$\y$};

\foreach \x / \y  in  {  -1/0,  0/0 , 1/ {0},  2/{0} ,  3/0,  4/ {0} , 5/{0},   6/{0} ,  7/{0},    8/{0} ,   9/{0},    10/0,   11/0, 12/0   }
\draw (\x,-6) node {$\y$};

\end{tikzpicture}
\end{array}
\eeq
Note that the red diamond is the same as in the $(1,1)$ theory. The other contributions
correspond to conformal blocks with two full rows, which therefore did not exist in the previous case with $m=1$. 

Finally let us look in the $(m,n)=(3,1)$ theory at the decomposition of $F_{\gamma=10,[3,1]}(x_1,x_2,x_3|y;\frac{1}{2})$. 
We find $(\cS_{\gamma=10})_{[3,1]}^{\umu \unu}=$
\beq
\begin{array}{c}
\!\!\!\!\!\!\!\!\!\!
\!\!\!\!\!\!\!\!\!\!\!\!\!\!\!
\begin{tikzpicture}[scale=.85]

\foreach \x / \y  in  { -1/$[1]$,  0/ $[2]$ , 1/ $[3]$,  2/$[4]$ ,  3/{$[1,1]$},  4/ {$[2,1]$} , 5/{$[3,1]$},   6/{$[4,1]$} ,  7/{$[2,2]$},   8/{$[3,2]$},    9/{$[4,2]$} ,   10/{$[3,3]$},    %
11/{$[1,1,1]$}, 12/{$[2,1,1]$} , 13/{$[3,1,1]$},  14/{$[4,1,1]$},  15/{$[2,2,1]$},  16/{$[3,2,1]$}, 17/ {$[4,2,1]$}  , 18/ {$[3,3,1]$} }
\draw (\x,0) node[scale=.7] {\y};

\foreach \y / \x  in  { -1/$[0]$,  -2/ $[1]$ , -3/ $[2]$,  -4/$[3]$ ,  -5/{$[4]$}  ,-6/{$[5]$}  }
\draw (-2,\y) node {\x};	

\draw (-2,-.5) -- (18.5,-.5);
\draw (-1.9, 0) node {$\umu$};

\draw (-1.5,.5) -- (-1.5,-6.5);
\draw (-2.5,-.3) node[scale=.9] {$\unu'$};

\foreach \x / \y  in  {  -1/0,  0/ 0 , 1/ 0,  2/0 ,  3/0,  4/ 0 , 5/{\color{blue} 1},   6/0 ,  7/0,   8/0,    9/0 ,   10/0,    %
11/0, 12/0, 13/0, 14/0, 15/0, 16/0, 17/0, 18/0    }
\draw (\x,-1) node {$\y$};

\foreach \x / \y  in  {  
-1/0,  0/ 0 , 1/ {\color{red} 1},  2/0 ,  3/0,  4/ {\color{blue} \frac{28}{45}} , 5/0,   6/{ \color{blue} \frac{1573}{53550}} ,  7/0,   8/{\color{blue} \frac{4096}{133875}},    9/0 ,   10/0,    %
11/0, 12/0, 13/{\frac{169}{4320}}, 14/0, 15/0, 16/0, 17/0, 18/0  }
\draw (\x,-2) node {$\y$};

\foreach \x / \y  in  {  
-1/0,  0/ {\color{red} \frac{7}{9}} , 1/ 0,  2/{\color{red} \frac{1573}{40320}} ,  3/0,  4/ 0 , 5/{ \color{blue} \frac{93428}{1461915}},   6/{0} ,  7/{\color{blue} \frac{64}{2025}},    8/0 ,   
9/{\color{blue} \frac{37752}{33632375}},    10/0,   %
11/0, 12/ {\frac{8}{243}},  13/0,  14/{\frac{1573}{1049580}},   15/0, 16/{\frac{1352}{1136025}}, 17/0, 18/0  }
\draw (\x,-3) node {$\y$};

\foreach \x / \y  in  {  -1/0,  0/0 , 1/ {\color{red} \frac{39}{1120}},  2/{0} ,  3/0,  4/ { \color{blue} \frac{8}{225}} , 5/{0},   6/{ \color{blue} \frac{1573}{937125}} ,  7/{0},    
8/{\color{blue} \frac{1664}{1561875}} ,   9/{0},    10/0,   %
11/0,    12/0,    13/{\frac{13}{4080} },   14/0,   15/{\frac{5}{3564}},  16/0,  17/{\frac{11583}{241076864}}, 18/ 0   }
\draw (\x,-4) node {$\y$};

\foreach \x / \y  in  {  -1/0,  0/0 , 1/ {0},  2/{0} ,  3/0,  4/ {0} , 5/{ \color{blue} \frac{13}{7875}},   6/{0} ,  7/{0},    8/{0} ,   9/{0},    10/0,   %
11/ 0, 12/{\frac{7}{4050}},  13/0, 14/{\frac{1573}{19992000}},   15/0, 16/{\frac{3328}{66268125}}, 17/0, 18/0   }
\draw (\x,-5) node {$\y$};

\foreach \x / \y  in  {  -1/0,  0/0 , 1/ {0},  2/{0} ,  3/0,  4/ {0} , 5/{0},   6/{0} ,  7/{0},    8/{0} ,   9/{0},    10/0,   %
11/0, 12/0, 13/{\frac{13}{129600}}, 14/0, 15/0, 16/ 0, 17/0, 18/0  }
\draw (\x,-6) node {$\y$};

\end{tikzpicture}
\end{array}
\eeq

The $m\otimes n$ decomposition generically will have a diamond/rhomboid structure w.r.t. 
to the products basis of conformal and internal blocks. %
However we should say that in order to see such a diamond structure, one has to considered a generic Young diagram $\ulmb$,  and a generic $\gamma$, 
otherwise parts of the diamond might be reflected on the boundaries. %

One interesting aspect that the above example illustrates is the appearance of the coefficients `1' at the top of each diamond. 
In the cases  $\theta=\frac{1}{2},1,2$, where there is a group theory interpretation, this `1' is easily explained by considering the corresponding decomposition in Minkowski superspace.\footnote{Minkowski superpsace corresponds to the coset space obtained by putting cross through all odd (white) nodes in the relevant super Dynkin diagram rather than the crossed through nodes of analytic superspace~\eqref{dynkin}.} 
There the decomposition simply corresponds to the standard superspace expansion of the corresponding superfield over Grassmann odd variables and the first term in the expansion (the term obtained by switching off all Grassmann odd variables) has corresponding coefficient `1'. We can view this superspace expansion  for different values of $(m,n)$, thus giving different `1' coefficients. So for example in the $(1,1)$ theory this gives the red `1' in~\eqref{11ex} and in the $(2,1)$ theory it gives the blue `1'. But of course stability means the red `1' remains there even for the $(2,1)$ theory. 

It is interesting to see how  to  recover the ``$1$ at the top of the diamond", i.e.~the leading conformal block in the decomposition of a superblock from the formula~\eqref{blockstructure} and show that it is true for all $\theta$ regardless of whether there is a group theory interpretation. 
This follows directly from the observation that the Jack structure constants are `1' whenever the large Young diagram is built from the smaller ones put on top of each other~\eqref{sis1}. A similar result then follows for the block structure constants 
\begin{align}\label{sis1}
	(\mathcal{S}_\gamma)^{\umu\unu}_{\ulmb}(\theta)=1 \qquad \text{if}\qquad  \ulmb=[\umu,\unu]\ .
\end{align}
since $\mathcal{S}$ and $\mathcal{S}_\gamma$  are simply related via a triangular change of (infinite) basis with 1's on the diagonal~\eqref{blockstructure} (recall $(T_\gamma)^\ulmb_\ulmb=1$).

Testing the conjecture~\eqref{conjecturestructure}  in general  turns into a sophisticated computation, 
because the Jack structure constants $\mathcal{S}$ themselves are non-trivial and explicit formulae are not known. 
The works \cite{Lassalle_2,Lassalle_3,Schlosser} obtain a recursive formulation which can be used to carry out the computation,  
but we are not aware of a more explicit formula. A naive approach we used, for generic number of rows, which nevertheless gives the desired result, 
is to start from the definitions, and use computer algebra. 
But we point out that the remarkable consequence of the conjecture~\eqref{conjecturestructure} is that even though ${\cal S}_\gamma$ involves infinite matrices $T_\gamma$, the result always  has only finitely many non zero values. In particular if we keep $\alpha, \beta, \gamma$ arbitrary in the recursion relation for $T_\gamma$ this truncation does not occur, but only takes place when $\beta$ (or $\alpha$) is an integer larger than the height of $\ulmb$. To help clarify these points we now give some explicit results for the two row case.

 \subsubsection*{Explicit formulae for two-rows}

For Young diagrams with two rows we can give alternative explicit (i.e.~non combinatorial) formulae for the building blocks of the block structure constants~\eqref{blockstructure}. 
Indeed, the Jack structure constants for two-row diagrams are known thanks to a result quoted in \cite{warnaar}, 
\beq\label{sumM_2rows}
{\cal C}^{[\mu_1,\mu_2]}_{[\kappa_1,\kappa_2]\ [\omega_1,\omega_2]}=\frac{ 
	(\theta,2\theta+\mu_-, 
	\omega_2+\kappa_1-\mu_2+1,  \omega_1+\kappa_2-\mu_2+1 )_{\mu_2-\kappa_2 -\omega_2} }{ 
	(1,1+\theta+\mu_-,
	\omega_2+\kappa_1-\mu_2+\theta,  \omega_1+\kappa_2-\mu_2+\theta   )_{\mu_2-\kappa_2-\omega_2}  }
\eeq
and
\beq
\mathcal{S}^{\underline{\kappa}\, \underline{\omega}}_{\umu }=
\frac{ \Pi_{\underline{\kappa}}(\theta) \Pi_{\underline{\omega}}(\theta) }{   \Pi_{\umu}(\theta)} \mathcal{C}^{ \umu}_{\underline{\kappa}\,\underline{\omega}\, }\ .
\eeq
(Where $(a,b,c)_n$ represents a product of Pochhammers $a_n b_nc_n$.)
Note this formula is manifestly symmetric in $\underline{\kappa}\leftrightarrow \underline{\omega}$. It is also clear that for $\theta=1$ the value of ${\cal C}$ collapse to unity, when it is non vanishing.

Now, we assemble the block structure constants $\mathcal{S}_\gamma$ in \eqref{blockstructure}.  The two-row \JtoB matrix $T_\gamma$ can be taken from \eqref{DO2rows_concorda}, 
and the inverse 
$T^{-1}_\gamma$ is given by the formula \eqref{inverse_T_formula}, 
in terms of the interpolation polynomials in \eqref{formula_Korn_interp}. Putting them together then gives the block structure constants $\mathcal{S}_\gamma$.

In the recursion we may leave
all variables $\alpha,\beta,\gamma$ unspecified in  $T_\gamma$ and $T^{-1}_\gamma$. We find that in this case there is actually no truncation and there are infinitely many non-zero structure constants 	$(\mathcal{S}_\gamma)^{\umu\unu}_{\ulmb}(\theta)$ for given $\ulmb$.
However one finds that for arbitrary $\alpha,\beta,\gamma$, for $\unu$ large enough they always appear with a factor of $(\alpha-2)(\beta-2)$. So for any $\ulmb,\umu$  exists a $\unu^*$ such that
\beq\label{crucial_truncation}
(\mathcal{S}_\gamma)^{\umu\unu}_{\ulmb}(\theta) = (\alpha-2)(\beta-2)\Big[\ldots \Big] 
\qquad \forall\  \unu\supset \unu^*
\eeq
where the terms understood in $\Big[\ldots \Big]$ might be cumbersome, 
but their knowledge is not needed to understand the truncation.
It is then clear from \eqref{crucial_truncation} that upon specifying 
either $\beta$ or $\alpha$ to be the value of the maximal number of rows of the Young diagram, 
in the case of this section $\beta=2$, the block structure constants $\cal{S}_\gamma$ will truncate. We checked explicitly that the very same mechanism  generalises to any number of rows.

The $m\otimes n$ decomposition then can be carried out in practice for arbitrary  Young diagrams and theories $(m,n;\theta)$. 
It depends on the Young diagrams and $\theta$ through rational functions. 
 For example, consider the $(2,2)$ theory and $B_{\gamma=4,[\lambda_1,\lambda_2]}$. Then, we know that
\beq
B_{\gamma=4,[\lambda_1,\lambda_2]}=\left(\frac{\prod_{i=1}^{2} x_i^\theta}{\prod_{j=1}^{2} y_j}\right)^{\!\!\frac{\gamma}{2}} \, \left(  F^{(2,0)}_{\gamma,[\lambda_1,\lambda_2]}({\bf x})\times   F^{(0,2)}_{\gamma,\varnothing}({\bf y})+ \ldots \right)\ .
\eeq
The coefficients of the next terms  beyond $F_{\gamma=4,[\lambda_1,\lambda_2]}\times F_{\gamma=4,\varnothing}$, with $p_{12}=p_{43}=0$, are
\beq\label{example_from_DO_mondeco}
\begin{array}{cccc}
	(\cS_{\gamma=4})^{[\lambda_1,\lambda_2-1],\Box}_{[\lambda_1,\lambda_2]}= &  \frac{\theta\lambda_2}{\lambda_2+\theta-1}\\[.2cm]
	(\cS_{\gamma=4})^{[\lambda_1-1,\lambda_2],\Box}_{[\lambda_1,\lambda_2]}= & \frac{\theta(\lambda_1+\theta)(\lambda_1-\lambda_2)(\lambda_1-\lambda_2+2\theta-1)}{(\lambda_1+2\theta-1)(\lambda_1-\lambda_2+\theta-1)(\lambda_1-\lambda_2+\theta)}\\[.2cm]
	(\cS_{\gamma=4})^{[\lambda_1+1,\lambda_2],\Box}_{[\lambda_1,\lambda_2]}= &\frac{\theta(\lambda_1+2\theta)(\lambda_1+3\theta-1)(\lambda_1+\lambda_2+2\theta)(\lambda_1+\lambda_2+4\theta-1) }{ 4 (2\lambda_1+4\theta-1)(2\lambda_1+4\theta+1)(\lambda_1+\lambda_2+3\theta-1)(\lambda_1+\lambda_2+3\theta)} \\[.2cm]
	(\cS_{\gamma=4})^{[\lambda_1,\lambda_2+1],\Box}_{[\lambda_1,\lambda_2]}= & \frac{ \theta(\lambda_1-\lambda_2)(\lambda_1-\lambda_2+2\theta-1)(\lambda_2+\theta)(\lambda_2+2\theta-1)(\lambda_1+\lambda_2+2\theta)(\lambda_1+\lambda_2+4\theta-1)}{
		4(\lambda_1-\lambda_2+\theta-1)(\lambda_1-\lambda_2+\theta)(\lambda_1+\lambda_2+3\theta-1)(3\theta+\lambda_1+\lambda_2)(2\lambda_2+2\theta-1)(2\lambda_2+2\theta+1)}
\end{array}
\eeq 
The complexity of these rational functions comes both from the complexity of the structure constants and the solution of the recursion, 
equivalently the complexity of interpolation polynomials evaluated at partitions.

\section{Conclusions and outlook}

Let us summarise our findings and conclude with an outlook.

In this paper we have uncovered a non trivial connection between
the theory of conformal and superconformal blocks for four-point correlators of scalar fields,
and the theory of symmetric functions (as well as Heckman Opdam hypergeometric functions, Calogero Sutherland Moser wave functions and supersymmetric generalisations of all of these). This connection generalises previous work done 
in \cite{Doobary:2015gia} for theories defined on a super Grassmannian 
$Gr(m|n,2m|2n)$, corresponding to the case $(m,n;\theta=1)$, 
where the connection between symmetric functions and superconformal blocks 
passed through the use of super Schur polynomials \cite{moens}. 
It also generalises the connection between bosonic external/internal blocks and 
$BC_2$ hypergeometric functions/Jacobi polynomials
uncovered in~
\cite{Dolan:2003hv,Isachenkov:2016gim,Chen:2016bxc,Isachenkov:2017qgn,Chen:2020ipe}.
The starting point has been to consider four-point functions of scalar operators on certain 
special generalised $(m,n)$ analytic superspaces in 3,4 and 6 dimensions following~\cite{Galperin:1984av,Howe:1995md, Hartwell:1994rp,hep-th/9408062,hep-th/0008048,
	Heslop:2001zm,Heslop:2002hp,Heslop:2004du,Doobary:2015gia,Howe:2015bdd}.
We then looked at the very 
concrete problem of solving a Casimir eigenvalue differential equation, and we 
have found that: \\

\begin{minipage}{15cm}
$\bullet$ The Casimir operator for these theories coincides with the $BC_{m|n}$ CMS operator for special values of a parameter $\theta$ ($\theta=\frac{1}{2},1,2$). This then immediately suggests to define
 $(m,n;\theta)$ analytic superspaces for more general $\theta$
as discussed in section \ref{summary_list_theory}.\\
\end{minipage}  \\

\begin{minipage}{15cm}
$\bullet$ The eigenfunctions of the $BC_{m|n}$ Casimir 
admit a representation as a multivariate series 
over super Jack polynomials,  
\begin{align}\label{Jexp_conclu}
B_{\gamma, \ulmb}({\bf z})=
	\left(\frac{\prod_i x_i^\theta}{\prod_j y_j}\right)^{\!\!\frac{\gamma}{2}} \,  F_{\gamma,\ulmb}({\bf z})\ , \qquad \qquad
	 F_{\gamma,\ulmb}({\bf z})=\sum_{\umu \supseteq \ulmb} { ({ T}_{\gamma})_\ulmb^\umu} \, P_{\umu}({\bf z})\ ,
\end{align}
with expansion coefficients $T_\gamma$ enjoying special properties.\\
\end{minipage}   \\

\begin{minipage}{15cm}
$\bullet$ The expansion coefficients $ ({ T}_{\gamma})_\ulmb^\umu$ do not depend on $m,n$ as long as $\ulmb,\umu$ are Young diagrams. Since super Jack polynomials are themselves stable, by construction, our superconformal blocks are stable!\\
\end{minipage}   \\

\begin{minipage}{15cm}
$\bullet$ Although the coefficients $T_\gamma$ do not depend on $(m,n)$ when $\ulmb,\umu$ are Young diagrams, the recursion (and its solution) can non-the-less be represented in various forms, in terms of the row lengths of the Young diagram, the column lengths, or a mixture of the two. 
All representations give the same result, as shown in section \ref{explicit_form_sec}, when the parameters coincide with those of a Young diagram. However, the different choices yield  inequivalent analytic continuations (section \ref{analytic_cont_sec}).  A mixed representation consistent with the  $(m,n)$ Young diagram structure is then most  appropriate for superblocks in an $(m,n;\theta$) theory and will give the correct superblocks for operators with non-integer (anomalous) dimensions.
Such an $(m,n)$  representation satisfies the shift symmetry~\eqref{deform_overview_sec} for arbitrary $\theta$ (which is a manifest symmetry for the  physical cases $\theta=\frac{1}{2},1,2$).\\
\end{minipage}   \\

\begin{minipage}{15cm}
$\bullet$ We   exemplified this $(m,n)$ analytic continuation  in the simplest $(1,1)$ theory, for arbitrary $\theta$. This shows, in a concrete case, that the recursion can be solved 
by using a mixed, or supersymmetric, representation where the Young diagrams are read both along rows and columns.\\ 
\end{minipage}

The above results find immediate application for superconformal theories 
not only in dimensions 3,4,6, but also in 1,2 and  possibly also 5 dimensions, as well as ordinary bosonic 
theories in any dimension, as explained in section \ref{summary_list_theory}. 
The superconformal block understanding is particularly important  for theories 
that have a gravity dual through the $AdS/CFT$ correspondence, where we expect the SCFT 
to provide important insights towards the quest for a theory of quantum 
gravity. The simplest examples which have made full use of  this $(m,n;\theta)$ formalism have been so far $AdS_5\times S^5$~\cite{Aprile:2017bgs,Aprile:2017qoy,Aprile:2019rep}. There has been some use of the formalism also for $AdS_7 \times S^4$ but restricted to the lowest charge correlators~\cite{Heslop:2004du,Abl:2019jhh}.%
\footnote{More recent work has considered superblocks for higher charge correlators in  the  $AdS_7 \times S^4$ case using the Ward identity approach discussed at the start of section~\ref{overv_conject} ~\cite{Beem:2015aoa,Alday:2020tgi}} 
It would be interesting to extend this further and consider other $AdS\times S$ backgrounds. 
An interesting case that has not been attacked yet is $AdS_3\times S^3\times S^3$, which realises a 
one-parameter family of two-dimensional superconformal 
theories with $D(2,1;\alpha)\times D(2,1;\alpha)$ symmetry.\footnote{In the context 
of integrability this has been studied in, e.g.~\cite{Babichenko:2009dk}. }

The connection between (super)conformal blocks and symmetric functions, that we have uncovered, yielded  in turn a 
number of mathematical results, providing further inspiration on the 
mathematics of $BC_{n|m}$ symmetric functions, as well as further insights for 
the superconformal blocks themselves.
In this context, we found that: \\ %

\begin{minipage}{15cm}
$\bullet$ The reduction our blocks from $BC_{m|n}$ to $BC_{n}$ produces the dual Jacobi polynomials 
$\tilde J_{\beta,\ulmb}(;\theta,p^{\pm})$ in $n$ variables, which are stable, and were not noticed before. 
From the knowledge of these polynomials (in particular their stability), we showed that they admit a natural supersymmetric extension 
which give superconformal blocks.\\
\end{minipage}   \\

\begin{minipage}{15cm}
$\bullet$ Jacobi polynomials $J_{\ulmb}(;\theta,p^{\pm})$ have a binomial formula due to Okounkov 
\cite{Okounkov_1,Rains_1,Koornwinder},  and so do the dual Jacobi polynomials. 
The difference is the way the 
Young diagrams enter the interpolation polynomials $P^*$. For the dual Jacobi polynomials, 
the Young diagrams are complemented and transposed w.r.t.~a given rectangle $R$. This manipulation stabilises the binomial formula for the Jacobi polynomials.\\
\end{minipage}   \\

\begin{minipage}{15cm}
$\bullet$ The dependence on the variables of the interpolation polynomials 
 $ P^*_{N^M\,\bslash\umu}(  N^M\,\bslash\ulmb 
;\,\theta,u{=}\tfrac{1}{2}{-}\theta\tfrac{\gamma}{2}{-}N )$, 
provides the highly non-trivial $\gamma$ dependence of 
$(T_{\gamma})_{\ulmb}^{\umu}$, 
which was quite challenging to understand from the point of view of the recursion. \\
\end{minipage}   \\

\begin{minipage}{15cm}
$\bullet$ Elaborating on the way complementation works on the \JtoB matrix, we showed that its 
inverse, the \BtoJ matrix, $T_\gamma$, has a natural representation in terms of interpolation polynomials, which we upgraded to super interpolation polynomials of Sergeev and Veselov \cite{Veselov_Macdonald}.  %
Thus, whenever an identity pairs a Young diagram with  its complement, we can use the \JtoB 
and the \BtoJ matrix to uplift that identity to an identity for superconformal blocks.\\  
 \end{minipage}   \\

\begin{minipage}{15cm}
$\bullet$ From the $(m,n)$ Cauchy identity for superJack polynomials we proved a superconformal Cauchy identity,
\beq
\prod_{r=1}^\beta \frac{ \prod_{j=1}^n (1-s_r y_j) }{ \, \prod_{i=1}^m (1-s_r x_i)^{\theta} }= 
\sum_{\ulmb} {J}_{\ulmb}({\bf s};\theta,  \tilde p^- , \tilde p^+ )\ 
{\Pi_{\ulmb}(\theta) }\, F_{\gamma,\ulmb}({\bf z};\theta,p_{12},p_{43}) 
\eeq
Then we showed that the LHS~becomes 
a propagator structure when ${\bf s}=(1^k,0^{\beta-k})$, up to the overall prefactor $(\prod_{i,j} x_i^{\theta}/y_j)^{\frac{\gamma}{2}}$, 
which distinguishes $B_{\gamma,\ulmb}$ from $F_{\gamma,\ulmb}$. It follows that, with no effort, the above Cauchy identity 
provides the decomposition of any (generalised) free theory diagram within our formalism.\\
\end{minipage}\\

\begin{minipage}{15cm}
	$\bullet$ The physics understanding of the $(m,n)$ super Cauchy identity allowed us to prove a 
	new doubly supersymmetric Cauchy identity ~\eqref{superCauchy2}
	involving superconformal blocks (dual super Jacobi functions) and the super Jacobi 
	polynomial of Sergeev and Veselov \cite{Veselov_Macdonald}.\\ 
\end{minipage}\\

\begin{minipage}{15cm}
	$\bullet$ Finally we could read off known results on higher order operators for Heckman Opdam hypergeometrics to obtain all higher order Casimirs in $\theta=1$ theories (in particular $\cN=4$ SYM).
\end{minipage}\\

Overall, we showed in section \ref{sec_guideline} and \ref{sc_Cauchy}  that the connection between superconformal blocks 
and symmetric functions comes with a non trivial exchange of information, which is fruitful for both sides. 
The superconformal Cauchy identity is a beautiful example of this exchange. On one side the understanding of stability clarifies the proofs greatly, on the other side, the Cauchy formula 
allows us to read off, with no effort, the superconformal block 
 decomposition of any free theory diagram. 

There are a number of topics one could investigate further. There is still more to be understood about the analytic properties of the superblocks. In~\cite{Isachenkov:2016gim,Isachenkov:2017qgn} bosonic blocks were identified with twisted versions of the Harish Chandra functions giving Heckman Opdam hypergeometric functions. We give here two new viewpoints on this. Both viewpoints start with the much simpler $BC_n$ Jacobi polynomials or more accurately the dual $BC_n$ Jacobi polynomials which are stable. This stability then leads immediately to a family of $(n,m)$ supersymmetric dual Jacobi functions, and then specialising to $n=0$ we obtain the conformal blocks (which are certain $S_n$ symmetric combinations of Harish Chandra functions). But the rank one considerations suggest another viewpoint, namely that the analytic continuation, regular at the origin,  of the dual Jacobi polynomials to negative row lengths  also yield the conformal blocks. We then expect similarly   that analytic continuation of the super dual Jacobi functions to negative row lengths for the infinite non compact direction will give the super Jacobi polynomials of~\cite{Veselov_Macdonald}.

Besides the immediate relevance of our work for theories that have a generalised free limit, 
the superconformal block expansion has a non perturbative nature,  and recent work in 
\cite{Penedones:2019tng,Sleight:2019ive,Carmi:2019cub,Caron-Huot:2020adz,Gopakumar:2021dvg} %
has shown that a fully non perturbative reconstruction of a four-point correlator can be given within the framework of the
inversion formula \cite{Caron-Huot:2017vep}. A prominent role  is played by the Polyakov-Regge block.%
 We think that this discussion should have a counterpart in the theory of symmetric functions 
worth investigating.\footnote{For similar reasons, it would be interesting to investigate the diagonal limit 
\cite{Hogervorst:2013kva} of the superconformal blocks, 
as well as the change of variables to the radial coordinate proposed in 
\cite{Hogervorst:2013sma}.}

It would be interesting to try to adapt our story beyond scalar 
representations, as well as for higher points, possibly by adapting the techniques 
of~\cite{Schomerus:2017eny,Buric:2019rms,Buric:2019dfk,Buric:2020buk}.

\subsection{Outlook on the $q$-deformed superblocks}

Finally, it would be fascinating to find a place for a possible $q$-deformation of the superblocks, 
since the various group theoretic objects that we have been discussing, i.e.~Jack, Jacobi, 
interpolation polynomials etc., have well known $q$-deformed generalisations.  
In particular, the expansion coefficients~\eqref{SCbinomial2} of the superblocks
\beq
(T_{\gamma})_{\ulmb}^{\umu}= 
(\mathcal{N}_{}\,)^{\umu}_{\ulmb}\times
\frac{ P^{*(M)}_{N^M\,\bslash\umu}(  N^M\,\bslash\ulmb ;\theta,u ) }{  P^{*(M)}_{ 
		N^M\,\bslash\ulmb}( N^M\,\bslash\ulmb ;\theta,u) 
}\Bigg|_{u=\tfrac{1}{2}-\theta\tfrac{\gamma}{2} -N } \ .
\label{SCbinomial22} 
\eeq
written as binomial coefficient has
such a $q$-deformation. A simple question  
 to ask is whether the $q$-deformed binomial coefficient plays a role in
physical theories.\footnote{In this paper, 
we used the limit $q\rightarrow 1$ and $t\rightarrow q^\theta$, 
	in the notation of \cite{Rains_1}. }  For example, is there a notion of $q$-deformed CFTs in any dimension? In these theories, how do $q$-deformed conformal blocks
compare to our multivariate series? For applications within the AdS/CFT correspondence, 
is it possible to investigate the field theory dual to the $q$-deformed world-sheet $\sigma$-model on 
$AdS_5\times S^5$ \cite{Delduc:2013qra,Delduc:2014kha,Arutyunov:2015mqj}.

In a little more detail, the $q$-deformed binomial coefficient can be used to define the Koornwinder polynomials $K_{\ulmb}({\bf z},q,t;a_{i=1,2,3,4})$, as shown by Okounkov \cite{Okounkov_1}.
The latter are $BC$ orthogonal (Laurent) polynomials in the bosonic variables ${\bf z}$, and
for a choice of parameters $\vec{a}$, the $q\rightarrow 1$ 
defines the $BC$ Jacobi polynomials (see e.g.~\cite{Koornwinder}):
\beq
\lim_{q\rightarrow 1} K_{\ulmb}({\bf z})= (-4)^{ |\ulmb| } J_{\ulmb}\left( -\tfrac{1}{4}(\sqrt{z_i}-\tfrac{1}{\sqrt{z_i}})^2\right).
\eeq
An obvious proposal to build the
$q$-deformed generalisation of the $(0,n)$ internal block would be to start from the Koornwinder polynomials.
Koornwinder polynomials have a binomial expansion, but the 
polynomials involved are themselves $q$-deformed $BC$ interpolation polynomials.\footnote{The 
	limit to the Jacobi polynomials is designed to degenerate the $BC$ interpolation polynomials to Jack polynomials
	and the $q$-deformed binomial coefficient to its classical counterpart \cite{Koornwinder}.}  
One might have expected the generalisation of the internal block would rather correspond to a series over  Macdonald polynomials, instead 
of $BC$ interpolation polynomials. After all the super Jack polynomials, which provided the 
relevant basis for the $(m,n)$ superconformal blocks, admit a natural generalisation to the super 
Macdonald operators of Sergeev and Veselov \cite{Veselov_Macdonald}. 
On the other hand, there is a fundamental reason why Koornwinder polynomials sum over $BC$ interpolation polynomials, and this is evaluation symmetry, i.e.~the property that evaluating $K_{\umu}$ 
at partition $\ulmb$ is invariant under $\umu\leftrightarrow \ulmb$.
This property is manifest in the binomial coefficient formalism \cite{Okounkov_1}, precisely because
the expansion coefficients on the basis of $BC$ interpolation polynomials
are again $BC$ interpolation polynomial evaluated at a partition, and therefore evaluation symmetry ``swaps" the two.

A proposal for the $q$-deformed generalisation of the $(0,n)$ internal block can be read from E.Rains' results in \cite{Rains_1}.\footnote{We thank Ole Warnaar for pointing out this to us.}
It is the virtual Koornwinder ``polynomial" $\hat{K}_{Q,\ulmb}$ (see Definition 7 and Theorem 7.13 in \cite{Rains_1}). This is a stable $S_n$ (rather than $BC_n$) \emph{function} 
which provides a 1-parameter, $Q$, generalisation of Koornwinder polynomials $K^{(n)}_{\ulmb}$ in the following sense,
\beq\label{final_Rains}
\hat{K}_{Q,\ulmb}(\ldots,z_n; q,t; \vec{a})\Big|_{Q=q^m}= \prod_i z_i^m\, K^{(n)}_{m^n\bslash \ulmb}({\bf z};q,t,\vec{a})\ .
\eeq
In more detail, $\hat{K}_{Q,\ulmb}$ is defined by a binomial formula that sums over Young diagrams $\umu$ such that $\ulmb\subseteq\mu$, 
and uses the equivalent of the $q$-deformed \JtoB matrix, as in our dual Jacobi function $\tilde J_{\beta,\ulmb}( {\bf y}|\,)$ in \eqref{sJ}, 
analytically continued in $\beta$ (see the discussion in section \ref{analytic_cont_sec} and appendix \ref{alphabetarevisited}.) 
The similarity is even more precise because when $Q=q^m$ 
the sum truncates to the polynomial on the r.h.s.~of \eqref{final_Rains}, which indeed has the same characteristics as dual Jacobi polynomials:
\beq
\tilde J_{\beta,\ulmb}({\bf y}|\,)\Big|_{\beta=m} \equiv   \prod_i y_i^m  
J_{m^n\bslash\,\ulmb}(\ldots ,\tfrac{1}{y_n}) \ .
\eeq
We can see now that by taking the limit from Koornwinder polynomials to Jacobi polynomials in \eqref{final_Rains}, and re-expressing the 
result in terms of our dual Jacobi polynomials we find that
\beq\label{rains_limit}
\lim_{q\rightarrow 1} \hat{K}_{Q,\ulmb}({\bf z})\Big|_{Q=q^m}= \prod_i(1-z_i)^{2m} \tilde J_{m,\ulmb}\left( \frac{-4 z_i}{(1-z_i)^2}\right) 
\eeq
The argument of $\tilde J_{}$ might be surprising at first, but  $Z(z)={-4 z}/{(1-z)^2}$ 
is just a conformal mapping which sends the unit disk in the $z$ plane to $Z\in \mathbb{C}\bslash(1,+\infty)$.
Stability of the r.h.s.~of \eqref{rains_limit} then follows from that of $\hat{K}_{Q,\ulmb}$, and coincides with 
the stability of $\tilde J_{\beta,\ulmb}$ which we discussed in this paper  (since that of $\prod_i (1-z_i)^{2m}$ is obvious). 
We thus conclude that $\hat{K}_{Q,\ulmb}$ is the $q$-generalisation of our $\tilde J_{\beta,\ulmb}({\bf y}|\,)$ where $Q=q^\beta$.

From stability,  following the same logic that led us from dual Jacobi polynomials to super Jacobi function $\tilde{ J}_{\beta}({\bf y}|{\bf x})$ in section \ref{dual_Jac_sec},
we infer that there is a supersymmetric uplift of the virtual $\hat{K}$ to a ``super virtual Koornwinder"  $\hat{K}_Q({\bf y}|{\bf x})$ obtained by supersymmetrising the 
expansion polynomials that define $\hat{K}$, called virtual $P^*$ for reference, but keeping the same expansion coefficients, i.e.~the $q$-deformed \JtoB matrix.   
It is crucial here that such virtual $P^*$ are again $S_n$ stable polynomials, the ``super virtual" $P^*$ will then generalise the super Macdonald polynomials  of \cite{Veselov_Macdonald}. 
The super virtual Koornwinder  $\hat{K}_Q({\bf y}|{\bf x})$ would then give the $q$-generalisation of our superconformal blocks. 
Following our general motivations, it would be fascinating to study in more detail the properties of the super $\hat{K}_Q({\bf y}|{\bf x})$ and understand their role in the context of $q$-deformed CFTs. 

\section*{Acknowledgements}
We would like to thank Ole Warnaar for very illuminating conversations about $BC$ symmetric functions, and guidance into the literature.
With great pleasure we would like to thank Reza Doobary for collaboration at an initial stage of this work, and James Drummond, Arthur Lipstein, and Hynek Paul, Michele Santagata 
for collaboration on related projects over the years.
We also thank Chiung Hwang, Gleb Arutunov, Pedro Vieira, Sara Pasquetti, Volker Schomerus and other members of the DESY theory group for discussions  on related topics. 
FA is supported by FAPESP grant 2016/01343-7, 2019/24277-8 and 2020/16337-8, and was partially supported by the ERC-STG grant 637844-HBQFTNCER.  PH is supported
by STFC Consolidated Grant ST/P000371/1 and the European Union's Horizon 2020 research and innovation programme
under the Marie Sklodowska-Curie grant agreement No.764850 ``SAGEX''.

\appendix

\section{ Super/conformal/compact groups of interest}\label{app_supecoset}

Here we give details about the supercoset construction of section~\ref{how_to_read}, 
and in particular show how the general formalism ties in with the more standard approach in the case of non-supersymmetric  conformal blocks.

\subsection{$\theta=1$: $SU(m,m|2n)$}

The $SU(m,m|2n)$ is the simplest family of theories %
with a supergroup interpretation for any value of $m,n$ positive integers. This corresponds to the case $\theta=1$.

The supercoset here is the most straightforward of the three general classes: Firstly view the complexified group $SU(m,m|2n)=SL(2m|2n;{\mathbb C})$ 
as the set of $(m|2n|m)\times (m|2n|m)$ matrices (this is a straightforward change of basis of the more standard $(2m|2n)\times (2m|2n)$
 matrices). Then  the supercoset space we consider has the $2\times2$ 
 block structure of~\eqref{hg} with each block being a $(m|n)\times (m|n)$ matrix 
 (or rearrangement thereof - see footnote~\ref{fbasis}.) These blocks are unconstrained beyond the overall unit superdeterminant condition:  $\sdet(G)=\sdet(H)=1$.
 So in particular the coordinates $X^{AA'}$ of~\eqref{cosetgen} are unconstrained $(m|n)\times(m|n)$ matrices. 
 Here $A$ and $A'$ are both superindices carrying the fundamental representation of two independent $SL(m|n)$ subgroups. 
This supercoset corresponds to the super Grassmannian $Gr(m|n,2m|2n)$. 
For more details see~\cite{Howe:1995md,Hartwell:1994rp,Heslop:2001zm,Heslop:2002hp,hep-th/0307210}.
 
Then a four point function (and in particular the superconformal blocks) can be written in terms of a function of the four points $X_1,X_2,X_3,X_4$ invariant under the action of $G$. This in turn boils down to a function of the $(m|n) \times (m|n)$ matrix $Z$ invariant under conjugation (see~\cite{Heslop:2002hp} for more details) 
\begin{align}\label{th1ev}
	Z=X_{12}X_{23}^{-1}X_{34}X_{41}^{-1} \sim \text{diag}(x_1,..,x_m|y_1,..,y_n)\ .
\end{align}  
The $m+n$ eigenvalues $x_i,y_i$ then yield the $m+n$ arguments of the superblocks $B_{\gamma,\ulmb}$.

Thus any function of $Z$  invariant under conjugation will automatically solve the superconformal Ward identities associated with the four point function.
As first pointed out in this context in~\cite{Heslop:2002hp} the simplest way to construct a  basis of all such polynomials associated with a Young diagram $\ulmb$ naturally yields  (super) Schur polynomials (which are just the Jack polynomials with $\theta=1$). The construction is as follows. Take $|\ulmb|$ copies of $Z_A^B$ and symmetrise the upper ($B$) indices according to the Young symmetriser of $\ulmb$. Then contract all upper and lower indices. For example
\Yboxdim{8pt}
\Yvcentermath1
\begin{align}
	\renewcommand*\arraystretch{0}
	\ulmb &= \yng(1) \quad &&\rightarrow &\quad Z_A^A &= \Tr(Z)&&=\quad P_{\ulmb}({\bf z};\theta{=}1) \notag\\
	\ulmb &= \yng(2) \quad &&\rightarrow &\quad Z^{(A_1}_{A_1} Z^{A_2)}_{A_2} &=\tfrac12\left( \Tr(Z)^2+\Tr(Z^2)\right)&&=\quad P_{\ulmb}({\bf z};\theta{=}1)\notag\\
	\ulmb &= \yng(1,1) \quad  &&\rightarrow \quad &\quad Z^{[A_1}_{A_1} Z^{A_2]}_{A_2} &=\tfrac12\left( \Tr(Z)^2-\Tr(Z^2)\right)&&=\quad P_{\ulmb}({\bf z};\theta{=}1)
\end{align}
Inputting the eigenvalues of $Z$~\eqref{th1ev} into these we obtain (up to an overall normalisation) precisely  the corresponding Schur polynomials i.e. Jack polynomials with $\theta =1$, $P_\ulmb({\bf z};\theta=1)$. This correspondence works for any Young diagram $\ulmb$.

\subsection{$\theta=2: OSp(4m|2n)$}
\label{appth2}

Details of this supercoset construction can be found (specialised to the $m=2$ case) in~\cite{hep-th/0008048,Heslop:2004du}. We summarise here.

When $\theta=2$ the relevant (complexified) supergroup is $OSp(4m|2n)$ which has bosonic subgroup 
$SO(4m) \times Sp(2n)$ with $SO(4m)$ non-compact and $Sp(2n)$ compact. 
Of physical interest is the case $m=2$ in which $SO(8)$ is the complexification 
of the 6d conformal group $SO(2,6)$ and $Sp(2n)$ is the internal symmetry group for  
$(0,n)$ superconformal field theories.

The supercoset space is defined by the marked Dynkin diagram~\eqref{dynkin_summary}b (or~\eqref{dynkin_summaryn0}b if $n=0$) and  can be realised in the   block $2\times 2$ matrix form of~\eqref{hg}.
To see this one needs to realise the group $osp(4m|2n)$ as the set of supermatrices orthogonal wrt the metric $J$:
\beq\label{osp1}
	\!\!\!\!\!\!\!\!osp(4m|2n)=\left\{M \in {\mathbb{C}}^{(2m|2n|2m)\times(2m|2n|2m)}: MJM^T=J %
	=\left(
			\addtolength{\tabcolsep}{-10pt}
		\begin{array}{c|cc|c}
			&&&1_{2m}  \\\hline
			&&1_n& \\
			&-1_n&& \\ \hline
			1_{2m}&&&
		\end{array}
	\right)\right\}\ .
\eeq
Inputting $M=H$ in the block form of~\eqref{hg} into this we  find that $a$ is related to  $d$ but is itself unconstrained. Thus the Levi subgroup (under which the operators  transform explicitly) is isomorphic to $GL(2m|n)$. 
Similarly, inputting $M=s(X)$ in the form of~\eqref{cosetgen}, we find that  the coordinates are $(2m|n)\times (2m|n)$  antisymmetric supermatrices 
\beq
X=-X^T\ .
\eeq 
Note that here and elsewhere $X^T$ denotes the supertranspose $(X^{AB})^T=X^{BA}(-1)^{AB}$.
Four-point functions are written in terms of a function of four such $X$s,  $X_1,X_2,X_3,X_4$ invariant under $OSp(4m|2n)$. We can use the symmetry to set $X_3 \rightarrow 0$ and then  $X_2^{-1} \rightarrow 0$. Then we can use the remaining symmetry  to set 
\begin{align}
	X_1 \rightarrow K=\left(\begin{array}{cc|c}
		&1_{m}&  \\
		-1_{m}&& \\
		\hline
		&&1_n
	\end{array}\right)\ .
\end{align}
Finally we end up with the problem of finding a function of $X_4$ which is invariant under $X_4 \rightarrow AX_4A^{T}$ where $AKA^T=K$. The group of matrices satisfying $AKA^T=K$ is $OSp(n|2m)$. So letting $Z=X_4K$ we seek  functions of $Z$  invariant under conjugation under $OSp(n|2m)\subset GL(2m|n)$. Ultimately, an invariant function of $X_1,X_2,X_3,X_4$ will be a function of the eigenvalues of this matrix $Z$. 
The  eigenvalues of the $m\times m$ piece of the matrix $Z$~\eqref{Zmat} are repeated
\begin{align}\label{th2ev}
	Z=X_{12}X_{23}^{-1}X_{34}X_{41}^{-1} \sim \text{diag}(x_1,x_1,x_2,x_2,..,x_m,x_m|y_1,..,y_n)\ .
\end{align}  
As always the independent ones yield the $m|n$ arguments of the superblocks~\eqref{evalues}.

This construction again provides a completely manifest way of solving the superconformal Ward identities. Fascinatingly the simplest way to construct a basis of functions solving these Ward identities yields the super Jack polynomials!
To construct a basis of such functions in 1-1 correspondence with Young diagrams $\ulmb$ proceed as follows.
Take $|\ulmb|$ copies of $W=X_4$, and symmetrise all indices according to the  Young symmetriser of $\widehat \ulmb$ the Young diagram obtained from $\ulmb$ by duplicating all rows. Then contract all indices with $|\ulmb|$ copies of $K$.

Let us illustrate this with the simplest examples:
\begin{align}
\renewcommand*\arraystretch{0}
	\ulmb &= \yng(1)  &&\Rightarrow&  \widehat \ulmb &=\young(<\scriptstyle A>,<\scriptstyle B) &&\rightarrow& W^{AB}K_{AB} &= \Tr(Z)= P_{\ulmb}({\bf z};\theta{=}2) \notag\\
		\ulmb &= \yng(2)  &&\Rightarrow& \widehat \ulmb &=\young(<\scriptstyle A ><\scriptstyle C>,<\scriptstyle B><\scriptstyle D>) &&\rightarrow& -W^{B(A}W^{C)D}K_{AB}K_{CD} &=\tfrac12\left( \Tr(Z)^2+\Tr(Z^2)\right)= P_{\ulmb}({\bf z};\theta{=}2)\notag\\
	\ulmb &= \yng(1,1)   &&\Rightarrow&  \widehat \ulmb &=\young(<\scriptstyle A>,<\scriptstyle B>,<\scriptstyle C>,<\scriptstyle D>)
&&\rightarrow&  W^{[AB}W^{CD]}K_{AB}K_{CD} &=\tfrac13\left( \Tr(Z)^2-2\Tr(Z^2) \right)= P_{\ulmb}({\bf z};\theta{=}2)
\end{align}
Inputting the eigenvalues of $Z$~\eqref{th2ev} into these we obtain (up to an overall normalisation) precisely  the corresponding Jack polynomial with $\theta =2$, $P_\ulmb({\bf z};\theta=2)$. This correspondence continues for any Young diagram $\ulmb$.

Note that this derivation of superblocks and super Jacks from group theory makes manifest the stability in $m,n$ of both, since they arise from formulae derived from matrices $Z$ of arbitrary dimensions. 
 A combinatoric description of (super)Jack polynomials with $\theta=2$ along these lines has also been  discussed in~\cite{bergeron1992zonal}.

\subsection{$\theta=\frac{1}{2}: OSp(4n|2m)$} %

The other value of $\theta$ for which there is a supergroup interpretation for any $m,n$ is $\theta=\frac{1}{2}$. 
Here the relevant (complexified) supergroup is $OSp(4n|2m)$, the same complexified group as previously 
but with the roles of $m,n$ reversed. This  has bosonic subgroup $ Sp(2m) \times SO(4n)$ but now with $Sp(2m)$ 
non-compact (for $m=2$ this is the complexified 3d conformal group $Sp(4) \sim SO(5) \sim SO(2,3)$) and $SO(4n)$ compact the internal subgroup.

The following is then essentially identical to the $\theta=2$ case but with the role of Grassmann odd and even exchanged.
The supercoset space defined by the marked Dynkin diagram~\eqref{dynkin_summary}c (or~\eqref{dynkin_summaryn0}c if $n=0$)  can also  be realised in the   block $2\times 2$ matrix form of~\eqref{hg} by realising  the group $osp(4n|2m)$ in the following form%
\footnote{This is a change of basis of the matrix form in~\eqref{osp1} }%
:
\beq
	osp(4n|2m)=\left\{M \in {\mathbb{C}}^{(m|4n|n)\times(m|4n|m)}: MJM^T=J=\left(
	\addtolength{\tabcolsep}{-10pt}
	\begin{array}{c|cc|c}
		&&&1_{m}  \\\hline
		&&1_{2n}& \\
		&1_{2n}&& \\ \hline
		-1_{m}&&&
	\end{array}
	\right) \right\}%
\eeq
The Levi subgroup (under which the operators  transform explicitly) is isomorphic to $GL(m|2n)$ and  the coordinates are $(m|2n)\times (m|2n)$ {\em symmetric} supermatrices  
\beq
X=X^{T}\ .
\eeq
Four-point functions are written in terms of a function of four such $X$s,  $X_1,X_2,X_3,X_4$ invariant under $OSp(4n|2m)$. We can use the symmetry to set $X_3 \rightarrow 0$ and then  $X_2^{-1} \rightarrow 0$. Then we can use the remaining symmetry to set 
\begin{align}
	X_1 \rightarrow K=\left(\begin{array}{c|cc}
		1_m&&\\\hline
		&&1_{n}  \\
		&-1_{n}&
	\end{array}\right)\ .
\end{align}
We then are left with a function of $X_4$ which invariant under $X_4 \rightarrow AX_4A^{T}$ where $AKA^T=K$ i.e. under $OSp(m|2n)$. Letting  $Z=X_4K$ we seek  functions of $Z$  invariant under conjugation under $OSp(m|2n)\subset GL(m|2n)$. Thus ultimately, an invariant function of $X_1,X_2,X_3,X_4$ will be a function of the eigenvalues of this matrix $Z$. 
The  eigenvalues of the internal $n\times n$ piece of the matrix $Z$~\eqref{Zmat} are repeated and the independent eigenvalues yield the $m|n$ arguments of the superblocks~\eqref{evalues}
\begin{align}\label{ev2}
	Z=X_{12}X_{23}^{-1}X_{34}X_{41}^{-1} \sim \text{diag}(x_1,..,x_m|y_1,y_1,y_2,y_2,..,y_n,y_n)\ .
\end{align}

As for $\theta=2$ this construction provides a completely manifest way of 
solving the superconformal Ward identities naturally giving $\theta=\frac{1}{2}$  super 
Jack polynomials!
Take $|\ulmb|$ copies of $W=X_4$, and symmetrise all indices according to the  Young symmetriser of $\widehat \ulmb$ which is this time the Young diagram obtained from $\ulmb$ by duplicating all {\em columns} rather than rows. Then contract all indices with $|\ulmb|$ copies of $K$.

In the simplest examples:
\begin{align}
	\renewcommand*\arraystretch{0}
	\ulmb &= \yng(1) \quad  &&\Rightarrow& \quad \widehat \ulmb &=\young(<\scriptstyle A><\scriptstyle B)\quad &&\rightarrow& \quad W^{AB}K_{AB} &= \Tr(Z) \notag\\
	\ulmb &= \yng(2)  \quad &&\Rightarrow& \quad \widehat \ulmb &=\young(<\scriptstyle A><\scriptstyle B><\scriptstyle C><\scriptstyle D>)
	\quad &&\rightarrow& \quad W^{(AB}W^{CD)}K_{AB}K_{CD} &=\tfrac13\left( \Tr(Z)^2+2\Tr(Z^2) \right)\notag\\
	\ulmb &= \yng(1,1)  \quad &&\Rightarrow& \quad \widehat \ulmb &=\young(<\scriptstyle A ><\scriptstyle B>,<\scriptstyle C><\scriptstyle D>)\quad &&\rightarrow& \quad W^{(B[A}W^{C]D}K_{AB}K_{CD} &=\tfrac12\left( \Tr(Z)^2-\Tr(Z^2)\right)
\end{align}
Inputting the eigenvalues of $Z$~\eqref{ev2} into these we obtain (up to an overall normalisation) precisely  the corresponding Jack polynomials with $\theta =\frac{1}{2}$, $P_\ulmb({\bf z};\theta=\frac{1}{2})$. This correspondence works for any Young diagram $\ulmb$.

As for $\theta=1,2$ this derivation of superblocks and super Jacks from group theory makes manifest the stability in $m,n$ of both, since they arise from formulae derived from matrices $Z$ of arbitrary dimensions.

\subsection{Non-supersymmetric conformal and internal blocks}
\label{m2n0}

We conclude our discussion by considering non-supersymmetric conformal and compact blocks.
These were first analysed in~\cite{Dolan:2003hv} using an embedding space formalism for Minkowski space. Here we see how an unusual  coset representation of Minkowski space relates  these cases to the general matrix  formalism we have presented.

\subsubsection*{Conformal blocks $(m,n)=(2,0)$ with any $\theta \in \mathbb{Z}^+/2$}\label{bosonic_groups}

Complexified Minkowski space in $d$ dimensions, $M_d$, can be viewed as a coset of the complexified 
conformal group $SO(d+2;\mathbb{C})$ divided by the subgroup consisting of Lorentz transformations, 
dilatations and special conformal transformations. It is both an orthogonal Grassmannian and a flag manifold 
and can be conveniently denoted by taking the Dynkin diagram of $SO(d+2;\mathbb{C})$ and putting 
a single cross through the first node (see \cite{bastoneastwood})
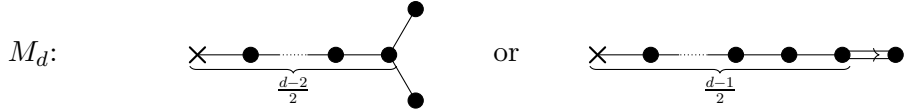
\begin{center}
	$M_{d}$: \qquad \qquad 	
	\begin{dynkinDiagram}{D}{x*.****}
		\dynkinBrace[\frac{d-2}2]{1}{4};
	\end{dynkinDiagram}
	\qquad or \qquad 
	\begin{dynkinDiagram}{B}{x*.****}
		\dynkinBrace[\frac{d-1}2]{1}{5};
	\end{dynkinDiagram}
\end{center}
This crossed-through node  represents the group of  dilatations and the remaining Dynkin diagram (with the crossed node omitted) represents the Lorentz subgroup $SO(d;\mathbb{C})$ with $d=2\theta+2$.

The coset construction for $\theta=\frac{1}{2},1,2$ in the previous section suggests to start now from the spinorial representation of $SO(d+2)$, with $d=2\theta+2$, rather than the fundamental representation.
This representation is $2^{d/2}$ dimensional for $d$ even (the Weyl representation) and $2^{(d+1)/2}$ dimensional for $d$ odd. 
In this representation, the coset space can be written in the block form~\eqref{cosetgen} and the coordinates $X$ are $2^{\lceil\theta\rceil} \times 2^{\lceil\theta\rceil}$ matrices. The matrices will be highly constrained in general. 
The cross-ratios  arise as eigenvalues of the matrix $Z$ which are now repeated (due to the constraints, this matches~\eqref{th2ev} for the case $\theta=2$ when $m=2,n=0$):
\begin{align}
	Z=X_{12}X_{23}^{-1}X_{34}X_{41}^{-1} \sim \text{diag}(x_1,..,x_1,x_2,..,x_2)\ .
\end{align}  
thus there are always just two independent cross-ratios $x_{1},x_2$ which become the variables of the conformal blocks in~\eqref{evalues}.

Representations of the conformal group are specified by placing Dynkin weights below the nodes, so for example 
a fundamental scalar is given by placing a -1 by the first node and zero's everywhere else. The scalars of 
dimension $\Delta$,  $\cO_\Delta$ appearing as external operators in the four-point function~\eqref{4pt}, have a $-\Delta$ by the first node 
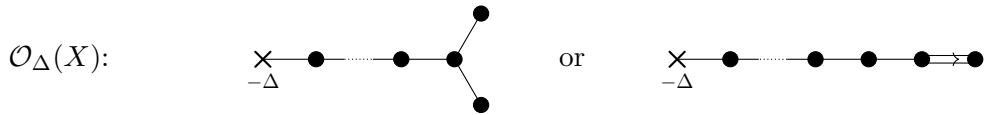
\begin{center}
	$\cO_\Delta(X)$: \qquad \qquad 	\dynkin[labels={-\Delta}]{D}{x*.****} \qquad or \qquad \dynkin[labels={-\Delta}]{B}{x*.****}
\end{center}
The fact that it has a negative Dynkin label is consistent with the fact that this corresponds to an infinite dimensional 
representation of the (complexified) conformal group (positive Dynkin labels give finite reps). One reads off from the 
diagram the representation as a field on Minkowski space: it has dilatation weight $\Delta$ 
and is a scalar under the Lorentz subgroup (since it has zero's under all nodes of the uncrossed 
part of the Dynkin diagram).

The only reps which occur in   the OPE of two such scalars has dimension  $\Delta$ and  Lorentz spin $l$. The corresponding Dynkin diagram is
\begin{center}
	$O_{\gamma,\ulmb}(X)$: \qquad \qquad 	\dynkin[labels={-b'\ ,l,,,,}]{D}{x*.****} \qquad or \qquad \dynkin[labels={-b'\ ,2l,,,,}]{B}{x*.****}
\end{center}
where the Young tableau $\ulmb$ is at most two row (the only shape consistent with $(m,n)=(2,0)$) and 
\begin{align}\label{dilbos}
	 b'=\tfrac{ \Delta+l}{\theta} =  \gamma+ \tfrac2\theta\lambda_1, \qquad;\qquad \Delta=\theta \gamma+\lambda_1+\lambda_2, \qquad;\qquad   l=\lambda_1-\lambda_2\ .
\end{align}
Notice again the redundancy in the description in terms of $\gamma$ and $\ulmb$. 
The two operators  $\cO_{\gamma,[\lambda_1,\lambda_2]}=\cO_{\gamma-2k,[\lambda_1+\theta k,\lambda_2+\theta k]}$ 
give the same representation for any $k$ as long as it leaves a valid Young tableau. 
We could use this to   set $\gamma=0$ in this case and just have a  description in terms of the Young tableau $\ulmb$ only.

Note that there are different ways of realising $M_d$ as a coset. The above way is a bit complicated for large $d$. Had we started instead from the 
fundamental representation of $SO(d+2)$, the coset construction would be equivalent to the 
embedding space formalism, in which $M_d$ is viewed as the space of null $d+2$ vectors in projective space $P_{d+1}$. 
This is the approach used in~\cite{Dolan:2003hv}.  This approach does not fit directly into the 
 general $(m,n,\theta)$ matrix formalism we worked out in  the previous cases with $\theta=\frac{1}{2},1,2$ however. In particular the coset 
is not in the $2\times 2$ block form of~\eqref{cosetgen}, and the independent cross-ratios have no natural origin arising from a diagonal matrix.

\subsubsection*{Internal blocks:  $SO(2 \theta+4)$ blocks: $(m,n)=(0,2)$ with $\theta \in 2/\mathbb{Z}^+$}
\label{m0n2}

A very similar case occurs for $m=0,n=2$, corresponding to internal blocks. 
Namely instead of four conformal scalars we have four finite dimensional reps of $SO(4+\frac{2}{\theta})$. 
Blocks for these were again discussed in Dolan and Osborn~\cite{Dolan:2003hv}.
Here the complexified coset space is the same as in the previous case with $\theta \leftrightarrow \frac{1}{\theta}$ 
(although the real forms will be different, previously $SO(2,2+2\theta)$, now $SO(4+\frac{2}{\theta})$). 
\begin{center}
 \qquad \qquad 	
	\begin{dynkinDiagram}{D}{x*.****}
		\dynkinBrace[\frac{1}\theta]{1}{4};
	\end{dynkinDiagram}
	\qquad or \qquad 
	\begin{dynkinDiagram}{B}{x*.****}
		\dynkinBrace[\frac{1}\theta+\frac{1}2]{1}{5};
	\end{dynkinDiagram}
\end{center}
The discussion of these spaces as explicit matrices  is also identical to the previous section.

The external states are a specific representation of $SO(4+2/\theta)$ specified by placing $p$ 
above the first node and zeros everywhere else. Note that this time the Dynkin label is positive as it is a finite dimensional representation.
\begin{center}
	$\cO_p(X)$: \qquad \qquad 	\dynkin[labels={p}]{D}{x*.****} 
	\quad or \quad \dynkin[labels={p}]{B}{x*.****}
\end{center}
The only reps which occur in   the `OPE' (in this case equivalent to  the tensor product) of two such reps have Dynkin labels 
\begin{center}
	$O_{\gamma,\ulmb}(X)$: \qquad \qquad 	\dynkin[labels={b,a,,,,}]{D}{x*.****} \quad or \quad  \dynkin[labels={b,2a,,,,}]{B}{x*.****}\ .
\end{center}
The Young tableau $\ulmb$ is at most two column (the only shape consistent with $(m,n)=(0,2)$ this is a rotation of a standard $SL(2)$ Young tableau) and we have
\begin{align}\label{p_dynkin_label}
	{b}={\gamma}-2\lambda_1' \qquad;\qquad a=\lambda_1'-\lambda_2'\ .
\end{align}
Again there is redundancy in the description in terms of $\gamma$ and $\ulmb$ with $\cO_{\gamma,[\lambda'_1,\lambda'_2]'}=\cO_{\gamma+2\delta,[\lambda'_1+\delta,\lambda'_2+\delta]'}$ for any $\delta$ as long as it leaves a valid Young tableau. Here we  could use this to  force  $\lambda_2'=0$, so only allow 1 column Young tableaux.

\section{From CMS Hamiltonians to the superblock 
Casimir}\label{app_differential_operators}

Jack polynomials (in suitable variables) are eigenfunctions of the Hamiltonian 
of the Calogero-Moser-Sutherland system for $A_n$ root sytem whilst Jacobi 
polynomials (as well as dual Jacobi functions/ blocks) are eigenfunctions of 
the $BC_n$ root system. 
Similarly {\em deformed} (i.e. supersymmetric) Calogero-Moser-Sutherland (CMS) 
Hamiltonians of the {\em generalised} root systems $A_{n|m}$ and $BC_{n|m}$ 
yield super Jack polynomials and  dual super Jacob fucntions (superblocks) as 
eigenfunctions.

In this section we review the {\em deformed} Calogero-Moser-Sutherland (CMS) 
Hamiltonian associated to {\em generalised} root systems and show explicitly 
how the $BC$ type relates to 
the super Casimir for superblocks in our theories.

\subsection {Deformed CMS Hamiltonians.}

The  (non-deformed) Calogero-Moser-Sutherland (CMS) operator for {\em any} (not necessarily 
reduced)  root system was 
first given in~\cite{Olshanetsky:1983wh}
via the Hamiltonian
\begin{align}\label{cmsop}
{\cal H}&=-\partial_I \partial^I + \sum_{\alpha\in\, 
R_{+}}\frac{k_{\alpha}(1+k_{\alpha}+2k_{2\alpha})\alpha^2}{\sin^2\alpha_Iu^I}\ ,
\end{align}
where
\begin{itemize}
	\item
	the $u^I$ are coordinates in a vector space, $\partial_I:=\partial/\partial 
	u^I$, and the indices are 
	raised and lowered using the Euclidean metric $g_{IJ}$. 
	\item
	the roots $\{\alpha_I\}$ are a set of covectors (the roots of the root 
	system) 
	and $R_+$ is the set of positive roots. $\alpha^2$ denotes the length 
	squared 
	of root $\alpha$, $\alpha^2= \alpha_I \alpha_Jg^{IJ}$.
	\item
	Finally to each root $\alpha$ is associated a parameter $k_{\alpha}$, which 
	is 
	constant under Weyl transformations, so 
	$k_{w\alpha}=k_{\alpha}$ for $w$ in the Weyl group. For any covector 
	$\alpha$  
	which  is not a root then we have $k_{\alpha}=0$.
\end{itemize}

In~\cite{Veselov_2} this story was generalised and {\em deformed} CMS operators 
were defined in terms of  {\em 
generalized} (or supersymmetric) root systems. The irreducible generalized root 
systems were first classified by Serganova~\cite{serganova}. The generalized 
root systems are no longer required to have a Euclidean metric but can have 
non-trivial signature, and they have the 
analogous relation to Lie superalgebras as root systems do to Lie algberas. In 
the classification there are just two infinite series, $A_{m,n}$ and $BC_{m,n}$ 
(which we will concentrate on) together with a handful of exceptional cases.

The deformed Hamiltonian has the same form as~\eqref{cmsop} but with some 
additional subtleties
\begin{itemize}
	\item The metric $g_{IJ}$ need not be
	Euclidean. 
	\item There are odd (also known as  imaginary) roots which must have 
	$k_\alpha=1$. 
	\item There will be relations between some of the parameters $k_\alpha$ 
	even when they are not related by Weyl reflections.
\end{itemize}
We will give the two main cases, $A_{n|m}$ and $BC_{n|m}$ explicitly  in the 
next subsections.

In all cases there is a special  eigenfunction  (the ground state) of
$\mathcal{H}$ which takes the universal form, 
\begin{align}\label{gs}
	\Psi_0( u_I,\{ k_{\alpha}\} )=\prod_{\alpha\in R^+}\sin^{-k_\alpha}\alpha_I 
	u^I\ .
\end{align}
This is automatic in the non-deformed case but in the deformed case it requires 
the additional relations between parameters $k_\alpha$.

It is useful to then write other eigenfunctions of $\mathcal{H}$  as the 
product 
$\Psi_0 f$ 
for some function $f(u_I,\{ k_{\alpha}\})$. 
The function $f$ is then an eigenfunction of the conjugate operator 
$\Psi_0^{-1} {\mathcal{H}} \Psi_0$. This is the one we shall relate to the 
super Casimir. This conjugate operator also has a universal (and indeed simpler 
) form for all such (generalised) root systems
\begin{align}\label{CMSgen2}
	{\mathcal L}:= \Psi_0^{-1} {\mathcal{H}} \Psi_0+c = -\partial_{I}\partial^I 
	+ 
	\sum_{\alpha\in R_{+}}{2k_{\alpha}}{\cot(\alpha_J u^J) }\alpha^I\partial_I\ 
\end{align}
for some constant $c$.

Our presentation in this appendix collects a number of known results and follows \cite{Veselov_2}.
In addition, we will rewrite the differential operators $\mathcal{L}$, by using the (orthogonal) 
measures associated to the root systems.

\subsection{$A$-type Hamiltonians and Jack polynomials}\label{A_type_op_append}

The $A_{n-1|m-1}$ generalised root system can be placed inside an $n+m$ 
dimensional vector space.  
The positive roots will be parametrised as%
\begin{align}
A_{n-1|m-1}: \qquad R^{+}_{} = \left\{\, \begin{array}{l}  { e}_i{-}{ e}_j \\  
{ e}_i{-}{ e}_{j'} 
\\ { e}_{i'}{-}{ e}_{j'}\end{array} 
\quad; \quad 1\leq i < j \leq n, \ n+1\leq i' < j' \leq m+n \,\right\}\ 
\end{align}
where $e_{I=1,\ldots m+n}$ give the basis of unit vectors.
The (inverse) metric is 
\begin{align}\label{inverse_metric}
	g^{IJ} &= \left\{ \begin{array}{ll}
	\delta^{ij} \qquad  &i,j=1,..,n\\[.2cm]
	-\theta \delta^{i'j'} &i',j'=n{+}1,..,m{+}n
	\end{array} \right.
\end{align}
$R^{+}_{}$ splits into  three separate Weyl orbits (therefore there will be 
three distinct $k_{\alpha}$)
and the three families of roots are
\begin{align}
\begin{array}{lcll}
\alpha={{e}_i{-}{e}_j} 		&:  &\quad\alpha^2=2,				&\quad 
k_\alpha=-\theta\\[.2cm]
\alpha={{e}_i{-}{e}_{j'}}	&:  &\quad \alpha^2=1-\theta, 			&\quad k_\alpha=1\\[.2cm]
\alpha={{e}_{i'}{-}{e}_{j'}}	&:  & \quad\alpha^2=-2\theta,			&\quad k_\alpha=-\frac1\theta
\end{array}
\end{align}

Plugging in these, the CMS operator~\eqref{cmsop} thus becomes
\begin{align}
{\cal H}^A_{}&=-\sum_{i}{\partial_{i}^2}+\theta\sum_{i'}^{}{\partial_{{i'}}^2} 
-\sum_{i<j} \frac{2\theta (1-\theta)}{\sin^{2}u_{ij}}+\sum_{i}\sum_{j'} 
\frac{2(1-\theta)}{\sin^{2}(u_{i}{-}u_{j'})}+\sum_{i'<j'} \frac{2 
(1-1/\theta)}{\sin^{2}u_{i'j'}}\notag\\
&= -\sum_{I=1}^{n+m} (-\theta)^{\pi_I} \partial_i^2 +2(1-\theta) \sum_{1\leq I<J\leq n+m} \frac{ (-\theta)^{1-\pi_I-\pi_J} }{ \sin^{2} (u_{I}-u_J) }\label{my_Hamilt_Anm}
\end{align}
where the parity assignment is $\pi_{i=1,\ldots n}=0$ and $\pi_{i'=n+1,\ldots n+m}=1$.

The ground state is given by 
\begin{align}
\Psi_0= \frac{ \prod_{ i<j} \sin^\theta (u_i{-}u_j) \prod_{ i'<j'} 
\sin^{\frac{1}{\theta}} (u_{i'}{-}u_{j'} ) }{ \prod_{i}\prod_{j'} 
\sin(u_i{-}u_{j'}) }  
\end{align}
and the conjugated operator~\eqref{CMSgen2} is
\begin{align}
\!\!\!\mathcal{L}^{A}_{}=\Psi_0^{-1}\left( {\cal H}^{A}_{}- |\rho_\theta|^2 \right)\Psi_0=&
-\sum_{I=1}^{n+m} (-\theta)^{\pi_I} \partial_I^2
+2\sum_{i,j'} {\cot(u_{i}{-}u_{j'})}(\partial_{i}+\theta\partial_{{j'} })\notag\\
& -{2\theta} \sum_{i<j} {\cot (u_{i}{-}u_j)}(\partial_{i}-\partial_{j})
+2\sum_{i'<j'} {\cot (u_{i}{-}u_j)}(\partial_{i'}-\partial_{j'}) \notag
\end{align}
where $\rho_{\theta}=\sum_{\alpha} k_{\alpha}\alpha=\sum_{I<J}(-\theta)^{1-\pi_I-\pi_J}(e_I-e_J)$, and $|\rho_\theta|^2$ is the norm of $\rho_{\theta}$ under $g^{IJ}$.

Notice that $\mathcal{L}^{A}_{}$ contains in principle all terms coming from 
$\sum_I (-\theta)^{\pi_I} \partial_I^2 \Psi_0$ and in particular the terms  
$\sum_{\beta} \sum_{\alpha\neq \beta}  k_{\alpha} k_{\beta}\, (\alpha_I  g^{IJ} \beta_J)\, \cot(\alpha^{}_I u^I) \cot(\beta_I u^I)$, 
which have mixed nature  and do not cancel immediately with the ones from $\mathcal{H}^A$.
However these can be replaced in favour of a $\theta$ dependent constant,  
$-\sum_{\beta} \sum_{\alpha\neq \beta}  k_{\alpha} k_{\beta}\, (\alpha_I  g^{IJ} \beta_J)$, 
because of the following relation
\begin{align}
\sum_{\beta} \sum_{\alpha\neq \beta}  k_{\alpha} k_{\beta}\, (\alpha_I  g^{IJ} \beta_J)\, \left( \cot(\alpha^{}_I u^I) \cot(\beta_I u^I)+1\right)=0
\end{align}
This condition is automatically satisfied by the $k_{\alpha}$ assignment of the roots, 
and can be understood as a consistency condition on $\Psi_0$ being the ground state. 

Finally, by changing variables to $z_I=e^{2i u_I}$, we find 
$\cot(u_I-u_J)=i(z_I+z_J)/(z_I-z_J)$
and $\partial_{u_I}=2i z_I \partial_{z_I}$. Therefore, 
\begin{align}\label{formula_Lamn_preB11}
\tfrac{1}{4}\mathcal{L}^{A}_{}&=\sum_{I=1}^{n+m} (-\theta)^{\pi_I} (z_I\partial_I)^2
+{\theta} \sum_{I<J}^{n+m} \frac{z_I+z_J}{z_I-z_J}( (-\theta)^{-\pi_J } 
z_I\partial_{I}-(-\theta)^{-\pi_I} z_J\partial_{J})\ .
\end{align}
and one can check that 
\begin{align}
	\label{rew_app_numero1}
	\tfrac{1}{4}\mathcal{L}^{A}_{} = \mathbf{H}_{}+ (\theta(m-1)-(n-1))\sum_I 
	z_I \partial_I
\end{align}
where $\mathbf{H}_{}$ is defined in~\eqref{hh} and is the defining operator of 
super Jacks~\eqref{ham} . %

At this point it is also interesting to note that
superJack operators are orthogonal in an $A_{(m,n)}$ measure \cite{1802.02016}
\begin{align}
\mathcal{S}_{(m,n)}({\bf z};\theta)= \prod_{I}\prod_{J\neq I} \left(1-\frac{z_I}{z_J}\right)^{ - (-\theta)^{1-\pi_I-\pi_J} }
\end{align}
(where the parity assignment is $\pi_{i=1,\ldots n}=0$ and $\pi_{i'=n+1,\ldots 
n+m}=1$) and the $A$-type CMS operator~\eqref{formula_Lamn_preB11} can be 
rewritten in a very simple way in terms of this measure as
\begin{align}\label{measure_superjackdiff_explanation}
\tfrac{1}{4}\mathcal{L}^{A}_{} = { \mathcal{S}_{} }^{-1}
\sum_I (-\theta)^{\pi_I} z_I \partial_{I}  [\mathcal{S}_{} z_I 
\partial_I]\ .
\end{align}

Finally, we point out that Jack polynomials are also eigenfunctions of the one-parameter 
family of Sekiguchi differential operators \cite{Okounkov_Olshanski}.

\subsection{$BC_{}$-type Hamiltonians and superblocks}\label{BCtoCas}

We now move on to the CMS operator for the generalised $BC$ root system and 
relate it to the Casimir which give superblocks.

The positive roots of the $BC_{n|m}$ root system live in an $m+n$ dimensional 
vector space and are as follows  
\begin{align}
R^{BC+}_{} = \left\{ 
 \begin{array}{l} { e}_i\\ { e}_{i'}\end{array} \,;\,
\begin{array}{l} 2{e}_i \\ 2 {e}_{i'} \end{array} \,;\,
\begin{array}{l}
{ e}_i \pm { e}_j\\
{ e}_{i} \pm { e}_{j'}\\
{ e}_{i'} \pm { e}_{j'}\end{array} \quad;\quad 1\leq i < j \leq n, \ n+1\leq i' 
< j' \leq m+n \right\}\ 
\end{align}
where again $e_{I=1,\ldots m+n}$ are the basis of unit vectors and the inverse 
metric is the same as in the $A_{n|m}$ case~\eqref{inverse_metric}.
The  parameters are assigned as follows,
\beq
\begin{array}{lcll}
\alpha={{ e}_i{\pm}{ e}_j}    	&:  		&\quad \alpha^2=2,			&\quad k_\alpha=-\theta  \\[.2cm]			%
\alpha={{ e}_i{\pm}{ e}_{j'}} 	 &:  		&\quad \alpha^2=1-\theta, 	&\quad k_\alpha=1 \\[.2cm] 		%
\alpha={{ e}_{i'}{\pm}{ e}_{j'}}  	&:  		&\quad \alpha^2=-2\theta 		&\quad k_\alpha=-\frac1\theta			%
\end{array}\qquad
\eeq
and
\beq\quad
\begin{array}{lcll}
\alpha={{ e}_i}		&:  &\quad\alpha^2=1,		&\quad k_\alpha=p\\[.2cm]
\alpha={{ e}_{i'}}	&:  &\quad\alpha^2=-\theta,  	&\quad k_\alpha=r  
\end{array}
\qquad;\qquad
\begin{array}{lcll}
\alpha={2{ e}_i}		&:  &\quad \alpha^2=4, 			&\quad k_\alpha=q  \\[.2cm]
\alpha={2{ e}_{i'}}	&:  &\alpha^2=-4\theta,  			&\quad k_\alpha=s
\end{array}
\eeq
Plugging these values into the general formula~\eqref{cmsop} we obtain the 
Hamiltonian 
\begin{align}
\mathcal{H}^{BC}_{}=& -\sum_{I=1}^{n+m} (-\theta)^{\pi_I} \partial_I^2 +2(1-\theta) \sum_{1\leq I<J\leq n+m} \frac{ (-\theta)^{1-\pi_I-\pi_J} }{ \sin^{2} (u_{I}\pm u_J) } \\
&+ \sum_{i=1}^{n} \left( \frac{p(p+2q+1)}{\sin^{2}u_{i}}+ \frac{4q(q+1)}{\sin^{2}2u_{i}} \right) -\theta \left( \sum_{i'=n+1}^{n+m} \frac{r(r+2s+1)}{\sin^{2}u_{i'}}+ \frac{4s(s+1)}{\sin^{2}2u_{i'}}\right)\notag
\end{align}
where the first line is a trivial modification of \eqref{my_Hamilt_Anm}.

The ground state~\eqref{gs} becomes 
\begin{align}
\Psi_0&= \frac{ \prod_{ i<j} \sin^\theta (u_i\pm u_j) \prod_{ i'<j'} 
\sin^{\frac{1}{\theta}} (u_{i'}\pm u_{j'} ) }{ 
\prod_{i=1}^n\prod_{j'=n+1}^{m+n} \sin(u_i\pm u_{j'})\prod_{i=1}^{n}  
(\sin^p(u_i) \sin^q(2u_i)) \prod_{i'=n+1}^{n+m} (\sin^r(u_{i'}) \sin^{s}(2 
u_{i'})) } 
\end{align}
and the conjugated operator is
\begin{align}
\!\!\!\mathcal{L}^{BC}_{}=\Psi_0^{-1}\left( {\cal H}^{BC}_{}- |\rho_\theta|^2 \right)\Psi_0=&
-\sum_{I=1}^{n+m} (-\theta)^{\pi_I} \mathcal{D}_I \partial_I
+2\sum_{i,j'} {\cot(u_{i}{\pm }u_{j'})}(\partial_{i}\mp \theta\partial_{{j'} })\notag\\
& -{2\theta} \sum_{i<j} \cot(u_{i}\pm u_{j}) (\partial_{i}\pm\partial_{j})
+2\sum_{i'<j'} \cot(u_{i'}\pm u_{j'})(\partial_{i'}\pm\partial_{j'}) \notag
\end{align}
where $\rho_{\theta}=\sum_{\alpha} k_{\alpha}\alpha$ 
and $\mathcal{D}_{I}=( \partial_{I}+ 2\cot(2u_I) )-2(-\theta)^{-\pi_I} ( p \cot u_I + (2q+1) \cot(2u_I) )$.

Going through the computation notice that for the ground state to be so, the following relation had to be satisfyied
\begin{align}\label{values_accessory_BC}
\sum_{\beta}\sum_{\beta \nsim \alpha}  k_{\alpha} k_{\beta}\, (\alpha_I  g^{IJ} \beta_J)\, \left( \cot(\alpha^{}_I u^I) \cot(\beta_I u^I)+1\right)=0\qquad;\quad
 \begin{array}{l} p=-\theta r\\[.2cm] 2q+1=-\theta(2s+1) \end{array}
\end{align}
This was not automatic for the $k_{\alpha}$ assignment but puts a constraint which we used
to solve for $r$ and $s$.  The summation over $\beta \nsim \alpha$ excludes roots which are parallel in the vector space.
The origin of this constraint again can be understood as a rewriting of the potential terms arising 
from $ \sum (-\theta)^{\pi_I} \partial_I^2 \Psi_0$ coming from mixed products which do not cancel immediately against the potential terms in $\mathcal{H}^{BC}_{}$.

We now change variables to exponential coordinates $\hat z_I = e^{2iu_I}$, %
and we get

\begin{align}
\tfrac{1}{4}\mathcal{L}^{BC}_{}=& \sum_{I=1}^{n+m} \left( (-\theta)^{\pi_I}\left( \hat z_I \hat \partial_{I}+ \frac{\hat z_I^2+1}{\hat z_I^2-1} \right)
							- \left( p \frac{\hat z_I+1}{\hat z_I-1} + (2q+1)\frac{\hat z_I^2+1}{\hat z_I^2-1} \right)\right)(\hat z_I \hat \partial_I)\notag\\
&\qquad +{\theta} \sum_{I<J}^{n+m} \frac{\hat z_I+\hat z^{\pm}_J}{\hat z_I-\hat z^{\pm}_J}( (-\theta)^{-\pi_J } \hat z_I \hat \partial_{I}\mp(-\theta)^{-\pi_I} \hat z_J \hat \partial_{J})
\end{align}
We can rearrange the sum over many-body interactions as a sum over $I\neq J$, 
then for a single derivative we can add up the two terms
$\frac{\hat z_I+\hat z^{\pm}_J}{\hat z_I-\hat z^{\pm}_J}$ 
to find $\frac{2\hat z_J(1-\hat z_I^2)}{(\hat z_I -\hat z_J)(1-\hat z_i \hat z_J)}$.

Finally we need to change variables again to%
\footnote{
Note the following useful relations 
	\begin{align*}%
		\left\{\begin{array}{cl} 
			z_I&=-\frac{1}{4} \left(\frac{1}{\sqrt{ \hat z_I} }-\sqrt{ \hat z_I 
			} \right)^{\!\! 2} \\[.2cm]
			z_I-1&=-\frac{1}{4} \left(\frac{1}{\sqrt{ \hat z_I} }+\sqrt{ \hat 
			z_I } \right)^{\!\! 2}  \end{array}\right.
		\qquad ;\qquad 
		\left\{\begin{array}{lll}  
			\hat z_I \frac{\partial z_I}{\partial \hat z_I}= &\frac{1}{4} 
			\frac{ (1-\hat z_I^2 ) }{ \hat z_I}& = \sqrt{ z_I(z_I-1)}  \\[.2cm] 
			&\frac{1}{4}\frac{ (1+\hat z_I^2 ) }{ \hat z_I} &= 
			-(z_I-\frac{1}{2})\end{array}\right.
		\end{align*}
	} 
\begin{align}
z_I=\frac{1}{2}-\frac{1}{4}\left(\hat z_I+\frac{1}{\hat z_I}\right)
\end{align}
and  we obtain
\begin{align}\label{formula_Lbcmn_preB21}
\tfrac{1}{4}\mathcal{L}^{BC}_{}({\bf z},\theta)=  & 
					\sum_{I=1}^{n+m} \Big( (-\theta)^{\pi_I}\partial_I z_I(z_I-1) \partial_I 
							- \left( p (z_I-1) + (2q+1)\left(z_I-\tfrac{1}{2}\right) \right)\partial_I \Big)  \notag\\
&
-2\theta \sum_{I\neq J} \frac{ \ \ (-\theta)^{-\pi_J}  }{ z_I-z_J} z_I(1-z_I)  \partial_I
\end{align}

At this point we can understand the relation between the super Casimir ${\bf 
C}$, defining the superblocks, and 
$\mathcal{L}^{BC}_{}$. 
One can check that they are closely related after conjugation and a shift
\begin{align}
	&
	\left(\frac{ \prod_{i=1}^{m} x_i^\theta }{ \prod_{j=1}^{n} y_{j} 
	}\right)^{\!\!-\beta } 
	\tfrac{1}{4}\mathcal{L}_{}^{BC} (\tfrac{1}{x_1}\ldots \tfrac{1}{x_m}; 
	\tfrac{1}{y_1}\ldots \tfrac{1}{y_n}; {\theta})\   
	\left(\frac{ \prod_{i=1}^{m} x_i^\theta }{ \prod_{j=1}^{n} y_{j} 
	}\right)^{\!\!+\beta } =  %
	\notag\\[.2cm]
	&
	\rule{1.2cm}{0pt} 
	\mathbf{C}^{(\theta,\frac12(\gamma-p_{12}),\frac12(\gamma-p_{43}),\gamma)}_{}
	  - 
	\big[\beta (n-m\theta)((n-1)-(m-1)\theta+(\gamma-\beta)\theta) \big]
\end{align}
with $\beta= \min\left( \tfrac 1 2 (\gamma {-}  p_{12}), \tfrac 1  2 (\gamma 
{-}  p_{43}) \right)$. 
We have thus shown that the differential operator corresponding to $BC_{}$ root 
system is equivalent 
to our Casimir operator.

Before concluding this section however, we point that there is an analogue of the measure based 
relation~\eqref{measure_superjackdiff_explanation} 
in the super BC case which we have not seen in the literature previously.
Macdonald used a measure to define $BC_n$ Jacobi polynomials as orthogonal  
(see for example  
\cite{Macdonald_3} page 52).
This measure has a natural generalisation to the $BC_{n|m}$ case
as follows
\begin{align}
\mathcal{S}_{}^{(p^-,p^+)}({\bf z};\theta)= \prod_{I=1}^{n+m}(z_{I})^{ 
p^-(-\theta)^{1-\pi_I} } (1-z_{I})^{p^+(-\theta)^{1-\pi_I} } \prod_{I<J} 
(z_{I}-z_{J})^{-2(-\theta)^{1-\pi_I-\pi_J}}\ .
\end{align}
We then find that the operator $\tfrac{1}{4}\mathcal{L}^{BC}$ has the simple 
form
\begin{align}
\tfrac{1}{4}\mathcal{L}^{BC}({\bf z};\theta,p,q)=-{ \mathcal{S}}^{-1}
\sum_{I=1}^{n+m} (-\theta)^{\pi_I} \partial_{z_{I}} \Big[ z_{I}(1{-}z_{I})\ 
\mathcal{S}\ \partial_{z_{I}}  \Big] 
\end{align}
where $p=\theta(p^- - p^+)$ and $q=-\frac{1}{2}+\theta p^+$. Recall that 
$p^{\pm}=\frac{|p_{43}\pm p_{12}|}{2}$ %
in terms of the external charges.

\section{Symmetric and supersymmetric polynomials } \label{symm_supersym_poly_sec}

In this section we review relevant background regarding 
Young diagrams and symmetric polynomials (Jacks and interpolation Jacks), both 
in the bosonic and supersymmetric case. 
 We shall view  multivariate Jack polynomials
and interpolation polynomials as fundamental, in the sense 
that they are homogeneous polynomials in certain variables. In particular they 
can be constructed as a sum over semistandard Young tableaux, or filling, as we will see.  
On the 
other hand, Jacobi polynomials (and blocks) we view as `composite': they  
naturally 
can be thought of as a sum of  Jacks as we have done 
throughout the paper. In this appendix we focus on the fundamental 
objects themselves.

\subsection{Symmetric polynomials}
\label{sec_sym}

A Young diagram is a collection of boxes drawn consecutively on rows and 
columns, with the number of boxes on each row decreasing as we go down, for 
example

  \beq
  \begin{array}{c}
             \begin{tikzpicture}

		\foreach \y / \x  in  {   0 /5,  0.5/ 5,  1 / 3, 1.5 /3  ,  2 / 2.5  , 2.5/1,  3/1  }
		\draw (0,-\y) -- (\x,-\y);

		\foreach \x / \y  in  {   0 /3,  0.5/ 3,  1 / 3, 1.5 /2  ,  2 / 2  , 2.5/2,  3/1.5, 3.5/0.5,  4/0.5,  4.5/0.5, 5 /0.5  }
		\draw (\x,0) -- (\x,-\y);

		\draw (7,	-2) 	node 	{ $\ulmb\,=[10,6,6,5,2,2]$};
		\draw (7,	-3) 	node 	{ $\ulmb'=[6,6,4,4,4,3,1,1,1,1]$};

	   \end{tikzpicture}
	   \end{array}
\eeq
By counting the number of boxes on the rows we define a representation of the Young diagram of the form $\ulmb=[\lambda_1,\ldots]$. 
Equivalently, by counting the number of boxes on the columns we define the transposed representation $\ulmb'=[\lambda_1',\ldots]$. 
Both $\ulmb$ and $\ulmb'$ are partitions of the total number of boxes. A box $\Box$ in the diagram has coordinates $(i,j)$ where for each row index $i$, $1\leq j\leq \lambda_i$, 
and for each column index $j$, $1\leq j\leq \lambda'_i$.

Physics-wise, Jack and Jacobi polynomials are best known as eigenfunctions of 
known differential operators, but more abstractly, the theory of symmetric polynomials associates 
polynomials to Young diagrams in such a way that polynomials are characterised by properties and 
uniqueness theorems \cite{Macdonald_3,Macdonald_2,Okounkov_1,Rains_1}, which are equivalent to 
solving the corresponding differential equations.\footnote{See \cite{Macdonald_book} for a comprehensive introduction.}

In this appendix we will highlight a combinatorial definition for Jack and interpolation polynomials, 
which is very efficient in actual computations. 
 We focus on bosonic polynomials first, for which there 
is \emph{no} distinction among $\{x_1,\ldots x_m\}$ variables, differently from 
the supersymmetric case discussed afterwards. 

For a polynomial $P$ of $m$ variables, the combinatorial formula  takes the 
form
\beq\label{combinat_form_poly}
P_{\ulmb}(x_1,\ldots x_m;\vec{s})=\sum_{\{ {\cal T}\} } \Psi_{\cal T}(\vec{s})  
\prod_{(i,j) \in \ulmb} {f}_{}\Big(x_{{\cal T}(i,j)};\vec{s}\Big)\ .
\eeq 
The   functions $\Psi$ and ${f}$ depend on the polynomials under 
consideration (i.e.~Jack or Interpolation polynomials).
Note that in some cases the function ${f}_{}$ may depend explicitly on the integers $(i,j)$ and the integer ${\cal T}(i,j)$, in addition to the variable $x_{{\cal T}(i,j)}$, but we will usually suppress  this dependence in order to avoid cluttering the notation.
Both $\Psi$ and $f$ may also depend on various external parameters, which we denoted collectively by 
$\vec{s}$.
As a concrete example to have in mind: Jack polynomials have simply ${f}_{}(x,i,j;\theta)= x$ 
and for the special case with $\theta=1$ 
(corresponding to Schur polynomials) the coefficient $\Psi_{\cal T}(\theta=1)=1$. 

The sum in~\eqref{combinat_form_poly} is over
all {\em fillings} (also known as semistandard Young tableaux) of the Young diagram  $\ulmb$, denoted here and after by 
$\{{\cal T}\}$. 
A filling $\mathcal{T}$ 
assigns to a box  with coordinates $(i,j)\in\ulmb$, a number in $\{1, \ldots 
m\}$ in such a way that
${\cal T}(i,j)$ is weakly decreasing in $j$, which means from left-to-right, and 
strongly decreasing in $i$, from top-to-bottom, which means ${\cal T}(i,j)>{\cal 
T}(i-1,j)$, ${\cal T}(i,j)\geq {\cal T}(i,j-1)$. 
%


Note that the fillings  precisely correspond to 
the independent states of the  $U(m)$ representation  $\ulmb$ 
which one typically views as a tensor with $|\ulmb|$ indices symmetrised via a 
Young symmetrizer.

For example,
if $\ulmb=[3,1]$ and $m=2$, the fillings $\{\mathcal{T}\}$ are
\beq
\label{exampleyt}
\begin{array}{c}
	\begin{tikzpicture}
		\def\stepx{.8};
		\def\stepy{.6};
		\draw (0,2*\stepy) -- (3*\stepx,2*\stepy) ;
		\draw (0,0*\stepy) -- (1*\stepx,0*\stepy);
		\draw (0,1*\stepy) -- (3*\stepx,1*\stepy) ;
		\draw (0,2*\stepy) -- (0,0*\stepy); \draw (1*\stepx,2*\stepy) 
		--(1*\stepx,0*\stepy);
		\draw (2*\stepx,2*\stepy) --(2*\stepx,1*\stepy);\draw 
		(3*\stepx,2*\stepy) --(3*\stepx,1*\stepy);
		
		\draw (.5*\stepx,1.5*\stepy) node {$2$};
		\draw (1.5*\stepx,1.5*\stepy) node {$2$};
		\draw (2.5*\stepx,1.5*\stepy) node {$2$};
		\draw (.5*\stepx,0.5*\stepy) node {$1$};
	\end{tikzpicture}
\end{array}
\qquad
\begin{array}{c}
	\begin{tikzpicture}
		\def\stepx{.8};
		\def\stepy{.6};
		\draw (0,2*\stepy) -- (3*\stepx,2*\stepy) ;
		\draw (0,0*\stepy) -- (1*\stepx,0*\stepy);
		\draw (0,1*\stepy) -- (3*\stepx,1*\stepy) ;
		\draw (0,2*\stepy) -- (0,0*\stepy); \draw (1*\stepx,2*\stepy) 
		--(1*\stepx,0*\stepy);
		\draw (2*\stepx,2*\stepy) --(2*\stepx,1*\stepy);\draw 
		(3*\stepx,2*\stepy) --(3*\stepx,1*\stepy);
		
		\draw (.5*\stepx,1.5*\stepy) node {$2$};
		\draw (1.5*\stepx,1.5*\stepy) node {$2$};
		\draw (2.5*\stepx,1.5*\stepy) node {$1$};
		\draw (.5*\stepx,0.5*\stepy) node {$1$};
	\end{tikzpicture}
\end{array}
\qquad
\begin{array}{c}
	\begin{tikzpicture}
		\def\stepx{.8};
		\def\stepy{.6};
		\draw (0,2*\stepy) -- (3*\stepx,2*\stepy) ;
		\draw (0,0*\stepy) -- (1*\stepx,0*\stepy);
		\draw (0,1*\stepy) -- (3*\stepx,1*\stepy) ;
		\draw (0,2*\stepy) -- (0,0*\stepy); \draw (1*\stepx,2*\stepy) 
		--(1*\stepx,0*\stepy);
		\draw (2*\stepx,2*\stepy) --(2*\stepx,1*\stepy);\draw 
		(3*\stepx,2*\stepy) --(3*\stepx,1*\stepy);
		
		\draw (.5*\stepx,1.5*\stepy) node {$2$};
		\draw (1.5*\stepx,1.5*\stepy) node {$1$};
		\draw (2.5*\stepx,1.5*\stepy) node {$1$};
		\draw (.5*\stepx,0.5*\stepy) node {$1$};
	\end{tikzpicture}
\end{array}
\eeq
which correspond to the three states in the corresponding rep of $U(2)$, the 
independent states in a tensor of the form $S_{abcd} = T_{(abc)d}-T_{(dbc)a}$ 
where the indices $a,b,c,d= 1,2$.
Then for example the Schur polynomial can be directly read off 
from~\eqref{combinat_form_poly} with  $f(x) =x$ and $\Psi=1$ as
\begin{align}\label{schurex}
	P_{[3,1]}(x_1,x_2;\theta=1)=x_1^3x_2+ x_1^2x_2^2 +  x_1 x_2^3\ . 
\end{align}
We immediately see from this definition that if  the number of rows of $\ulmb$ 
is greater than 
the number of variables $m$ then $P_\ulmb =0$ since it is not possible 
to construct a valid semistandard tableaux and the sum is empty (just as for 
$U(m)$ reps).

In order to define $\Psi$ it is first useful to note that there is a simple way 
to generate all the fillings $\{{\cal T}\}$ given a Young 
diagram $\ulmb$, via recursion in the number of variables $m$. (See for example the discussion in \cite{Okounkov_Olshanski}.)
This recursion in fact gives a way of generating the polynomial itself also, equivalent to the combinatorial formula~\eqref{combinat_form_poly}. It 
reads
\begin{align}\label{most_gen_rec}
P_{\ulmb}(x_1,\ldots x_{m},x_{m+1};\vec{s}) = &\sum_{\underline{\kappa} 
\prec\ulmb}\ 
\psi^{}_{\ulmb,\underline{\kappa}}(\vec{s})\ \left(\prod_{(i,j)\in 
\ulmb/\underline{\kappa}} f_{}(x_{m+1};\vec{s})\right)
P_{\underline{\kappa}}(x_1,\ldots x_{m};\vec{s})\notag\\
P_{[\varnothing]}=&\,1
\end{align} 
where $\ulmb / \underline{\kappa}$ is the skew Young diagram obtained by taking 
the Young diagram of $\ulmb$ and deleting the boxes of the sub Young diagram 
$\underline{\kappa}$ (see section~\ref{sec_skew}).
Here $\psi^{}$ is closely related to $\Psi$ mentioned above (we will give the 
precise relation shortly), and  
the symbol $\underline{\kappa} \prec\ulmb$ means that 
$\underline{\kappa}$ belongs to the following set,\footnote{The skew diagram $\ulmb/\underline{\kappa}$ is sometimes known as a {\em horizontal strip}.}  
\bea\label{horiz_strips}
\{\ [\kappa_1,\ldots ,\kappa_{m} ]:\ \ \lambda_{m+1}\leq 
\kappa_{m}\leq \lambda_{m}\,,\ \ldots\ ,\,\lambda_2\leq\kappa_1\leq\lambda_1\ 
\}\ .
\eea
In this formula, if $\ulmb$ is a partition with less than $m+1$ rows, it is extended with trailing zeros. 
The recursion generates sequences of Young diagrams
of the form 
\beq\label{sequence_hor_strip}
[\varnothing]\equiv \underline{\kappa}^{(0)}\prec \underline{\kappa}^{(1)} \prec\ldots \prec \underline{\kappa}^{(m)}\prec \underline{\kappa}^{(m+1)}\equiv \ulmb,
\eeq
with a strict inclusion, i.e. $\underline{\kappa}^{(i-1)}\subset \underline{\kappa}^{(i)}$. 
This is the same as considering the set of fillings 
$\{{\cal T}\}$. 
For example, for the above case
with $\ulmb=[3,1]$ in~\eqref{exampleyt}
the sum in~\eqref{most_gen_rec} would be over  $\underline{\kappa}\in 
\{[3],[2],[1]\}$. These sub Young diagrams are then filled with $x_1$'s and the 
remaining bits $\ulmb/\underline{\kappa}$ filled with $x_2$ 
reproducing~\eqref{exampleyt}. 

It follows that for a filling ${\cal T}$ corresponding to a sequence 
\eqref{sequence_hor_strip},
\beq
\Psi_{\cal T}(\vec{s})   = 
\prod_{i=1}^{m+1} \psi_{\underline{\kappa}^{(i)}\!,\,\underline{\kappa}^{(i-1)} } 
\,,
\qquad  \qquad \prod_{(i,j)\in \ulmb} {f}_{}(x_{{\cal T}(i,j)};\vec{s})=\prod_{l=1}^{m+1} \prod_{(i,j)\in 
\kappa^{(l)}/\kappa^{(l-1)}}
 f(x_l;\vec{s})
\eeq
and the recursion and the combinatorial formula \eqref{combinat_form_poly} are 
the same.  In particular $l$ is the level of nesting in the recursive formula. In the example above with $m+1$ variables,  the relation between $l$ and ${\cal T}(i,j)$ is
\beq\label{level_vs_calT}
l=(m+1)-{\cal T}(i,j)+1.
\eeq

Given the above facts it is enough for us to 
define $\psi^{}_{\ulmb,\underline{\kappa}}$ and the function $f(x;\vec{s})$   in order to fully 
define the symmetric function. 

We will now examine the various specific cases, beginning with the Jack polynomials.

\subsection{Jack polynomials}

Jack polynomials depend on one parameter, $\theta$. 
The defining function is  (for example \cite{Okounkov_Olshanski}) 
\begin{align}
\psi_{\ulmb,\underline{\kappa}}(\theta)= 
\prod_{1\leq i \leq j <m+1 } \frac{ ( \kappa_i -\lambda_{j+1} +1 + \theta (j-i) )_{\lambda_i-\kappa_i } }{   (\kappa_i-\lambda_{j+1}+\theta(j-i+1))_{\lambda_i-\kappa_i } } 
 \frac{ (\kappa_i-\kappa_j+\theta(j-i+1) )_{ \lambda_i-\kappa_i} }{ ( 
 \kappa_i-\kappa_j+1+\theta(j-i) )_{\lambda_i-\kappa_i} }\ .
\label{Jackpsi}
\end{align}
with,
\begin{align}
	f(x;\theta)= x 
\end{align}

Considering the top-left Pochhammer symbol, notice that 
$\psi_{\ulmb,\underline{\kappa}}$ 
vanishes when,
\begin{align}
 ( \kappa_i -\lambda_{i+1} +1 )_{\lambda_i-\kappa_i }=
(\kappa_i+1-\lambda_{i+1})\ldots (\lambda_i-\lambda_{i+1})=0
\end{align}
i.e.~when one of the terms vanishes.
This happens precisely when $\underline{\kappa}\nprec\ulmb$. 

As a simple example then consider $P_{[2,1]}(x_1,x_2;\theta)$. The case with $\theta=1$ (Schur polynomial) is given in~\eqref{schurex}. 
For arbitrary $\theta$ we need the coefficient $\Psi_{\cal T}(\theta)$ for the three fillings $\cal T$ in~\eqref{exampleyt}. 
The first and third filling  have $\Psi=1$ whereas the second has $\Psi=\tfrac{2\theta}{1+\theta}$ and we thus obtain
\begin{align} 
P_{[3,1]}(x_1,x_2;\theta)=x_1^3 x_2 +\tfrac{2\theta}{1+\theta}  x_1^2 x_2^2 + x_1 x_2^3\ .
\end{align}

Notice that Jack polynomials are stable, i.e.~$P_{\ulmb}(x_1,\ldots 
x_m,0;\theta)=P_{\ulmb}(x_1,\ldots x_m;\theta)$ as can be shown directly from the 
combinatorial formula (as well as from their definition as  the unique polynomial eigenfunctions of the differential 
equation~\eqref{ham} with eigenvalues and differential operator independent of 
$m$, and  with the same normalisation).

Also note that Jack polynomials have the following property 
\begin{align}\label{jackshift}
	P^{}_{\ulmb+\tau^m} = (x_1 \ldots x_m)^{\tau} P^{}_{\ulmb}\ .
\end{align}
It is instructive to prove \eqref{jackshift} by showing again that both 
RHS~and~LHS~are eigenfunctions of the $A_{n}$ CMS operator ${\bf H}$ in 
\eqref{ham}.
Applying ${\bf H}$ on the LHS~we simply find the corresponding eigenvalue $h_{\ulmb+\tau^m}$. On the RHS~we need to consider what happens upon conjugation,
\beq
(x_1 \ldots x_m)^{-\tau}\cdot{\bf H}\cdot(x_1 \ldots x_m)^{\tau} = {\bf H}+2\tau \sum_i x_i \partial_i + h^{(\theta)}_{\tau^m}
\eeq 
Thus applying  ${\bf H}\!\cdot\!$ to \eqref{jackshift} we find
\beq\label{bosonic_galilean}
h^{(\theta)}_{\ulmb+\tau^m}=h^{(\theta)}_{\ulmb}+ 2\tau |\ulmb|+ h^{(\theta)}_{\tau^m}
\eeq
which is an identity, and proves \eqref{jackshift}, since both RHSand l.h.s have the same small variable expansion.\\ 

\emph{Remark.}
When there is an ambiguity in the notation, in relation to the supersymmetric case, 
we will specify $P^{(m,0)}({\bf z};\theta)$ to mean the bosonic Jack polynomial.

\subsubsection*{Dual Jack polynomials} 

{Jack polynomials are orthogonal but not orthonormal (except when $\theta=1$ 
where they reduce to Schur polynomials) under the Hall inner product.
 The 
\emph{dual} Jack polynomials (where dual here denotes the usual vector space 
dual 
under the Hall inner product) are thus simply a normalisation of the Jacks, 
defined so that they have unit inner product with the corresponding Jack. 	}
Given a Jack polynomial, the \emph{dual} Jack polynomial has the 
 form~\cite{Macdonald_book}
\begin{align}\label{dual_jack}
Q_{\underline{\kappa} }({\bf x};\theta) =  \frac{ C^{-}_{\underline{\kappa} }(\theta;\theta) }{ C^{-}_{\underline{\kappa} }(1;\theta) } P_{ \underline{\kappa} }({\bf x};\theta)\quad;\quad 
C_{\underline{\kappa}}^-(t;\theta)=\prod_{(ij)\in\underline{\kappa} } \left( \kappa_i{-}j+\theta(\kappa'_j{-}i)+t \right)
\end{align}
where, as throughout
\begin{align}\label{def_PI}
\Pi_{\underline{\kappa} }(\theta)=   \frac{ C^{-}_{\underline{\kappa} 
}(\theta;\theta) }{ C^{-}_{\underline{\kappa} }(1;\theta) } 
\qquad;\qquad \Pi_{\underline{\kappa}}(\tfrac{1}{\theta} )= 
\big(\Pi_{\underline{\kappa}' }(\theta) \big)^{-1}\ .
\end{align}

\subsubsection*{Skew Jack polynomials} 
\label{sec_skew}

For one way of defining super Jack polynomials shortly we will also need the 
concept of skew Jack polynomials.

A skew Young diagram $\ulmb/\umu$, where $\umu\subseteq \ulmb$
is obtained by erasing $\umu$ from $\ulmb$.  
The Figure below gives a simple example,\\

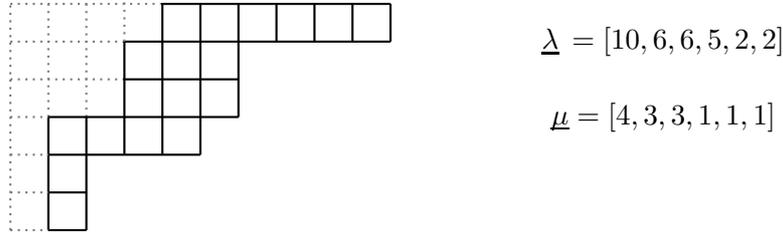
        \begin{center}
             \begin{tikzpicture}
		\def\xuno{2}
		\def\yuno{2}
	
		\def\ydue{5}
		\def\xtre{7}
		
		\def\ytre{4.5}
		\def\xquattro{5}
		
		\def\yquattro{3.5}
		\def\xcinque{4.5}
		
		\def\ycinque{3}
		\def\xsei{3}
		
		\def\step{.5}

		\draw[thick,gray,dotted] (\xuno,	\yuno) -- 			(\xuno,	\ydue);
		
		\draw[thick] (\xuno+4*.495,	\ydue) -- 			(\xtre,	\ydue);
		\draw[thick,gray,dotted] (\xuno,	\ydue) -- 			(\xuno+4*.495,	\ydue);
		
		\foreach \x in {4, ..., 10}
		\draw[thick] (\xuno+\x*.5,	\ydue) -- 			(\xuno+\x*.5,	\ydue-\step);
		
		\draw[thick] (\xquattro,	\ydue-\step) -- 			(\xtre,	\ydue-\step);
		
		\draw[thick] (\xuno+3*.495,	\ytre) -- 			(\xquattro,	\ytre);
		\draw[thick,gray,dotted] (\xuno,	\ytre) -- 			(\xuno+3*.495,	\ytre);
		\draw[thick,gray,dotted] (\xuno,	\ytre-\step) -- 			(\xuno+3*.495,	\ytre-\step);
	
		\draw[thick,gray,dotted] (\xuno+3*\step,	\ydue-\step) -- 			(\xuno+3*\step,	\ydue+.025);
		\draw[thick,gray,dotted] (\xuno+\step,	\ytre-2*\step) -- 			(\xuno+\step,	\ydue+.025);
		\draw[thick,gray,dotted] (\xuno+2*\step,	\ytre-2*\step) -- 			(\xuno+2*\step,	\ydue+.025);

		\foreach \x in {3, ..., 6}
		\draw[thick] (\xuno+\x*.5,	\ytre) -- 			(\xuno+\x*.5,	\ytre-2*\step);
		
		\draw[thick] (\xuno+3*.495,	\ytre-\step) -- 			(\xquattro,	\ytre-\step);
		\draw[thick] (\xuno+1*.495,	\ytre-2*\step) -- 			(\xquattro,	\ytre-2*\step);

		\foreach \x in {1, ..., 5}
		\draw[thick] (\xuno+\x*.5,	\yquattro) -- 			(\xuno+\x*.5,	\yquattro-\step);
		
		\draw[thick] (\xsei,	\yquattro-\step) -- 			(\xcinque,	\yquattro-\step);

		\draw[thick] (\xuno+1*.495,	\ycinque) -- 			(\xsei,	\ycinque);
		
		\foreach \x in {1, ..., 2}
		\draw[thick] (\xuno+\x*.5,	\ycinque) -- 			(\xuno+\x*.5,	\ycinque-2*\step);
		
		\draw[thick] (\xuno+1*.495,	\ycinque-\step) -- 			(\xsei,	\ycinque-\step);
		\draw[thick] (\xuno+1*.495,	\ycinque-2*\step) -- 			(\xsei,	\ycinque-2*\step);

		\draw[thick,gray,dotted] (\xuno,	\ycinque-2*\step) -- 			(\xuno+\step,	\ycinque-2*\step);
		\draw[thick,gray,dotted] (\xuno,	\ycinque-1*\step) -- 			(\xuno+\step,	\ycinque-1*\step);
		\draw[thick,gray,dotted] (\xuno,	\ycinque) -- 			(\xuno+\step,	\ycinque);
		\draw[thick,gray,dotted] (\xuno,	\ycinque+\step) -- 			(\xuno+\step,	\ycinque+\step);

		\draw (\xtre+3.6,	\ytre) 	node 	{ $\ulmb\,=[10,6,6,5,2,2]$};
		\draw (\xtre+3.6,	\yquattro) 	node 	{ $\umu=[4,3,3,1,1,1]$};

	   \end{tikzpicture}
	\end{center}
Skew Jack polynomials are then defined by a similar combinatorial formula but 
where one sums only over semi standard skew Young tableaux. Or equivalently 
by the recursion formula
\begin{align}\label{most_gen_rec_jack_skew}
P_{\ulmb/\umu}(x_1,\ldots x_{m},x_{m+1};\theta) = \sum_{\umu \, \preceq\,  \underline{\kappa}\, \prec\, \ulmb}\ 
\psi_{\ulmb,\underline{\kappa}}(\theta)\ 
x_{m+1}^{|\ulmb|-|\underline{\kappa}|} P^{}_{\underline{\kappa}/\umu}(x_1,\ldots
 x_{m};\theta)\ .
\end{align}
Notice that $m+1\ge \lambda'_1-\mu'_1$.
The number of variables here is the number of variables that can fill in the Young diagram 
according to recursion in \eqref{horiz_strips}.
The recursion goes on as long as
$\umu\prec \underline{\kappa}^{(1)} \prec\ldots \prec \underline{\kappa}^{(m)}\prec \underline{\kappa}^{(m+1)}\equiv \ulmb$
and the condition $\umu\prec \underline{\kappa}^{(1)}$ is non trivial, 
since it implies that a skew Jack polynomial has the vanishing property
\begin{align}\label{skew_vanishing_cond}
P_{\ulmb/\umu}(x_1,\ldots x_m)=0\qquad {\rm if}\quad  \lambda_{m+i}> \mu_i\ .
\end{align}
For example, imagine a $\ulmb$ with very long rows, and take a very small $\umu$, then no horizontal strip of $\ulmb$ will contain $\umu$. 
Indeed the minimal diagram $\underline{\kappa}$ generated by the table \eqref{horiz_strips} at each 
step of the recursion is given by components of $\ulmb$ properly shifted upwards. 

An example of a skew Jack polynomial is
\begin{align}
P_{[3,1,1]/[1]}(x_1,x_2;\theta)=x_1^3 x_2 + x_1 x_2^3 + \frac{2\theta}{1+\theta} x_1^2 x_2^2
\end{align}

Skew polynomials have the property that
\begin{align}\label{property}
P_{\ulmb}({\bf x} ;\theta)= \sum_{\umu\subset \ulmb} P_{\umu}(x_1,\ldots x_{m-n};\theta) P_{\ulmb/\umu}(x_{m-n+1},\ldots x_m;\theta)
\end{align}

\subsubsection*{Structure constants and decomposition formulae for Jacks}

The Jack structure constants $\cC_{\ulmb \umu}^\unu(\theta)$ are defined as follows 
\begin{align}\label{structure}
	P_\ulmb P_{\umu} =\sum_{\unu} \cC_{\ulmb \umu}^\unu(\theta) P_{\unu}\ .
\end{align}
For $\theta=1$ they are just the Littlewood-Richardson coefficients.

Then there are related coefficients ${\cal S}_{\unu}^{\underline{\ulmb}\,\umu}(\theta)$ obtained from decomposing skew Jack polynomials into Jack polynomials 
\begin{align}\label{skewinjack}
	P_{\unu/\ulmb} = \sum_{\umu} {\cal S}_{\unu}^{\ulmb\,\umu}(\theta) P_{\umu}\ .
\end{align}
The property~\eqref{property} then yields the  decomposition formula for decomposing higher  dimensional Jacks into sums of products of lower dimensional Jacks
 \begin{align}\label{bosdecomp}
 	P_{\ulmb}(x_1,..,x_{m+n})= \sum_{\umu,\unu} P_{\umu}(x_1,..,x_m) \cS_\ulmb^{\umu \unu}(\theta) P_{\unu}(x_{m+1},..,x_{m+n})\ .
 \end{align}

For $\theta=1$ these decomposition coefficients are also the Littlewood-Richardson coefficients,  $\cC_{\ulmb \umu}^\unu(1)={\cal S}_{\unu}^{\underline{\ulmb}\,\umu}(1)$, but for general $\theta$ they are related via normalisation:
\beq\label{from_C_to_S}
\mathcal{S}^{\umu\unu}_{\ulmb}(\theta)=
\frac{ \Pi_{\umu}(\theta) \Pi_{\unu}(\theta) }{   \Pi_{\lmb}(\theta)} \mathcal{C}^{\ulmb}_{\umu \unu}(\theta)\ .
\eeq

Note that the structure constants $\cC_{\ulmb \umu}^\unu(\theta)$ and ${\cal S}_{\unu}^{\underline{\ulmb}\,\umu}(\theta)$ do not depend on the dimensions of the Jack polynomials in any of the above formulae.

Also note that if the Young diagram $\ulmb$ is built from two Young diagrams $\umu,\unu$ on top of each other then the corresponding structure constant is 1:
\begin{align}\label{sis1}
	\mathcal{S}^{\umu\unu}_{\ulmb}(\theta)=1 \qquad \text{if}\qquad  \ulmb=[\umu,\unu]\ .
\end{align}
This can be seen by~\eqref{skewinjack} noting that the Jack polynomial of dimension equal to the height of $\umu$ is equal to the skew Jack 
\begin{align}
	P_\umu(x_1,..,x_{\mu'_1}) = P_{\ulmb /\unu }(x_1,..,x_{\mu'_1})\qquad \text{if}\qquad  \ulmb=[\umu,\unu]\ .
\end{align}
This can be verified from the respective combinatoric formulae.

\subsection{Interpolation polynomials} 
\label{sec_interpolation}

The interpolation polynomials of relevance here  (i.e.~the $BC$-type which appear 
in section~\ref{sec_guideline}) 
depend on two parameters, $\theta$, and a new one denoted here by $u$. 
They can be defined in a very similar way to the Jack polynomials and we will 
denote them  $P^{ip}_{}({\bf x};\theta,u)$. 
They are symmetric polynomials in the variables ${\bf x}=(x_1,..,x_m)$.
Generically $P^{ip}_{}$ is a complicated, non factorisable polynomial and 
it 
is uniquely 
defined by the following vanishing property 
\begin{align}\label{vanishing}
	P^{ip}_{\underline{\kappa}}(\underline{\nu} 
	+\theta\underline{\delta}+u;\theta,u)=0 \qquad \text{if} \qquad \ulmb 
	\subset \umu 
\end{align}
where $\udelta=(m{-}1,..,1,0)$.
This idea generalises what happens for the Pochhammer symbol 
$(-z)_\lambda=(-z)(-z+1)\ldots (-z+\lambda-1)$ which vanishes if $0\leq 
z<\lambda$.%
\footnote{Indeed the Pochhammer symbol $(-z)_k$ is precisely a case of interpolating 
polynomial, giving the standard one-variable binomial expansion $(1+x)^n = 
\sum_k \begin{psmallmatrix}
		n\\k
	\end{psmallmatrix}x^k = \sum_k \frac{(-n)_k}{(-k)_k}x^k$.} 

The interpolation polynomial is defined via the combinatorial 
formula~\eqref{combinat_form_poly} (or the equivalent 
recursion~\eqref{most_gen_rec}) with $\psi$ exactly the same as for the 
Jacks~\eqref{Jackpsi}, but with a more complicated $x$ dependence arising from 
a 
modified $f$ 
	\begin{align}\label{BCpsi}
		\psi_{\ulmb,\underline{\kappa}}(\theta,u)&=\psi_{\ulmb,\underline{\kappa}}
		 (\theta)\\	
		 f(x_l,i,j,l;\theta,u)&=  x_l^2-( (j{-}1){-}\theta(i-1)+\theta(l-1)+u)^2\label{BCf}
	\end{align}
where $\psi_{\ulmb,\underline{\kappa}}(\theta)$ is the defining function for 
Jack polynomials~\eqref{Jackpsi}. Instead,
note that $f$ here depends explicitly on $(i,j)$ and $l$. As in \eqref{sequence_hor_strip} the index $l$ labels the nesting and is related to ${\cal T}(i,j)$ via \eqref{level_vs_calT}.
 
One can see an example of the vanishing property from the above definition: take the last variable $x_{m}$
and the last row $i=m$ in \eqref{BCf}, this contributes with terms 
which vanish for all $x_{m}=j-1+u$, where $j=1,\ldots \kappa_{m}$.

Because of the various shifts in the vanishing property~\eqref{vanishing}, it is useful to define the 
non-symmetric version of the interpolation polynomial $P^{*}({\bf x};\theta,u)$ 
by
\begin{align}
	P^{*}({\bf x};\theta,u)\equiv P^{ip}({\bf x}+\theta \udelta+u;\theta,u) \qquad 
	\udelta=(m{-}1,..,1,0)\ 
\end{align}
So the vanishing property~\eqref{vanishing} takes the form
\begin{align}
	P^*_{\umu}(\ulmb;\theta,u)=0 \qquad \text{if} \quad \ulmb \subset \umu \ .
\end{align}
(Recalling however that $P^*_{\umu}$ here is no longer symmetric in its variables $\bf{z}$.)

This interpolation polynomial is $\mathbb{Z}_2$ invariant under $x_i\leftrightarrow -x_i$.
It was introduced by Okounkov \cite{Okounkov_1}, and re-obtained by Rains 
\cite{Rains_1}, using a different approach.

\subsection{Supersymmetric polynomials}

The general combinatorial formulation of symmetric polynomials outlined in 
section~\ref{sec_sym} has a natural supersymmetric 
generalisation~\cite{Veselov_1}. This allows in particular to define super Jack 
polynomials and super interpolation Jack polynomials.  The only real 
modification in the general story is that the definition 
of a filling (semistandard tableaux) is modified to become a supersymmetric 
filling (or bitableau)  since it has to 
take into account two alphabets, 
$x_1,\ldots x_m$ and $y_1,\ldots y_n$.  Written in terms of the letters ${\bf 
	z}$, the labelling is $z_i=x_i$ 
for $i=1,\ldots m$, and $z_{m+j}=y_j$ for $j=1,\ldots n$ and  the  
combinatorial formula then looks the same as in~\eqref{combinat_form_poly} 
\beq\label{combinat_form_poly_susy}
P_{\ulmb}(z_1,\ldots z_{m+n};\vec{s})=\sum_{\{ {\cal T}\} } \Psi_{\cal 
T}(\vec{s})  
\prod_{(i,j)\in \ulmb} f(z_{ {\cal T}(i,j)};\vec{s}) 
\eeq 
but with  the sum now over all \emph{supersymmetric} fillings.

The supersymmetric filling assigns to a box of coordinate $(i,j)$ a number 
$\{1,\ldots n+m\}$ such that
${\cal T}(i,j)$ is weakly decreasing%
\footnote{ Conventionally the 
	supersymmetric 
	filling is also reversed, hence decreasing rather than increasing.} in $j$, i.e.~from left-to-right, and 
is weakly decreasing in $i$, i.e.~from top-to-bottom. 
But, if  ${\cal T}(i,j)\in\{1,\ldots m\}$, then ${\cal T}(i,j)$ is 
strictly decreasing in $i$, i.e.~from top-to-bottom, and 
if  ${\cal T}(i,j)\in\{n+1,\ldots n+m\}$, then ${\cal T}(i,j)$ is 
strictly decreasing in $j$, i.e. from left-to-right. 

Note that the above notion of a supersymmetric filling has a direct relation with 
states
of the supergroup $U(m|n)$ just as the ordinary fillings relate to states  of 
$U(m)$.
This can again be clearly seen by representing the $U(m|n)$ irrep $\ulmb$ as a  
tensor $T_{A_1 A_2 .. A_{|\ulmb|}}$ with 
superindices $A\in\{1,..,m|m+1,..,m+n\}$ and symmetrising the indices in the 
standard fashion via a Young symmetriser. The only  
caveat is that when the index $A \in \{m+1,..,m+n\}$ the index is viewed as 
`fermionic' and so symmetrising two of them  
becomes anti-symmetrising and vice versa. 
The resulting independent states obtained in this way will have a precise 
correspondence with 
the supersymmetric fillings.

\subsubsection*{A first example, $\ulmb=[3,1]$ with $(z_1,z_2|z_3)$.} This has 
two rows, so we can 
fill it in as in \eqref{exampleyt}, namely
\beq
\begin{array}{c}
	\begin{tikzpicture}[scale=.9]
		\def\stepx{.8};
		\def\stepy{.6};
		\draw (0,2*\stepy) -- (3*\stepx,2*\stepy) ;
		\draw (0,0*\stepy) -- (1*\stepx,0*\stepy);
		\draw (0,1*\stepy) -- (3*\stepx,1*\stepy) ;
		\draw (0,2*\stepy) -- (0,0*\stepy); \draw (1*\stepx,2*\stepy) 
		--(1*\stepx,0*\stepy);
		\draw (2*\stepx,2*\stepy) --(2*\stepx,1*\stepy);\draw 
		(3*\stepx,2*\stepy) --(3*\stepx,1*\stepy);
		
		\draw (.5*\stepx,1.5*\stepy) node {$2$};
		\draw (1.5*\stepx,1.5*\stepy) node {$2$};
		\draw (2.5*\stepx,1.5*\stepy) node {$2$};
		\draw (.5*\stepx,0.5*\stepy) node {$1$};
	\end{tikzpicture}
\end{array}
\qquad
\begin{array}{c}
	\begin{tikzpicture}[scale=.9]
		\def\stepx{.8};
		\def\stepy{.6};
		\draw (0,2*\stepy) -- (3*\stepx,2*\stepy) ;
		\draw (0,0*\stepy) -- (1*\stepx,0*\stepy);
		\draw (0,1*\stepy) -- (3*\stepx,1*\stepy) ;
		\draw (0,2*\stepy) -- (0,0*\stepy); \draw (1*\stepx,2*\stepy) 
		--(1*\stepx,0*\stepy);
		\draw (2*\stepx,2*\stepy) --(2*\stepx,1*\stepy);\draw 
		(3*\stepx,2*\stepy) --(3*\stepx,1*\stepy);
		
		\draw (.5*\stepx,1.5*\stepy) node {$2$};
		\draw (1.5*\stepx,1.5*\stepy) node {$2$};
		\draw (2.5*\stepx,1.5*\stepy) node {$1$};
		\draw (.5*\stepx,0.5*\stepy) node {$1$};
	\end{tikzpicture}
\end{array}
\qquad
\begin{array}{c}
	\begin{tikzpicture}[scale=.9]
		\def\stepx{.8};
		\def\stepy{.6};
		\draw (0,2*\stepy) -- (3*\stepx,2*\stepy) ;
		\draw (0,0*\stepy) -- (1*\stepx,0*\stepy);
		\draw (0,1*\stepy) -- (3*\stepx,1*\stepy) ;
		\draw (0,2*\stepy) -- (0,0*\stepy); \draw (1*\stepx,2*\stepy) 
		--(1*\stepx,0*\stepy);
		\draw (2*\stepx,2*\stepy) --(2*\stepx,1*\stepy);\draw 
		(3*\stepx,2*\stepy) --(3*\stepx,1*\stepy);
		
		\draw (.5*\stepx,1.5*\stepy) node {$2$};
		\draw (1.5*\stepx,1.5*\stepy) node {$1$};
		\draw (2.5*\stepx,1.5*\stepy) node {$1$};
		\draw (.5*\stepx,0.5*\stepy) node {$1$};
	\end{tikzpicture}
\end{array}
\eeq
Then we introduce the variable $z_3=y_1$, once
\beq
\begin{array}{c}
	\begin{tikzpicture}[scale=.9]
		\def\stepx{.8};
		\def\stepy{.6};
		\draw (0,2*\stepy) -- (3*\stepx,2*\stepy) ;
		\draw (0,0*\stepy) -- (1*\stepx,0*\stepy);
		\draw (0,1*\stepy) -- (3*\stepx,1*\stepy) ;
		\draw (0,2*\stepy) -- (0,0*\stepy); \draw (1*\stepx,2*\stepy) 
		--(1*\stepx,0*\stepy);
		\draw (2*\stepx,2*\stepy) --(2*\stepx,1*\stepy);\draw 
		(3*\stepx,2*\stepy) --(3*\stepx,1*\stepy);
		
		\draw (.5*\stepx,1.5*\stepy) node {$3$};
		\draw (1.5*\stepx,1.5*\stepy) node {$2$};
		\draw (2.5*\stepx,1.5*\stepy) node {$2$};
		\draw (.5*\stepx,0.5*\stepy) node {$2$};
	\end{tikzpicture}
\end{array}
\
\begin{array}{c}
	\begin{tikzpicture}[scale=.9]
		\def\stepx{.8};
		\def\stepy{.6};
		\draw (0,2*\stepy) -- (3*\stepx,2*\stepy) ;
		\draw (0,0*\stepy) -- (1*\stepx,0*\stepy);
		\draw (0,1*\stepy) -- (3*\stepx,1*\stepy) ;
		\draw (0,2*\stepy) -- (0,0*\stepy); \draw (1*\stepx,2*\stepy) 
		--(1*\stepx,0*\stepy);
		\draw (2*\stepx,2*\stepy) --(2*\stepx,1*\stepy);\draw 
		(3*\stepx,2*\stepy) --(3*\stepx,1*\stepy);
		
		\draw (.5*\stepx,1.5*\stepy) node {$3$};
		\draw (1.5*\stepx,1.5*\stepy) node {$2$};
		\draw (2.5*\stepx,1.5*\stepy) node {$2$};
		\draw (.5*\stepx,0.5*\stepy) node {$1$};
	\end{tikzpicture}
\end{array}
\
\begin{array}{c}
	\begin{tikzpicture}[scale=.9]
		\def\stepx{.8};
		\def\stepy{.6};
		\draw (0,2*\stepy) -- (3*\stepx,2*\stepy) ;
		\draw (0,0*\stepy) -- (1*\stepx,0*\stepy);
		\draw (0,1*\stepy) -- (3*\stepx,1*\stepy) ;
		\draw (0,2*\stepy) -- (0,0*\stepy); \draw (1*\stepx,2*\stepy) 
		--(1*\stepx,0*\stepy);
		\draw (2*\stepx,2*\stepy) --(2*\stepx,1*\stepy);\draw 
		(3*\stepx,2*\stepy) --(3*\stepx,1*\stepy);
		
		\draw (.5*\stepx,1.5*\stepy) node {$3$};
		\draw (1.5*\stepx,1.5*\stepy) node {$2$};
		\draw (2.5*\stepx,1.5*\stepy) node {$1$};
		\draw (.5*\stepx,0.5*\stepy) node {$2$};
	\end{tikzpicture}
\end{array}
\
\begin{array}{c}
	\begin{tikzpicture}[scale=.9]
		\def\stepx{.8};
		\def\stepy{.6};
		\draw (0,2*\stepy) -- (3*\stepx,2*\stepy) ;
		\draw (0,0*\stepy) -- (1*\stepx,0*\stepy);
		\draw (0,1*\stepy) -- (3*\stepx,1*\stepy) ;
		\draw (0,2*\stepy) -- (0,0*\stepy); \draw (1*\stepx,2*\stepy) 
		--(1*\stepx,0*\stepy);
		\draw (2*\stepx,2*\stepy) --(2*\stepx,1*\stepy);\draw 
		(3*\stepx,2*\stepy) --(3*\stepx,1*\stepy);
		
		\draw (.5*\stepx,1.5*\stepy) node {$3$};
		\draw (1.5*\stepx,1.5*\stepy) node {$2$};
		\draw (2.5*\stepx,1.5*\stepy) node {$1$};
		\draw (.5*\stepx,0.5*\stepy) node {$1$};
	\end{tikzpicture}
\end{array}
\
\begin{array}{c}
	\begin{tikzpicture}[scale=.9]
		\def\stepx{.8};
		\def\stepy{.6};
		\draw (0,2*\stepy) -- (3*\stepx,2*\stepy) ;
		\draw (0,0*\stepy) -- (1*\stepx,0*\stepy);
		\draw (0,1*\stepy) -- (3*\stepx,1*\stepy) ;
		\draw (0,2*\stepy) -- (0,0*\stepy); \draw (1*\stepx,2*\stepy) 
		--(1*\stepx,0*\stepy);
		\draw (2*\stepx,2*\stepy) --(2*\stepx,1*\stepy);\draw 
		(3*\stepx,2*\stepy) --(3*\stepx,1*\stepy);
		
		\draw (.5*\stepx,1.5*\stepy) node {$3$};
		\draw (1.5*\stepx,1.5*\stepy) node {$1$};
		\draw (2.5*\stepx,1.5*\stepy) node {$1$};
		\draw (.5*\stepx,0.5*\stepy) node {$2$};
	\end{tikzpicture}
\end{array}
\
\begin{array}{c}
	\begin{tikzpicture}[scale=.9]
		\def\stepx{.8};
		\def\stepy{.6};
		\draw (0,2*\stepy) -- (3*\stepx,2*\stepy) ;
		\draw (0,0*\stepy) -- (1*\stepx,0*\stepy);
		\draw (0,1*\stepy) -- (3*\stepx,1*\stepy) ;
		\draw (0,2*\stepy) -- (0,0*\stepy); \draw (1*\stepx,2*\stepy) 
		--(1*\stepx,0*\stepy);
		\draw (2*\stepx,2*\stepy) --(2*\stepx,1*\stepy);\draw 
		(3*\stepx,2*\stepy) --(3*\stepx,1*\stepy);
		
		\draw (.5*\stepx,1.5*\stepy) node {$3$};
		\draw (1.5*\stepx,1.5*\stepy) node {$1$};
		\draw (2.5*\stepx,1.5*\stepy) node {$1$};
		\draw (.5*\stepx,0.5*\stepy) node {$1$};
	\end{tikzpicture}
\end{array}\notag
\eeq
and twice
\beq
\begin{array}{c}
	\begin{tikzpicture}[scale=.9]
		\def\stepx{.8};
		\def\stepy{.6};
		\draw (0,2*\stepy) -- (3*\stepx,2*\stepy) ;
		\draw (0,0*\stepy) -- (1*\stepx,0*\stepy);
		\draw (0,1*\stepy) -- (3*\stepx,1*\stepy) ;
		\draw (0,2*\stepy) -- (0,0*\stepy); \draw (1*\stepx,2*\stepy) 
		--(1*\stepx,0*\stepy);
		\draw (2*\stepx,2*\stepy) --(2*\stepx,1*\stepy);\draw 
		(3*\stepx,2*\stepy) --(3*\stepx,1*\stepy);
		
		\draw (.5*\stepx,1.5*\stepy) node {$3$};
		\draw (1.5*\stepx,1.5*\stepy) node {$2$};
		\draw (2.5*\stepx,1.5*\stepy) node {$2$};
		\draw (.5*\stepx,0.5*\stepy) node {$3$};
	\end{tikzpicture}
\end{array}
\
\begin{array}{c}
	\begin{tikzpicture}[scale=.9]
		\def\stepx{.8};
		\def\stepy{.6};
		\draw (0,2*\stepy) -- (3*\stepx,2*\stepy) ;
		\draw (0,0*\stepy) -- (1*\stepx,0*\stepy);
		\draw (0,1*\stepy) -- (3*\stepx,1*\stepy) ;
		\draw (0,2*\stepy) -- (0,0*\stepy); \draw (1*\stepx,2*\stepy) 
		--(1*\stepx,0*\stepy);
		\draw (2*\stepx,2*\stepy) --(2*\stepx,1*\stepy);\draw 
		(3*\stepx,2*\stepy) --(3*\stepx,1*\stepy);
		
		\draw (.5*\stepx,1.5*\stepy) node {$3$};
		\draw (1.5*\stepx,1.5*\stepy) node {$1$};
		\draw (2.5*\stepx,1.5*\stepy) node {$1$};
		\draw (.5*\stepx,0.5*\stepy) node {$3$};
	\end{tikzpicture}
\end{array}
\
\begin{array}{c}
	\begin{tikzpicture}[scale=.9]
		\def\stepx{.8};
		\def\stepy{.6};
		\draw (0,2*\stepy) -- (3*\stepx,2*\stepy) ;
		\draw (0,0*\stepy) -- (1*\stepx,0*\stepy);
		\draw (0,1*\stepy) -- (3*\stepx,1*\stepy) ;
		\draw (0,2*\stepy) -- (0,0*\stepy); \draw (1*\stepx,2*\stepy) 
		--(1*\stepx,0*\stepy);
		\draw (2*\stepx,2*\stepy) --(2*\stepx,1*\stepy);\draw 
		(3*\stepx,2*\stepy) --(3*\stepx,1*\stepy);
		
		\draw (.5*\stepx,1.5*\stepy) node {$3$};
		\draw (1.5*\stepx,1.5*\stepy) node {$2$};
		\draw (2.5*\stepx,1.5*\stepy) node {$1$};
		\draw (.5*\stepx,0.5*\stepy) node {$3$};
	\end{tikzpicture}
\end{array}
\eeq

These correspond to the states of 
the $U(2|1)$ rep $[3,1]$ as can be seen  via Young symmetrising as described 
above.

\subsubsection*{A second example, $\ulmb=[3,1]$ with $(z_1,z_2|z_3,z_4)$.} 
Again 
we can fill $\ulmb$ 
as in the previous example, with both $3$ and $3\rightarrow 4$. 
Then we have new fillings in which both $z_3$ and $z_4$ appear. These are
\beq
\begin{array}{c}
	\begin{tikzpicture}[scale=.9]
		\def\stepx{.8};
		\def\stepy{.6};
		\draw (0,2*\stepy) -- (3*\stepx,2*\stepy) ;
		\draw (0,0*\stepy) -- (1*\stepx,0*\stepy);
		\draw (0,1*\stepy) -- (3*\stepx,1*\stepy) ;
		\draw (0,2*\stepy) -- (0,0*\stepy); \draw (1*\stepx,2*\stepy) 
		--(1*\stepx,0*\stepy);
		\draw (2*\stepx,2*\stepy) --(2*\stepx,1*\stepy);\draw 
		(3*\stepx,2*\stepy) --(3*\stepx,1*\stepy);
		
		\draw (.5*\stepx,1.5*\stepy) node {$4$};
		\draw (1.5*\stepx,1.5*\stepy) node {$3$};
		\draw (2.5*\stepx,1.5*\stepy) node {$2$};
		\draw (.5*\stepx,0.5*\stepy) node {$3$};
	\end{tikzpicture}
\end{array}
\
\begin{array}{c}
	\begin{tikzpicture}[scale=.9]
		\def\stepx{.8};
		\def\stepy{.6};
		\draw (0,2*\stepy) -- (3*\stepx,2*\stepy) ;
		\draw (0,0*\stepy) -- (1*\stepx,0*\stepy);
		\draw (0,1*\stepy) -- (3*\stepx,1*\stepy) ;
		\draw (0,2*\stepy) -- (0,0*\stepy); \draw (1*\stepx,2*\stepy) 
		--(1*\stepx,0*\stepy);
		\draw (2*\stepx,2*\stepy) --(2*\stepx,1*\stepy);\draw 
		(3*\stepx,2*\stepy) --(3*\stepx,1*\stepy);
		
		\draw (.5*\stepx,1.5*\stepy) node {$4$};
		\draw (1.5*\stepx,1.5*\stepy) node {$3$};
		\draw (2.5*\stepx,1.5*\stepy) node {$1$};
		\draw (.5*\stepx,0.5*\stepy) node {$3$};
	\end{tikzpicture}
\end{array}
\
\begin{array}{c}
	\begin{tikzpicture}[scale=.9]
		\def\stepx{.8};
		\def\stepy{.6};
		\draw (0,2*\stepy) -- (3*\stepx,2*\stepy) ;
		\draw (0,0*\stepy) -- (1*\stepx,0*\stepy);
		\draw (0,1*\stepy) -- (3*\stepx,1*\stepy) ;
		\draw (0,2*\stepy) -- (0,0*\stepy); \draw (1*\stepx,2*\stepy) 
		--(1*\stepx,0*\stepy);
		\draw (2*\stepx,2*\stepy) --(2*\stepx,1*\stepy);\draw 
		(3*\stepx,2*\stepy) --(3*\stepx,1*\stepy);
		
		\draw (.5*\stepx,1.5*\stepy) node {$4$};
		\draw (1.5*\stepx,1.5*\stepy) node {$3$};
		\draw (2.5*\stepx,1.5*\stepy) node {$2$};
		\draw (.5*\stepx,0.5*\stepy) node {$4$};
	\end{tikzpicture}
\end{array}
\
\begin{array}{c}
	\begin{tikzpicture}[scale=.9]
		\def\stepx{.8};
		\def\stepy{.6};
		\draw (0,2*\stepy) -- (3*\stepx,2*\stepy) ;
		\draw (0,0*\stepy) -- (1*\stepx,0*\stepy);
		\draw (0,1*\stepy) -- (3*\stepx,1*\stepy) ;
		\draw (0,2*\stepy) -- (0,0*\stepy); \draw (1*\stepx,2*\stepy) 
		--(1*\stepx,0*\stepy);
		\draw (2*\stepx,2*\stepy) --(2*\stepx,1*\stepy);\draw 
		(3*\stepx,2*\stepy) --(3*\stepx,1*\stepy);
		
		\draw (.5*\stepx,1.5*\stepy) node {$4$};
		\draw (1.5*\stepx,1.5*\stepy) node {$3$};
		\draw (2.5*\stepx,1.5*\stepy) node {$1$};
		\draw (.5*\stepx,0.5*\stepy) node {$4$};
	\end{tikzpicture}
\end{array}
\eeq
and
\beq
\begin{array}{c}
	\begin{tikzpicture}[scale=.9]
		\def\stepx{.8};
		\def\stepy{.6};
		\draw (0,2*\stepy) -- (3*\stepx,2*\stepy) ;
		\draw (0,0*\stepy) -- (1*\stepx,0*\stepy);
		\draw (0,1*\stepy) -- (3*\stepx,1*\stepy) ;
		\draw (0,2*\stepy) -- (0,0*\stepy); \draw (1*\stepx,2*\stepy) 
		--(1*\stepx,0*\stepy);
		\draw (2*\stepx,2*\stepy) --(2*\stepx,1*\stepy);\draw 
		(3*\stepx,2*\stepy) --(3*\stepx,1*\stepy);
		
		\draw (.5*\stepx,1.5*\stepy) node {$4$};
		\draw (1.5*\stepx,1.5*\stepy) node {$1$};
		\draw (2.5*\stepx,1.5*\stepy) node {$1$};
		\draw (.5*\stepx,0.5*\stepy) node {$3$};
	\end{tikzpicture}
\end{array}
\
\begin{array}{c}
	\begin{tikzpicture}[scale=.9]
		\def\stepx{.8};
		\def\stepy{.6};
		\draw (0,2*\stepy) -- (3*\stepx,2*\stepy) ;
		\draw (0,0*\stepy) -- (1*\stepx,0*\stepy);
		\draw (0,1*\stepy) -- (3*\stepx,1*\stepy) ;
		\draw (0,2*\stepy) -- (0,0*\stepy); \draw (1*\stepx,2*\stepy) 
		--(1*\stepx,0*\stepy);
		\draw (2*\stepx,2*\stepy) --(2*\stepx,1*\stepy);\draw 
		(3*\stepx,2*\stepy) --(3*\stepx,1*\stepy);
		
		\draw (.5*\stepx,1.5*\stepy) node {$4$};
		\draw (1.5*\stepx,1.5*\stepy) node {$2$};
		\draw (2.5*\stepx,1.5*\stepy) node {$2$};
		\draw (.5*\stepx,0.5*\stepy) node {$3$};
	\end{tikzpicture}
\end{array}
\
\begin{array}{c}
	\begin{tikzpicture}[scale=.9]
		\def\stepx{.8};
		\def\stepy{.6};
		\draw (0,2*\stepy) -- (3*\stepx,2*\stepy) ;
		\draw (0,0*\stepy) -- (1*\stepx,0*\stepy);
		\draw (0,1*\stepy) -- (3*\stepx,1*\stepy) ;
		\draw (0,2*\stepy) -- (0,0*\stepy); \draw (1*\stepx,2*\stepy) 
		--(1*\stepx,0*\stepy);
		\draw (2*\stepx,2*\stepy) --(2*\stepx,1*\stepy);\draw 
		(3*\stepx,2*\stepy) --(3*\stepx,1*\stepy);
		
		\draw (.5*\stepx,1.5*\stepy) node {$4$};
		\draw (1.5*\stepx,1.5*\stepy) node {$2$};
		\draw (2.5*\stepx,1.5*\stepy) node {$1$};
		\draw (.5*\stepx,0.5*\stepy) node {$3$};
	\end{tikzpicture}
\end{array}
\eeq
\beq
\begin{array}{c}
	\begin{tikzpicture}[scale=.9]
		\def\stepx{.8};
		\def\stepy{.6};
		\draw (0,2*\stepy) -- (3*\stepx,2*\stepy) ;
		\draw (0,0*\stepy) -- (1*\stepx,0*\stepy);
		\draw (0,1*\stepy) -- (3*\stepx,1*\stepy) ;
		\draw (0,2*\stepy) -- (0,0*\stepy); \draw (1*\stepx,2*\stepy) 
		--(1*\stepx,0*\stepy);
		\draw (2*\stepx,2*\stepy) --(2*\stepx,1*\stepy);\draw 
		(3*\stepx,2*\stepy) --(3*\stepx,1*\stepy);
		
		\draw (.5*\stepx,1.5*\stepy) node {$4$};
		\draw (1.5*\stepx,1.5*\stepy) node {$3$};
		\draw (2.5*\stepx,1.5*\stepy) node {$1$};
		\draw (.5*\stepx,0.5*\stepy) node {$1$};
	\end{tikzpicture}
\end{array}
\
\begin{array}{c}
	\begin{tikzpicture}[scale=.9]
		\def\stepx{.8};
		\def\stepy{.6};
		\draw (0,2*\stepy) -- (3*\stepx,2*\stepy) ;
		\draw (0,0*\stepy) -- (1*\stepx,0*\stepy);
		\draw (0,1*\stepy) -- (3*\stepx,1*\stepy) ;
		\draw (0,2*\stepy) -- (0,0*\stepy); \draw (1*\stepx,2*\stepy) 
		--(1*\stepx,0*\stepy);
		\draw (2*\stepx,2*\stepy) --(2*\stepx,1*\stepy);\draw 
		(3*\stepx,2*\stepy) --(3*\stepx,1*\stepy);
		
		\draw (.5*\stepx,1.5*\stepy) node {$4$};
		\draw (1.5*\stepx,1.5*\stepy) node {$3$};
		\draw (2.5*\stepx,1.5*\stepy) node {$2$};
		\draw (.5*\stepx,0.5*\stepy) node {$2$};
	\end{tikzpicture}
\end{array}
\
\begin{array}{c}
	\begin{tikzpicture}[scale=.9]
		\def\stepx{.8};
		\def\stepy{.6};
		\draw (0,2*\stepy) -- (3*\stepx,2*\stepy) ;
		\draw (0,0*\stepy) -- (1*\stepx,0*\stepy);
		\draw (0,1*\stepy) -- (3*\stepx,1*\stepy) ;
		\draw (0,2*\stepy) -- (0,0*\stepy); \draw (1*\stepx,2*\stepy) 
		--(1*\stepx,0*\stepy);
		\draw (2*\stepx,2*\stepy) --(2*\stepx,1*\stepy);\draw 
		(3*\stepx,2*\stepy) --(3*\stepx,1*\stepy);
		
		\draw (.5*\stepx,1.5*\stepy) node {$4$};
		\draw (1.5*\stepx,1.5*\stepy) node {$3$};
		\draw (2.5*\stepx,1.5*\stepy) node {$1$};
		\draw (.5*\stepx,0.5*\stepy) node {$2$};
	\end{tikzpicture}
\end{array}
\
\begin{array}{c}
	\begin{tikzpicture}[scale=.9]
		\def\stepx{.8};
		\def\stepy{.6};
		\draw (0,2*\stepy) -- (3*\stepx,2*\stepy) ;
		\draw (0,0*\stepy) -- (1*\stepx,0*\stepy);
		\draw (0,1*\stepy) -- (3*\stepx,1*\stepy) ;
		\draw (0,2*\stepy) -- (0,0*\stepy); \draw (1*\stepx,2*\stepy) 
		--(1*\stepx,0*\stepy);
		\draw (2*\stepx,2*\stepy) --(2*\stepx,1*\stepy);\draw 
		(3*\stepx,2*\stepy) --(3*\stepx,1*\stepy);
		
		\draw (.5*\stepx,1.5*\stepy) node {$4$};
		\draw (1.5*\stepx,1.5*\stepy) node {$3$};
		\draw (2.5*\stepx,1.5*\stepy) node {$2$};
		\draw (.5*\stepx,0.5*\stepy) node {$1$};
	\end{tikzpicture}
\end{array}
\eeq
for a total of $3+2\times 9+4+7$ fillings. These correspond to the states of 
the $U(2|2)$ rep $[3,1]$ as can be seen  via Young symmetrising.

The variables  ${\bf z}=(x_1,\ldots x_m|y_{m+1},\ldots y_{m+n})$ 
are defined to have a parity $\pi_{i}=0$ if $x_i$ and $\pi_i=1$ if $y_i$.\\

A Young diagram $\ulmb$ endowed with an $(m,n)$ structure
is a Young diagram $\ulmb$ that has to satisfy the condition $\lambda_{m+1}\leq 
n$. 
One can easily check that only then will the diagram allow any supersymmetric 
filling, and  correspondingly only then can it yield a non vanishing $U(m|n)$ rep. 
This condition implies that the Young diagram has at most a \emph{hook} shape, 
with only $m$ arbitrarily long rows and only $n$ arbitrarily long columns. 
The reps split into two cases, 
\begin{itemize}
\item[$\bullet$] \emph{typical} Young diagram, i.e. $\lambda_{m+1}\leq n$ such that it contains 
the rectangle $n^m$. These correspond to long representations of $U(m|n)$.

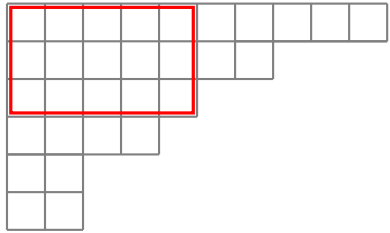
        \begin{equation}\label{typical}
             \begin{tikzpicture}
		\def\xuno{2}
		\def\yuno{2}
	
		\def\ydue{5}
		\def\xtre{7}
		
		\def\ytre{4.5}
		\def\xquattro{5}
		
		\def\yquattro{3.5}
		\def\xcinque{4}
		
		\def\ycinque{3}
		\def\xsei{3}
		
		\def\step{.5}

		\draw[thick,gray] (\xuno,	\yuno) -- 			(\xuno,	\ydue);
		
		\draw[thick,gray] (\xuno,	\ydue) -- 			(\xtre,	\ydue);
		
		\foreach \x in {1, ..., 10}
		\draw[thick,gray] (\xuno+\x*.5,	\ydue) -- 			(\xuno+\x*.5,	\ydue-\step);
		
		\draw[thick,gray] (\xuno,	\ydue-\step) -- 			(\xtre,	\ydue-\step);
		
		\draw[thick,gray] (\xquattro,	\ydue-\step) -- 			(\xtre,	\ydue-\step);
		
		\foreach \x in {1, ..., 5}
		\draw[thick,gray] (\xuno+\x*.5,	\ytre) -- 			(\xuno+\x*.5,	\ytre-2*\step);
		
		\foreach \x in {6, ..., 7}
		\draw[thick,gray] (\xuno+\x*.5,	\ytre) -- 			(\xuno+\x*.5,	\ytre-1*\step);
		
		\draw[thick,gray] (\xuno,	\ytre-\step) -- 			(\xquattro+0.5,		\ytre-\step);
		\draw[thick,gray] (\xuno,	\ytre-2*\step) -- 			(\xquattro-.5,	\ytre-2*\step);

		\foreach \x in {1, ..., 4}
		\draw[thick,gray] (\xuno+\x*.5,	\yquattro) -- 			(\xuno+\x*.5,	\yquattro-\step);
		
		\draw[thick,gray] (\xuno,	\yquattro-\step) -- 			(\xcinque,	\yquattro-\step);

		\draw[thick,gray] (\xuno,	\ycinque) -- 			(\xsei,	\ycinque);
		
		\foreach \x in {1, ..., 2}
		\draw[thick,gray] (\xuno+\x*.5,	\ycinque) -- 			(\xuno+\x*.5,	\ycinque-2*\step);
		
		\draw[thick,gray] (\xuno,	\ycinque-\step) -- 			(\xsei,	\ycinque-\step);
		\draw[thick,gray] (\xuno,	\ycinque-2*\step) -- 			(\xsei,	\ycinque-2*\step);

		\draw (\xtre+3,	\ytre) 	node 	{ ${\color{red} (m,n)=(3,5)}$};
		\draw (\xtre+4,	\yquattro+.25) 	node 	{ $\ulmb_{}=[10,7,5,4,2,2]$};

		\draw[very thick, red]  (\xuno+.05,\ydue-.05) rectangle 
		(\xcinque+.5-.05,\yquattro+.05);
			
	   \end{tikzpicture}
	\end{equation}

\item[$\bullet$] \emph{atypical} Young diagram, i.e. $\lambda_{m+1}\leq n$ such that it \emph{does not}
contain $n^m$. These correspond to short representations of $U(m|n)$.
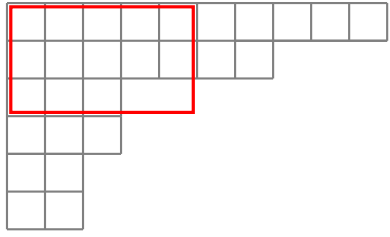
        \begin{equation}\label{atypical}
             \begin{tikzpicture}
		\def\xuno{2}
		\def\yuno{2}
	
		\def\ydue{5}
		\def\xtre{7}
		
		\def\ytre{4.5}
		\def\xquattro{5}
		
		\def\yquattro{3.5}
		\def\xcinque{4}
		
		\def\ycinque{3}
		\def\xsei{3}
		
		\def\step{.5}

		\draw[thick,gray] (\xuno,	\yuno) -- 			(\xuno,	\ydue);
		
		\draw[thick,gray] (\xuno,	\ydue) -- 			(\xtre,	\ydue);
		
		\foreach \x in {1, ..., 10}
		\draw[thick,gray] (\xuno+\x*.5,	\ydue) -- 			(\xuno+\x*.5,	\ydue-\step);
		
		\draw[thick,gray] (\xuno,	\ydue-\step) -- 			(\xtre,	\ydue-\step);
		
		\draw[thick,gray] (\xquattro,	\ydue-\step) -- 			(\xtre,	\ydue-\step);
		
		\foreach \x in {1, ..., 3}
		\draw[thick,gray] (\xuno+\x*.5,	\ytre) -- 			(\xuno+\x*.5,	\ytre-2*\step);
		
		\foreach \x in {4, ..., 7}
		\draw[thick,gray] (\xuno+\x*.5,	\ytre) -- 			(\xuno+\x*.5,	\ytre-1*\step);
		
		\draw[thick,gray] (\xuno,	\ytre-\step) -- 			(\xquattro+0.5,		\ytre-\step);

		\foreach \x in {1, ..., 3}
		\draw[thick,gray] (\xuno+\x*.5,	\yquattro) -- 			(\xuno+\x*.5,	\yquattro-\step);
		
		\draw[thick,gray] (\xuno,	\yquattro-\step) -- 			(\xcinque-.5,	\yquattro-\step);

		\draw[thick,gray] (\xuno,	\ycinque) -- 			(\xsei,	\ycinque);
		
		\foreach \x in {1, ..., 2}
		\draw[thick,gray] (\xuno+\x*.5,	\ycinque) -- 			(\xuno+\x*.5,	\ycinque-2*\step);
		
		\draw[thick,gray] (\xuno,	\ycinque-\step) -- 			(\xsei,	\ycinque-\step);
		\draw[thick,gray] (\xuno,	\ycinque-2*\step) -- 			(\xsei,	
		\ycinque-2*\step);
		\draw[thick,gray] (\xuno,	\ycinque+.5) -- 			(\xsei+.5,	
		\ycinque+.5);

		\draw (\xtre+3,	\ytre) 	node 	{ \color{red} $(m,n)=(3,5)$};
		\draw (\xtre+4,	\yquattro+.25) 	node 	{ $\ulmb_{\color{red}}=[10,7,3,3,2,2]$};

		\draw[very thick, red]  (\xuno+.05,\ydue-.05) rectangle 
		(\xcinque+.5-.05,\yquattro+.05);
			
	   \end{tikzpicture}
	\end{equation}

\end{itemize}
Typical representations are such that the box with coordinates $(m,n)$ lies in 
the diagram.
The simplest example of an atypical representation is the empty diagram.

In both cases it can be useful to define two sub Young diagrams, by essentially 
cutting open the diagram vertically after the $n$th column. We then define the 
Young diagram obtained from the first $n$ columns and transposing as $\ulmb_s$ 
and the remaining diagram as $\ulmb_e$. So in the above atypical example:
    
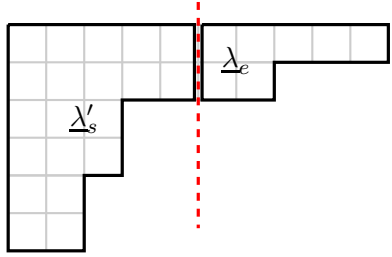
\begin{align}
	\begin{tikzpicture}
		\def\xuno{2}
		\def\yuno{2}
		\def\ydue{5}
		\def\xtre{7}
		\def\ytre{4.5}
		\def\xquattro{5}
		\def\yquattro{3.5}
		\def\xcinque{4}
		\def\ycinque{3}
		\def\xsei{3}
		\def\step{.5}
		\draw[thick,black!20] (\xuno,	\yuno) -- 			(\xuno,	\ydue);
		\draw[thick,black!20] (\xuno,	\ydue) -- 			(\xtre,	\ydue);
		\foreach \x in {1, ..., 10}
		\draw[thick,black!20] (\xuno+\x*.5,	\ydue) -- 			(\xuno+\x*.5,	
		\ydue-\step);
		\draw[thick,black!20] (\xuno,	\ydue-\step) -- 			(\xtre,	
		\ydue-\step);
		\draw[thick,black!20] (\xquattro,	\ydue-\step) -- 			(\xtre,	
		\ydue-\step);
		\foreach \x in {1, ..., 3}
		\draw[thick,black!20] (\xuno+\x*.5,	\ytre) -- 			(\xuno+\x*.5,	
		\ytre-2*\step);
		\foreach \x in {4, ..., 7}
		\draw[thick,black!20] (\xuno+\x*.5,	\ytre) -- 			(\xuno+\x*.5,	
		\ytre-1*\step);
		\draw[thick,black!20] (\xuno,	\ytre-\step) -- 			
		(\xquattro+0.5,		\ytre-\step);
		\foreach \x in {1, ..., 3}
		\draw[thick,black!20] (\xuno+\x*.5,	\yquattro) -- 			
		(\xuno+\x*.5,	
		\yquattro-\step);
		\draw[thick,black!20] (\xuno,	\yquattro-\step) -- 			
		(\xcinque-.5,	\yquattro-\step);
		\draw[thick,black!20] (\xuno,	\ycinque) -- 			(\xsei,	
		\ycinque);
		\foreach \x in {1, ..., 2}
		\draw[thick,black!20] (\xuno+\x*.5,	\ycinque) -- 			
		(\xuno+\x*.5,	
		\ycinque-2*\step);
		\draw[thick,black!20] (\xuno,	\ycinque-\step) -- 			(\xsei,	
		\ycinque-\step);
		\draw[thick,black!20] (\xuno,	\ycinque-2*\step) -- 			(\xsei,	
		\ycinque-2*\step);
		\draw[thick,black!20] (\xuno,	\ycinque+.5) -- 			(\xsei+.5,	
		\ycinque+.5);
		\draw (\xtre+4,	\ytre+.25) 	node 	{ $(m,n)=(3,5)$};
		\draw (\xtre+4,	\yquattro+.25) 	node 	{ 
			$\ulmb_{\color{red}}=[10,7,3,3,2,2]$};
		\draw (\xtre+4,	\yquattro-.5) 	node 	{ 
			$\ulmb_{s}=[6,6,4,2,2]$};
		\draw (\xtre+4,	\yquattro-1.25) 	node 	{ 
			$\ulmb_{e}=[5,2]$};
		\draw (\xtre-4,	\yquattro+.25) 	node 	{ 
			$\ulmb'_{s}$};
		\draw (\xtre-2,	\yquattro+.9) 	node 	{ 
			$\ulmb_{e}$};
		\draw[very thick, black]  (\xuno,\ydue) -- 
		(\xcinque+.5-.05,\ydue) -- (\xcinque+.5-.05,\yquattro+.5)-- 
		(\xcinque-.5,\yquattro+.5)-- (\xcinque-.5,\yquattro-.5)-- 
		(\xcinque-1,\yquattro-.5)-- 
		(\xcinque-1,\yquattro-1.5)-- 
		(\xcinque-2,\yquattro-1.5)-- (\xuno,\ydue);
		\draw[very thick, black]	(\xcinque+.5+.05,\ydue) -- 
		(\xcinque+.5+.05,\yquattro+.5)-- (\xcinque+1.5,\yquattro+.5)-- 
		(\xcinque+1.5,\yquattro+1)-- (\xcinque+3,\yquattro+1)-- 
		(\xcinque+3,\ydue)--	(\xcinque+.5+.05,\ydue)
		;
		\draw[very thick, dashed,red]	(\xcinque+.5,\ydue+.3) -- 	
		(\xcinque+.5,\ydue-2.7);
	\end{tikzpicture}
\label{southeast}
\end{align}
These sub Young tableaux have a direct interpretation in terms of the corresponding $U(m|n)$ rep. 
They are simply the corresponding representations of the $U(n)$ and $U(m)$ 
subgroups of the highest 
weight state. Or put another way, the highest term in the supersymmetric filling 
will have only $y$ entries in $\ulmb_s$ and only $x$ entries in $\ulmb_e$.

\subsection{Super Jack polynomials}
\label{intro_sjack_sec}

Having outlined the general structure of supersymmetric polynomials let us now 
specify to the main interest, the super Jack polynomials. We just need to 
define $\Psi$ and $f$ in~\eqref{combinat_form_poly_susy}. 
To define $\Psi$ we  split the superfilling $\cal T$ of $\ulmb$ into ${\cal T}_1$, the 
part containing $m+1,..,m+n$, and the rest ${\cal T}_0={\cal T}/{\cal T}_1$. Say that $\umu$ is the shape of ${\cal T}_1$.
Then 
$\Psi({\cal 
	T};\theta)$ is defined in terms of the Jack $\Psi$ as~\cite{Veselov_1}
\begin{align}
	\Psi_{\cal T}(\theta) & = (-)^{|\umu|} \Pi_{\umu'}(\tfrac{1}{\theta} )
	\Psi_{{\cal T}'_1}(\tfrac{1}{\theta}) 	\Psi_{{\cal T}_0}(\theta) \\
	{f}_{} (z;\theta)&=z\ .
\end{align}

According to this definition the superJack is given as a sum over all 
decompositions of 
the form 
 \beq
  P_{\ulmb}({\bf z};\theta) = \!\! 
  \sum_{
  \umu\subseteq\ulmb } (-)^{|\umu|} 
  Q_{\umu'}(y_1,\ldots y_n,\tfrac{1}{\theta})\,  P_{ \ulmb / \umu } (x_1,\ldots 
  x_{m};\theta)\ .
  \label{theoremVeselov}
 \eeq
where the sum can also be restricted to $\umu$ such that ${\rm max}(\lambda'_j-m,0) \leq \mu'_{j}\leq \lambda'_j $ with $j=1,\ldots n$.

Note also that we could construct the superJacks directly from Jack 
polynomials via their 
decomposition into $U(m)$ and $U(n)$ reps following \cite{Veselov_1}  (see 
also \cite{1802.02016} 
for a nice review). Quite nicely \cite{Veselov_1}  showed that this decomposition can be brought to \eqref{theoremVeselov}. 
From this point of view it might be useful to compare this with the bosonic decomposition~\eqref{property}. 

\emph{Remark.} A super Jack polynomial is denoted by $P_{\ulmb}^{(m,n)}({\bf z} 
;\theta)$. 
However, the $(m,n)$ dependence can be read off the variables ${\bf z}$, when
there are no ambiguities in the notation. 
When this is the case, we will simply use  $P_{\ulmb}^{}({\bf z} ;\theta)$.\\

Some simple observations which can be seen directly from this formula: \\

\begin{minipage}{14.5cm}
{\bf 1)}~If $\ulmb$ fails to have an $(m,n)$ structure, so  $(i,j)=(m{+}1,n{+}1)$ is in the Young diagram $\ulmb$,  then the superJack 
vanishes. For any 
$\umu$ in the summation, either the box  $(m+1,n+1)$ is in $\umu$, in which case   $Q_{\mu'}=0$ as $\umu'$ will have $n+1$ rows, or
$(m+1,n+1)$ is in ${\ulmb/ \umu}$ in which case  $P_{\ulmb/ \umu}=0$ as $\ulmb/ \umu$ will have a full column with $m+1$ elements. \\

{\bf 2)}~When $m=0$ the sum localises on $\umu=\ulmb$, and since 
$P_{\ulmb/\ulmb}=P_{\varnothing}=1$,
 the superJack polynomial reduces to a (normalised) dual Jack polynomial $(-)^{|\ulmb'|}Q_{\ulmb'}( {\bf y};\tfrac{1}{\theta})$. 
When $n=0$ the sum localises on $\umu=\varnothing$ and the superJack polynomials reduces to $P_{\ulmb}({\bf x};\theta)$.
\end{minipage}


\subsection{Properties of Super Jack polynomials}


\subsubsection*{Stability}

The combinatorial formula makes stability of superJacks manifest. 
Alternatively, from the fact that the eigenvalue of the $A$-type CMS 
differential depends only 
on the Young diagram, and the uniqueness of the polynomial solution for given Young diagram,  
we also infer that the super Jack polynomials are stable.

\subsubsection*{The $m$ and $n$ switch}
From the bosonisation of the eigenvalue and a property of the differential 
operator ${\bf H}$, namely
\beq
h_{\ulmb}^{(\theta)}=-\theta\sum_{j=1} \lambda'_j ( \lambda'_j-1-\tfrac{2}{\theta} (j-1) )=-\theta h_{\ulmb'}^{(\frac{1}{\theta})}
\qquad;\qquad {\bf H}^{(\frac{1}{\theta})}({\bf y}|{\bf x})= -\tfrac{1}{\theta} {\bf H}^{({\theta})}({\bf x}|{\bf y})
\eeq
we deduce that $P_{\ulmb'}({\bf y}|{\bf x}; \frac{1}{\theta})$ has to be 
proportional to $P_{\ulmb}({\bf x}|{\bf y};{\theta})$.  
Looking at the precise normalisations we arrive at
\beq
\label{sJduality}
P^{(m,n)}_{\lambda}({\bf x}|{\bf y};\theta)=(-1)^{|\ulmb'|} \Pi_{\ulmb'}(\tfrac{1}{\theta})P^{(n,m)}_{\ulmb'}({\bf y}|{\bf x}; \tfrac{1}{\theta})
\eeq

We can also show this more directly from either the combinatorial formula or 
the decomposition formula (where we further decompose the skew Jacks into 
Jacks via structure constants, and use some known properties of the structure 
constants).

\subsubsection*{Shift transformation of superJacks}
\label{prop_undershift_app}

We have emphasised that blocks should be invariant under the shift symmetry~\eqref{deform_overview_sec}. In appendix~\ref{sec_shift} we will show that the coefficients $T_\gamma$ in the expansion of blocks over Jacks possesses this symmetry, but we thus also require the symmetry for the Jacks themselves  in order that the blocks have this symmetry. We consider this here.

Let $\ulmb$ be a typical $(m,n)$ representation,  i.e.~$\lambda_m \geq n,\lambda_{m+1}\leq n$. 
Then we consider: 1) the Young diagram obtained by a horizontal shift of $\theta \tau'$ to
 the first $m$ rows of $\ulmb$ (on the east),  and 2) the Young diagram obtained by a vertical  
 shift of $\tau'$ to the first $n$ columns of $\ulmb'$ (on the south).  We will show that
\beq\label{property_sjack_shift}
P^{(m,n)}_{\ulmb+(\theta \tau')^m}({\bf x}|{\bf y};\theta)\qquad;\qquad     \left(\frac{\prod_i x_i^\theta}{\prod_j y_j}\right)^{\!\!\tau'} P^{(m,n)}_{(\ulmb'+\tau'^n)'}({\bf x}|{\bf y};\tfrac{1}{\theta})
\eeq
are proportional to each other, then we will find the proportionality factor. 

For the bosonic $n=0$ Jack polynomials, the above claim follows from \eqref{bosonic_galilean} 
with $\tau=\theta\tau'$.  We will thus use the same argument, and show that both the LHS~and r.h.s 
in \eqref{property_sjack_shift} are eigenfunctions of ${\bf H}$ with both $m$ and $n$ turned on.

So let us see what happens when we apply ${\bf H}$ on 
\eqref{property_sjack_shift}. 
On the LHS~we simply find the eigenvalue $h_{\ulmb+(\tau)^m}^{(\theta)}$. On 
the RHS~we use
\beq
  \left(\frac{\prod_i x_i^\theta}{\prod_j y_j}\right)^{\!\!-\tau'}\!\!\!\!\cdot{\bf H}\cdot   \left(\frac{\prod_i x_i^\theta}{\prod_j y_j}\right)^{\!\!\tau'} =
  {\bf H}+2\theta\tau' \sum_I z_I \partial_I + \tau'(m\theta-n)(n-1-\theta(m-1)+\tau'\theta)
\eeq
Note also that the constant term can be written in terms of the same eigenvalue of ${\bf H}$, 
\beq
\tau'(m\theta-n)(n-1-\theta(m-1)+\tau'\theta)= +2\theta\tau' |m^n| + h_{(\theta\tau')^m}^{(\theta)}-\theta h_{(-\tau')^n}^{(\frac{1}{\theta})}
\eeq
(In the bosonic case, $n=0$, this was $h^{(\theta)}_{\tau^m}$). 
Then, the action of ${\bf H}$ on \eqref{property_sjack_shift} matches 
because of the identity
\beq
h^{(\theta)}_{\ulmb+(\theta \tau')^m}=+h_{(\theta\tau')^m}^{(\theta)}-\theta h_{\ulmb'+\tau'^n}^{(\frac{1}{\theta})}-\theta h_{(-\tau')^n}^{(\frac{1}{\theta})}
+ 2\theta\tau'\Big( |m^n|+|\lambda'| + |(\tau')^n| \Big)\ .
\eeq
Note that this is true \emph{only} if $\lambda_m \geq n,\lambda_{m+1}\leq n$ ie 
only for typical, long representations. For example it is clearly false 
for $\ulmb=[\varnothing]$, which would read``$0=2\theta\tau' 
|m^n|$".\footnote{We can also rewrite the above equality as
\beq
\frac{ h^{(\theta)}_{\ulmb+(\theta \tau')^m}-h_{(\theta\tau')^m}^{(\theta)}}{2\theta\tau'} +\frac{ h_{\ulmb'+\tau'^n}^{(\frac{1}{\theta})}+ h_{(-\tau')^n}^{(\frac{1}{\theta})}}{2\tau'}=
|m^n|+|\lambda'|+|(\tau')^n|
\eeq
In the limit $\tau'\rightarrow 0$ the RHS~is simply $\sum_{i=1}^m \lambda_i + \sum_{j=1}^n \lambda'_j$. }

Having established that the polynomials in \eqref{property_sjack_shift} are proportional 
to each other, we only need to fix the proportionality. 
By considering the first term in the decomposition 
formula~\eqref{theoremVeselov} the only change to 
take into account is 
the $\Pi$ function inside $Q$ for the lowest representation. Following this through we arrive at the 
shift formula~\eqref{shiftsJack} 
\beq
 \Pi^{}_{(\ulmb_s-m^n)'}({\theta}) P^{(m,n)}_{\ulmb+(\theta \tau')^m}({\bf x}|{\bf y};\theta) =
 \Pi^{}_{(\ulmb_s-m^n+\tau'^n)'}({\theta}) \left(\frac{\prod_i x_i^\theta}{\prod_j y_j}\right)^{\!\!\tau'} P^{(m,n)}_{(\ulmb'+\tau'^n)'}({\bf x}|{\bf y};{\theta})
\eeq
where we are using $\Pi_{\unu}^{-1}(\frac{1}{\theta})=\Pi_{\nu'}(\theta)$, and we introduced
\beq
\ulmb_s=[\lambda'_1,\ldots \lambda'_n]
\eeq
as in~\eqref{southeast}.

\subsubsection*{ Super Jacks and the superconformal Ward identity}

The uplift of Jack to superJack polynomials follows from a 
{characterisation theorem}, cf. Theorem 2 of \cite{Veselov_1}.
In particular, a superJack polynomial $P_{\ulmb}({\bf x}|{\bf y};\theta)$ 
is generated by the map $\varphi_{(m,n)}$ which acts on the power sum decomposition of $P_{\ulmb}({\bf z};\theta)$ as 
\bea
\varphi_{(m,n)}\left(  p_r\equiv \sum_{j} z^r \right) = \sum_{j=1}^m x_j^r 
-\tfrac{1}{\theta} 
\sum_{j=1}^n y_j^r\quad\rightarrow\quad P_{\ulmb,(m,n)}(;\theta) = 
\varphi_{(m,n)}( P_{\ulmb}({\bf z};\theta))
\label{theoremVeselov1}
\eea

Note that this map is very easy to understand  in the cases $\theta=1,2,\frac{1}{2}$ 
where there is a group theoretic interpretation. For example, when $\theta=1$ 
we consider the symmetric polynomials as functions of the $n\times n$ matrix 
$Z$, invariant under conjugation $f(Z)=f(G^{-1}Z G)$, with $z_i$ the 
eigenvalues of $Z$. Then the supersymmetric case just corresponds to the case 
where $Z$ is a $(m|n)\times (m|n)$ supermatrix and the map $\varphi$ is just 
taking the supertrace (see~\eqref{th1ev} and the following discussion). So for 
example the power sums
\begin{align}
	p_r=\text{tr}(Z^r)=\sum_j z^r \qquad \rightarrow \qquad 
	\text{str}(Z^r)=\sum_{i=1}^m 
	x_i^r -	\sum_{j=1}^n y_j^r\ .
\end{align}
A similar discussion  also follows in the cases $\theta=2$ and $\theta=\frac{1}{2}$ 
(which is dual under the $m,n$ swap) with the only real difference as far as 
power sums is concerned being that  there are 
repeated eigenvalues of $Z$ in these cases, which accounts for the factors of 
$\theta$ appearing. (See~\eqref{th2ev} and the following for the $\theta=2$ 
case 
and~\eqref{ev2} for the $\theta=\frac{1}{2}$ case.)

The characterisation \eqref{theoremVeselov1} implies that superJack polynomials satisfy the condition
\begin{align}\label{Ward_id}
\left[ \left(\frac{\partial}{\partial{x_i}} + \theta\frac{\partial}{\partial {y_i} }\right) P({\bf z};\theta) \right]_{x_i=y_i}=0
\end{align}
From the point of view of the four point functions invariant under the 
superconformal group, this condition is a consequence of the super-conformal 
Ward identity. 
We understand in this way that our construction of the superconformal blocks, as given by the series over super Jack polynomials,
automatically satisfy the super-conformal Ward identity. 
Our approach here is alternative to various other approaches which instead use the 
superconformal Ward identity to fix the superblock, given an ansatz of $m\otimes n$ bosonic blocks. We will consider this approach in appendix~\ref{section_mn_deco}.

\subsubsection*{Structure constants and decomposition formulae for super Jacks}

The super Jacks (from stability) have exactly the same structure constants as the Jacks. So~\eqref{structure} is true for superJacks  
\begin{align}
	P_\ulmb P_{\umu} = \cC_{\ulmb \umu}^\unu(\theta) P_{\unu}\ .
\end{align}
and furthermore so is the decomposition formula for decomposing higher  dimensional super Jacks into sums of products of lower dimensional super Jacks
\begin{align}\label{superdecomp}
	P^{(m+m'|n+n')}_{\ulmb}= \sum_{\umu,\unu} P_{\umu}^{(m|n)} \cS_\ulmb^{\umu \unu}(\theta) P_{\unu}^{(m'|n')}\ .
\end{align}
Here the arguments of the superJack on the LHS  are split between the two superJacks on the RHS e.g. $ P_{\umu}^{(m|n)}(x_1,..,x_m|y_1,..,y_n)$ and $ P_{\unu}^{(m'|n')}(x_{m+1},..,x_{m+m'}|y_{n+1},..,y_{n+n'})$. 
The coefficients  $\cS_\ulmb^{\umu \unu}(\theta)$ are completely independent of $m,m',n,n'$ and are related to the structure constants just as before~\eqref{from_C_to_S}
\beq
\mathcal{S}^{\umu\unu}_{\ulmb}(\theta)=
\frac{ \Pi_{\umu}(\theta) \Pi_{\unu}(\theta) }{   \Pi_{\lmb}(\theta)} \mathcal{C}^{\ulmb}_{\umu \unu}(\theta)\ .
\eeq

Indeed the definition of superJacks itself in terms of Jacks~\eqref{theoremVeselov} is just an example of this decomposition with $n=m'=0$ (after using~\eqref{sJduality} with $n=0$).

Note that in the $\theta=1$ case this corresponds to the decomposition of $U(m+m'|n+n') \rightarrow  U(m|n)\otimes U(m'|n')$ for which the coefficients are the Littlewood Richardson coefficients.

\subsection{Super interpolation polynomials} \label{rewriting_Veselov}

The last polynomials we review are the super interpolation polynomials which we first discuss below~\eqref{tildepstar}. 
These were first introduced in~\cite{Veselov_3}, 
and we will repeat below the definition given in \cite{Veselov_3} following our conventions. 

Introduce the bosonic polynomial,
\begin{align}
\hat{I}^{(m)}_{\ulmb}(x_1,\ldots x_m;\theta,h)&=P^{ip}_{\ulmb}(\ldots,x_i -\theta i + h,\ldots; \theta, h-\theta m)\\
&=P_{\ulmb}^*({\bf x};\theta,h-\theta m)
\end{align}
where the corresponding function $f$ follows from \eqref{BCf}, i.e.~
\beq
{f}_{}(x_l,i,j,l;\theta,h)= (x_{ l}-\theta l +h)^2  - ( (j-1)-\theta(i-1) + h -\theta {\cal T}(i,j) )^2
\eeq
where ${\cal T}(i,j)=1+m-l$.
In the main text, see \eqref{tildepstar}, we wrote
\beq
{\tilde P}_{\ulmb}^{*(m) }({\bf x};\theta,h)\equiv \hat{I}^{(m)}_{\ulmb}({\bf x};\theta,h)=P^{*(m)}_{\ulmb}({\bf x};\theta,h-\theta m)\,.
\eeq

The supersymmetrised version of ${\hat I}$ has the same structure as \eqref{theoremVeselov}, simply reflecting the underlying sum over superfillings, and it is simply
\beq
\!\!\!\hat{I}_{\ulmb}^{(m,n)}({\bf x}|{\bf y};\theta,h)=\sum_{\mu\subseteq\ulmb} 
(-)^{|\umu|}\left[ (\theta^2)^{|\umu'|}\Pi_{\mu'}(\tfrac{1}{\theta})\hat{I}_{\umu'}(y_1,\ldots y_n; \tfrac{1}{\theta}, \tfrac{1}{2}+\tfrac{1}{2\theta}-\tfrac{h}{\theta} + m) \right]  \hat{I}_{\ulmb/\umu}(x_1,\ldots x_m;\theta, h)
\eeq
In particular, we find
$
\hat{I}^{(0,n)}_{\ulmb}(|{\bf y};\theta,h)= (-\theta^2)^{|\ulmb|} \Pi_{\ulmb'}(\frac{1}{\theta}) {\tilde P}^{*(n)}_{\ulmb'}(\ldots y_n;\frac{1}{\theta} ,\tfrac{1}{2}+\tfrac{1}{2\theta}-\tfrac{h}{\theta} )
$, and for $n=0$ we obvious recover $\hat{I}_{\ulmb}^{(m)}$.
Note that if we take the scaling limit $z\rightarrow \epsilon z$ and we look at the leading contribution as $\epsilon\rightarrow \infty$, that polynomial is a superJack, i.e.
\beq
{\it lead.\, contr.}\Big[\hat{I}_{\ulmb}^{(m,n)}({\bf x}|{\bf y};\theta,h)\Big]= P^{(m,n)}_{\ulmb}( \ldots ,x^2_m| \ldots, \theta^2 y^2_n;\theta)
\eeq

Following Veselov and Sergeev, we now define
\beq
{I}^{(m,n)}_{\ulmb}({\bf x}|{\bf y};\theta, h)=(-\theta^2)^{|\ulmb|} \Pi_{\ulmb'}(\tfrac{1}{\theta}) \hat{I}^{(n,m)}_{\ulmb'}({\bf y}|{\bf x}; \tfrac{1}{\theta}, \tfrac{1}{2}+\tfrac{1}{2\theta}-\tfrac{h}{\theta}) 
\eeq
This object is such that
\begin{align}
I^{(m,0)}_{\ulmb}={\tilde P}^{*(m)}_{\ulmb}(\ldots x_m;\theta,h) \qquad;\qquad \
{I}^{(m,n)}_{\ulmb}(\umu_e|\umu_s;\theta,h)=0\qquad{\rm if}\qquad \umu\subset\ulmb
\end{align}
where $\umu_s=[\mu'_1,.., \mu'_{{N}}]$ and $\umu_e\equiv\umu/\umu'_s=[(\mu_1 {-} {N})_+,..,(\mu_{{M}} {-} {N})_+]$ where $(x)_+\equiv\max(x,0)$. 
Note also that the leading contribution of $I$ now gives a superJack polynomial through the relation in \eqref{sJduality}. 
In the main text, see \eqref{our_veselov}, we wrote
\beq
\tilde P^{(m,n)}_{\ulmb}({\bf z};\theta,h) \equiv I^{(m,n)}_{\ulmb}({\bf z};\theta h)
\eeq
with $I$ being the supersymmetrisation we were looking for. Finally
\beq
{I}^{(m,n)}_{\ulmb}(\ulmb_e|\ulmb_s;\theta,h)=\prod_{(i,j)\in\ulmb} (1+\lambda_i-j+\theta(\lambda'_j-i))(2h-1+\lambda_i+j-\theta(\lambda'_j+i))
\eeq
The r.h.s.~of course does not depend on $(m,n)$ and it becomes \eqref{evaluationBCinterp} upon matching the parameters.

\section{More properties of analytically continued superconformal blocks}
\label{sec_shift}

\subsection{Shift symmetry of the supersymmetric form of the recursion} \label{sec_shift_analitic}

We would like the supersymmetric recursion~\eqref{super_recursion2} to be
invariant under the supersymmetric shift:
\beq\label{deform_preview}
\begin{array}{c} \lambda_i \rightarrow \lambda_i-\theta\tau'   \\[.2cm] \mu_i \rightarrow \mu_i-\theta\tau'  \\[.2cm] \!i=1,\ldots m\end{array} \qquad;\qquad 
\begin{array}{c} \lambda'_j \rightarrow \lambda'_j+\tau' \\[.2cm] \mu'_j \rightarrow \mu'_j+\tau'  \\[.2cm] \!j=1,\ldots n\end{array}  \qquad;\qquad            
\gamma \rightarrow \gamma+2\tau'\ .
\eeq
This is {\em almost} the case for~\eqref{super_recursion2}. The issue is to do with  normalisation  and we will fix it below. Let us emphasise first that
this shift  symmetry arises directly from the group theory  in all cases with a group theory interpretation, 
$\theta=\frac{1}{2},1,2$ (see~\eqref{shift}) but we expect it for any $\theta$. Note that in writing \eqref{deform_preview}, 
we are implicitly assuming that the Young diagram, prior to analytic continuation, contains the box with coordinates $(m,n)$.   
Thus in a supersymmetric theory we are looking at typical or long representations.

We claim that the following normalisation of $T_\gamma$ gives a  shift invariant  result: 
\beq\label{def_Tlong_def}
(T^{\tt long}_{\gamma})^{[\umu_s;\,\umu_e]}_{[\ulmb_s;\,\ulmb_e]} =  \frac{ \Pi_{\umu_s-m^n}(\tfrac{1}{\theta}) }{\Pi_{\ulmb_s-m^n}(\tfrac{1}{\theta}) }(T_{\gamma})_{\ulmb}^{\umu}
\qquad;\qquad
\begin{array}{ccc}  
	\ulmb_s=[\lambda'_1,\ldots \lambda'_n] & ; & \ulmb_e=[\lambda_1,\ldots \lambda_m]\\[.2cm]
	\umu_s=[\mu'_1,\ldots \mu'_n] & ; & \umu_e=[\mu_1,\ldots \mu_m]
\end{array}
\eeq
where we split the diagrams into east and south components by taking the corresponding first $m$ rows and first $n$ columns. 
We might consider $\ulmb_e\rightarrow \ulmb_e-n^m$, however this is not important here.

So the claim is that $T^{\tt long}_{\gamma}$ is  invariant under the shift~\eqref{deform_preview}. In particular, it solves the following recursion
\begin{mdframed}
	\begin{align}
		&
		\!\!\!\Big({h}_{\umu}{-}{h}_{\ulmb} {+} \theta \gamma\, (|\umu|{-}|\ulmb| )\Big)  (T^{\tt long}_{\gamma})^{[\umu_s;\,\umu_e]}_{[\ulmb_s;\,\ulmb_e]} 
		=  \notag\\[.2cm]
		&
		\rule{1cm}{0pt}
		\sum_{i=1}^{m } \left(\mu_i{-}1{-} \theta(i{-}1{-}\alpha ) \right)\left(\mu_i{-}1{-} \theta(i{-}1{-}\beta) \right) 
		\mathbf{f}^{(i)}_{\umu_e{-}\Box_i}(\theta) (T^{\tt long}_{\gamma})^{[\umu_s;\,\umu_e-\Box_i]}_{[\ulmb_s;\,\ulmb_e]} {+}%
		\notag
		\\
		& 
		\rule{.5cm}{0pt}
		{+}\theta\sum_{j=1}^{n }\left(\mu'_j {-}1{-}\alpha{-} \tfrac{j{-}1}{\theta} \right)\left(\mu'_j {-}1{-}\beta{-}\tfrac{j{-}1}{\theta}\right) 
		\mathbf{f}^{(j)}_{\umu_s{-}\Box_j}(\tfrac1\theta)  {\bf c}^{(j)}_{\umu-\Box_j}(\tfrac{1}{\theta}) 
		(T^{\tt long}_{\gamma})^{[\umu_s-\Box_j;\,\umu_e]}_{[\ulmb_s;\,\ulmb_e]}%
		\label{super_recursion_analitic} 
	\end{align}
	where
	\beq
	\ \ {\bf c}^{(j)}_{\umu}(\tfrac{1}{\theta})=\prod_{i=1}^m  \frac{ 
		\left(\mu_i-j+1+\theta{(\mu'_j-i)}{} \right)\left(\mu_i-j+\theta{(\mu'_j-i+2)}{}\right) }{ 
		\left(\mu_i-j+1+\theta{(\mu'_j-i+1)}{}\right)\left(\mu_i-j+\theta{(\mu'_j-i+1)}{}\right)}
	\eeq
\end{mdframed}
Note that ${\bf c}^{(j)}_{\umu}(\theta)$ is built out of arm and leg length symbols, $
c_{\unu}(i,j)=\nu_i-j+(\nu'_j-i+1)/\theta$ and $c'_{\unu}(i,j)=\nu_i-j+1+(\nu'_j-i)/\theta$ for $\unu=\umu,(\umu'+\Box_j)'$.
This ${\bf c}^{(j)}_{\umu}(\theta)$ is now invariant under \eqref{deform_preview} since it mixes rows and columns coherently, i.e. 
\beq
\mu_i+\theta\mu'_j \qquad {\rm is\ invariant\ under\ \eqref{deform_preview}\ if} \qquad 1\leq i\leq m\quad;\quad 1\leq j\leq n.
\eeq

The origin of the normalisation  in \eqref{def_Tlong_def} can be understood  starting from the Casimir differential equations and the super Jack polynomials.
Indeed, a quick way to see the invariance of the superblocks under \eqref{deform_preview} is to consider that the eigenvalue of the 
original Casimir operator on $B_{\gamma,\ulmb}$ \eqref{intro_eigenv} 
\beq
{ E}_{\gamma,\ulmb}^{(m,n;\theta)}={h}_{\ulmb}^{(\theta)}+\theta \gamma |\ulmb|
+\Big[ \gamma\theta|m^n|+ h_{[e^m]}^{(\theta)} - \theta h_{[s^n]}^{(\frac{1}{\theta})}\Big]
\eeq
with $e=+\frac{\theta\gamma}{2}$ and $s=-\frac{\gamma}{2}$, 
is indeed invariant under \eqref{deform_preview}.  To see this, the hybrid form of the RHS~is useful, 
therefore $|\gamma|=\sum_{i=1}^m \lambda_i + \sum_{j=1}^n \lambda'_j -nm$,
and for $h_{\ulmb}(\theta)$ the expression given in  \eqref{evboth}.

The precise statement about shift invariance of the long superconformal blocks is
\beq\label{shift_superconformal}
\!\!\!\Pi_{ (\ulmb_s -m^n)'}(\theta)\, B_{\gamma,\ulmb+(\theta\tau')^m}=\Pi_{ (\ulmb_s-m^n+(\tau')^n)'}(\theta)\, B_{\gamma+2\tau',(\ulmb'+(\tau')^n)'}
\eeq
The $\Pi$ factors follow from the fact that we normalise our blocks 
so that the coefficient of the leading superJack $P_{\ulmb}$ is unity, i.e.~$B_{\gamma,\ulmb}=(\prod x_i^\theta\big/\prod_j y_j)^{\frac{\gamma}{2}} \big( P_{\ulmb}+\ldots \big)$, 
and from the analogous relation proved in appendix \ref{prop_undershift_app} for super Jack polynomials, namely
\beq\label{shiftsJack}
\ \  \Pi^{}_{(\ulmb_s-m^n)'}({\theta})\, P^{(m,n)}_{\ulmb+(\theta \tau')^m} =
\Pi^{}_{(\ulmb_s+(\tau')^n-m^n)'}({\theta}) \left(\frac{\prod_i x_i^\theta}{\prod_j y_j}\right)^{\!\!\tau'}\, P^{(m,n)}_{(\ulmb'+(\tau')^n)'}\ .
\eeq

From \eqref{shift_superconformal} and~\eqref{shiftsJack} we find the corresponding transformation on the coefficients
\beq\label{altra_formula}
(T_{\gamma})^{\umu+(\theta\tau')^m}_{\ulmb+(\theta\tau')^m}=(T_{\gamma+2\tau'})^{(\umu'+(\tau')^n)'}_{(\ulmb'+(\tau')^n)'} \times
\frac{ \Pi_{\lambda_s-m^n}(\frac{1}{\theta} )  }{   \Pi_{\lambda_s-m^n+(\tau')^n}(\frac{1}{\theta} )    }   \frac{ \Pi_{\umu_s-m^n+(\tau')^n}(\frac{1}{\theta} )  }{  \Pi_{\umu_s-m^n}(\frac{1}{\theta} )  }\ .
\eeq
This gives the normalisation of  $T^{\tt long}_{\gamma}$ in~\eqref{def_Tlong_def}, so that it is invariant under the shift
\beq
(T^{\tt long}_{\gamma})^{[\umu_s;\,\umu_e+(\theta\tau')^m]}_{[\ulmb_s;\,\ulmb_e+(\theta\tau')^m]}=(T^{\tt long}_{\gamma})^{[\umu_s+(\tau')^n;\,\umu_e]}_{[\ulmb_s+(\tau')^n;\,\ulmb_e]}\ .
\eeq
We could then define normalised long super Jack polynomial $P_{\ulmb}^{\tt long}$ so that $(T^{\tt long}_{\gamma})$ gives the expansion coefficients and all formulae are manifestly shift symmetric.

Finally, let us point out that the ratio of $\Pi$ functions in \eqref{altra_formula} should simplify because it does not depend on $\gamma$. Indeed, 
by using a property of $C^-$ under shifts\footnote{This is $$
	C^-_{\underline{\kappa}+(\tau')^n}(w;\theta)\big/ C^-_{\underline{\kappa}}(w;\theta)=C^-_{(\tau')^n}(w;\theta) \times C^0_{\underline{\kappa}}(w+\tau'+\theta(n-1);\theta)\big/ C^0_{\underline{\kappa}}(w+\theta(n-1);\theta)
	$$} and rearranging the expression we obtain
\beq
(T_{\gamma})^{\umu+(\theta\tau')^m}_{\ulmb+(\theta\tau')^m}=(T_{\gamma+2\tau'})^{(\umu'+(\tau')^n)'}_{(\ulmb'+(\tau')^n)'} \times
\frac{ C^0_{\umu_s/\ulmb_s}(\tau'+\frac{n}{\theta}-m)C^0_{\umu_s/\ulmb_s}(1+\frac{n-1}{\theta}-m) }{ 
	C^0_{\umu_s/\ulmb_s}(\frac{n}{\theta}-m) C^0_{\umu_s/\ulmb_s}(\tau'+1+\frac{n-1}{\theta}-m)}
\eeq
As expected the ratio of $\Pi$ functions above is more explicitly just a function of the skew diagrams.

\subsection{Truncations  of the superconformal block} \label{alphabetarevisited}

An $(m,n)$ superconformal block $B_{\gamma,\ulmb}$ has been so far defined as the multivariate series 
\beq\label{recap_sum_alphabeta}
B_{\gamma,\ulmb}=\left(\frac{\prod_i x_i^\theta}{\prod_j y_j}\right)^{\!\!\frac{\gamma}{2}} \, 
\sum_{\umu \supseteq \ulmb} { ({ T}_{\gamma})_\ulmb^\umu} \, P_{\umu}({\bf z};\theta)\ .
\eeq
In this section we discuss how the infinite sum over $\umu$ depends effectively on the external parameters $\alpha,\beta$ and $\ulmb$. 
Let us recall indeed that the $\alpha$ and $\beta$ dependence of $T_{\gamma}$ is solved by the $C^0$ factors, 
as we pointed out already in section \ref{example_Acoeff}.  Thus %
\beq\label{vanishgT}
(T_{\gamma})_{\ulmb}^{\umu} = {C^0_{\umu/\ulmb}(\theta\alpha;\theta) C^0_{\umu/\ulmb}(\theta\beta;\theta) }\times (T^{\tt\, rescaled}_\gamma)_{\ulmb}^{\umu} %
\eeq
where $T^{\tt\, rescaled}_\gamma$ is $\alpha,\beta$ \emph{independent.} The observation we will elaborate on is that the $C^0$ factors 
have certain vanishing properties which in practise truncate the sum.

Given the various forms of the recursions, spelled out in previous sections, it is useful to put the $C^0$ factors accordingly. 
To do so consider the equivalent rewritings\footnote{For the last rewriting consider that $((\mu_s-m^n)')_i=\mu_{m+i}$ $\forall i\ge 1$.}
\begin{align}
\label{C0forthesouth}
C^0_{\umu/\ulmb}(w;\theta)&=(-\theta)^{|\umu|-|\ulmb|} C^0_{\umu'/\ulmb'}(-\tfrac{w}{\theta};\tfrac{1}{\theta}) \\
\label{C0forthemixed}
&=(-\theta)^{|\umu_s|-|\ulmb_s|}C^0_{\umu_e/\ulmb_e}(w;\theta)C^0_{\umu'_s/\ulmb'_s}(-\tfrac{w}{\theta};\tfrac{1}{\theta})
\end{align}
which follow from the definition in \eqref{C0symbol_sec5}. 
Then, let us note that $C^0$ has the following analytic continuation, 
\beq
C^0_{[\kappa_1,\ldots ,\kappa_{\ell} ]}(w;\theta)=\prod_{i=1}^{\ell} (w-\theta(i-1))_{\kappa_i}
\eeq
therefore, if we consider the $(m,0)$ analytic continuation of section \ref{sec_various_analcont}, 
we are keeping fixed the number of rows of $\ulmb,\umu$, as for the Young diagrams, and we find that $C^0_{\umu/\ulmb}(w;\theta)$ 
has the correct $(m,0)$ analytic continuation.
If instead we consider the $(0,n)$ analytic continuation, we are keeping fixed the number of columns of $\ulmb,\umu$, and thus it is 
$C^0_{\umu'/\ulmb'}(-\tfrac{w}{\theta};\tfrac{1}{\theta})$ on the RHS~of \eqref{C0forthesouth} the one with  the correct $(0,n)$ analytic continuation.
Finally, the rewriting on the RHS~of \eqref{C0forthemixed} is the one with the correct $(m,n)$ analytic continuation.

To appreciate the vanishing properties of the $C^0$ factors in \eqref{vanishgT}, consider %
for example $C^0_{\umu/\ulmb}(\theta \alpha;\theta)$. It will vanish when %
\beq
(\lambda_i+ \theta\alpha-\theta(i-1))\ldots (\mu_i-1+ \theta\alpha-\theta(i-1)) =0%
\eeq
and similarly for  $C^0_{\umu/\ulmb}(\theta \beta;\theta)$.
In particular, it is enough that only one factor becomes zero.
This of course depends on the values of $\alpha$, $\beta$,  $\theta$, as well as $\ulmb$.  
The general situation is summarised by the following tables. On the east
\beq\label{table_truncationeast}
\rule{.5cm}{0pt}\begin{array}{cc|c}
	& & {\tt truncation } \\
	\hline & & \\
	{\rm if}\ \exists\,i\leq m\ {\rm such\ that}\  {(i-1)-w}{}\ge \frac{ \lambda_i}{\theta}\in \mathbb{Z} & &  \mu_i= {\theta(i-1-w)}{}
\end{array}
\eeq
with $w$ here being either $\beta$ or $\alpha$. This means that when the condition is satisfied on a row index $i$, by progressively  increasing $\mu_i=\lambda_i+n_i$ by integers $n_i$ we will hit a vanishing point. 
Similarly on the south %
\beq\label{table_truncationsouth}
\begin{array}{cc|c}
	& & {\tt truncation } \\
	\hline & & \\
	{\rm if}\ \exists\,j\leq n\ {\rm such\ that}\  {(j-1)}{}+\theta w\ge \theta \lambda'_j\in \mathbb{Z}& & \quad \mu'_j= \frac{(j-1)}{\theta}+w
\end{array}
\eeq
where again $w$ is either $\beta$ or $\alpha$.

Consider now a situation with a group theory interpretation, thus relevant for superconformal blocks. Assume first that $\ulmb$ is a Young diagram, then  
$\ulmb$ has at most $\beta\in \mathbb{N}$ rows (since in our conventions $\beta\leq \alpha$). In particular, $\lambda'_j-\beta\leq 0$. %
From \eqref{table_truncationsouth}
with $j=1$ we thus find $\mu'_1\leq\beta$,
and we  conclude that there is a vertical cut off on the Young diagrams $\umu$ over which we sum.
On the contrary, note that by construction  
$\lambda_i+\theta \beta\ge \theta(i-1)$, $\forall$ $i\leq m\leq \beta$, i.e.~the opposite of the condition in \eqref{table_truncationeast},  
precisely because $\beta$ sets the value for the maximal number of rows.  Therefore there is no truncation on the horizontal east directions. 
The picture to have in mind for the superconformal blocks is thus
\beq\label{example_picture_block}
\begin{tikzpicture}[scale=.8]

\def\shift{2.5}
\def\xuno{2-\shift}
\def\yuno{2}

\def\ydue{5}
\def\xtre{6-\shift}

\def\ytre{4.5}
\def\xquattro{5-\shift}

\def\yquattro{3.5}
\def\xcinque{4-\shift}

\def\ycinque{3}
\def\xsei{3-\shift}

\def\step{.5}
\def\stepp{.1}

\draw[thick] (\xuno,	\yuno+\step) -- 			(\xuno,	\ydue);
\draw[thick] (\xuno,	\ydue) -- 				(\xtre,	\ydue);
\draw[thick] (\xtre,	\ytre) -- 				(\xquattro,	\ytre);
\draw[thick] (\xquattro,	\yquattro) -- 		(\xcinque,	\yquattro);
\draw[thick] (\xcinque,	\ycinque) -- 		(\xsei,	\ycinque);
\draw[thick] (\xsei,	\yuno+\step) -- 			(\xuno, \yuno+\step);

\draw[thick](\xtre,	\ydue)-- 			(\xtre,	\ytre);
\draw[thick](\xquattro,	\ytre)-- 		(\xquattro,	\yquattro);
\draw[thick](\xcinque,	\yquattro)-- 	(\xcinque,	\ycinque);
\draw[thick](\xsei,	\ycinque)-- 		(\xsei,	\yuno+\step);

\draw[thick] (\xtre+\stepp,	\ytre-\stepp) -- 				(\xquattro+\stepp,	\ytre-\stepp);
\draw[thick] (\xquattro+\stepp,	\yquattro-\stepp) -- 		(\xcinque+\stepp,	\yquattro-\stepp);
\draw[thick] (\xcinque+\stepp,	\ycinque-\stepp) -- 		(\xsei+ \stepp,	\ycinque-\stepp);
\draw[thick] (\xsei+\stepp,	\yuno+\step-\stepp) -- 			(\xuno, \yuno+\step-\stepp);

\draw[thick](\xtre+\stepp,	\ydue)-- 			(\xtre+\stepp,	\ytre-\stepp);
\draw[thick](\xquattro+\stepp,	\ytre-\stepp)-- 		(\xquattro+\stepp,	\yquattro-\stepp);
\draw[thick](\xcinque+\stepp,	\yquattro-\stepp)-- 	(\xcinque+\stepp,	\ycinque-\stepp);
\draw[thick](\xsei+\stepp,	\ycinque-\stepp)-- 		(\xsei+\stepp,	\yuno+\step-\stepp);

\filldraw[red!30!white]  (\xtre+\stepp+.025,\ydue) rectangle (\xtre+4-\stepp,\ycinque);
\filldraw[red!30!white]  (\xquattro+\stepp+.04,	\ytre-\stepp-.025) rectangle(\xtre+1-\stepp,\yuno-.05);
\filldraw[red!30!white]  (\xcinque+\stepp+.04,	\yquattro-\stepp-.025) rectangle(\xtre+1-\stepp,\yuno-.05);
\filldraw[red!30!white]  (\xsei+\stepp+.04,	\ycinque-\stepp-.025) rectangle(\xtre+1-\stepp,\yuno-.05);
\filldraw[red!30!white]  (\xuno+.04,	\yuno+\step-\stepp-.04) rectangle(\xtre+1-\stepp,\yuno-.1);

\draw[thick]  (\xtre+\stepp,\ydue) -- (\xtre+4-\stepp,\ydue);
\draw[thick]  (\xuno,\yuno-\stepp) -- (\xuno,\yuno+\step-\stepp);	

\draw[thick]  (\xuno,\yuno-\stepp) -- (\xtre+1-\stepp,\yuno-\stepp);	
\draw[thick]  (\xtre+1-\stepp,\yuno-\stepp) -- (\xtre+1-\stepp,\ycinque);
\draw[thick]  (\xtre+1-\stepp,\ycinque) --  (\xtre+4-\stepp,\ycinque);

\draw[thick,dashed] (\xtre+4-\stepp,\ydue)--(\xtre+5-\stepp,\ydue);
\draw[thick,dashed] (\xtre+4-\stepp,\ycinque)--(\xtre+5-\stepp,\ycinque);

\draw[stealth-stealth]  (\xtre+2-\stepp,	 \ydue-.2)  --     (\xtre+2-\stepp,	 \ycinque+.2) ;
\draw[]  (\xtre+2.3-\stepp,	 \ycinque+1) 	node [rotate=0]	{ \color{blue} $m$ };

\draw[latex-latex] (\xuno-.5,\yuno-.1) -- (\xuno-.5,\ydue+.1);
\draw[latex-latex] (\xuno,\ydue+.5) -- (\xtre+1,\ydue+.5);

\draw (\xuno-.8,	  3.5) 	node 	{ \color{blue} $\beta$ };
\draw (4.5-\shift,	  \ydue+.8) 	node 	{ \color{blue} $n$ };

\draw[] (3.5-\shift,	 3.75) 	node [rotate=0]	{ \color{blue} $\ulmb$ };

\draw (\xuno-4.5,3.5) node {$\displaystyle F_{\gamma,\ulmb}({\bf x}|{\bf y})=\sum_{{\color{red!60!white}\umu}:  \ulmb\subseteq \umu } { ({ T}_{\gamma})_\ulmb^\umu} \, P_{\umu}({\bf x}|{\bf y})\qquad\qquad$};

\draw[-stealth]  (3-\shift,4.5) -- ( 4-\shift,4.5) ;
\draw[-stealth]  (3-\shift,4.35) -- ( 4-\shift,4.35);

\draw[] (8.8,	 2.5) 	node [rotate=0]	{  $\phantom{cazzo}$ };	

\end{tikzpicture}
\eeq
A posteriori, the cut-off on the south is expected, because the internal subgroup of the superconformal algebra is compact.
Note also that for Young diagrams, since $T$ only depends on diagrams, we can see that a solution $(T_{\gamma})_{\ulmb}^{\umu}$ where $\umu$ has more than $\beta$ rows does not exist, 
by considering an equivalent reasoning on the east. In fact, when $i=\beta+1$ and $\lambda_{i=\beta+1}=0$, we find from \eqref{table_truncationeast} that $\mu_i=0$. %

Consider now a long block with analytically continued $\ulmb$, according to an $(m,n)$ structure, in such a way to describe an anomalous dimension. 
In this case we still require finiteness of the sum  \eqref{recap_sum_alphabeta} on the south, as expected for a compact group. From \eqref{table_truncationsouth} we see then that even 
when $\beta$ and $\lambda'_j$ are generic, if however $0\leq \beta-\lambda'_{j=1,\ldots n}\in \mathbb{Z}^+$, the condition on the truncation can be satisfied.
Note instead that our Young diagram argument on the east \eqref{table_truncationeast} now will not work, precisely because $\beta$ is not integer for cases of physical interest where $\lambda_i>0$. 
We notice however that taking the $\lambda_i<0$ it woud be possible to have truncation on the east and the south simultaneously and perhaps this reproduces the Sergeev-Veselov super Jacobi polynomials.

The generic `shape' of a physical superconformal block is thus the one illustrated in \eqref{example_picture_block}.

\section{Revisiting known blocks with the binomial coefficient}\label{know_solu}

Our formula for the coefficients $(T_{\gamma})_{\ulmb}^{\umu}$ in \eqref{SCbinomial2}, written in terms of interpolation polynomials, 
is valid for any Young diagram.  We find instructive to test \eqref{SCbinomial2} against 
known analytic solutions of the recursion, such as the rank-one and rank-two bosonic blocks. Then, we will 
repeat a similar check for the determinantal solution found in \cite{Doobary:2015gia} for $\theta=1$.

We will use the {\tt rescaled} version of it, where we omit the dependence on $\alpha$ and $\beta$.  
Therefore, 
\beq\label{app_SCbinomial2}
(T^{\tt rescaled}_{\gamma})_{\ulmb}^{\umu}= 
(\mathcal{N}^{\tt rescaled}\,)^{\umu}_{\ulmb}\times
\frac{ P^*_{N^M\,\bslash\umu}(  N^M\,\bslash\ulmb ;\theta,u ) }{  P^{*}_{ 
N^M\,\bslash\ulmb}( N^M\,\bslash\ulmb ;\theta,u) 
}\Bigg|_{u=\tfrac{1}{2}-\theta\tfrac{\gamma}{2} -N } 
\eeq
where  
\begin{gather}\label{app_nn}
(\mathcal{N}^{\tt rescaled}\,)^{\umu}_{\ulmb}=
\frac{(-)^{|\umu|} \Pi_{\umu}(\theta) }{(-)^{|\ulmb|} \Pi_{\ulmb}(\theta)}\ 
\frac{C^0_{\umu/\ulmb}(1{-}\theta{+}M\theta;\theta)}{C^0_{\umu/\ulmb}(M\theta;\theta)
 } 
\end{gather}
 As pointed out already,
the interpolation polynomials encode the non-factorisable $\gamma$ dependence of $(T_{\gamma})_{\ulmb}^{\umu}$, which we 
saw emerging experimentally from the recursion. However, the way the interpolation polynomials are evaluated is quite different compared with the recursion. In fact,  in order to match
$(T_{\gamma})_{\ulmb}^{\umu}$ we will need to compute $P^*$ of $N^M\bslash\umu$, 
rather than looking recursing over  $\umu/\ulmb$, as we do in the recursion. Our exercise here will show in simple cases how these two combinatorics actually produce the same result.  

Through this section, we shall take the minimal choice for $N$ and $M$ in the above formulae, i.e.~
\beq
N=\mu_1\qquad;\qquad M=\beta
\eeq
where $\beta$ as usual fixes the maximal height of both $\ulmb$ and $\umu$.

\subsection{The half-BPS solution}

The simplest case to start with is the derivation from~\eqref{app_SCbinomial2} of the half-BPS solution in \eqref{solu_halfBPS}, 
\beq\label{solu_halfBPS_again}
(T^{\tt rescaled}_{\gamma})^{\umu}_{[\varnothing]}= \frac{ 1 }{ C^0_{\umu}(\theta \gamma;\theta) }\frac{1}{C^-_{\umu}(1;\theta)}\ . %
\eeq
To obtain the $\gamma$ dependence  we need consider the ratio 
of interpolation polynomials in \eqref{app_SCbinomial2} and since $\ulmb=[\varnothing]$, the numerator $P^*$ is 
evaluated on $\mu_1^\beta$, which is a ``constant" Young diagram.
This evaluation is known. 
From \cite{Rains_1} we find
\beq\label{evalu_at_const}
P^{ip}_{\underline{\kappa}}(c+\theta \delta_\ell+u;\theta,u)=(-)^{|\ulmb|}  C^0_{\underline{\kappa}} (c+2u+\theta(\ell-1),-c;\theta) 
												\frac{C^0_{\underline{\kappa} }(\ell \theta;\theta)}{C^-_{\underline{\kappa}} (\theta;\theta) }
\eeq
which we need here for $c=\mu_1$ and $\ell=\beta$. 
For what concerns the denominator instead, this is always factorised for any $\ulmb$ 
and it is given by \eqref{evaluationBCinterp}.  Putting together we find, 
\beq
\begin{array}{ccc}
\frac{ P^*_{\mu_1^\beta\bslash\umu}(  \mu_1 + \theta \delta_{\beta} +u;\theta,u ) }{ C^+_{\mu_1^\beta}(-2\mu_1-\theta\gamma+2\theta\beta) C^-_{\mu_1^\beta}(1;\theta) }&=&
\frac{ C^0_{\mu_1^\beta\bslash\umu}(-\theta\gamma-(\mu_1-1)+\theta(\beta-1);\theta)  }{ C^0_{\mu_1^{\beta}} (-\theta\gamma-(\mu_1-1)+\theta(\beta-1);\theta) }\times
\frac{ C^0_{\mu_1^\beta \bslash\umu}(-\mu_1;\theta)\, C^0_{\mu_1^\beta \bslash\umu}(\beta\theta;\theta) }{  
(-)^{|\umu|-\beta\mu_1}C^-_{\mu_1^\beta}(1;\theta) C^-_{\mu_1^\beta\bslash\umu}(\theta;\theta)} %
\end{array}
\eeq
where we rewrote $C^+_{}$ in terms of $C^0$ using the explicit definition \eqref{recap_Csymbol}, and the fact that the Young diagram involved is a rectangle $\mu_1^{\beta}$. 
At this point, the $1/C^0(\theta\gamma;\theta)$ comes out thanks to the identity, 
\beq\label{ident_uno}
\frac{C^0_{\umu} (w-(\mu_1-1)+\theta(\beta-1);\theta) C^0_{\mu_1^\beta-\umu}(-w;\theta) }{ C^0_{\mu_1^\beta}(-w;\theta)}= {(-1)^{|\umu|} }\quad;\quad \forall\, w
\eeq
taken with $w=-\theta\gamma$, and the $(-1)^{|\umu|}$ cancels against  the one in \eqref{app_nn}.  %
Finally, the contribution $1/C^-_{\umu}(1;\theta)$ comes from
$\Pi_{\umu}(\theta)$ in  \eqref{app_nn}. The remaining terms 
in the binomial coefficient necessarily have to simplify to unity. When we collect 
them all,\footnote{These terms are
$$
\frac{(-)^{|\umu|+\beta\mu_1} C^-_{\umu}(\theta;\theta) }{ C^0_{\umu}(\beta\theta;\theta)C^-_{\mu_1^\beta\bslash\umu}(\theta;\theta) }\times 
\frac{C^0_{\mu_1^\beta\bslash\umu}(\beta\theta;\theta)  }{  C^-_{\mu_1^\beta}(1;\theta) }\times
{ C^0_{\mu_1^\beta \bslash\umu}(-\mu_1;\theta)  C^0_{\umu}(1+\theta(\beta-1);\theta) }{}=1
$$
using $w=\theta$ in \eqref{ident_due} and $w=\mu_1$ in \eqref{ident_uno}, then again $w=-\beta\theta$ in  \eqref{ident_uno}.
Finally $(-)^{\beta\mu_1}C^0_{\mu_1^\beta}(-\mu_1;\theta)/C^-_{\mu_1^\beta}(1;\theta)=1$ and $C^0_{\mu_1^\beta}(\beta\theta;\theta)/C^-_{\mu_1^\beta}(\theta;\theta)=1$, directly from their definitions.}
upon using another identity
\beq\label{ident_due}
\frac{C^-_{\umu}(w;\theta) }{C^0_{\mu_1^{\beta}}( \theta(\beta-1)+w;\theta) C^-_{\mu_1^\beta\bslash\umu}(w;\theta)} =\frac{ C^0_{\umu}(-w-(\mu_1-1);\theta) }{  (-)^{|\umu|} C^-_{\mu_1^\beta}(w;\theta) }
\quad;\quad \forall\, w
\eeq
we finally obtain our desired result. 

Already in this simple example we needed several identities between $C^{\pm,0}$ coefficients. This shows that the rewriting of $T$ in terms of interpolation polynomials is quite non trivial.

\subsection{Rank-one and rank-two}

After the half-BPS solution, we will consider the rank-one cases, since these are fully factorised.
Next, the rank-two case, which involves a$~_4F_3$ and has non trivial dependence on $\gamma$. Thus it is the most interesting case for our purpose. Our task will be to revisit the solution
found by Dolan and Osborn in \cite{Dolan:2003hv}.

\subsubsection*{Rank-one $(1,0)$}
For the single row case $\beta=1$, with $\umu=[\mu]$ and $\ulmb=[\lambda]$, we find
\beq
\begin{array}{rccc}
(\mathcal{N}^{\tt reduced}\,)^{\umu}_{\ulmb}=&
				\frac{(-)^{|\umu|} \Pi_{\umu}(\theta) }{(-)^{|\ulmb|} \Pi_{\ulmb}(\theta)} &\times& 
				\frac{C^0_{\umu/\ulmb}(1;\theta)}{C^0_{\umu/\ulmb}(\theta;\theta) } \\[.3cm]  %
=&
					\frac{(-)^{\mu} (\theta)_\mu (1)_{\lambda} }{(-)^{\lambda} (1)_{\mu} (\theta)_\lambda} &\times& 
					\ \ \frac{(\lambda+1)_{\mu-\lambda} }{(\lambda+\theta)_{\mu-\lambda} }%
\end{array}
\eeq
which simplifies to a sign. Then, 
for the interpolation polynomials we find 
\beq
\frac{ P^*_{[\varnothing]}(  \mu-\lambda;\theta,u ) }{  P^{*}_{ [\mu-\lambda]}( \mu-\lambda;\theta,u) } =
\frac{1}{(-)^{\mu-\lambda} (2\lambda+\theta\gamma)_{\mu-\lambda}(1)_{\mu-\lambda} }
\eeq
in particular, the numerator is trivial, and the denominator follows straightforwardly from 
\eqref{evaluationBCinterp} and \eqref{recap_Csymbol}.  Putting together the two results 
above we obtain the formula
\beq
(T^{\tt rescaled}_{\gamma})^{[\mu]}_{[\lambda]}=
	\frac{1}{(\mu{-}\lambda)!(2\lambda+{\theta \gamma})_{\mu-\lambda}}
\eeq
which coincides with the one derived in section \ref{sec_rank1}. 

This case was quite immediate and the reason is that the formula for the $P^*$ is oriented on the east, as the $(1,0)$ theory. We will see now what happens in the $(0,1)$ theory.

\subsubsection*{Rank-one $(0,1)$} 

For the single column case, the diagrams are $\ulmb=[1^{\lambda'}]$ and $\umu=[1^{\mu'}]$. 
This is a case in which $\lambda',\mu'\leq \beta$.  
Let us proceed with a  direct computation starting with~ \eqref{app_SCbinomial2}. 
The normalisation~\eqref{app_nn} becomes
\beq
\begin{array}{rccc}
(\mathcal{N}^{\tt rescaled}\,)^{\umu}_{\ulmb}=&
				\frac{(-)^{|\umu|} \Pi_{\umu}(\theta) }{(-)^{|\ulmb|} \Pi_{\ulmb}(\theta)} &\times& 
				\frac{C^0_{\umu/\ulmb}(1+(\beta-1)\theta;\theta)}{C^0_{\umu/\ulmb}(\beta\theta;\theta) } \\[.3cm]  %
=&
					\frac{(-)^{\mu'} (1)_{\mu'} (\frac{1}{\theta})_{\lambda'}  }{ (-)^{\lambda'}  (\frac{1}{\theta})_{\mu'} (1)_{\lambda'} } &\times& 
					\ \ \frac{(\frac{1}{\theta}+\beta-\mu')_{\mu'-\lambda'} }{(1+\beta-\mu')_{\mu'-\lambda'} }
\end{array}
\eeq
For the ratio of interpolation polynomials
the diagrams are $\mu_1^\beta\bslash\umu=[1^{\beta-\mu'}]$ and $\mu_1^\beta\bslash\ulmb=[1^{\beta-\lambda'}]$ 
and the polynomials have $\beta$ variables.  This time the numerator involves a non trivial Young diagram, which however has less than $\beta$ rows. 
To simplify the result we use the non-trivial reducibility property  for the interpolation polynomials, which reads
\beq
P_{[\kappa_1,\ldots \kappa_{s}]}^{ip}(\underbrace{ \ldots, w_{s},u,u+\theta,\ldots u+(r-1)\theta}_{\rm variables};\theta,u)=P_{[\kappa_1,\ldots \kappa_{s}]}^{ip}(\ldots,w_{s};\theta, u+r\theta)
\eeq
where $r$ is counting the excess between the number of variables $\beta$ and the non zero 
parts of $\underline{\kappa}$.\footnote{Of course the same is true for a symmetric permutation 
of all variables on the LHS}  In our case, the above result gives back an evaluation formula to
\beq
P^{ip}_{[1^{\beta-\mu'}]}(1^{\beta-\lambda'}+\delta_{\beta-\lambda'}+u';\theta,u')\Bigg|_{u'=u+\lambda'\theta}
=\tfrac{(\mu'-\lambda'+1)_{\beta-\mu'} }{ (1)_{\beta-\mu'} }(\theta)^{2(\beta-\mu')} (\tfrac{1}{\theta})_{\beta-\mu'} (\lambda'+\mu'-\gamma)_{\beta-\mu'}
\eeq
and in fact the RHS~follows from \eqref{evalu_at_const}. %
Thus, the ratio of interpolation polynomials contributes as, 
\beq
\qquad\quad
\frac{ P^*_{[1^{\beta-\mu'}]}(  1^{\beta-\lambda'} ;\theta,u ) }{  P^{*}_{ [1^{\beta-\lambda'}]}( 1^{\beta-\lambda'} ;\theta,u) } =
\frac{
(\theta)^{2(\beta-\mu')} (\lambda'+\mu'-\gamma)_{\beta-\mu'} (\tfrac{1}{\theta})_{\beta-\mu'}
}{ (\theta)^{2(\beta-\lambda')} (2\lambda'-\gamma)_{\beta-\lambda'}(\frac{1}{\theta})_{\beta-\lambda'} }\ .
\eeq
Putting it all together, the final result for $T^{\tt rescaled}_{\gamma}$ (defined in~\eqref{attempt_gen_sol}) is
\beq
(T^{\tt rescaled}_{\gamma})_{[1^{\lambda'}]}^{[1^{\mu'}]}=\frac{(\theta)^{2(\lambda'-\mu')} %
}{(\mu'-\lambda')!(2\lambda'{-}\gamma)_{\mu'-\lambda'}}\ 
	\frac{ \mu'! (\frac{1}{\theta})_{\lambda'}}{\lambda'! (\frac{1}{\theta})_{\mu'} }\, (-1)^{ \mu'-\lambda'}.
\eeq
This coincides precisely with the one derived in section \ref{sec_rank1} for $(T_{\gamma})_{[1^{\lambda'}]}^{[1^{\mu'}]}$ when we re-insert  
\beq
C^0_{\umu/\ulmb}\left(\tfrac{\theta(\gamma-p_{12})}{2},\tfrac{\theta(\gamma-p_{43})}{2};\theta\right)=(-\theta)^{2(\mu'-\lambda')}\big(\lambda'-\tfrac{(\gamma-p_{12})}{2},\lambda'-\tfrac{(\gamma-p_{43})}{2}\big)_{\mu'-\lambda'}
\eeq
properly oriented towards the south.

\subsubsection*{Rank-two $(2,0)$}

In the rank-one cases there is no room for any non trivial polynomial in $\gamma$. The first non 
trivial case in this sense is then rank-two, corresponding to $T_\gamma$ for two row Young diagrams. 
The recursion~\eqref{rec_section_scblocks} for this case  was solved explicitly by Dolan and Osborn~\cite{Dolan:2003hv}.
They did this first on a case by case basis in dimensions $d=2,4,6$, then for general $\theta$
by using properties of the$~_4F_3$ function, and manipulations inspired by those in 
\cite{DOcopypaste}.\footnote{In \cite{Dolan:2003hv}, see formulae  (3.11), (3.18) and (3.19).} 
We will show here how this relates to the interpolation polynomial of~\eqref{SCbinomial2}.

In our conventions the solution of~\cite{Dolan:2003hv} reads
\beq\label{DO2rows_concorda}
(T^{\tt rescaled}_{\gamma})_{\ulmb}^{\umu}=
\frac{ (\theta)_{\theta}}{(\mu_-+\theta)_{\theta}}
\times (D^{\tt rescaled})^{\umu+\frac{\theta}{2} \gamma}_{\ulmb+ \frac{\theta}{2} \gamma}
\eeq
with
\begin{align}\label{erre_DO}
	({D}^{\tt rescaled})_{\ulmb}^{\umu}=&\ 
	\frac{
		(2\theta)_{\lambda_-}(2\theta)_{\mu_-} }{(\theta)_{\lambda_- } (\theta)_{\mu_- } } \frac{
		\frac{(\mu_-+1)_{\theta} }{(\mu_1-\lambda_2+1)_{\theta} } 
	}{(\mu_1-\lambda_1)!(\mu_2-\lambda_2)!  }\times \\
	&\ \frac{
		\frac{(\lambda_+-2\theta)_{\theta} }{(\lambda_2+\mu_2-2\theta)_{\theta} }
	}{ (2\lambda_1)_{\mu_1-\lambda_1}(2\lambda_2-2\theta)_{\mu_2-\lambda_2}} ~_4F_3\left[\begin{array}{c}
		-\mu_-\, ,\, \theta\, ,\, -\lambda_-\, ,\,  \lambda_+-1 \\ 
		-(\mu_1-\lambda_2)\,,\, \lambda_2+\mu_2-\theta \,,\, 2\theta \end{array};1\right]\notag
\end{align}
where  $\ulmb=[\lambda_1,\lambda_2]$, $\umu=[\mu_1,\mu_2]$, and $\kappa_{\pm}=\kappa_1\pm\kappa_2$, 
for $\underline{\kappa}=\ulmb,\umu$.

The contribution denoted by $D$ is the translation to our conventions of the result of \cite{Dolan:2003hv}, 
where we implemented the shift symmetry and slightly rewrote some Pochhammers.
Note that the$~_4F_3$ has a series expansion which truncates at $\min(\lambda_-, \mu_-)$ and can thus be written as the explicit finite sum
\begin{align}
({D}^{\tt rescaled})_{\ulmb}^{\umu}=& \sum_{n=0}^{\min(\lambda_-,\mu_-)} \frac{(-\lambda_-)_n}{n!} \frac{
(2\theta)_{\lambda_-}(\theta)_n(2\theta)_{\mu_-} }{(\theta)_{\lambda_- }(2\theta)_n (\theta)_{\mu_- } }\frac{ (\mu_--n+1)_{n+\theta} }{ (\mu_1-\lambda_2-n+1)_{n+\theta} }  \\
&\rule{1cm}{0pt}\  \frac{ (\lambda_+-1)_n (\lambda_+-2\theta)_{\theta} }{(\mu_1-\lambda_1)!(\mu_2-\lambda_2)! (2\lambda_1)_{\mu_1-\lambda_1}(2\lambda_2-2\theta)_{\mu_2-\lambda_2+n+\theta}  }\ .
\notag
\end{align}

Now compare this expression with our expression for $T_{\gamma}$ written in terms of  interpolation polynomials~\eqref{SCbinomial2} (with $N=\mu_1$, $M=2$):
\beq\label{TBC2_ex}
\qquad\quad
(T_{\gamma})_{\ulmb}^{\umu}=(\mathcal{N}_{}\,)^{\umu}_{\ulmb}\times 
\frac{ P^{ip}_{[\mu_1-\mu_2]}( w_1,w_2;\theta,\frac{1}{2}-\theta\frac{\gamma}{2}-\mu_1 ) }{  
	P^{ip}_{[ \mu_1-\lambda_2,\mu_1-\lambda_1]}(  w_1,w_2;\theta,\frac{1}{2}-\theta\frac{\gamma}{2}-\mu_1) } \Bigg|_{w_i=\frac{1}{2}-\theta(\frac{\gamma}{2}-i+1)-\lambda_i}
\eeq
(Note that the shift symmetry shows up nicely in this formula since we can always put together combinations 
of the form $\kappa_i+\frac{\theta}{2}\gamma$, and $\mathcal{N}$ is automatically invariant.) 
All contributions 
in \eqref{TBC2_ex} coming from the denominator and the normalisation, can be straightforwardly computed, 
and written as product of Pochhammers or Gamma functions. 
The $BC_2$ interpolation polynomial in the numerator 
is also known explicitly, and described by a $~_4F_3$~\cite{Koornwinder}, 
\begin{align}
P^{ip}_{\underline{\kappa}}(w_1,w_2;\theta,u)=& (-)^{|\kappa|} (u\pm w_1,u\pm w_2)_{\kappa_2} 
(\kappa_2+\theta+u\pm w_1)_{\kappa_{-} }\times\notag\\
& \rule{2cm}{0pt}
\times~_4F_3\left[\begin{array}{c}
 -\kappa_-\, ,\, \theta\, ,\, \kappa_2+u\pm w_2\,  \\ 
1-\theta-\kappa_-\,,\, \kappa_2+\theta+u\pm w_1  \end{array};1\right]
\label{formula_Korn_interp}
\end{align}
Although similar,  this $~_4F_3$ is not identical to that of~\eqref{erre_DO}. Also note that it is not manifestly $w_1 \leftrightarrow w_2$ 
invariant even though the  polynomial is symmetric under this interchange. 
We thus have two possibilities leading to the same result.  
Using $P^{ip}_{[\mu_1-\mu_2]}( w_2,w_1)$ we find
\beq
{\rm numerator\, of\,}(T_{\gamma=0})_{\ulmb}^{\umu}\ \rightarrow\ 
~_4F_3\left[\begin{array}{c}
 -\mu_-\, ,\, \theta\, ,\, 1-\lambda_1-\mu_1\, ,\, \lambda_1-\mu_1  \\ 
\lambda_2-\mu_1\,,\, 1-\theta-\mu_1+\mu_2,1+2\theta-\lambda_2-\mu_1  \end{array};1\right]
\eeq
which  becomes the same hypergeometric as in \eqref{erre_DO} upon using the Whipple identity
\beq
_4F_3\left[\begin{array}{c}
-n\,,\, a\,,\,b\,,\,c \\ 
d\,,\,e\,,\,f \end{array};1 \right]= \frac{(e-a)_n(f-a)_n}{(e)_n(f)_n}
~_4F_3\left[\begin{array}{c}
-n\,,\, a\,,\,d-b\,,\,d-c \\ 
d\,,\,a-e+1-n\,,\,a-f+1-n \end{array} ;1\right]
\eeq
Similarly had we chosen $P^*_{[\mu_1-\mu_2]}( w_1,w_2)$ a different ${}_4F_3$ identity will give the same final result. 
Thus we see how the$~_4F_3$ of~\cite{Dolan:2003hv} arises directly from the interpolation polynomial.

It is quite spectacular to implement on a computer the above representation  of $T_{\gamma}$ and the one arising directly from the recursion and check that they agree over many examples.

\subsection{Revisiting the $\theta=1$ case and determinantals}\label{D1508}

The superconformal blocks for the case $\theta=1$ and any $(m,n)$ were obtained 
in \cite{Doobary:2015gia}.  One of the main outcomes of that derivation is an explicit expression 
for the coefficients in an expansion over super Schur polynomials, that we quote here below, 
\begin{align}
({R}_{\gamma})_{\ulmb}^{\umu} =  & \sum_{\sigma} (-)^{|\sigma|} \prod_{i=1}^{m} \frac{ 
\left(\lambda_i-i+1+\frac{\gamma-p_{12}}{2}\right)_{\mu_{\sigma(i)}+i-\sigma(i) -\lambda_i } \left(\lambda_i-i+1+\frac{\gamma-p_{43}}{2}\right)_{\mu_{\sigma(i)}+i-\sigma(i) -\lambda_i } }{ 
(\mu_{\sigma(i) }+i-\sigma(i)-\lambda_i)! (2\lambda_i-2i+2+\gamma)_{\mu_{\sigma(i)}+i-\sigma(i) -\lambda_i }}  \notag
\end{align}
We can also conveniently rewrite the above formula for the coefficients as a determinant
\beq
\quad\ \ ({R}_{\gamma})_{\ulmb}^{\umu} = \det\left(
\frac{ (\lambda_i-i+\frac{\gamma-p_{12}}{2}+1)_{\mu_j-j-\lambda_i+i}
(\lambda_i-i+\frac{\gamma-p_{43}}{2}+1)_{\mu_j-j-\lambda_i+i} }{(\mu_{j }-j-\lambda_i+i)!  (2\lambda_i-2i+2+\gamma)_{\mu_{j}-j-\lambda_i+i  }  }  \right)_{1\leq i,j\leq \beta}
\eeq
In this section we want to directly show how $({R}_{\gamma})_{\ulmb}^{\umu}$ also arises from the interpolation polynomials via~\eqref{SCbinomial2}, namely we want to show
\beq
(R_{\gamma})_{\ulmb}^{\umu} = (T_{\gamma})_{\ulmb}^{\umu}\Bigg|_{\theta=1}=
(\mathcal{N}_{}\,)^{\umu}_{\ulmb}\times
\frac{ P^*_{N^M\,\bslash\umu}(  N^M\,\bslash\ulmb ;\theta,u ) }{  P^{*}_{ 
N^M\,\bslash\ulmb}( N^M\,\bslash\ulmb ;\theta,u) 
}\Bigg|_{u=\tfrac{1}{2}-\theta\tfrac{\gamma}{2} -N } \Bigg|_{\theta=1} 
\label{we_want_to_prove}
\eeq
 Notice that for $\theta=1$ the normalisation simplifies massively, 
\beq
(\mathcal{N}_{}\,)^{\umu}_{\ulmb}\Bigg|_{\theta=1}=C^0_{\umu/\ulmb}\left(\tfrac{(\gamma-p_{12})}{2},\tfrac{(\gamma-p_{43})}{2};1\right)
\eeq
and only the ratio of interpolation polynomials is important.

Let us begin by improving the expression of the determinant in such a way as to extract the 
same normalisation. Using $(a)_x=\Gamma[a+x]/\Gamma[a]$,  it is simple to realise that  various 
contributions depend solely on either row or column index, therefore can be factored out,
and rearranged. We arrive at the expression
\begin{align}
({R}_{\gamma})_{\ulmb}^{\umu} &= (\mathcal{N}_{}\,)^{\umu}_{\ulmb}\times ({R}^{\tt rescaled}_\gamma)_{\ulmb}^{\umu}\\[.2cm]
({R}^{\tt rescaled}_{\gamma})_{\ulmb}^{\umu}&=\left(
\prod_{i=1}^{\beta} \Gamma[ 2-2i+\gamma +2\lambda_i ]
 \right)\det\left( 
 \frac{\ \Gamma[1+i-j-\lambda_i+\mu_j] ^{-1}}{ \Gamma[2-i-j+\gamma+\lambda_i+\mu_j]}\right)_{ 1\leq i ,j\leq \beta  }  \label{Paul_theta_equal1}
 \end{align}

Our formula for $T_{\gamma}$ in terms of interpolation polynomials can be manipulated quite 
explicitly when $\theta=1$, since the $BC$ interpolation polynomials for $\theta=1$ 
themselves also have a determinantal representation \cite{Okounkov_FactoShur},
\begin{align}
 P^{ip}_{[\kappa_1,\ldots \kappa_{\ell} ]}({\bf w};\theta=1,u)&= \frac{ \det \left( P^{ip}_{[\kappa_j+\ell-j]}(w_i;u) \right)_{ 1\leq i,j \leq \ell } }{ \prod_{i<j} (w_i^2-w_j^2) }
 \label{determinantal_Pinterp2}
 \end{align}
The entries of the matrix are single-variable interpolation polynomials $P^{ip}_{\kappa}(w_i,u)$. These do not depend on $\theta$ and are given by a pair of Pochhammers, 
\begin{align}
 P^{ip}_{\kappa}(w;u)=(-)^k (u+w)_{k}(u-w)_k &= (-w-u-k+1)_{k} (u-w)_k \notag\\
 &= (-)^k (-w-u-k+1)_{k} (w-u-k+1)_k\label{BC1Pstar_useful}
\end{align}
The denominator of~\eqref{determinantal_Pinterp2} is the $\mathbb{Z}_2$ invariant Vandermonde determinant.

We can now proceed and compute $(T_{\gamma})_{\ulmb}^{\umu}$ for $\theta=1$.
Since we already identified the normalisation, we 
will focus on $({R}^{\tt rescaled}_{\gamma})_{\ulmb}^{\umu}$. 
This has to agree with
\bea
({T}^{\tt rescaled}_{\gamma})_{\ulmb}^{\umu} \equiv \frac{(-)^{|\mu|} }{(-)^{|\ulmb|} } \frac{ P^{ip}_{\umu_1^\beta-\umu}(\tfrac{1}{2}- \tfrac{\gamma}{2}+i-1 -\lambda_{i} ;1,\tfrac{1}{2}-\tfrac{\gamma}{2}-\mu_1) }{ 
P^{ip}_{\mu_1^\beta-\ulmb}(\tfrac{1}{2}- \tfrac{\gamma}{2}+i-1 -\lambda_{i} ;1,\tfrac{1}{2}-\tfrac{\gamma}{2}-\mu_1) }
\eea
The idea is simple: we need to recognise the matrix
in \eqref{Paul_theta_equal1}. To do so, we first use   the expression of the $BC_1$ interpolation 
polynomials given in \eqref{BC1Pstar_useful}, and pass from Pochhammers to $\Gamma$ functions. 
The result is 
\begin{align}
({T}^{\tt rescaled}_{\gamma})_{\ulmb}^{\umu} =  \frac{(-)^{|\mu|} }{(-)^{|\ulmb|} } \frac{  (-)^{\beta\mu_1-|\umu| +\frac12\beta(\beta-1)}
\det \left( \frac{ \Gamma[i-\lambda_i+\mu_1 ] \Gamma[1-i+\gamma+\lambda_i-\mu_1] }{  \Gamma[1+i-J-\lambda_i+\mu_J]\Gamma[2-i-J+\gamma+\lambda_i+\mu_J ] } \right)_{1\leq i,j\leq \beta} }{
(-)^{\beta\mu_1-|\ulmb| +\frac12\beta(\beta-1)}
\det \left( \frac{ \Gamma[i-\lambda_i+\mu_1 ] \Gamma[1-i+\gamma+\lambda_i-\mu_1] }{  \Gamma[1+i-J-\lambda_i+\lambda_J]\Gamma[2-i-J+\gamma+\lambda_i+\lambda_J ] } \right)_{1\leq i,j\leq \beta} } 
\end{align}
where $J=\beta+1-j$ and the signs have been factored out since only depended on the column index. 
For the same reason we can factor out  $\Gamma[i-\lambda_i+\mu_1 ] \Gamma[1-i+\gamma+\lambda_i-\mu_1]$ 
and cancel it between numerator and denominator. Moreover we can reverse the columns 
and switch $j\leftrightarrow J$.  We arrive at the simple formula
\begin{align}
({T}^{\tt rescaled}_{\gamma})_{\ulmb}^{\umu}  = \frac{  \det\left( \frac{ \ \Gamma[1+i-j-\lambda_i+\mu_j]^{-1} }{ \Gamma[2-i-j+\gamma+\lambda_i+\mu_j]}\right)_{1\leq i,j\leq \beta} }{
  \det\left( \frac{ \ \Gamma[1+i-j-\lambda_i+\lambda_j]^{-1} }{ \Gamma[2-i-j+\gamma+\lambda_i+\lambda_j]}\right)_{1\leq i,j\leq \beta} } 
\end{align}
Comparing this result with \eqref{Paul_theta_equal1} we are left with the statement
\begin{align}
\prod_{i=1}^{\beta} \Gamma[ 2-2i+\gamma +2\lambda_i ]=
\frac{1}{ \det\left( \frac{ \ \Gamma[1+i-j-\lambda_i+\lambda_j]^{-1} }{ \Gamma[2-i-j+\gamma+\lambda_i+\lambda_j]}\right)_{1\leq i,j\leq \beta} }
\end{align}
But notice that $\Gamma[1+i-j-\lambda_i+\lambda_j]^{-1}$, because $\lambda_{i}\ge \lambda_{i+1}$, makes the matrix on the r.h.s 
a triangular matrix, and the identity follows. Thus $(T^{\tt rescaled}_{\gamma})_{\ulmb}^{\umu} =({R}^{\tt rescaled})_{\ulmb}^{\umu}$ and our proof is concluded.

\end{document}